\documentclass[11pt, a4paper, oneside]{Thesis} 

\graphicspath{{Pictures/}} 

\usepackage[square,sort,compress,numbers]{natbib}

\usepackage{amsmath,latexsym} 
\usepackage{float}
\usepackage[dvipsnames]{xcolor}

\newcommand{\mycirc}[1][black]{\normalsize{\textcolor{#1}{\ensuremath\bullet}}\normalsize}
\usepackage{hyperref}
\hypersetup{colorlinks,citecolor=red}
\hypersetup{colorlinks=true,urlcolor=black}
\title{\ttitle} 
\DeclareUnicodeCharacter{2212}{-}
\DeclareUnicodeCharacter{2212}{+}

\begin{document}

\setstretch{1.3} 

\fancyhead{} 
\rhead{\thepage} 
\lhead{} 

%


\thesistitle{Cosmological Evolution of the Universe in Torsion-based Modified Gravity}
\documenttype{\textbf{THESIS}}
\supervisor{\textbf{Prof. Pradyumn Kumar Sahoo}}
\supervisorposition{\textbf{Professor}}
\supervisorinstitute{\textbf{BITS-Pilani, Hyderabad Campus}}
\examiner{}
\degree{Ph.D. Research Scholar}
\coursecode{\textbf{DOCTOR OF PHILOSOPHY}}
\coursename{Thesis}
\authors{\textbf{Sai Swagat Mishra}}
\IDNumber{\textbf{2021PHXF0458H}}
\addresses{}
\subject{}
\university{\texorpdfstring{\href{http://www.bits-pilani.ac.in/} 
                {Birla Institute of Technology and Science, Pilani}} 
                {Birla Institute of Technology and Science, Pilani}}
\UNIVERSITY{\texorpdfstring{\href{http://www.bits-pilani.ac.in/} 
                {\textbf{BIRLA INSTITUTE OF TECHNOLOGY AND SCIENCE, PILANI}}} 
                {\textbf{BIRLA INSTITUTE OF TECHNOLOGY AND SCIENCE, PILANI}}}


\department{\texorpdfstring{\href{http://www.bits-pilani.ac.in/pilani/Mathematics/Mathematics} 
                {Mathematics}} 
                {Mathematics}}
\DEPARTMENT{\texorpdfstring{\href{http://www.bits-pilani.ac.in/pilani/Mathematics/Mathematics} 
                {Mathematics}} 
                {Mathematics}}
\group{\texorpdfstring{\href{Research Group Web Site URL Here (include http://)}
                {Research Group Name}} 
                {Research Group Name}}
\GROUP{\texorpdfstring{\href{Research Group Web Site URL Here (include http://)}
                {RESEARCH GROUP NAME (IN BLOCK CAPITALS)}}
                {RESEARCH GROUP NAME (IN BLOCK CAPITALS)}}
\faculty{\texorpdfstring{\href{Faculty Web Site URL Here (include http://)}
                {Faculty Name}}
                {Faculty Name}}
\FACULTY{\texorpdfstring{\href{Faculty Web Site URL Here (include http://)}
                {FACULTY NAME (IN BLOCK CAPITALS)}}
                {FACULTY NAME (IN BLOCK CAPITALS)}}


\maketitle

\clearpage
\frontmatter 
\Certificate
\setstretch{1.3} 

\Declaration

\addtocontents{toc}{\vspace{2em}}

\begin{acknowledgements}
It is a great privilege to express my heartfelt gratitude to everyone whose encouragement, inspiration, and support have guided me throughout my doctoral journey.  

First and foremost, I am profoundly thankful to my supervisor, \textbf{Prof. Pradyumn Kumar Sahoo}, Professor in the Department of Mathematics, BITS-Pilani, Hyderabad Campus. His perseverance, unwavering enthusiasm, and constant guidance have been invaluable throughout my PhD. Working under his supervision has been a deeply enriching experience, and his exceptional mathematical expertise, insightful suggestions, and inspiring mentorship have shaped not only this dissertation but also my outlook as a researcher.  

I am also grateful to my Doctoral Advisory Committee members, \textbf{Prof. Bivudutta Mishra} and \textbf{Prof. Prasant Samantray}, for their insightful questions, constructive feedback, and continuous support, which have greatly improved the quality of my research. I would also like to thank the Head of the Department, the DRC Convener, and all faculty and staff of the Department of Mathematics at BITS-Pilani, Hyderabad Campus, for their assistance and encouragement. My sincere thanks also go to AGRSD, BITS-Pilani, Hyderabad Campus, for their support.  

I gratefully acknowledge the financial assistance and facilities provided by \textbf{BITS-Pilani, Hyderabad Campus}, as well as the Council of Scientific \& Industrial Research \textbf{(CSIR)}, Govt. of India, for awarding me the Junior and Senior Research Fellowship (E-Certificate No.: JUN21C05815), which enabled the successful realization of this research.  

I also acknowledge the financial support received through multiple international travel grants, including the SERB, CSIR, BITS, and sponsorship from the University of Oslo. These grants enabled me to present my work and participate in major international events, including the Galileo Galilei Institute (Italy), the University of Oslo (Norway), the NuDM conference (Egypt), and the CosmoGravitas (Thailand). Such opportunities have been invaluable in broadening my academic exposure and fostering collaborations within the global research community.

I also gratefully acknowledge the \textbf{Inter-University Centre for Astronomy and Astrophysics (IUCAA), Pune}, for providing me the opportunity to visit twice --- once in December 2024 and again in October 2025 --- and to engage in fruitful discussions with researchers there. These visits significantly enriched the scientific depth of my doctoral work.

My warmest thanks to \textbf{Dr. Kavya N.~S.} from Christ University, Bangalore, for being the closest friend and collaborator. Most of my research has been carried out in collaboration with her, and her constant encouragement, insightful discussions, and steady support have played a pivotal role in shaping this thesis. Beyond academic guidance, her presence has been a source of strength and motivation throughout this journey.

I am also fortunate to have worked closely with \textbf{Ameya Kolhatkar}, who has been both a collaborator and a very good friend. I have spent most of my PhD days alongside him, and our discussions, academic and beyond, have been an inseparable part of my research life. His companionship has made this journey not only productive but also deeply enjoyable.  

I extend my thanks to all my collaborators for their valuable contributions and stimulating discussions. I am especially grateful to my senior, \textbf{Dr. Sanjay Mandal} (Fukushima University, Japan), for guiding me in the early years of my PhD and for his constant encouragement. I also thank my friends \textbf{Soumyakanta Bhoi}, \textbf{Mayuri Verma}, \textbf{Santanu Dash}, and \textbf{Kailash Swami} for their unwavering support and encouragement during my doctoral studies.  

Finally, I owe my deepest gratitude to my parents, sisters, and my entire family. Their unconditional love, constant encouragement, and support in every high and low of life have been my greatest strength. This thesis is as much theirs as it is mine.  
\end{acknowledgements}

\clearpage

\begin{abstract}
General Relativity, despite its century-long success, faces both conceptual and observational challenges, including singularities, incompatibility with quantum mechanics, and the need to postulate dark matter and dark energy. In addition, precision cosmology has revealed persistent tensions, most notably the $H_0$ and $S_8$ discrepancies, which cast doubt on the completeness of the concordance $\Lambda$CDM model. These issues provide strong motivation to explore alternative frameworks of gravity and cosmology.

This thesis investigates cosmological applications of teleparallel gravity and its extensions, with a particular focus on $f(T)$ and $f(T,\mathcal{T})$ theories. Beginning with the motivation from cosmological tensions, in \autoref{chap2} we show that torsion-based modifications can effectively shift late-time expansion rates and matter clustering, providing a natural alleviation of the $H_0$ and $S_8$ discrepancies. Using datasets including cosmic chronometers (CC), baryon acoustic oscillations (BAO), Type Ia supernovae, Pantheon+SH0ES, Union3, DESI, and gravitational wave standard sirens, we conduct detailed Markov Chain Monte Carlo (MCMC) analyses to constrain model parameters.

To establish model-independent diagnostics, we employ cosmography in the teleparallel framework, demonstrating that cosmographic expansions can successfully constrain extended coupled theories such as $f(T,\mathcal{T})$ in the \autoref{chap3}. Further, in \autoref{chap4}, we implement Padé approximants of the luminosity distance alongside direct dynamical reconstructions from modified Friedmann equations, finding consistent results across both approaches. These methods not only confirm the viability of $f(T)$ models but also outperform $\Lambda$CDM in light of the latest DESI and Union3 data. Finally, we propose a novel connection between late-time cosmic acceleration and early-Universe baryogenesis within the teleparallel framework in \autoref{chap5}. By linking baryon-to-entropy asymmetry with present-day observations, we demonstrate that torsional gravity can successfully reproduce the observed baryon asymmetry while remaining consistent with the late-time expansion history. Finally, the conclusions and future perspectives are discussed in \autoref{Chapter6}.

Overall, the results of this thesis show that teleparallel gravity and its extensions constitute a robust and predictive alternative to the $\Lambda$CDM paradigm. By alleviating key cosmological tensions, offering consistent observational fits, and bridging early- and late-Universe physics, torsional theories provide fertile ground for advancing our understanding of gravity and the evolution of the Universe.
\end{abstract}

\clearpage
\pagestyle{empty} 
\Dedicatory{\textbf{Dedicated to}\\ \vspace{0.1 in}
My beloved grandfather, the late Shri Arun Kumar Mishra, and my family members}


\addtocontents{toc}{\vspace{1em}}
\tableofcontents 
\addtocontents{toc}{\vspace{1em}}
\lhead{\emph{List of Tables}}
\listoftables 
\addtocontents{toc}{\vspace{1em}}
\lhead{\emph{List of Figures}}
\listoffigures 
\addtocontents{toc}{\vspace{1em}}



\lhead{\emph{List of Symbols and Abbreviations}}
\listofsymbols{ll}{
$G_{ij}:$\,\,\,\, Einstein tensor\\
$g_{ij}:$ \,\,\,\,\,\,\,Lorentzian metric\\
$g:$ \,\,\,\,\,\,\,\,\,\,\,Determinant of $g_{ij}$\\
$\overset{w}{\Gamma}{}^{\lambda}_{\nu \mu}:$ \,\,\,\,Weitzenb\"ock  connection\\
$\square$\,:\,\,\,\,\,\,\,\,\,\, D'Alembert operator\\
$R^{\sigma}_{\,\,\lambda ij}:$ \,\,Riemann tensor \\
$R_{ij}:$\,\,\,\,\,\,\,\,Ricci tensor \\
$R:$ \,\,\,\,\, \,\,\,Ricci scalar \\
$z:$\,\,\,\,\, \,\,\,\,\,\,\,Redshift\\
$q:$\,\,\,\,\, \,\,\,\,\,\,\,Deceleration parameter\\
$s:$\,\,\,\,\, \,\,\,\,\,\,\,Snap parameter\\
$j:$\,\,\,\,\, \,\,\,\,\,\,\,Jerk parameter\\
$l:$\,\,\,\,\, \,\,\,\,\,\,\,Lerk parameter\\
$T:$\,\,\,\,\, \,\,\,\,\,\,Torsion\\
$Q:$\,\,\,\,\, \,\,\,\,\,\,Non-metricity\\
$\Lambda:$\,\,\,\,\, \,\,\,\,\,\,Cosmological constant\\
$T_{ij}:$ \,\,\,\,\,\,\,\,Stress-energy tensor\\
$G:$\,\,\,\,\, \,\,\, \,Newton's gravitational constant\\
$\Omega:$\,\,\,\,\, \,\,\,\,\,\,Density parameter\\
$\chi^{2}:$\,\,\,\,\, \,\,\,\,\,Chi-Square\\
GR:\,\,\,\,\, \,\,\,\,General Relativity\\
$\Lambda$CDM:\,\,$\Lambda$ Cold Dark Matter\\
ECs:\,\,\,\,\,\,\,\,\,\,Energy Conditions\\
EoS:\,\,\,\,\,\,\,\,\,\,Equation of State\\
SNeIa:\,\,\,\,\,Type Ia Supernovae\\
CMB: \,\,\,\,\,Cosmic Microwave Background\\
BAO: \,\, \,\,Baryon Acoustic Oscillations\\
MCMC:\,\,Markov Chain Monte Carlo\\
LSS:\,\,\,\,\, \,\,\,\,\,Large Scale Structure\\
SPH:\,\,\,\,\, \,\,\,\,Smoothed Particle Hydrodynamics\\
RSD:\,\,\,\,\, \,\,\,\,Redshift Space Distortion\\
DE:\,\,\,\,\,\,\,\,\,\,\,\,\,\,Dark Energy\\
DM:\,\,\,\,\,\,\,\,\,\,\,\,\,Dark Matter\\
MG:\,\,\,\,\,\,\,\,\,\,\,\,Modified Gravity\\
TEGR:\,\,\,\,\,Teleparallel Equivalent to GR\\
STEGR:\,\,\,Symmetric Teleparallel Equivalent to GR\\
}


\addtocontents{toc}{\vspace{1em}}
 
\addtocontents{toc}{\vspace{1em}}





%
%


\clearpage 

\lhead{\emph{Glossary}} 

 


\mainmatter 

\pagestyle{fancy} 


\newpage
\thispagestyle{empty}

\vspace*{\fill}
\begin{center}
    {\Huge \color{RedViolet} \textbf{CHAPTER 1}}\\
    \
    \\
    {\Large\color{Emerald}\textsc{\textbf{Introduction}}}\\
    \end{center}
\vspace*{\fill}

\def\baselinestretch{1}
\markboth{}{\it\small{{Chapter 1 \hfill Introduction}}}
\chapter{\textsc{Introduction}}\label{ch:intro}
\def\baselinestretch{1.5}
\pagestyle{fancy}
\fancyhead[R]{\textit{Chapter 1}}
\lhead{\emph{Chapter 1. Introduction}}
\rhead{\thepage}
\markboth{Introduction}{}
\setstretch{1.5}
\section{Historical Background and Motivation}
The quest to understand the Universe and its governing laws has captivated humanity since antiquity. Early cosmological models were deeply intertwined with philosophy and empirical observations, evolving from Ptolemy’s geocentric worldview to Copernicus’s revolutionary heliocentric paradigm, which fundamentally reshaped our cosmic perspective. The pioneering efforts of Kepler and Galileo established a firm mathematical and observational foundation for celestial mechanics, culminating in Sir Isaac Newton’s \textit{Philosophiæ Naturalis Principia Mathematica} \cite{Newton1687}. Newton unified terrestrial and celestial dynamics through his law of universal gravitation, providing a framework that successfully explained a wide array of phenomena.

Despite its enduring success, Newtonian gravity exhibited limitations as new phenomena emerged. The anomalous perihelion precession of Mercury, the invariance of the speed of light, and the advent of special relativity \cite{Einstein1905} exposed the need for a more fundamental understanding of gravity. This culminated in Einstein’s formulation of General Relativity (GR) in 1915 \cite{Einstein1916}, which reinterpreted gravity as the curvature of spacetime induced by mass-energy. Beyond resolving previous anomalies, GR predicted novel gravitational phenomena such as light bending and gravitational waves, which have since been empirically confirmed.

Simultaneously, cosmology underwent a transformative evolution. Edwin Hubble’s discovery of the expanding Universe \cite{Hubble1929}, alongside the formulation of relativistic cosmological models by Friedmann \cite{Friedmann1922}, Lemaître \cite{Lemaitre1927}, Robertson \cite{Robertson1935}, and Walker \cite{Walker1937}, laid the groundwork of modern physical cosmology. Landmark observational breakthroughs, the discovery of the Cosmic Microwave Background (CMB)  radiation \cite{Penzias1965} and the success of Big Bang Nucleosynthesis (BBN) \cite{Wagoner1967}, provided compelling evidence for the hot Big Bang paradigm.

However, despite these remarkable achievements, profound mysteries persist. The discovery of the accelerated expansion of the Universe in the late 1990s, led by the seminal works of Riess et al. (1998) \cite{SupernovaSearchTeam:1998fmf} and Perlmutter et al. (1999) \cite{SupernovaCosmologyProject:1998vns}, radically altered our understanding of cosmology. It introduced the concept of dark energy, a mysterious component that now dominates the cosmic energy budget. Within the prevailing Lambda Cold Dark Matter ($\Lambda$CDM) model, this accelerated expansion is attributed to a cosmological constant $\Lambda$, which accounts for approximately 70\% of the total energy density as reported by the Planck Collaboration (2020) \cite{Planck:2018vyg}. Despite its observational success, $\Lambda$CDM faces unresolved theoretical issues, including the cosmological constant problem \cite{Weinberg:1988cp}, where quantum field theory overpredicts the vacuum energy by $\sim20$ orders of magnitude, and the coincidence problem \cite{Steinhardt:1999nw}, which puzzles over why dark energy began to dominate so recently in cosmic time.

In addition to these theoretical concerns, observational discrepancies continue to raise doubts about the completeness of the $\Lambda$CDM model. The most prominent is the Hubble tension \cite{DiValentino:2021izs, Verde:2019ivm}, a persistent and statistically significant ($\sim 4 -6\sigma$) discrepancy between early-Universe estimates of the Hubble constant $H_0$ (from CMB measurements) \cite{Planck:2018vyg} and local, late-Universe determinations (based on supernovae and Cepheid variable stars) \cite{Riess:2021jrx}. If this tension cannot be explained by systematic errors, it may indicate the need for new physics, such as evolving dark energy, modifications to GR, or the presence of extra relativistic species.

Therefore, this thesis is motivated by the need to deepen our understanding of gravity and cosmology beyond the standard paradigms. It focuses on investigating modified gravity scenarios as plausible frameworks to address existing theoretical and observational challenges. The following chapters systematically develop these models, analyze their cosmological implications, and assess their consistency with current observational data, thus contributing to the broader effort of uncovering the fundamental workings of our Universe.

\section{Fundamental Aspects of Gravity}
Gravity is one of the four fundamental interactions in nature, essential in shaping the large-scale structure and dynamics of the Universe. Unlike the electromagnetic, weak, and strong forces, which are described as gauge interactions in the framework of quantum field theory, gravity influences the very fabric of spacetime itself. Its universal and long-range character makes it unique: it is always attractive, cannot be shielded, and extends to infinite range. While weaker than other forces at microscopic scales, gravity dominates astrophysical and cosmological dynamics.  

In Newtonian mechanics, gravity is modeled as an instantaneous action-at-a-distance, governed by the inverse-square law. This framework provided accurate descriptions of planetary orbits, cometary trajectories, and tidal effects, establishing Newton’s theory as the foundation of classical mechanics \cite{Newton1687}. Despite its success, Newtonian gravity relied on the concept of absolute space and time, which eventually proved incompatible with developments in relativity and electromagnetism.  

A profound shift occurred in the early 20th century with the advent of Einstein’s special relativity \cite{Einstein1905}, which abolished absolute simultaneity and imposed the constancy of the speed of light as a universal principle. This reformulation of kinematics laid the groundwork for a deeper understanding of gravitation. A cornerstone in this transition is the \textit{equivalence principle}, which states that inertial and gravitational mass are indistinguishable. This principle implies that a freely falling observer cannot locally distinguish between acceleration and a gravitational field, bridging the gap between Newtonian concepts and Einstein’s later reformulation of gravity.  

Thus, while Newton’s law unified terrestrial and celestial mechanics, the recognition of its limitations and the emergence of relativity set the stage for Einstein’s geometric theory of gravitation. These fundamental aspects of gravity, ranging from its classical formulation to the insights of relativity, form the essential foundation for GR, discussed in the next section.

\section{General Relativity}
\label{sec:gr}
Einstein’s General Theory of Relativity, published in 1916 \cite{Einstein1916}, revolutionized the description of gravity by interpreting it as a manifestation of spacetime curvature rather than a force. In GR, mass and energy determine the geometry of spacetime, represented by a four-dimensional pseudo-Riemannian manifold, and particles follow geodesics of this curved geometry.  

The central dynamical equations of GR are the Einstein field equations, which relate the curvature of spacetime to the matter–energy content:
\begin{equation}\label{eq:efe}
G_{\mu\nu} + \Lambda g_{\mu\nu} = \frac{8\pi G}{c^4} T_{\mu\nu},
\end{equation}
where $G_{\mu\nu}$ is the Einstein tensor, $T_{\mu\nu}$ the stress-energy tensor, $g_{\mu\nu}$ the metric tensor, $G$ Newton’s gravitational constant, $c$ the speed of light, and $\Lambda$ the cosmological constant. These equations encode the fundamental relationship between geometry and physics.  

\subsection*{Newtonian Limit}
In the weak-field and low-velocity limit, GR reduces to Newtonian gravity. If the metric is written as a perturbation of flat space, $g_{\mu\nu} = \eta_{\mu\nu} + h_{\mu\nu}$ with $|h_{\mu\nu}| \ll 1$, the time-time component (00-component) of the field equations recovers Poisson’s equation for the gravitational potential $\Phi$:
\begin{equation}
\nabla^2 \Phi = 4\pi G \rho.
\end{equation}
This correspondence guarantees consistency between GR and Newtonian gravity in appropriate regimes.  

\subsection*{Cosmological Framework}
A key cosmological application of GR comes from assuming a homogeneous and isotropic Universe, described by the Friedmann–Lemaître–Robertson–Walker (FLRW) metric:
\begin{equation}\label{eq:flrw}
ds^2 = -c^2 dt^2 + a^2(t)\left[\frac{dr^2}{1 - k r^2} + r^2(d\theta^2 + \sin^2 \theta\, d\phi^2)\right],
\end{equation}
where $a(t)$ is the scale factor and $k$ determines the spatial curvature ($k=0,\pm 1$). Substituting this metric into the Einstein field equations \eqref{eq:efe} yields the Friedmann equations:
\begin{align}
H^2 &= \left(\frac{\dot{a}}{a}\right)^2 = \frac{8\pi G}{3}\rho - \frac{k}{a^2} + \frac{\Lambda}{3}, \label{eq:grfriedmann1}\\
\frac{\ddot{a}}{a} &= -\frac{4\pi G}{3}\left(\rho + \frac{3p}{c^2}\right) + \frac{\Lambda}{3}, \label{eq:grfriedmann2}
\end{align}
where $H$ is the Hubble parameter, $\rho$ the energy density, and $p$ the pressure. These equations govern the dynamics of cosmic expansion and provide the backbone of modern cosmology.  

\subsection*{Principles and Empirical Successes}
Two guiding principles underpin GR:  
\begin{itemize}
    \item \textbf{General covariance}: the laws of physics take the same form in all coordinate systems.  
    \item \textbf{Equivalence principle}: locally, gravitational effects are indistinguishable from those of acceleration, ensuring consistency with special relativity.  
\end{itemize}
GR has been confirmed across a vast array of experimental and observational tests. Classical confirmations include the perihelion precession of Mercury, the gravitational redshift, and the deflection of starlight by the Sun, first verified during Eddington’s 1919 eclipse expedition. More modern tests include the Shapiro time delay, frame-dragging measurements, binary pulsar timing, and most notably, the direct detection of gravitational waves by the LIGO-Virgo collaboration \cite{LIGOScientific:2016aoc}, which opened a new window into the strong-field regime of gravity.  

Cosmologically, GR underpins the FLRW framework, providing the theoretical basis for the expanding Universe. It successfully explains the dynamics of cosmic expansion, the anisotropies of the CMB \cite{Penzias1965,Planck:2018vyg}, and large-scale structure formation.  
\section{Limitations of GR}
Despite its elegance and remarkable success across a wide range of phenomena, GR is not a complete theory of gravity. Several conceptual, theoretical, and observational challenges highlight its limitations and motivate the search for alternative frameworks. These limitations can be broadly categorized as follows.

\subsection{Incompatibility with Quantum Mechanics}
GR is a classical field theory and does not incorporate the principles of quantum mechanics. Attempts to quantize gravity using standard perturbative techniques lead to a non-renormalizable theory, meaning that an infinite number of counterterms are required to absorb divergences at higher orders. This incompatibility prevents GR from being a consistent fundamental theory at the Planck scale ($\sim 10^{19}$ GeV). While approaches such as string theory and loop quantum gravity attempt to unify gravity with quantum mechanics, no fully accepted quantum theory of gravity has yet emerged \cite{Rovelli:1997yv,Ashtekar:2004eh,Polchinski:1998rq}.

\subsection{Singularity Problems}
The Einstein field equations predict the formation of spacetime singularities, regions where curvature invariants diverge, and classical physics breaks down. Notable examples include the Big Bang singularity in cosmology and the central singularities of black holes. The Hawking--Penrose singularity theorems formalize the inevitability of singularities under general conditions \cite{Hawking1970}. Since physical quantities become undefined at singularities, GR ceases to provide a predictive description, suggesting the need for new physics beyond classical relativity.

\subsection{Dark Matter and Dark Energy}
\begin{figure}[H]
    \centering
    \includegraphics[width=0.85\linewidth]{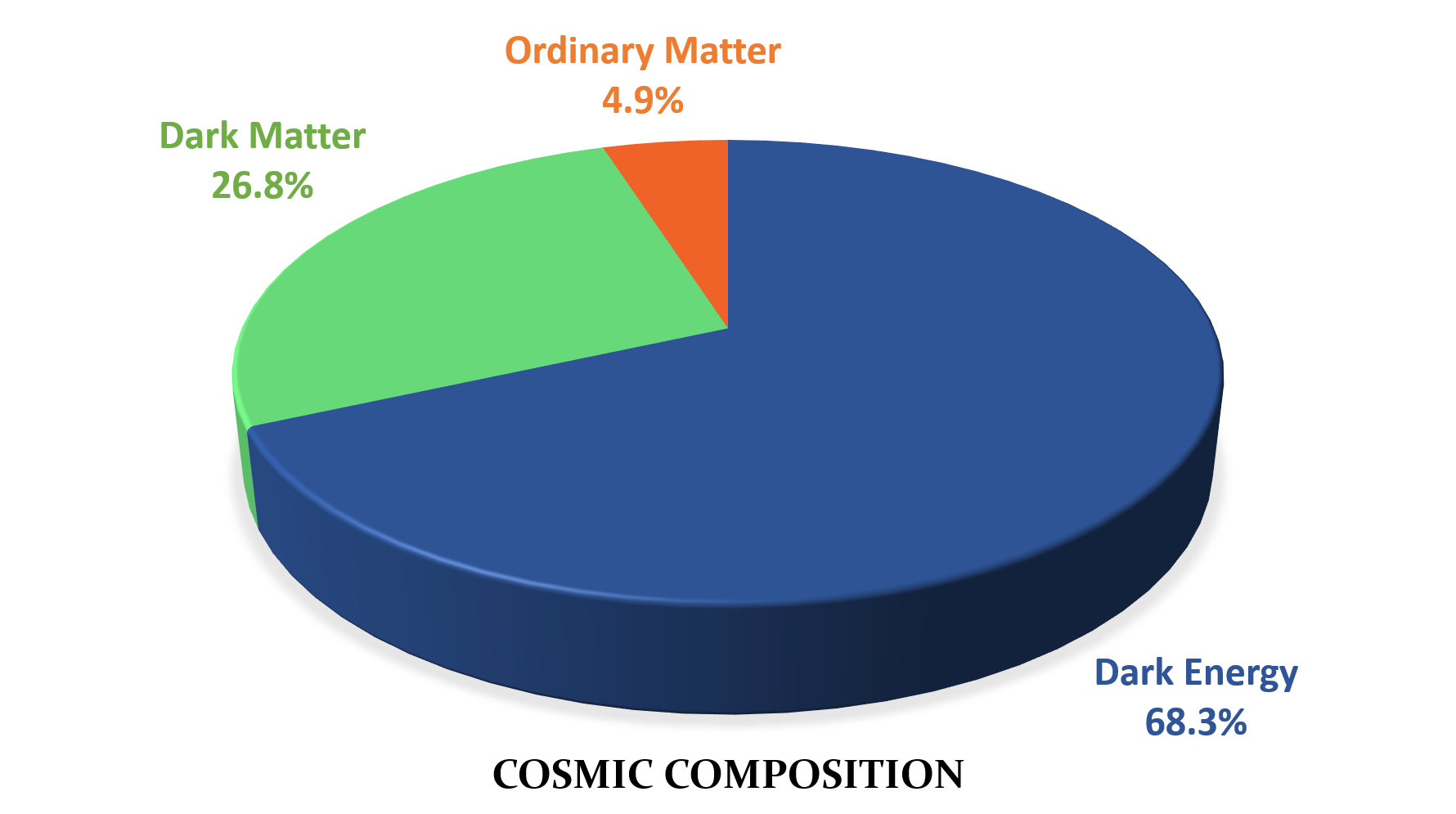}
\caption{The cosmic energy budget.}
    \label{chap1/fig:prop}
\end{figure}

GR, when applied to cosmology, requires the postulation of two unknown components: dark matter and dark energy. Observations of galactic rotation curves, gravitational lensing, and large-scale structure formation demand non-luminous matter \cite{Rubin1970,Zwicky1933}, while Type Ia supernovae and CMB anisotropies indicate an accelerated cosmic expansion driven by dark energy \cite{SupernovaCosmologyProject:1998vns,SupernovaSearchTeam:1998fmf,Planck:2018vyg}. Together, these components constitute about 95\% of the energy budget of the Universe in the $\Lambda$CDM model. However, neither dark matter particles nor the true nature of dark energy have been directly detected, raising the possibility that the apparent effects attributed to them may instead arise from modifications of gravity on cosmological scales. As shown in \autoref{chap1/fig:prop}, the total cosmic energy budget is composed of approximately 68.3\% dark energy, 26.8\% dark matter, and 4.9\% ordinary (baryonic) matter \cite{Planck:2018vyg}.

\subsection{Cosmological Constant Problem}
When interpreted as a cosmological constant, dark energy introduces the so-called ``cosmological constant problem''. Quantum field theory predicts vacuum energy densities that are up to 120 orders of magnitude larger than the observed value inferred from cosmology \cite{Weinberg:1988cp, Martin2012}. This represents the largest known fine-tuning problem in physics. GR itself provides no mechanism for explaining the extreme smallness or stability of $\Lambda$, which remains one of the central puzzles of modern theoretical physics.

\subsection{Cosmic Coincidence Problem}
In the $\Lambda$CDM model, the energy densities of matter and dark energy evolve very differently with cosmic expansion: while matter dilutes as $a^{-3}$, the cosmological constant $\Lambda$ remains unchanged. Despite this, observations indicate that the two contributions are of the same order of magnitude today, with $\Omega_\Lambda \simeq 0.7$ and $\Omega_m \simeq 0.3$ \cite{Planck:2018vyg}. This raises the so-called \textit{cosmic coincidence problem}: why do we live at the special epoch in which the densities of matter and dark energy are comparable, given that their ratio changes rapidly over cosmic history \cite{Steinhardt:1999nw, Velten:2014nra}? GR itself provides no explanation for this apparent fine-tuning, which motivates consideration of modifications of gravity that may alleviate the coincidence.

\subsection{Hubble Tension and Other Observational Discrepancies}
Recent precision cosmology has revealed inconsistencies between early- and late-Universe measurements. The most prominent is the Hubble tension: local distance-ladder measurements of the Hubble constant $H_0$ \cite{Riess:2020fzl} differ significantly from values inferred from CMB data under the $\Lambda$CDM model \cite{Planck:2018vyg,Verde:2019ivm, DiValentino:2021izs}. Additional tensions, such as discrepancies in the amplitude of matter clustering ($S_8$ tension), challenge the universality of GR+$\Lambda$CDM. These anomalies may point to systematic errors, new astrophysics, or the need for modifications of GR.

\subsection{Non-Renormalizability and Effective Theory Status}
From a field-theoretic perspective, GR can be treated as a low-energy effective field theory, valid below the Planck scale. While this approach provides predictive power at accessible energies, it implies that GR is not a fundamental theory but an approximation to a more complete framework that incorporates quantum effects \cite{Donoghue1994,Burgess2004}. This limitation motivates efforts to embed GR into a consistent ultraviolet-complete theory.

\subsection{Challenges in Explaining Cosmic Inflation}
The inflationary paradigm, proposed to resolve the horizon, flatness, and monopole problems, requires the introduction of an additional scalar field or exotic matter component. GR alone does not naturally explain inflation without invoking such extensions, again pointing to incompleteness at early-Universe scales \cite{Guth1981,Linde1982,Lyth1999}. 

\subsection{Strong-Field and Astrophysical Puzzles}
Although GR has passed stringent tests in the weak-field regime (solar system, binary pulsars), the strong-field regime is far less explored. Astrophysical observations such as supermassive black holes, galaxy core-cusp problems, and the dynamics of galaxy clusters sometimes present challenges to GR combined with standard dark matter models \cite{Bull:2015stt}. Future gravitational wave astronomy will provide critical tests in this regime.

\bigskip

In summary, while GR stands as the cornerstone of modern gravitational physics, its incompatibility with quantum mechanics, the existence of singularities, reliance on undetected dark sectors, and growing observational tensions strongly suggest that GR is not the final theory of gravity. These limitations provide compelling motivation for exploring modified and extended theories of gravity, which aim to address both theoretical inconsistencies and empirical anomalies.

\section{Modifications of General Relativity}\label{sec:grmodifications}

The limitations of GR, discussed in the previous section, have motivated the proposal of numerous modified and extended theories of gravity. These frameworks can be broadly classified into several categories: (i) extensions of the underlying spacetime geometry, (ii) modifications of the action by higher-order invariants and non-minimal couplings, (iii) the introduction of additional scalar, vector, or tensor fields, and (iv) structural or phenomenological modifications involving massive gravitons, extra dimensions, or non-local effects. The schematic diagrams in \autoref{fig:geom}--\ref{fig:field} summarize the main directions.

\subsection{Geometrical Extensions}
\begin{figure}[t]
    \centering
    \includegraphics[width=.85\linewidth]{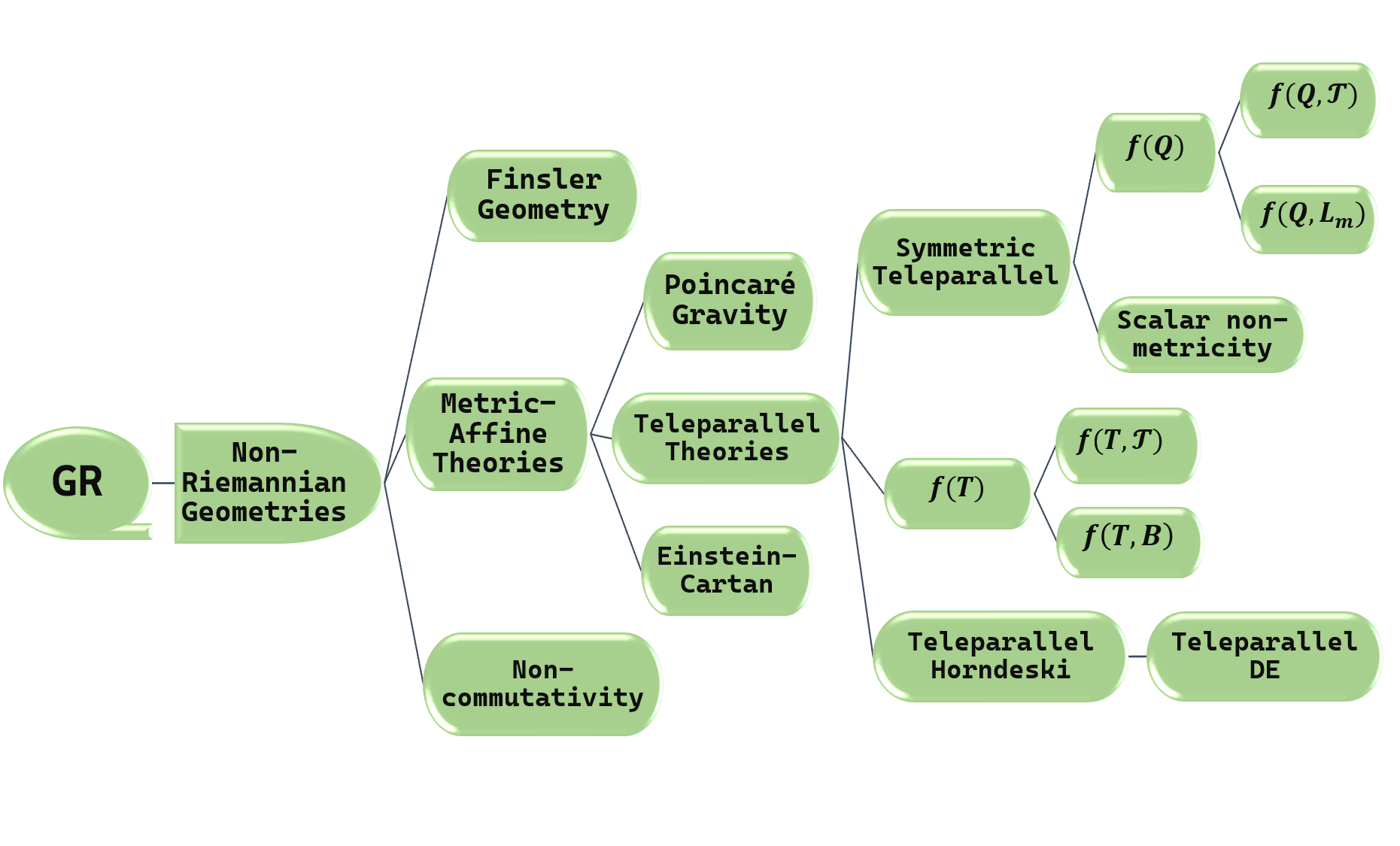}
    \caption{Extensions of GR by modifying the underlying spacetime geometry.}
    \label{fig:geom}
\end{figure}

A first approach is to generalize the geometry underlying GR, moving beyond the pseudo-Riemannian framework. A schematic classification is shown in \autoref{fig:geom}.

\paragraph{Metric-affine and non-Riemannian frameworks:\\}
In a broader geometrical setting, one may relax the assumptions of Riemannian geometry by allowing both curvature and torsion to coexist, leading to the general framework of metric-affine gravity. Einstein--Cartan theory incorporates torsion sourced by spin \cite{Hehl1976}, while Poincaré gauge theory unifies translations and Lorentz transformations \cite{Blagojevic2012}. More exotic generalizations include non-commutative geometry and Finsler geometry, which introduce intrinsic anisotropies in spacetime structure.

\paragraph{Teleparallel and symmetric teleparallel theories:\\}
An alternative approach to the geometric description of gravity eliminates curvature altogether, attributing gravitation to torsion or non-metricity. The Teleparallel Equivalent of GR (TEGR) replaces curvature with torsion, while its extension, $f(T)$ gravity, promotes the torsion scalar to a function \cite{Cai:2015emx}. Similarly, symmetric teleparallel gravity employs the non-metricity scalar $Q$, leading to generalizations such as $f(Q)$ gravity \cite{BeltranJimenez:2019tme, Kolhatkar:2025ubm} and further couplings like $f(Q,\mathcal{T})$ \cite{Kavya:2024kdi} and $f(Q,L_m)$ \cite{Rana:2025qjw}. These formulations provide novel geometric interpretations of gravitational interaction and open new avenues in cosmology.

\subsection{Higher-order Invariants and Non-minimal Couplings}
\begin{figure}[t]
    \centering
    \includegraphics[width=.85\linewidth]{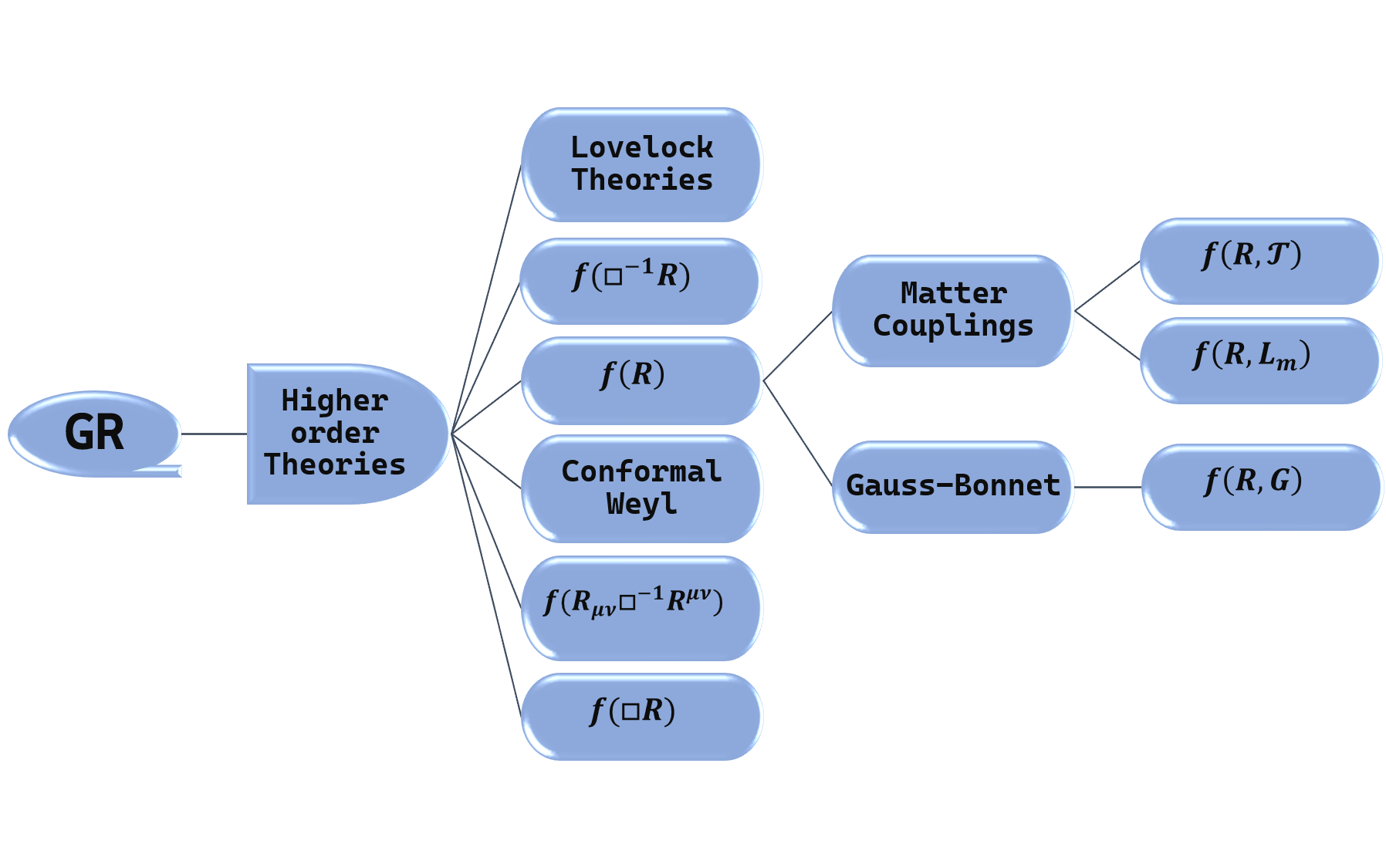}
    \caption{Modifications of GR through higher-order invariants and non-minimal couplings.}
    \label{fig:invariant}
\end{figure}

Another strategy is to generalize the Einstein--Hilbert action by replacing or augmenting the Ricci scalar with more general invariants. A schematic diagram is presented in \autoref{fig:invariant}.

\paragraph{\texorpdfstring{$f(R)$}{f(R)} and Lovelock theories:\\}
The simplest extension, $f(R)$ gravity, promotes the Ricci scalar to an arbitrary function \cite{Starobinsky1980,Nojiri2011}, which can mimic dark energy and inflation. In higher dimensions, Lovelock gravity introduces higher-curvature terms that maintain second-order equations of motion \cite{Lovelock1971}.
\paragraph{Gauss--Bonnet and string-inspired models:\\}
The Gauss--Bonnet invariant $G = R^2 - 4R_{\mu\nu}R^{\mu\nu}+R_{\mu\nu\alpha\beta}R^{\mu\nu\alpha\beta}$ arises naturally in string corrections. Generalizations such as $f(G)$ and $f(R,G)$ have been widely explored in cosmology \cite{Nojiri2005}.

\paragraph{Non-minimal matter couplings:\\}
Non-minimal couplings between curvature invariants and matter fields have been proposed, such as $f(R,\mathcal{T})$ gravity \cite{Harko2011}, where $\mathcal{T}$ is the trace of the energy--momentum tensor, or $f(R,L_m)$ couplings to the matter Lagrangian density \cite{Kavya:2022dam,Kavya:2023tjf}. These theories lead to non-conservation of $T_{\mu\nu}$ and modified geodesic motion.

\paragraph{Conformal and non-local models:\\}
Weyl (conformal) gravity is based on invariance under local rescalings of the metric \cite{Mannheim1989}. Non-local theories, e.g., $R \Box^{-1}R$ or $f(\Box^{-1}R)$, capture effective infrared modifications and are motivated by quantum corrections \cite{Deser2007,Maggiore2014}.

\subsection{Adding Extra Fields}
\begin{figure}[t]
    \centering
    \includegraphics[width=.85\linewidth]{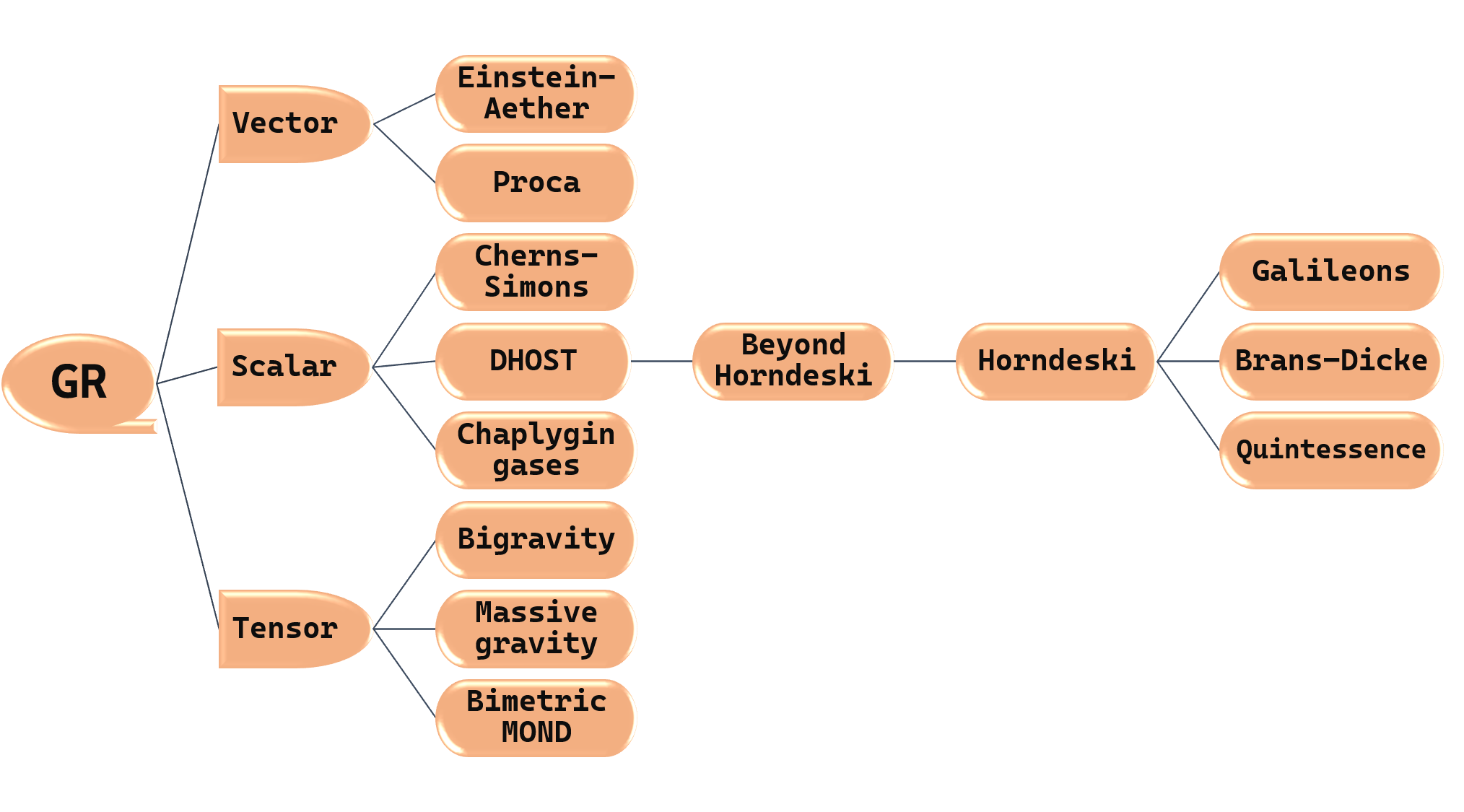}
    \caption{Modifications of GR by introducing additional fields (scalar, vector, tensor).}
    \label{fig:field}
\end{figure}

A third approach supplements the metric with additional dynamical fields, enriching the gravitational sector with extra degrees of freedom. These are classified into scalar, vector, and tensor extensions, as depicted in \autoref{fig:field}.

\paragraph{Scalar fields:\\}
Scalar--tensor theories generalize GR by coupling scalar fields to curvature. The earliest is Brans--Dicke theory \cite{Brans1961}, followed by quintessence \cite{Caldwell1998}, $k$-essence \cite{ArmendarizPicon2000}, and generalized Galileon/Horndeski models \cite{Horndeski1974,Deffayet2011}. DHOST theories \cite{Langlois2016} extend these while avoiding Ostrogradsky instabilities. Exotic fluids like Chaplygin gas have also been studied as scalar field analogues.

\paragraph{Vector fields:\\}
Einstein--Aether theory introduces a dynamical timelike vector field \cite{Jacobson2001}, while generalized Proca theories \cite{Heisenberg2014} extend massive vector fields consistently with Lorentz symmetry breaking. Chern--Simons gravity couples a pseudoscalar to the Pontryagin density, leading to parity-violating effects.

\paragraph{Tensor fields:\\}
Tensorial extensions include bigravity, where two dynamical metrics interact \cite{Hassan2012}, and massive gravity, where the graviton acquires a small mass \cite{deRham2011}. Bimetric MOND-type models also fall under this category.

\subsection{Structural and Phenomenological Modifications}
Some proposals alter the very structure of spacetime or interpret gravity as emergent.

\paragraph{Extra dimensions:\\}
In braneworld scenarios, our four-dimensional Universe is embedded in a higher-dimensional bulk. The Randall--Sundrum models \cite{Randall1999a,Randall1999b} and the DGP model \cite{Dvali2000} are representative examples, modifying gravity on cosmological scales.

\paragraph{Emergent gravity:\\}
Jacobson derived Einstein’s equations from thermodynamic arguments \cite{Jacobson1995}, while Verlinde proposed gravity as an entropic force \cite{Verlinde2011}. These scenarios suggest gravity may not be fundamental but emergent.

\bigskip

In summary, modifications of GR span a broad landscape: geometrical extensions beyond Riemannian geometry, higher-order invariants and matter couplings, additional scalar/vector/tensor fields, and structural or emergent models. Each category addresses specific theoretical or observational shortcomings of GR and provides testable predictions, making them essential candidates in the quest for a deeper theory of gravity. This thesis specifically focuses on the geometric modifications of GR.

\section{The Geometric Trinity of Gravity}
\begin{figure}[H]
    \centering
    \includegraphics[width=0.6\linewidth]{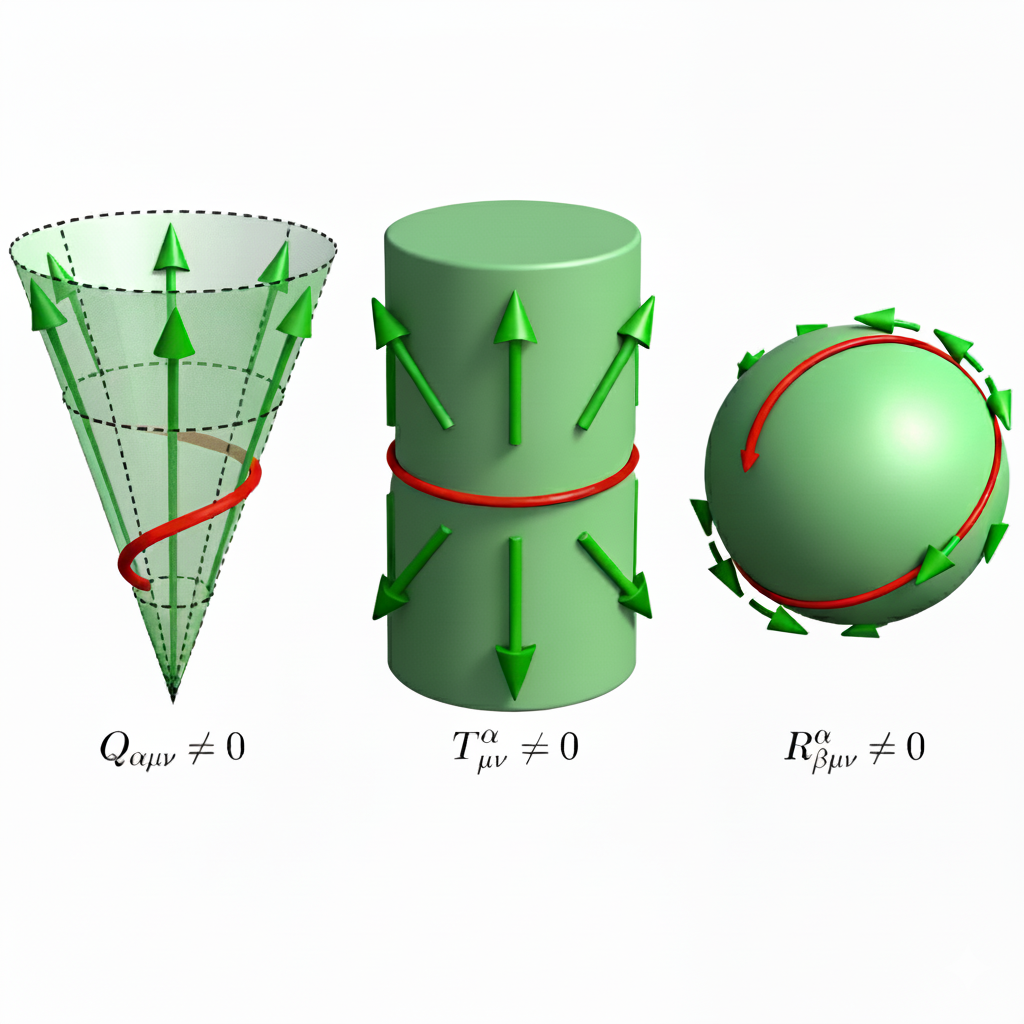}
    \caption{Visualization of the geometric trinity in metric-affine gravity: Pure non-metricity (left), pure torsion (center), and pure curvature (right).}
    \label{fig:trinity}
\end{figure}
General Relativity, in its conventional formulation, describes gravitation as the manifestation of spacetime curvature. However, it is not the only consistent geometric description of gravity. A remarkable result in modern theoretical physics is that gravity admits three equivalent formulations, often referred to as the \textit{geometric trinity of gravity} \cite{BeltranJimenez:2019esp}. These formulations are based on different geometric attributes of spacetime: curvature, torsion, and non-metricity. Each provides a distinct perspective on the gravitational interaction and yet all of them give an equivalent description of the underlying gravitational phenomenon. 

The \autoref{fig:trinity} visually represents the geometric trinity in metric-affine gravity theories, illustrating the three fundamental geometrical features of spacetime--non-metricity ($Q_{\alpha\mu\nu} \neq 0$), torsion ($T^\alpha_{\mu\nu} \neq 0$), and curvature ($R^\alpha_{\beta\mu\nu} \neq 0$)--and their respective manifestations.
\begin{itemize}
        \item The left panel (cone) corresponds to non-metricity, where the change of vector length (green arrows) along parallel transport produces a cone-like deformation, signifying the deviation from metric compatibility.
        \item The middle panel (cylinder) depicts torsion, where the parallel transport of vectors does not close a loop, visualized as a translation when completing a circuit (red path wrapped around), characteristic of a space with torsion but zero curvature and non-metricity.
        \item The right panel (sphere) illustrates curvature, where the failure of vectors to return to their original direction after parallel transport around a closed loop demonstrates spacetime curvature with vanishing torsion and non-metricity.
\end{itemize}
\subsection{Curvature formulation: General Relativity}
In the standard Einstein-Hilbert approach, gravity is described in terms of spacetime curvature encoded in the Levi-Civita connection, which is torsion-free and metric-compatible. The action reads
\begin{equation}\label{eq:graction}
    S_{\text{GR}} = \frac{1}{2\kappa^2} \int d^4x \, \sqrt{-g}\, R + S_m ,
\end{equation}
where $R$ is the Ricci scalar constructed from the Levi–Civita connection, $\kappa^2 = 8\pi G/c^4$, and $S_m$ is the matter action. The resulting field equations are the Einstein equations, which constitute the foundation of GR (already introduced in the \autoref{sec:gr}).

\subsection{Torsion formulation: Teleparallel Gravity}
An alternative but dynamically equivalent formulation arises when the gravitational interaction is attributed to spacetime torsion rather than curvature. This is achieved by employing the Weitzenböck connection, which is curvature-free but exhibits torsion. The action is given by
\begin{equation}\label{eq:tegraction}
    S_{\text{TEGR}} = \frac{1}{2\kappa^2} \int d^4x \, \sqrt{-g}\, T + S_m ,
\end{equation}
where $T$ is the torsion scalar. This theory is known as the TEGR \cite{Aldrovandi:2013wha,Cai:2015emx}. The field equations derived from this action are equivalent to those of GR, though the underlying geometrical interpretation is different. TEGR serves as the starting point for modifications such as $f(T)$ gravity.

\subsection{Non-metricity formulation: Symmetric Teleparallel Gravity}
A third equivalent representation of gravity arises from non-metricity, characterized by the failure of the connection to preserve the metric under parallel transport. By employing the coincident gauge, one constructs a connection that is flat and torsionless but possesses non-metricity. The corresponding action is
\begin{equation}\label{eq:stegraction}
    S_{\text{STEGR}} = -\frac{1}{2\kappa^2} \int d^4x \, \sqrt{-g}\, Q + S_m ,
\end{equation}
where $Q$ is the non-metricity scalar \cite{BeltranJimenez:2019tme}. This framework is called the Symmetric Teleparallel Equivalent of General Relativity (STEGR). Like GR and TEGR, it yields the same field equations, but with a distinct geometric underpinning. It also provides a natural starting point for generalizations such as $f(Q)$ gravity.

\subsection{Equivalence and Extensions}
The three formulations, GR, TEGR, and STEGR, constitute the geometric trinity of gravity. Despite their different geometric approaches, they are dynamically equivalent at the level of Einstein’s equations. However, once generalized beyond the linear scalar invariants $R$, $T$, or $Q$, they lead to inequivalent modified gravity theories: $f(R)$, $f(T)$, and $f(Q)$ respectively. These modifications predict distinct cosmological dynamics, offering a fertile ground for addressing the puzzles of dark energy, cosmic acceleration, and potential observational tensions.

\bigskip
In summary, the geometric trinity illustrates that the gravitational interaction can be equivalently described through curvature, torsion, or non-metricity. This insight not only broadens the conceptual foundations of gravity but also motivates diverse modified gravity models, which will be central to the investigations in this thesis.

\section{TEGR and its extensions}
\subsection{Geometric Description}
In contrast to GR, which is formulated using the torsionless and metric-compatible Levi--Civita connection, teleparallel gravity employs the Weitzenb\"ock connection, which has vanishing curvature but non-zero torsion. 
Working in the so-called \textit{Weitzenb\"ock gauge}, where the spin connection $\omega^{A}{}_{B\mu}$ is set to zero, the Weitzenb\"ock connection is constructed directly from the tetrad fields as
\begin{equation}
\label{connection}
\overset{w}{\Gamma}{}^{\lambda}{}_{\mu\nu} = e^\lambda_{A}\,\partial_\nu e^A_{\mu}.
\end{equation}
Here, $e^A_{\mu}$ are the tetrad fields forming an orthonormal basis in the tangent space, and $e^\mu_{A}$ are their inverses, satisfying 
$g_{\mu\nu} = \eta_{AB} e^A_{\mu} e^B_{\nu}$. 
The superscript $w$ distinguishes the Weitzenb\"ock connection from the Levi--Civita one $\{^{\,\lambda}_{\mu\nu}\}$ used in GR.

 The torsion tensor is defined as
\begin{equation}
    T^{\lambda}_{\ \mu \nu} = \overset{w}{\Gamma}{}^{\lambda}_{\nu \mu} - \overset{w}{\Gamma}{}^{\lambda}_{\mu \nu} = e^{\lambda}_{\ A} (\partial_{\mu} e^A_{\ \nu} - \partial_{\nu} e^A_{\ \mu}).
\end{equation}
The spacetime metric is related to the tetrads through
\begin{equation}
    g_{\mu \nu} = \eta_{\alpha \beta} e^{\alpha}_{\ \mu} e^{\beta}_{\ \nu},
\end{equation}
with \( \eta_{\alpha \beta} = \text{diag}(1,-1,-1,-1) \) representing the Minkowski metric of the local inertial frame.
The torsion scalar, which serves a role analogous to the Ricci scalar in GR, is defined as
\begin{equation}
    T \equiv {S_{\lambda}}^{\ \mu \nu} T^{\lambda}_{\ \mu \nu},
\end{equation}
where \( {S_{\lambda}}^{\ \mu \nu} \) is the superpotential tensor, expressed as
\begin{equation}
    {S_{\lambda}}^{\ \mu \nu} = \frac{1}{2} \left( {K^{\mu \nu}}_{\ \lambda} + \delta^\mu_\lambda {T^{\alpha \nu}}_{\ \alpha} - \delta^\nu_\lambda {T^{\alpha \mu}}_{\ \alpha} \right).
\end{equation}
The contorsion tensor \( {K^{\mu \nu}}_{\ \lambda} \), which measures the difference between the Weitzenb\"ock and Levi-Civita connections, is given by
\begin{equation}
    {K^{\mu \nu}}_{\ \lambda} = -\frac{1}{2} \left( {T^{\mu \nu}}_{\ \ \lambda} - {T^{\nu \mu}}_{\ \ \lambda} - {T_{\lambda}}^{\ \mu \nu} \right).
\end{equation}
It is well known that the source of modification for teleparallel action \eqref{eq:tegraction} is GR action \eqref{eq:graction} where Ricci scalar is present. Using the Riemann curvature tensor and contorsion tensor, one can obtain the relation between the Ricci scalar and torsion scalar as $R=-T+B$. Here, $B$ is the total divergence which acts as a boundary term in the TEGR action. The boundary term can be ignored as it does not contribute to the field equations.

\subsubsection{\texorpdfstring{$f(T)$}{f(T)} gravity}
A natural extension of the TEGR involves promoting the torsion scalar \( T \) to a general function, giving rise to the framework of \( f(T) \) gravity. The teleparallel action \eqref{eq:tegraction} extends GR by allowing the Lagrangian to depend on a general function of the torsion scalar:
\begin{equation}\label{eq:ftaction}
    S_{f(T)} = \frac{1}{2k^2} \int d^4x\, e \left[ T + f(T) \right] + \int d^4x\, e\, \mathcal{L}_m,
\end{equation}
where \( e = \det(e_\mu^A) = \sqrt{-g} \) and \( k = \sqrt{8\pi G} \) is the gravitational coupling constant.

Varying the action \eqref{eq:ftaction} with respect to the tetrad fields yields the modified field equations of \( f(T) \) gravity:
\begin{equation}\label{eq:ftfield}
    (1 + f_T) \left[ e^{-1} \partial_\mu \left(e\, e^\lambda_A S_\lambda^{\ \nu \mu} \right) 
    - e^\alpha_A T^\lambda_{\ \mu \alpha} S_\lambda^{\ \mu \nu} \right] 
    + e^\lambda_A S_\lambda^{\ \nu \mu} \partial_\mu T\, f_{TT}
    + \frac{1}{4} e^\nu_A (T + f) 
    = 4\pi G\, e^\lambda_A \, \overset{em}{T}{}_\lambda^{\ \nu}.
\end{equation}
Here, \( f_T = \frac{df}{dT} \) and \( f_{TT} = \frac{d^2 f}{dT^2} \) represent the first and second derivatives of the function \( f(T) \), while \( \overset{em}{T}{}_\lambda^{\ \nu} \) denotes the energy-momentum tensor of matter.

For a spatially flat FLRW background, the line element (cartesian coordinates) takes the form  
\begin{equation}\label{eq:linelement}
    ds^2 = dt^2 - a^2(t)\, \delta_{ij} dx^i dx^j,
\end{equation}
where \( a(t) \) is the scale factor. The corresponding diagonal tetrad choice is  
\begin{equation}
    e_\mu^A = \text{diag}(1,\, a(t),\, a(t),\, a(t)),
\end{equation}
which leads to a simple expression for the torsion scalar  
\begin{equation}\label{eq:ts}
    T = -6H^2,
\end{equation}
with \( H = \dot{a}/a \) denoting the Hubble parameter.

By inserting this ansatz into the field equations of \( f(T) \) gravity \eqref{eq:ftfield}, one can obtain the modified Friedmann equations
\begin{gather}
    12 H^2 (1 + f_T) + T + f = 2k^2 \rho, \label{eq:ftmot1} \\
    48 H^2 \dot{H} f_{TT} - (1 + f_T)(4\dot{H} + 12H^2) + f - T = 2k^2 p. \label{eq:ftmot2}
\end{gather}
Here, \( \rho = \rho_m + \rho_r \) and \( p = p_m + p_r \) represent the total matter and radiation energy densities and pressures, respectively.

To cast these equations into a form reminiscent of standard GR, an effective dark energy sector is introduced
\begin{gather}
    3H^2 = k^2 (\rho + \rho_{\text{DE}}), \\
    3H^2 + 2\dot{H} = -k^2 (p + p_{\text{DE}}),
\end{gather}
where the effective dark energy density and pressure are defined as
\begin{gather}\label{eq:ftrhode}
    \rho_{\text{DE}} = \frac{3}{k^2} \left( \frac{T f_T}{3} - \frac{f}{6} \right), \\\label{eq:ftpde}
    p_{\text{DE}} = \frac{1}{2k^2} \frac{f - T f_T + 2 T^2 f_{TT}}{1 + f_T + 2T f_{TT}}.
\end{gather}
 These modified field equations demonstrate that \( f(T) \) gravity inherently introduces an effective energy-momentum contribution. This additional component is capable of explaining the observed late-time cosmic acceleration, thereby offering a geometrical alternative to the cosmological constant paradigm.

It is well known that in the pure tetrad formulation of $f(T)$ gravity, local Lorentz invariance is not preserved, which may lead to ambiguities in the choice of tetrad. This issue, however, is resolved in the covariant formulation of $f(T)$ gravity, where the tetrad is supplemented by the spin connection, thereby restoring local Lorentz symmetry \cite{Krssak:2015oua}. Throughout this thesis, we adopt the standard FLRW tetrad, which corresponds to a covariant choice, and thus the lack of invariance in the pure tetrad formalism does not affect our background-level cosmological analysis or results.
\subsubsection{\texorpdfstring{$f(T,\mathcal{T})$}{f(T,T)} gravity}
A further extension of $f(T)$ gravity, involving a coupling between the torsion scalar $T$ and the trace of the energy-momentum tensor $\mathcal{T}$, known as $f(T, \mathcal{T})$ gravity, is introduced by Harko et al. \cite{Harko:2014aja}. The action for this theory is given by \cite{Ganiou:2015lvv, Pace:2017dpu}
\begin{equation}
\label{eq:fTTaction}
S_{f(T, \mathcal{T})} = \frac{1}{2 k^2} \int d^4x e \left[T + f(T,\mathcal{T})\right] + \int d^4x e \mathcal{L}_m.
\end{equation}
It is very evident that the TEGR action can be recovered with the generalized functional form $f(T, \mathcal{T})$ being zero. Varying the action \eqref{eq:fTTaction} with respect to the vierbeins leads to the field equation \cite{Farrugia:2016pjh}
\begin{multline}
 \label{eq:fttfield}
2(1+f_T)\left[e^{-1}\partial_\alpha(e\,{e_B}^A S_A^{\gamma\alpha})-e_B^A T^{\alpha}_{\beta A} S_{\alpha}^{\beta \gamma} \right]+
   2\left(f_{TT}\, \partial_{\alpha} T+ f_{T \mathcal{T}}\,\partial_{\alpha} \mathcal{T}\right)e\,e_{B}^{A} S_{A}^{\gamma \alpha}+\\e_{B}^\gamma\left(\frac{f+T}{2}\right)-f_{\mathcal{T}} \left(e^{A}_B \stackrel{em}{T}_A^{\,\ \gamma} +p\, e_B^{\gamma} \right)
= k^2\, e^{A}_B \stackrel{em}{T}_A^{\,\ \gamma}.
\end{multline}
Here, $f$ denotes the functional of both variables $T$ and $\mathcal{T}$
and ${\stackrel{em}{T}_A}^\gamma$ is the energy-momentum tensor. 

For the spatially flat FLRW metric \eqref{eq:linelement}, the field equation \eqref{eq:fttfield} gives rise to the motion equations for $f(T,\mathcal{T})$,
\begin{equation}
\label{eq:fttmot1}
  {3H^2=k^2{\rho_m}-\frac{1}{2}(f+12H^2f_T)+f_\mathcal{T}(\rho_m+p_m)},
 \end{equation}
  \begin{equation}
  \label{eq:fttmot2}
  {2\dot{H}=-k^2(\rho_m+p_m)}-2\dot{H}(f_T-12H^2f_{TT})-2H(\dot{\rho_m}-3{\dot{p_m}})f_{T\mathcal{T}}-f_\mathcal{T}(\rho_m+p_m).
  \end{equation}
  Analogous to GR, with the convention $k=1$, one can rewrite the motion equations as
   \begin{gather}
    3H^2=\rho_m + \rho_{DE}, \label{fgr1}\\
    -2\dot{H}=\rho_m + p_m+ \rho_{DE} + p_{DE},\label{fgr2}
\end{gather}
where 
\begin{gather}
\rho_{DE}=-6 H^2 f_T +f_{\mathcal{T}}(\rho_m + p_m)-\frac{f}{2},\label{rd}
\end{gather}
\begin{equation}
p_{DE}= (6H^2+ 2\dot{H})f_T+ \frac{f}{2}-24 H^2 \dot{H} f_{TT}+2H(\dot{\rho}_m-3\dot{p}_m)f_{T \mathcal{T}}. \label{pd}
\end{equation}

\subsection{Growth structure description}\label{sec:growth}
After a brief discussion about the background-level equations, we now move to the perturbation-level description of the theory. As we probe the Redshift Space Distortion data, it is convenient to begin by defining the measurable quantity $f\sigma_8(a) =f(a) \, \sigma_8(a)$. The growth rate function $f(a)$ and the root-mean-square mass fluctuation $\sigma_8(a)$ are defined in terms of the matter density perturbation $\delta_m$, respectively as,
\begin{equation}\label{eq:fa}
    f(a) \equiv \frac{d ln \delta_m(a)}{da},
\end{equation}
\begin{equation}\label{eq:s8}
    \sigma_8(a)= \sigma_{8,0} \frac{\delta_m(a)}{\delta_{m0}},
\end{equation}
where $\sigma_{8,0}$ and $\delta_{m0}$ are the values for the present time $(a=1)$. Further, the structure growth parameter $S_8$ can be defined as
\begin{equation}
    \label{chap1/eq:s8}
    S_8=\sigma_8\sqrt{\frac{\Omega_{m0}}{0.3}}.
\end{equation}

In the subhorizon regime, the evolution of matter perturbation for the $f(T)$ theory satisfies the following differential equation
\begin{equation}\label{eq:mp}
    {\delta_m}''(a)+\left(\frac{H'(a)}{H(a)}+\frac{3}{a}\right){\delta_m}'(a)=\frac{3}{4}\frac{\Omega_m G_{eff}}{ a^2}\left(\frac{H_0}{H(a)}\right)^2 {\delta_m}(a),
\end{equation}
where the effective Newton's constant $G_{eff}$ is a function of scale factor $a$ and cosmic wave factor $\mathrm{k}$ in general, and prime represents the differentiation with respect to $a$. Since the considered data is not sensitive to $\mathrm{k}$-dependence, $G_{eff}$ is merely a function of scale factor in this analysis and defined as $G_{eff}=1/(1+f')$. It is very evident from observations that the LSS of the Universe was formed during the matter-dominated era, specifically when $z \in [1,10000]$. Hence, for $a<<1$, the growth should appear as $\delta(a)\sim a$ at an initial time deep of the matter phase.

For flat $\Lambda$CDM Universe, the evolution can be obtained as follows
\begin{equation}
    {\delta_m}''(a)+\left(\frac{H'(a)}{H(a)}+\frac{3}{a}\right){\delta_m}'(a)=\frac{3}{2}\frac{\Omega_m}{ a^2} {\delta_m}(a).
\end{equation}

 To achieve the theoretical prediction of $f \sigma_8$, one has to first solve Eq.~\eqref{eq:mp} by employing $H(a)$ for the underlying model. Further, the obtained solution can be incorporated into Eq.~\eqref{eq:fa} and Eq.~\eqref{eq:s8}.

 Progressing onward, there are several works in the literature that qualify the significance of the TEGR approach \cite{Maluf:2002zc, Obukhov:2002tm, Ferraro:2006jd, Linder:2010py, Kavya:2024ssu, Mishramnras, Wu:2010mn,Chen:2010va,Dent:2010nbw,Sudharani:2023hss,Bahamonde:2021srr,Moreira:2021xfe,Mishra:2023khd,Kofinas:2014daa,Mishraptep, Gonzalez-Espinoza:2021mwr}. 
 
\section{The Standard Cosmological Model}
\label{sec:lcdm}
\begin{figure}
    \centering
    \includegraphics[width=0.5\linewidth]{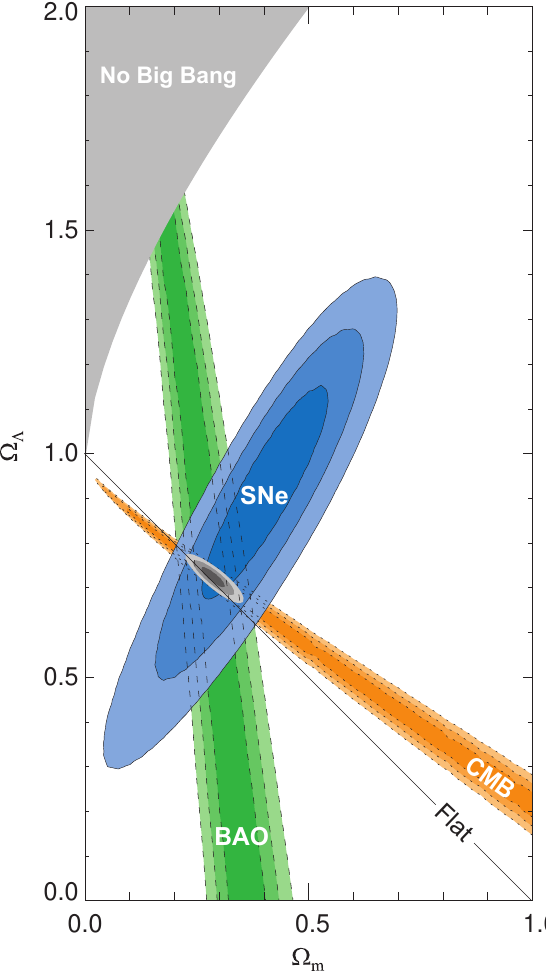}
    \caption{Concordance diagram of the $\Lambda$CDM model. Credit: Fig. 15 of \cite{SupernovaCosmologyProject:2008ojh}.}
    \label{fig:concordance}
\end{figure}
Having discussed the foundations of GR and its reformulations, we now turn to the cosmological framework that has become the prevailing paradigm: the $\Lambda$CDM model. This concordance cosmology represents the minimal and most successful realization of Einstein’s equations on large scales. Despite the theoretical challenges motivating modifications of gravity, $\Lambda$CDM remains the benchmark against which all competing scenarios are evaluated. The \autoref{fig:concordance} illustrates the joint observational constraints in the $(\Omega_m, \Omega_\Lambda)$ parameter space from Type~Ia supernovae, BAO, and CMB. Each probe independently constrains a different combination of cosmological parameters, yet their contours overlap in a common region around $\Omega_m \simeq 0.3$ and $\Omega_\Lambda \simeq 0.7$, consistent with a spatially flat Universe. This remarkable agreement among independent datasets is the reason the $\Lambda$CDM framework is often referred to as the \textit{concordance cosmological model}.

As introduced in \autoref{sec:gr}, assuming large-scale homogeneity and isotropy leads to the FLRW metric \eqref{eq:flrw} and the corresponding Friedmann equations (\ref{eq:grfriedmann1},\ref{eq:grfriedmann2}). In the $\Lambda$CDM model, the cosmic fluid consists of radiation, matter (baryonic and cold dark matter), and a cosmological constant $\Lambda$, with spatial flatness $k\simeq 0$ strongly supported by Planck observations \cite{Planck:2018vyg}.  

The Hubble expansion rate then takes the form
\begin{equation}\label{eq:Hlcdm}
    H(z) = H_0 \sqrt{\Omega_{m0}(1+z)^3 + \Omega_{r0}(1+z)^4 + \Omega_{\Lambda 0}} ,
\end{equation}
where $H_0$ is the present-day Hubble constant, $\Omega_{m0}$, $\Omega_{r0}$, and $\Omega_{\Lambda 0}$ are the current density parameters for matter, radiation, and dark energy, respectively. The radiation contribution, $\Omega_{r0} \sim 10^{-4}$, is negligible today but governed the early-Universe expansion.  

The latest Planck results (TT,TE,EE+lowE+lensing; base-$\Lambda$CDM) give \cite{Planck:2018vyg}
\[
H_0 = 67.4 \pm 0.5\ \mathrm{km\,s^{-1}\,Mpc^{-1}},\quad
\Omega_{m0} = 0.315 \pm 0.007,
\]
and, assuming flatness, \(\Omega_{\Lambda0}=1-\Omega_{m0}=0.685\pm0.007\) \cite{Planck:2018vyg}. The present radiation density (photons + standard neutrinos with \(N_{\rm eff}=3.046\)) is
\[
\Omega_{r0} = (9.21 \pm 0.14)\times 10^{-5},
\]
computed from \(T_0=2.72548\pm0.00057\) K \cite{Fixsen:2009ug} and the above $H_0$.

In summary, $\Lambda$CDM offers a minimal yet predictive description of the Universe, built directly upon the Einstein field equations in their FLRW specialization. Throughout this thesis, $\Lambda$CDM will serve as the reference model: the performance of modified gravity scenarios will be assessed relative to its well-established predictions.

\section{The Teleparallel Models} 
\subsection{Log-square-root Model}
We consider a successful model from the literature, namely the Log-square-root (LSR) model, which has the following form
\begin{equation}\label{model;log}
    f(T)=\epsilon T_0 \sqrt{\frac{T}{\eta T_0}} ln\left(\frac{\eta T_0}{T}\right),
\end{equation} 
where the model parameter $\epsilon$ is independent, and $\eta>0$. The logarithmic factor encodes an effective quantum (loop-like) correction, while the square-root 
structure regularizes the high- and low-$|T|$ behaviour, yielding controlled deviations from TEGR 
in the cosmological domain. As square root and logarithmic terms contain positive quantities $T/T_0$ and $T_0/T$, this imposes two reality requirements
\[
\frac{T}{\eta\,T_0}>0 \quad\text{(square root real)}, 
\qquad 
\frac{\eta\,T_0}{T}>0 \quad\text{(log argument $>0$)}.
\]
Since $T/T_0>0$, these two inequalities are simultaneously satisfied if and only if 
\[
\eta>0.
\]
Hence, $\eta$ must be strictly positive to keep the action real on the cosmological background.

The model exhibits a notable dynamical feature, phantom divide crossing, originally analyzed in the context of late-time cosmology by Bamba et al. \cite{Bamba:2010wb}. Its functional form has also proven viable within the framework of symmetric teleparallel gravity, where it successfully replicates the conditions necessary for consistent Big Bang nucleosynthesis, as shown in \cite{Anagnostopoulos:2022gej}.

In the upcoming chapters, we aim to explore the implications of the LSR model for various cosmological predictions. 
First, we will demonstrate that the LSR framework can serve as a promising candidate for addressing longstanding cosmological tensions, particularly by yielding a higher inferred value of the Hubble constant and a lower amplitude of matter clustering
compared to the predictions of the $\Lambda$CDM model. Furthermore, we will employ the model to establish a comparison between $f(T)$ gravity theories and cosmographic approaches in one of the chapters. Finally, we will investigate how this framework can provide a unified connection between the early-time baryogenesis mechanism and the late-time accelerated expansion of the Universe.

The Lagrangian \eqref{model;log}, along with the first motion equation \eqref{eq:ftmot1}, gives rise to
\begin{equation}
   \label{eq;hubblelsr} E^2(z)=\frac{2 \epsilon\sqrt{  \epsilon^2 + \eta \,{\Omega_{m0}}\, (1 + z)^3} + 2 \epsilon^2 + \eta \,{\Omega_{m0}}\, (1 + z)^3}{\eta}.
\end{equation}
 
 \subsection{Hybrid \texorpdfstring{$f(T)$}{f(T)} Model}
The Hybrid model has the following functional form
\begin{equation}\label{mod;hybrid}
    f(T)=\alpha T+\beta \frac{T_0^2}{T},
\end{equation}
where $\alpha$ and $\beta$ are the dimensionless model parameters. The model, in a similar line of \cite{Kolhatkar:2024oyy}, is considered because of its hybrid nature to unify the primordial and late time eras. The GR can be recovered from this model with the model parameters being zero. Incorporating the functional form \eqref{mod;hybrid}, in the first motion equation \eqref{eq:ftmot1} gives rise to
\begin{equation}
\label{eq;hubblehyb}
    E^2(z)=\frac{\sqrt{12 (\alpha +1) \beta +{\Omega_{m0}}^2 (z+1)^6}+ {\Omega_{m0}} (z+1)^3}{2(\alpha +1)}.
\end{equation}
In the above calculation, the matter density $\rho_m$ is considered as $3 \Omega_{m0}\, {H_0}^2\,(1+z)^3$ and $E(z)=H(z)/H_0$ is the normalized Hubble parameter. 

\subsection{Hybrid \texorpdfstring{$f(T,\mathcal{T})$}{f(T,T)} Model}
We consider the functional form 
\begin{equation}\label{mod;hybfTT}
    f(T,\mathcal{T})=\frac{a T_0^2}{T}
    +b \mathcal{T},
\end{equation}
where $a$ and $b$ are dimensionless free model parameters. The model is chosen because of its hybrid nature; specifically, presence of the $H^2$ term in both numerator ($\mathcal{T}=\rho_m\sim H^2$) and denominator makes it viable to explain both early and late time scenarios. These features of this class of models can be seen in the work \cite{Kolhatkar:2024oyy}. In addition, in \cite{Mishra:2025rhi}, the model has shown its suitability to explain the tensions. We have extended it with the inclusion of the trace of the energy-momentum tensor term. Employing this model in the first Friedmann equation \eqref{eq:fttmot1} gives rise to
\begin{equation}\label{eq:hub}
    {H^2(z)}=\frac{\rho_{m0}}{6}(2+b)(1+z)^3+ \frac{3a {H_0}^4}{{H^2(z)}}.
\end{equation}
Here, $H_0$ and $\rho_{m0}$ are the present time ($z=0$) Hubble parameter and matter density, respectively. The model parameters dependency can be achieved conveniently by assuming the present scenario in the above equation. It reads 
\begin{equation}
    a = \frac{1}{3}\left(1- \frac{\Omega_{m0}}{2}\left(2+b\right)\right).
\end{equation}
The relation $\rho_{m0}=3\Omega_{m0}{H_0}^2$ is applied in the above equation, where $\Omega_{m0}$ is the present matter density parameter. Now, our Hubble expression remains with the cosmological parameters along with an additional parameter $b$.
\section{Cosmological Parameters}\label{sec:cosmologicalparams}
The study of cosmic evolution is facilitated by a set of cosmological parameters that encapsulate the dynamics of expansion, the composition of the cosmic fluid, and the geometry of spacetime. Below we summarize the most relevant parameters for background cosmology.

\subsubsection*{Deceleration Parameter}
The deceleration parameter characterizes the acceleration or deceleration of the cosmic scale factor. It is defined as
\begin{equation}
    q(t) \equiv -\frac{1}{a(t)H^2(t)}\frac{d^2 a(t)}{dt^2} 
    = -1 - \frac{\dot{H}}{H^2}.
\end{equation}
Using the relation $\frac{d}{dt} = -(1+z)H(z)\frac{d}{dz}$, the deceleration parameter can be expressed as a function of redshift
\begin{equation}\label{eq:dec}
    q(z) = -1 + (1+z)\frac{1}{H(z)}\frac{dH(z)}{dz}.
\end{equation}
A negative value of $q$ indicates accelerated expansion, whereas a positive value signals deceleration.

\subsubsection*{Equation of State Parameter}
The effective equation of state (EoS) parameter (sometimes also referred to as total EoS $w_{tot}$) of the Universe is given by
\begin{equation}\label{eq:eos}
    w_{\mathrm{eff}} \equiv \frac{p_{\mathrm{eff}}}{\rho_{\mathrm{eff}}}
    = \frac{p_m + p_r + p_{\mathrm{DE}}}{\rho_m + \rho_r + \rho_{\mathrm{DE}}}.
\end{equation}
The dark energy equation of state can also be defined separately,
\begin{equation}
    w_{\mathrm{DE}} \equiv \frac{p_{\mathrm{DE}}}{\rho_{\mathrm{DE}}}.
\end{equation}
In $\Lambda$CDM, $w_{\mathrm{DE}}=-1$, whereas in modified gravity, $w_{\mathrm{DE}}$ may evolve with time.

\subsubsection*{Density Parameters}
The fractional contribution of each component to the total energy density is expressed in terms of density parameters,
\begin{equation}\label{eq:densityparam}
    \Omega_i(z) \equiv \frac{\rho_i(z)}{\rho_c(z)}, \qquad
    \rho_c(z) = \frac{3H^2(z)}{8\pi G},
\end{equation}
where $\rho_c(z)$ is the critical density. At present, $z=0$, these reduce to the standard $\Omega_{i0}$ parameters. The total energy density parameter satisfies
\begin{equation}
    \Omega_{\mathrm{tot}} = \sum_i \Omega_i = 1 + \frac{k}{a^2H^2}.
\end{equation}
Thus, $\Omega_{\mathrm{tot}}=1$ corresponds to spatial flatness.

\subsubsection*{Distance Measures}
Several cosmological observables are expressed in terms of distance measures derived from the transverse comoving distance
\begin{equation}\label{eq:dm}
    D_M(z) = c\int_0^z \frac{dz'}{H(z')}.
\end{equation}
From this, one can define
\begin{align}\label{eq:dh}
    D_H(z) &= \frac{c}{H(z)}  \qquad &\text{(Hubble distance)}, \\\label{eq:da}
    D_A(z) &= \frac{D_M(z)}{1+z} \qquad &\text{(angular diameter distance)}, \\ \label{eq:dl}
    D_L(z) &= (1+z) D_M(z) \qquad &\text{(luminosity distance)}.
\end{align}
The volume-averaged distance, commonly used in BAO analyses, is
\begin{equation}\label{eq:dv}
    D_V(z) = \left[ D_M^2(z)\, \frac{cz}{H(z)} \right]^{1/3}.
\end{equation}

\subsubsection*{Comoving Sound Horizon}
A key parameter for BAO and CMB analyses is the comoving sound horizon, defined as
\begin{equation}
    r_s(z) = \int_z^\infty \frac{c_s(z')}{H(z')} dz',
\end{equation}
where $c_s(z)$ is the sound speed of the photon–baryon fluid. In particular, $r_s(z_\ast)$ at recombination and $r_s(z_d)$ at the baryon drag epoch are standard reference scales.

\bigskip

In summary, the cosmological parameters encompass both background-level quantities ($H$, $q$, $w$, $\Omega_i$, distance measures) and structure formation indicators ($f\sigma_8$, $S_8$ [discussed earlier in \autoref{sec:growth}]). These parameters together provide a comprehensive description of cosmic evolution and constitute the key observables against which modified gravity scenarios will be tested in this thesis.

\section{Data Description}
\label{sec:mcmcmethodology}

The accurate estimation of cosmological parameters plays a crucial role in understanding the structure, composition, and evolution of the Universe. The framework for parameter estimation is fundamentally rooted in Bayesian statistics, which allows one to update prior beliefs about model parameters based on observed data. This approach is especially powerful when dealing with high-dimensional models such as those describing cosmic expansion and dark energy. The central equation governing this inference is Bayes’ theorem
\begin{equation}
    P(\theta | Data) = \frac{\mathcal{L}(Data | \theta)\, \pi(\theta)}{P(Data)},
\end{equation}

where $\theta$ denotes the set of cosmological parameters, $'Data'$ represents the observed data, $\mathcal{L}(Data|\theta)$ is the likelihood function describing the probability of the data given the model, $\pi(\theta)$ is the prior distribution encoding previous knowledge or constraints on the parameters, and $P(Data)$ is the Bayesian evidence, serving as a normalization constant across all parameter values.

In most cosmological applications, the evidence $P(Data)$ is difficult to compute directly but can be ignored when the goal is to determine the shape of the posterior distribution $P(\theta | Data)$, since it only acts as a normalization factor. Therefore, parameter estimation typically proceeds by sampling from the unnormalized posterior
\begin{equation}
    P(\theta | Data) \propto \mathcal{L}(Data | \theta)\, \pi(\theta).
\end{equation}
To evaluate this posterior distribution in a computationally feasible manner, especially in high-dimensional spaces where direct integration is intractable, we employ the MCMC technique. MCMC enables efficient sampling from complex posterior distributions by constructing a Markov chain, a sequence of samples where each sample depends only on the previous one, that eventually converges to the desired distribution.

The most widely used MCMC algorithm is the Metropolis-Hastings algorithm. It begins with an initial guess $\theta_0$ and generates a new candidate point $\theta'$ from a proposal distribution $q(\theta'|\theta_n)$. The new point is accepted with a probability given by
\begin{equation}
    \alpha = \min\left(1, \frac{\mathcal{L}(\theta') \pi(\theta') q(\theta_n | \theta')}{\mathcal{L}(\theta_n) \pi(\theta_n) q(\theta' | \theta_n)}\right).
\end{equation}
If the proposal distribution is symmetric, as is often the case, the acceptance probability simplifies to
\begin{equation}
    \alpha = \min\left(1, \frac{\mathcal{L}(\theta') \pi(\theta')}{\mathcal{L}(\theta_n) \pi(\theta_n)}\right).
\end{equation}
Repeated application of this process results in a Markov chain whose stationary distribution is the posterior $P(\theta | Data)$. In practice, an initial segment of the chain, known as the burn-in period, is discarded to ensure that the samples represent the true posterior distribution and are not influenced by the arbitrary starting point.

The likelihood function in cosmology is frequently assumed to be Gaussian in form, especially when observational uncertainties are well-characterized. The likelihood is then expressed as
\begin{equation}
    \mathcal{L}(Data | \theta) \propto \exp\left[-\frac{1}{2} \chi^2(\theta)\right],
\end{equation}
where the chi-square function is given by
\begin{equation}
    \chi^2(\theta) = \sum_{i,j} \left(D_i - T_i(\theta)\right) C^{-1}_{ij} \left(D_j - T_j(\theta)\right).
\end{equation}
Here, $D_i$ represents the observed data point, $T_i(\theta)$ is the theoretical prediction for that observable given the parameters $\theta$, and $C$ is the covariance matrix encapsulating uncertainties and correlations in the data.

Priors $\pi(\theta)$ play an important role in Bayesian analysis. Uniform priors are commonly used when minimal prior information is assumed, while Gaussian priors are appropriate when incorporating constraints from previous experiments. In some cases, log-flat or Jeffreys priors are employed to maintain scale invariance.

To implement MCMC, we used the Python-based \texttt{emcee} package \cite{Foreman-Mackey:2012any}, which is particularly well-suited for cosmological applications due to its affine-invariant ensemble sampler. This method uses multiple walkers that explore the parameter space simultaneously and is highly effective at dealing with degenerate or anisotropic posteriors, which are common in cosmology. The posterior samples produced by \texttt{emcee} can be analyzed to obtain marginalized distributions, mean values, credible intervals, and joint parameter constraints. For visualization and diagnostics, packages such as \texttt{GetDist} \cite{Lewis:2019xzd} are typically used to generate corner plots. A comprehensive overview of all the observational datasets employed throughout this thesis, along with their sources and statistical characteristics, is provided in the Appendices for reference and completeness.

\section{Model Selection Criteria}
\label{sec:modelselection}

When exploring cosmological models beyond the standard $\Lambda$CDM framework, it is essential not only to obtain the best-fit parameters but also to evaluate the relative performance of competing models. Since models often differ in the number of free parameters, a fair comparison cannot rely solely on the minimum chi-squared value. To address this, we employ statistical information criteria that balance the trade-off between goodness-of-fit and model complexity. Among these, the Akaike Information Criterion (AIC) and the Bayesian Information Criterion (BIC) are the most widely used in cosmology and other fields of data analysis.

The AIC is defined as
\begin{equation}
    \mathrm{AIC} = \chi^2_{\mathrm{min}} + 2k,
\end{equation}
where $\chi^2_{\mathrm{min}}$ is the minimum chi-squared value obtained from the fit, and $k$ denotes the number of free parameters in the model. The $2k$ term penalizes models with excessive complexity, ensuring that additional parameters are only favored if they yield a significant improvement in the fit. Lower values of AIC indicate a preferred model, but only relative differences $\Delta \mathrm{AIC}$ between models are meaningful. The standard interpretation is \cite{Burnham2002}
\begin{itemize}
\item $\Delta \mathrm{AIC} \leq 2$: substantial evidence in favor for the model,
\item $4 \leq \Delta \mathrm{AIC} \leq 7$: moderate support,
\item $\Delta \mathrm{AIC} \geq 10$: essentially no support.
\end{itemize}

Similarly, the BIC, motivated by Bayesian model comparison, is expressed as
\begin{equation}
    \mathrm{BIC} = \chi^2_{\mathrm{min}} + k \ln N,
\end{equation}
where $N$ is the total number of data points used in the analysis. The penalty term $k \ln N$ is typically stronger than that of AIC, especially when the dataset is large, making BIC more conservative in introducing extra parameters. Its interpretation follows the scale introduced by Kass and Raftery \cite{Kass1995}
\begin{itemize}
\item $0 \leq \Delta \mathrm{BIC} < 2$: strong evidence in favor,
\item $2 \leq \Delta \mathrm{BIC} < 6$: moderate evidence in favor,
\item $6 \leq \Delta \mathrm{BIC} < 10$: strong evidence against,
\item $\Delta \mathrm{BIC} \geq 10$: very strong evidence against.
\end{itemize}
Both AIC and BIC serve as tools for comparative model assessment rather than absolute measures of model validity. For a given dataset, the model with the lowest AIC or BIC is statistically preferred. However, their sensitivities differ: AIC tends to favor models that improve fit quality, even with moderate complexity, while BIC is stricter and tends to penalize additional parameters more heavily. In practice, using both criteria provides a balanced perspective on whether extended or modified gravity models are genuinely favored over $\Lambda$CDM.

In this thesis, AIC and BIC are employed in selected analyses where model comparison plays a central role. Their use allows us to assess, in a statistically consistent manner, whether the additional complexity introduced by modified gravity scenarios is justified by the data. By applying these criteria alongside parameter estimation, we aim to strike a balance between fit quality and theoretical economy, ensuring that the proposed models are evaluated on equal footing with the concordance $\Lambda$CDM framework.

\section{Conclusion}

In this chapter, we have traced the historical evolution of gravitational theory from Newtonian mechanics to Einstein’s GR, highlighting the profound shift in our understanding of gravity as a manifestation of spacetime geometry. We discussed the fundamental aspects and principles underlying GR, its remarkable successes across astrophysical and cosmological scales, and the formulation of the standard $\Lambda$CDM model as its most successful cosmological realization.  

Despite its achievements, we examined several theoretical and observational limitations of GR, including its incompatibility with quantum mechanics, the presence of singularities, and the growing tensions in modern cosmological data. These challenges naturally motivate the development of modified and extended theories of gravity. Accordingly, we provided a comprehensive overview of various extensions of GR, spanning geometrical modifications, higher-order invariants, additional dynamical fields, and structural generalizations and introduced the concept of the geometric trinity of gravity, which unifies the curvature, torsion, and non-metricity based formulations under a common framework.  

Finally, we outlined the basic cosmological framework, introduced the relevant dynamical parameters, and discussed the statistical and model-selection tools that will serve as the foundation for the analyses in the forthcoming chapters.  

Having thus established the theoretical groundwork and methodological framework, we are now prepared to apply these tools to investigate specific teleparallel gravity models, examine their cosmological consequences, and confront their predictions with current observational data. The analyses that follow aim to shed light on the nature of cosmic acceleration, the resolution of cosmological tensions, and the possible extensions of gravity beyond the standard paradigm.

\newpage
\thispagestyle{empty}

\vspace*{\fill}
\begin{center}
    {\Huge \color{RedViolet} \textbf{CHAPTER 2}}\\
    \
    \\
    {\Large\color{Emerald}\textsc{\textbf{Cosmological tensions in the teleparallel framework}}}\\
    \end{center}

\begin{center} 
\textbf{\color{RedViolet}PUBLICATIONS}\\
\textbf{Impact of teleparallalism on addressing current tensions and exploring the GW cosmology}\\
 Sai Swagat Mishra et al. 2025 \textit{The Astrophysical Journal} \textbf{981} 13.\\ \vspace{0.2 in}
\includegraphics[width=0.1\linewidth]{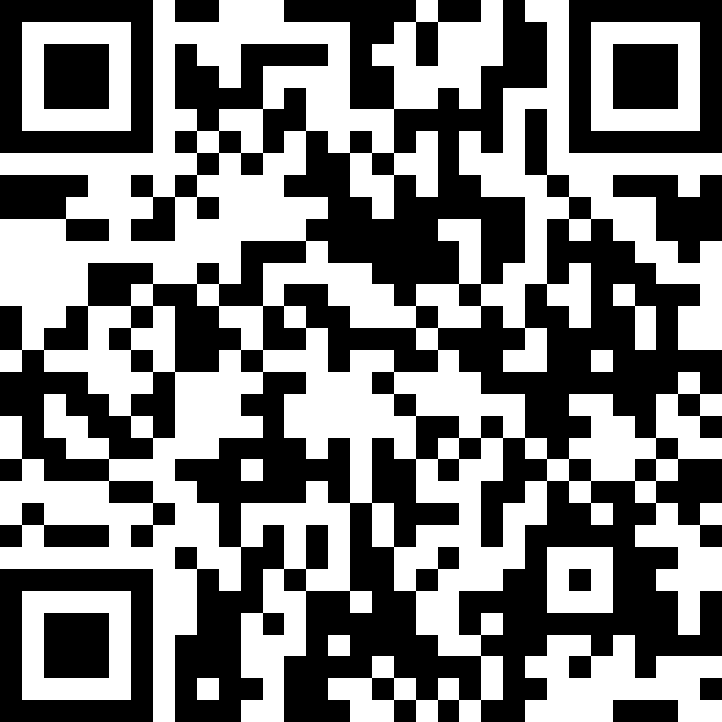} \\ DOI: \href{https://iopscience.iop.org/article/10.3847/1538-4357/adabc6}{10.3847/1538-4357/adabc6}\\
\textbf{Hubble Constant, $S_8$, and Sound Horizon Tensions: A Study within the Teleparallel Framework}\\
 Sai Swagat Mishra et al. 2025 \textit{Progress of Theoretical and Experimental Physics} \textbf{2025} 103E03\\ \vspace{0.2 in}
\includegraphics[width=0.1\linewidth]{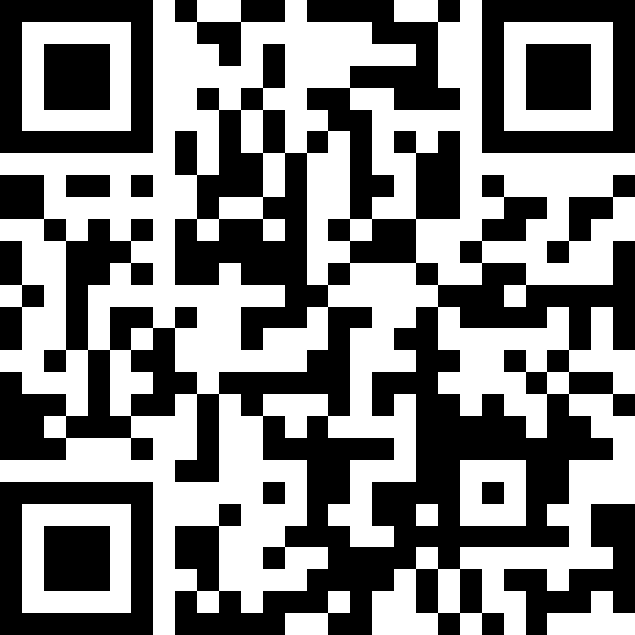} \\ DOI: \href{https://doi.org/10.1093/ptep/ptaf134}{10.1093/ptep/ptaf134}
\end{center}
\vspace*{\fill}

\pagebreak

\def\baselinestretch{1}
\chapter{\textsc{Cosmological tensions in the teleparallel framework}}\label{chap2}
\def\baselinestretch{1.5}
\pagestyle{fancy}
\lhead{\emph{Chapter 2. Cosmological tensions in the teleparallel framework}}
\rhead{\thepage}
\noindent\textbf{Highlights}
\begin{itemize}
    \item[$\star$] Investigates key cosmological tensions (\(H_0\), \(S_8\), and \(r_d\)) beyond the standard \(\Lambda\)CDM framework.  
    \item[$\star$] Tests two \(f(T)\) gravity models with datasets including CC, BAO, GRB, Pantheon+SH0ES, and GW, showing potential to alleviate both \(H_0\) and \(S_8\) tensions.  
    \item[$\star$] Extends the analysis to an \(f(T,\mathcal{T})\) model, where inverse torsion and matter coupling impact late-time expansion and structure growth.  
    \item[$\star$] Joint analyses with DESI, GW, CC, and Union3 confirm the model’s ability to statistically compete with \(\Lambda\)CDM (\(\Delta\chi^2 \lesssim 2\)).  
    \item[$\star$] Demonstrates that teleparallel extensions can provide consistent explanations of late-time acceleration while offering resolution to the cosmological tensions.  
\end{itemize}

\section{Introduction}

In modern cosmology, the Hubble parameter is fundamental to our understanding of the expansion and evolution of the Universe. It quantifies the rate at which galaxies recede from us, providing the relationship between their velocities and distances. This parameter is also central to constructing the cosmic distance ladder, a systematic method for determining astronomical distances that underpins much of observational cosmology. The Hubble constant,  $H_0$, is crucial for determining the age of the Universe. The accurate measurement of $H_0$ is essential, and the precision of these measurements has improved with the increasing number of observational methods. Generally, $H_0$ can be estimated using the cosmological model based on early-Universe observations or directly measured from the local Universe. 

$\Lambda$CDM has been the successful model in explaining the present-day acceleration in the expansion of the Universe, as highlighted by recent observations \cite{SupernovaCosmologyProject:1998vns,SupernovaSearchTeam:1998fmf}. For decades, $\Lambda$CDM model has been a cornerstone for understanding the Universe's dynamics, providing explanations for various cosmological phenomena with only a few fundamental parameters, thus establishing itself as a vital framework in cosmology. This is considered the underlying cosmological model for CMB measurements, allowing for the subsequent measurement of the value of $H_0$. 

\begin{figure}[h!]
    \centering
    \includegraphics[width=0.8\linewidth]{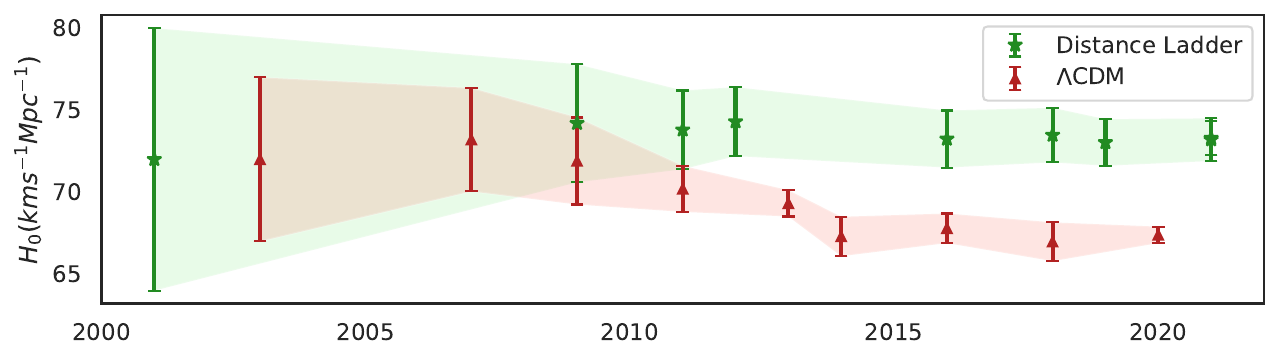}
    \caption{Tension between early and late time observations. (credit: Figure 12 of \citep{Perivolaropoulos:2021jda})}
    \label{fig:H0_tension}
\end{figure}

Notably, there is a discrepancy between the $H_0$ values obtained by these CMB measurements and the late-time surveys. Currently, numerous methods exist within the local Universe to estimate the Hubble constant using the distance-redshift relationship. These methods typically involve constructing a ``local distance ladder", with the most common approach being the use of geometric techniques to calibrate the luminosity of certain types of stars. Cepheid variables are frequently used to measure distances ranging from 10 to 40 Mpc. The green measurement band in \autoref{fig:H0_tension} represents the trend in the precision of values obtained from some significant surveys from the early 20s to the present-day \cite{HST:2000azd, Riess:2020fzl, Riess:2021jrx, Riess:2019qba, Riess:2018uxu, Riess:2016jrr, Riess:2011yx, Riess:2009pu, Freedman:2012ny}. Like-wise for early-time surveys \cite{WMAP:2003elm, WMAP:2006bqn, WMAP:2008rhx, WMAP:2010qai, WMAP:2012fli, Planck:2013pxb, Planck:2015fie, Planck:2018vyg, Addison:2017fdm}, the improved observations are represented by the red band. It is evident that the \autoref{fig:H0_tension} represents the recently growing degree of discrepancies in the value with these two estimations.

To address these issues with the background cosmological model and to build upon the loopholes of $\Lambda$CDM \cite{Perivolaropoulos:2021jda}, various approaches have been explored, either by extending GR or by modifying it (see \autoref{sec:grmodifications}). In this chapter, we focus on teleparallel formalism to modify the foundational geometric description of the spacetime manifold.

Another issue modern cosmology is facing is the $S_8$ tension, with \( S_8 \) being \( \sigma_8 \sqrt{\Omega_{m0} / 0.3} \), where \( \Omega_{m0}\) represents the current matter density, and \( \sigma_8 \) is the linearly evolved variance of the matter density field today, smoothed on an 8 Mpc/\( h \) scale. This parameter, \( S_8 \), is particularly useful as it encapsulates the combination of \( \sigma_8 \) and \( \Omega_{m0} \) to which cosmic shear (weak lensing) observations are most sensitive.
 Recent low-redshift observations of large-scale structure (LSS), including cosmic shear analyses and other measurements, yield best-fit values of \( S_8 \) that show a mild discrepancy with the predictions of the standard cosmological model. These predictions are based on parameter estimates derived from the CMB, BAO, and, CMB lensing. 

In this chapter, we investigate the persistent cosmological tensions, particularly the discrepancies in $H_0$ and $S_8$, within the framework of modified teleparallel gravity theories. The former part of the chapter is devoted to analyzing these tensions in the context of $f(T)$ gravity, exploring how the torsional modifications to General Relativity can potentially alleviate or explain the observed discrepancies. In the latter part, we extend our analysis to the $f(T,\mathcal{T})$ gravity framework, allowing for a richer coupling between matter and geometry. This section also incorporates the comoving sound horizon at the drag epoch ($r_d$) as a key observational parameter, offering further insight into the evolution of cosmic structures and the impact of modified gravity on cosmological observations.

The chapter is organized as follows. In \autoref{chap2/tensions}, we present several prominent cosmological surveys that highlight the observed tensions in parameters such as $H_0$ and $S_8$. The subsequent section, \autoref{chap2/tensionfT}, discusses the role of $f(T)$ gravity in addressing these tensions, providing theoretical interpretations and observational insights. In \autoref{chap2/tensionfTT}, we extend the analysis to the generalized teleparallel framework, namely $f(T,\mathcal{T})$ gravity, to further explore its potential in resolving the aforementioned cosmological discrepancies. Finally, we conclude our findings in \autoref{chap2/conclusion}.

\section{The Tension}\label{chap2/tensions}
 Two experiments are typically expected to yield consistent results for measured cosmological parameters, within the margin of reported errors. A discrepancy between these measurements (referred to as a tension) could suggest errors in one or both analyses, overlooked systematic uncertainties, or possibly reveal new physical phenomena. Despite the remarkable success of the standard cosmological model in explaining a wide range of observations, several persistent tensions have emerged in recent years, challenging its completeness. Notably, discrepancies in the measurements of the Hubble constant $H_0$
  and the matter fluctuation amplitude $S_8$
  between early and late-universe probes has attracted significant attention. These tensions are not only statistically significant but also potentially indicative of new physics beyond the $\Lambda$CDM paradigm. In this context, we explore how the teleparallel model addresses these anomalies and assess its effectiveness in alleviating or resolving the tensions by comparing it with the latest observations. Firstly, we report some major surveys and the discrepancies among them. Following that, how much our models fit/deviate from the landmark results will be discussed in the upcoming sections. 
  
  Beginning with the ``gold standard" experimental result from the Planck 2018 observations under a flat $\Lambda$CDM framework, the Hubble constant is determined to be $H_0 = 67.27 \pm 0.60$ km s$^{-1}$ Mpc$^{-1}$ at 68\% confidence level \cite{Planck:2018vyg}. When four trispectrum measurements are incorporated into the Planck data, the value slightly shifts to $H_0 = 67.36 \pm 0.54$ km s$^{-1}$ Mpc$^{-1}$ for the combined Planck 2018 + CMB lensing dataset \cite{Planck:2018vyg}. The Hubble constant parameter measurement from the South Pole Telescope (SPT) collaboration, as reported by Dutcher et al. \cite{SPT-3G:2021eoc}, is $H_0 = 68.8 \pm 1.5$ km s$^{-1}$ Mpc$^{-1}$. This measurement is independent of Planck observations and consistent with other CMB measurements. The $H_0$ measurement derived by D'Amico et al. \cite{DAmico:2019fhj} using BOSS DR12 data combined with Big Bang Nucleosynthesis constraints, without relying on CMB, is $H_0 = 68.5 \pm 2.2$ km s$^{-1}$ Mpc$^{-1}$. All the surveys we have discussed so far are indirect measurements. Next, we will focus on some important direct measurements in the literature.

  One of the most prominent model-independent direct measurement methods is the ``distance ladder". It allows us to measure $H_0$ locally. By this method, the Type Ia supernovae (SNIa) cepheid measurement, performed by the SH0ES team, reported the $H_0$ value $73.04 \pm 1.04$ km s$^{-1}$ Mpc$^{-1}$ at 68\% CL \cite{Riess:2021jrx}. A significant $5\sigma$ discrepancy between this measurement and that inferred from the CMB challenges the completeness of the standard cosmological model. Using the Tip of the Red Giant Branch (TRGB) is another independent method to calculate $H_0$, which is based on the calibration of SNIa. With this method, Anand et al. \cite{Anand:2021sum} reported $H_0 = 71.5 \pm1.8$ km s$^{-1}$ Mpc$^{-1}$ at 68\% CL. The HII galaxy observations can serve as reliable distance indicators, independent of Type Ia supernovae, offering an alternative method to investigate the Universe's background evolution. In \cite{FernandezArenas:2017dux}, the authors' best estimate of the Hubble constant is \( H_0 = 71.0 \pm 2.8 \,{\text{(random)}} \pm 2.1 \,{\text{(systematic)}} \, \text{km\,s}^{-1}\,\text{Mpc}^{-1} \). The latest Gravitational wave standard sirens measurements have come up as revolutionary in the cosmological context. Abbott et al. \cite{KAGRA:2021vkt}, the LIGO-VIRGO-KAGRA collaboration (GWTC-3) achieved $H_0 = 68_{-7}^{+12}$ km s$^{-1}$ Mpc$^{-1}$. Further, the CC method (flat $\Lambda$CDM with systematics) \cite{Moresco:2022phi} and TDCOSMO+SLACS \cite{Birrer:2020tax} yield, $H_0=66.5 \pm 5.4$ and $H_0=67.4_{-3.2}^{+4.1}$, respectively. Baxter et al. \cite{Baxter:2020qlr} combined CMB lensing measurements with supernova data to obtain the value $H_0=73.5 \pm 5.3$. This approach is notable because it does not rely on the sound horizon scale.

 A similar issue exists in the context of matter fluctuation amplitude, known as $S_8$ tension. Particularly, the results from the Planck satellite's measurements of the primary CMB anisotropies reveal a $2-3\sigma$ discrepancy in the amplitude of matter clustering compared to estimates from lower-redshift observations, such as weak gravitational lensing and galaxy clustering surveys. Here, we note down some major findings from early Universe followed by late Universe surveys.
  \begin{itemize}
    \item \textbf{Planck 2018 results} \hspace{2.2 in} Aghanim et al. \cite{Planck:2018vyg}
    \begin{itemize}
        \item \textit{TT, TE, EE+LowE+Lensing+BAO}: $S_8=0.825 \pm0.011$
        \item \textit{TT, TE, EE+LowE+Lensing}: $S_8=0.832 \pm 0.013$
        \item \textit{TT, TE, EE+LowE}: $S_8=0.834 \pm 0.016$
    \end{itemize}
    \item \textbf{CMB ACT+WMAP}: $S_8=0.84 \pm0.011$\hspace{0.92 in} Aiola et al. \cite{ACT:2020gnv}
      \item \textbf{WL KiDS-1000}: $S_8=0.759 \pm0.022$\hspace{1.23 in} Asgari et al. \cite{KiDS:2020suj}
      \item \textbf{WL+GC DES-Y1}: $S_8=0.773 \pm0.023$\hspace{1.03 in} Abbott et al. \cite{DES:2017myr}
      \item \textbf{WL+GC HSC+BOSS}: $S_8=0.795 \pm0.045$\hspace{0.73 in} Miyatake et al. \cite{Miyatake:2020uhg}
      \item \textbf{CC AMICO KiDS-DR3}: $S_8=0.78 \pm0.03$\hspace{0.73 in} Lesci et al. \cite{Lesci:2020qpk}
      
  \end{itemize}

 In recent years, there has been a growing focus on resolving the \( H_0 \) tension \cite{Efstathiou:2013via, Vagnozzi:2019ezj, Krishnan:2020vaf, Hazra:2022rdl, Efstathiou:2021ocp}, with various approaches proposed to address it.  
One potential solution involves early dark energy, a form of energy that mimics the behavior of the cosmological constant (\( \Lambda \)) at redshifts \( z \geq 3000 \) but diminishes over time, decaying as radiation or at an even faster rate during later stages of the evolution of the Universe. These models predict a smaller sound horizon at decoupling, resulting in a higher value of \( H_0 \) inferred from CMB data \cite{Poulin:2018cxd, Niedermann:2020dwg, Krishnan:2020obg}. Another avenue is the consideration of interacting dark energy, which presents a potential resolution to the \( H_0 \) tension between values derived from CMB anisotropies and recent direct measurements, as discussed in \cite{DiValentino:2017iww, Pan:2019gop}. Furthermore, several dark energy models address the discrepancies in both \( H_0 \) and \( S_8 \) measurements, as explored in \cite{DiValentino:2017oaw, Pandey:2019plg, Lyu:2020lps}. A model-independent approach has also been proposed, focusing on deriving necessary conditions that can be applied to generic late time cosmological extensions. This framework allows for systematic evaluation of possible solutions to the \( H_0 \) tension without relying on specific theoretical assumptions. The use of standard sirens, as discussed in \cite{Feeney:2018mkj}, represents another promising method, with ongoing and future surveys expected to provide valuable data for analysis in the coming decades.  

Additionally, modifying the underlying gravity models, where the fundamental Lagrangian description is altered, has been explored as a viable approach. Several studies demonstrate that such modifications can reduce or even eliminate the tension \cite{Schiavone:2022wvq, SolaPeracaula:2019zsl, Basilakos:2023kvk, Banerjee:2022ynv, Petronikolou:2021shp}. Specifically, the teleparallel formalism, a gravity framework employed in \cite{Wang:2020zfv, Mandal:2023bzo}, has shown promise in resolving the \( H_0 \) tension effectively. While this discrepancy could potentially arise from systematic errors, it is worth exploring the possibility of new physics beyond the standard model to address the issue. Hence, we use the framework of teleparallelism to address both tensions.

The \(n\sigma\) measures of tension are typically interpreted as probabilities based on a one-dimensional normal distribution, where \(1\sigma\) corresponds to a 68\% CL that the measurements are discrepant, \(2\sigma\) to 95\%, and so on. The challenge lies in converting constraints from two datasets into a probabilistic metric that quantifies their tension. Several methods are available for this purpose and are actively used within the scientific community. While these `tension metrics' tend to provide consistent insights when the datasets clearly agree or disagree, marginal cases can yield differing results. Factors such as the choice of priors, assumptions about posterior Gaussianity, and the higher-dimensional shape of the posterior distribution can influence the assessment of whether the datasets align. 

In the context of currently available surveys, we aim to conduct a comprehensive comparison with our theoretically proposed models. Our objective is to evaluate these results against our expectations of what the metrics should reveal regarding agreement or discrepancies between measurements. For this analysis, we classify tension as follows: \(0\)-\(2\sigma\) indicates negligible tension, \(2\)-\(3\sigma\) represents mild tension, and \(3\sigma\) or higher is considered significant inconsistency.

\section{Tensions in the \texorpdfstring{$f(T)$}{f(T)} framework}\label{chap2/tensionfT}

To explore cosmological tensions within the framework of $f(T)$ gravity, we consider the LSR model\footnote{The LSR model reads $f(T)=\epsilon T_0 \sqrt{\frac{T}{\eta T_0}} ln\left(\frac{\eta T_0}{T}\right)$.} \eqref{model;log} and the Hybrid model\footnote{The Hybrid model reads $f(T)=f(T)=\alpha T+\beta \frac{T_0^2}{T}$.} \eqref{mod;hybrid}. The corresponding Hubble parameter expressions, given in Eq.~\eqref{eq;hubblelsr} and Eq.~\eqref{eq;hubblehyb}, are used to carry out an MCMC analysis following the methodology described in \autoref{sec:mcmcmethodology}. The resulting constrained parameter spaces are subsequently examined in various cosmological contexts, as discussed in the following subsection.

\subsection{Results}\label{chap2.1/results} 
It is indeed a curiosity that among several modifications of GR, which theory can be considered a suitable alternative? However, no candidates are able to challenge the standard cosmological model in the context of describing all aspects of gravity till now. In order to this probe, we present two successful teleparallel models that can possibly alleviate the current tensions and be considered as perfect contenders to examine further. In this section, we intend to discuss the outcomes yielded from both Lagrangians. The first subsection is dedicated to review the results obtained from MCMC analysis and their further consequences. Subsequently, a complete subsection is presented to investigate the impact of our models on the well-known cosmological tensions.

\subsubsection{Observational Constraints}
\begin{figure}
    \centering
    \includegraphics[width=0.8\textwidth]{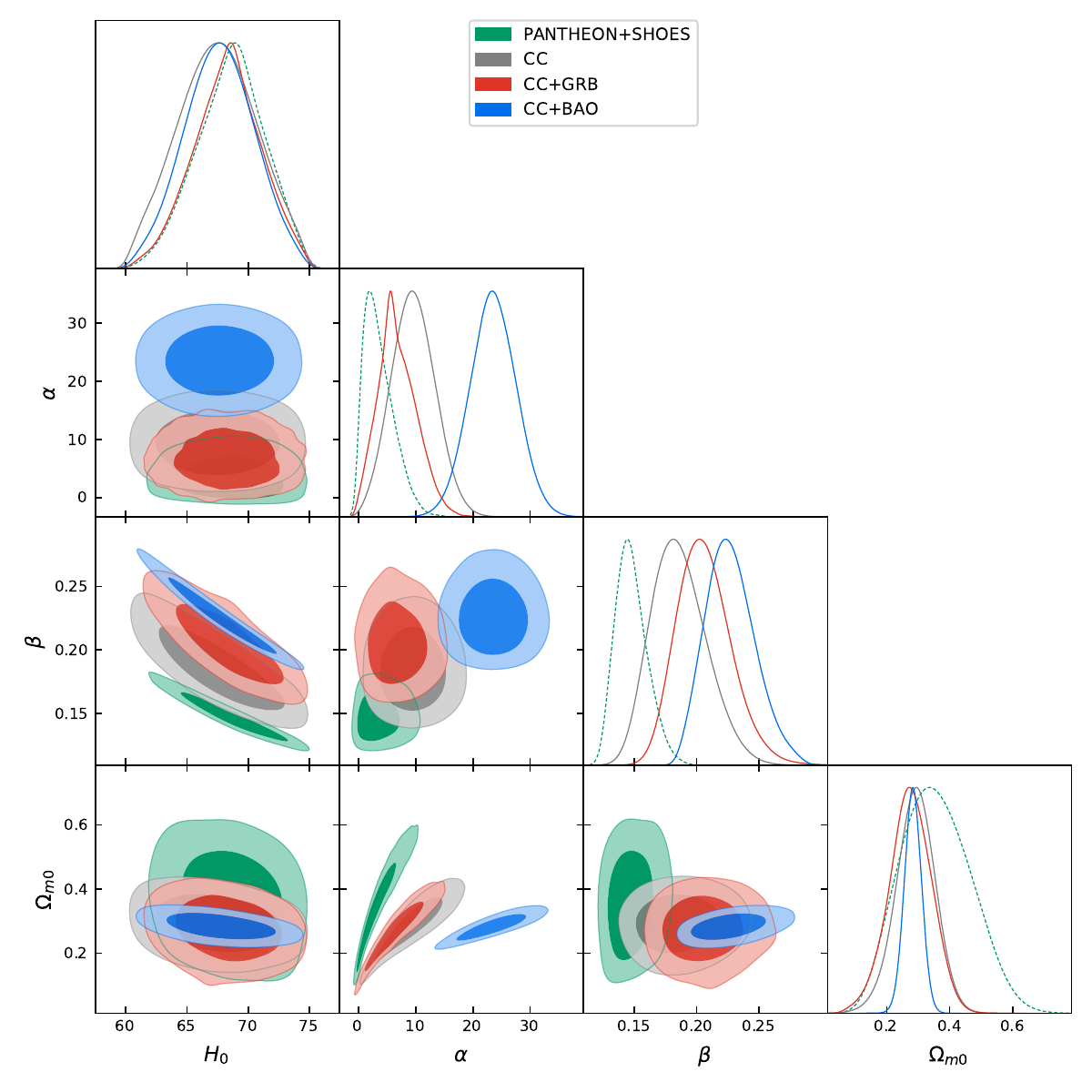}
    \caption{2D contours and Likelihood probability distributions for the Hybrid model.}
    \label{fig:c1}
\end{figure}

\begin{figure}
    \centering
    \includegraphics[width=0.8\textwidth]{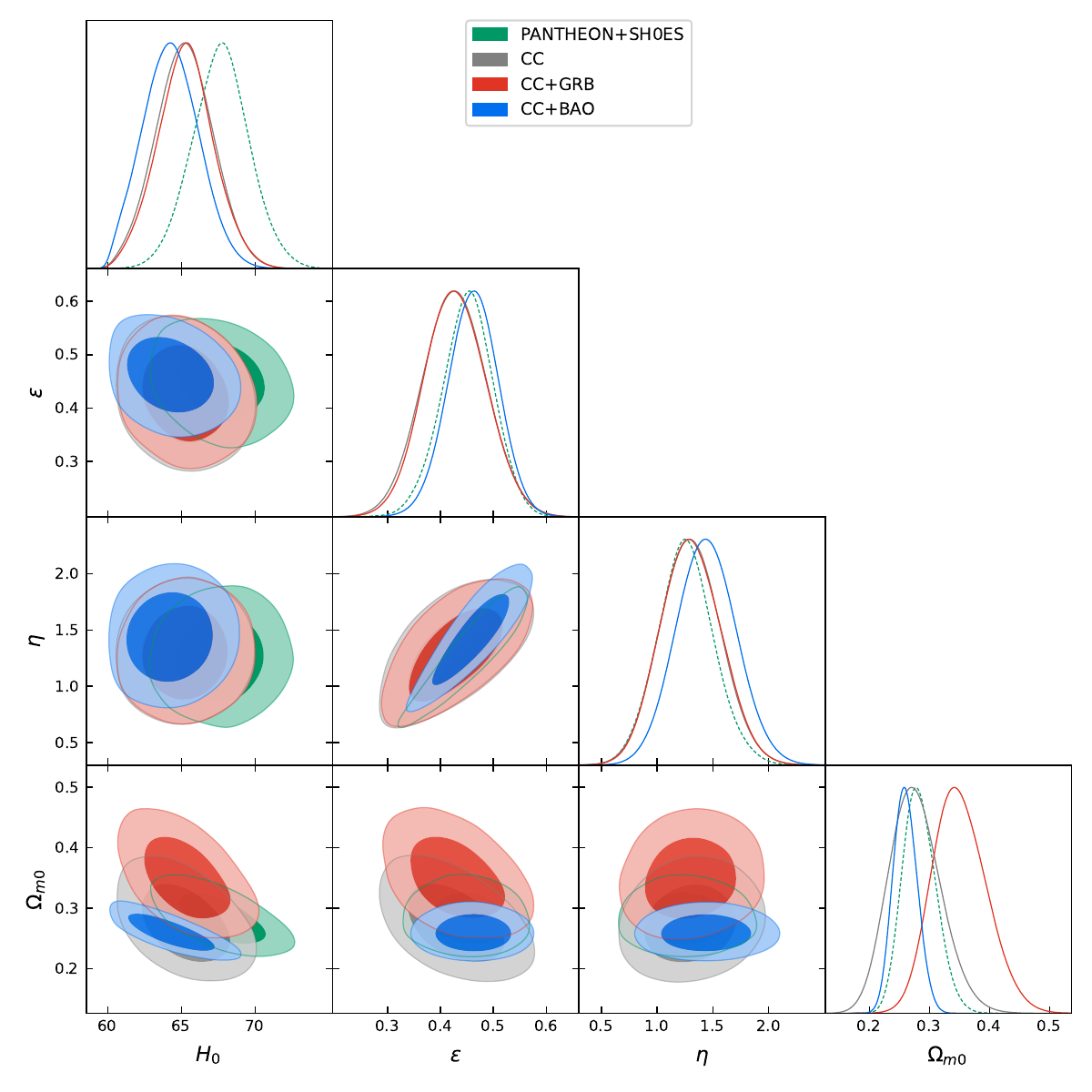}
    \caption{2D contours and Likelihood probability distributions for the LSR model.}
    \label{fig:c2}
\end{figure}

\begin{figure}
    \centering
    \includegraphics[width=0.8\textwidth]{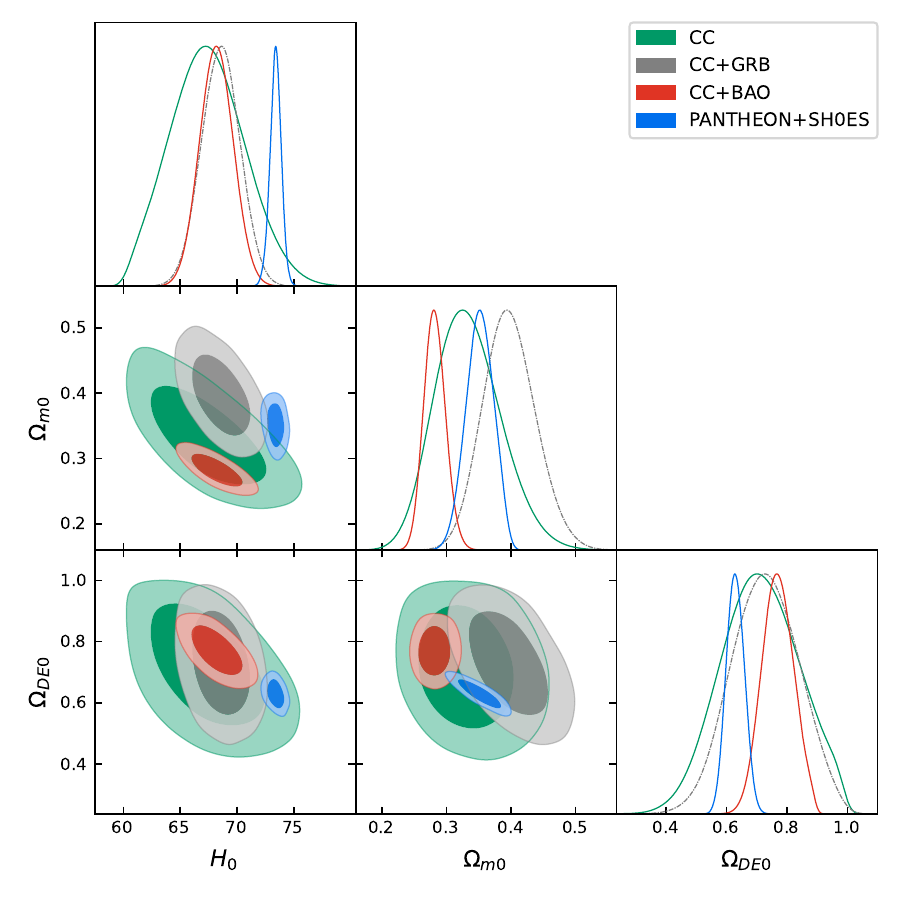}
    \caption{2D contours and Likelihood probability distributions for the $\Lambda$CDM model.}
    \label{fig:LCDM}
\end{figure}

To constrain both the models through the MCMC technique, the theoretical Hubble expressions, along with four combinations of datasets, are considered. The 2D likelihood profiles for both the models are presented in \autoref{fig:c1} and \autoref{fig:c2}, respectively. The dark-shaded regions represent the $1-\sigma \,\, (68 \%)$ CL, and the light-shaded region represents $2-\sigma \,\, (95.4 \%)$ CL. The exact numerical ranges for the former are summarized in the \autoref{chap2/table1} and \autoref{chap2/table2}. One can observe the overlapping ranges for all the parameters from the tables, especially the $H_0$ parameter. Regarding this, a detailed discussion is presented in the next subsection. Further, we repeat the statistical procedure involving the same combinations of data for the standard cosmological model. The corresponding 2D likelihood contour is represented in 
 \autoref{fig:LCDM} and the summarized result is given in \autoref{chap2/tab:LCDM}.

\begin{figure}
    \centering
        \centering
        \includegraphics[width=0.49\linewidth]{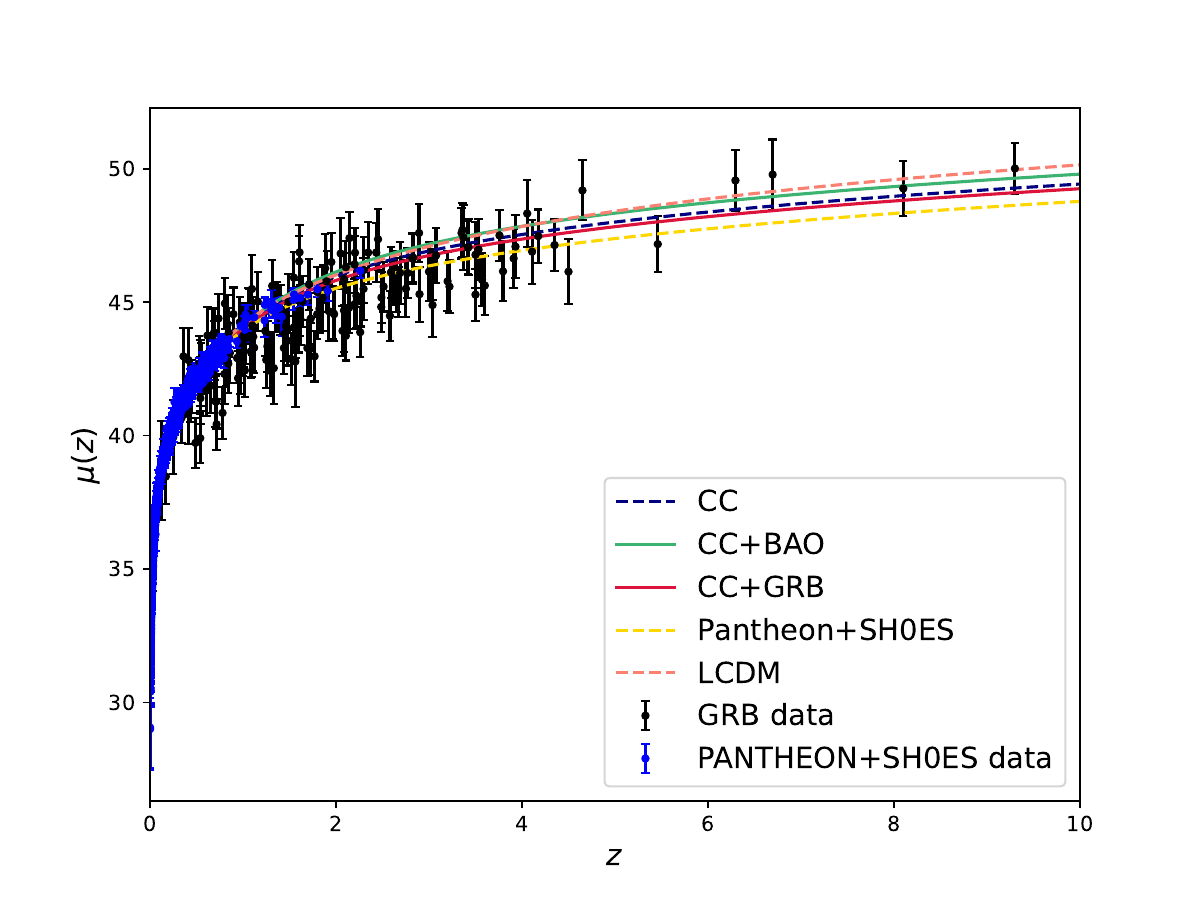}
        \includegraphics[width=0.49\linewidth]{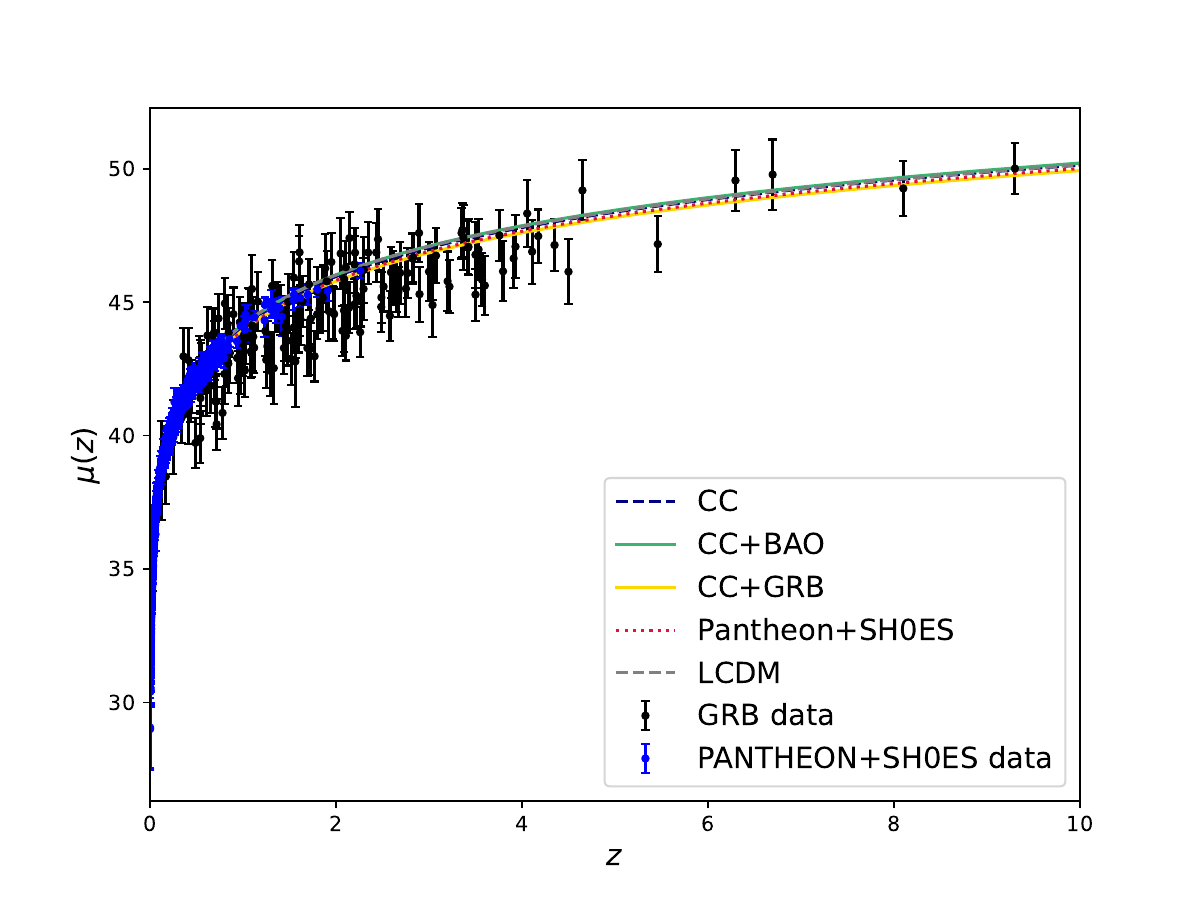}
    \caption{Distance modulus profile of the Hybrid model (left panel) and LSR model (right panel) 
    against the error bars of GRB and Pantheon+SH0ES.}
    \label{chap2.1/fig:mu}
\end{figure}

\begin{figure}
    \centering
        \includegraphics[width=0.49\linewidth]{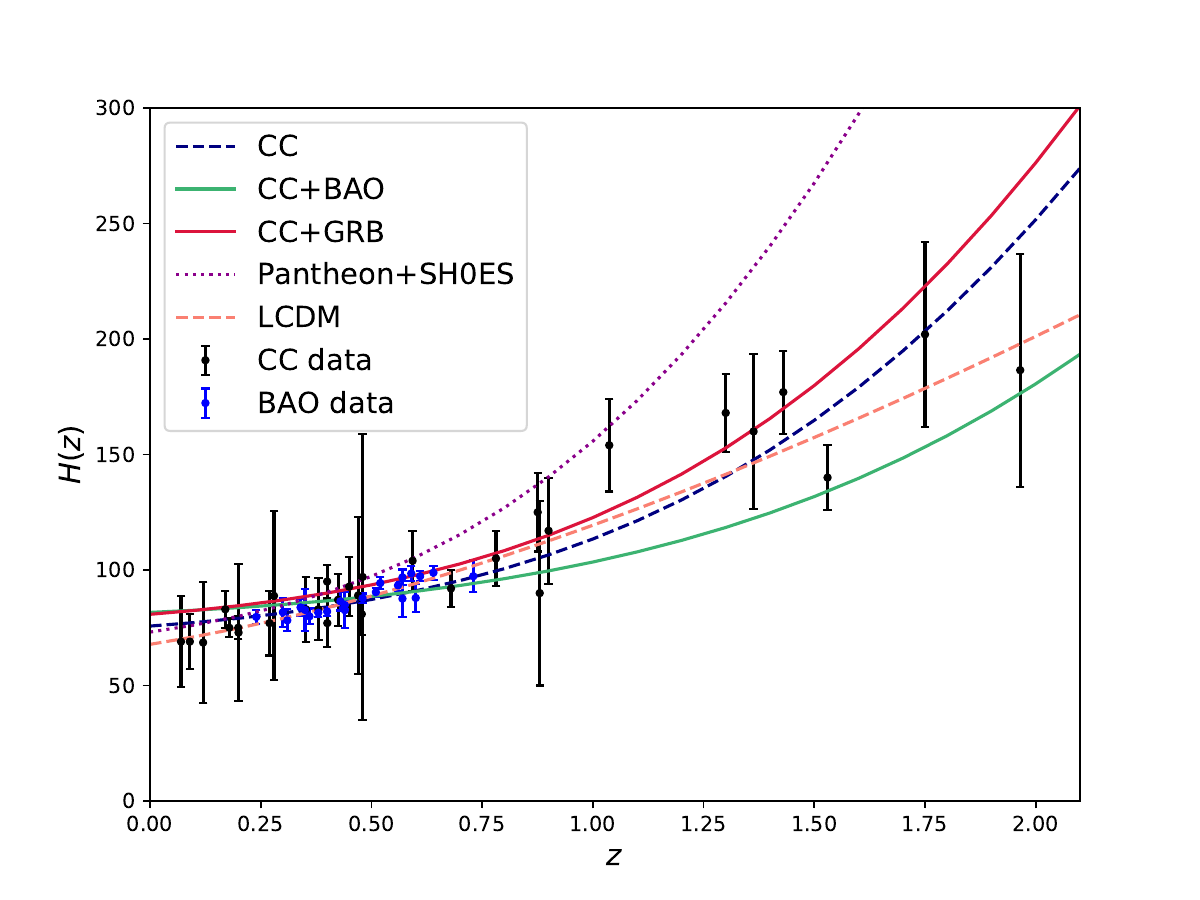}
    \includegraphics[width=0.49\linewidth]{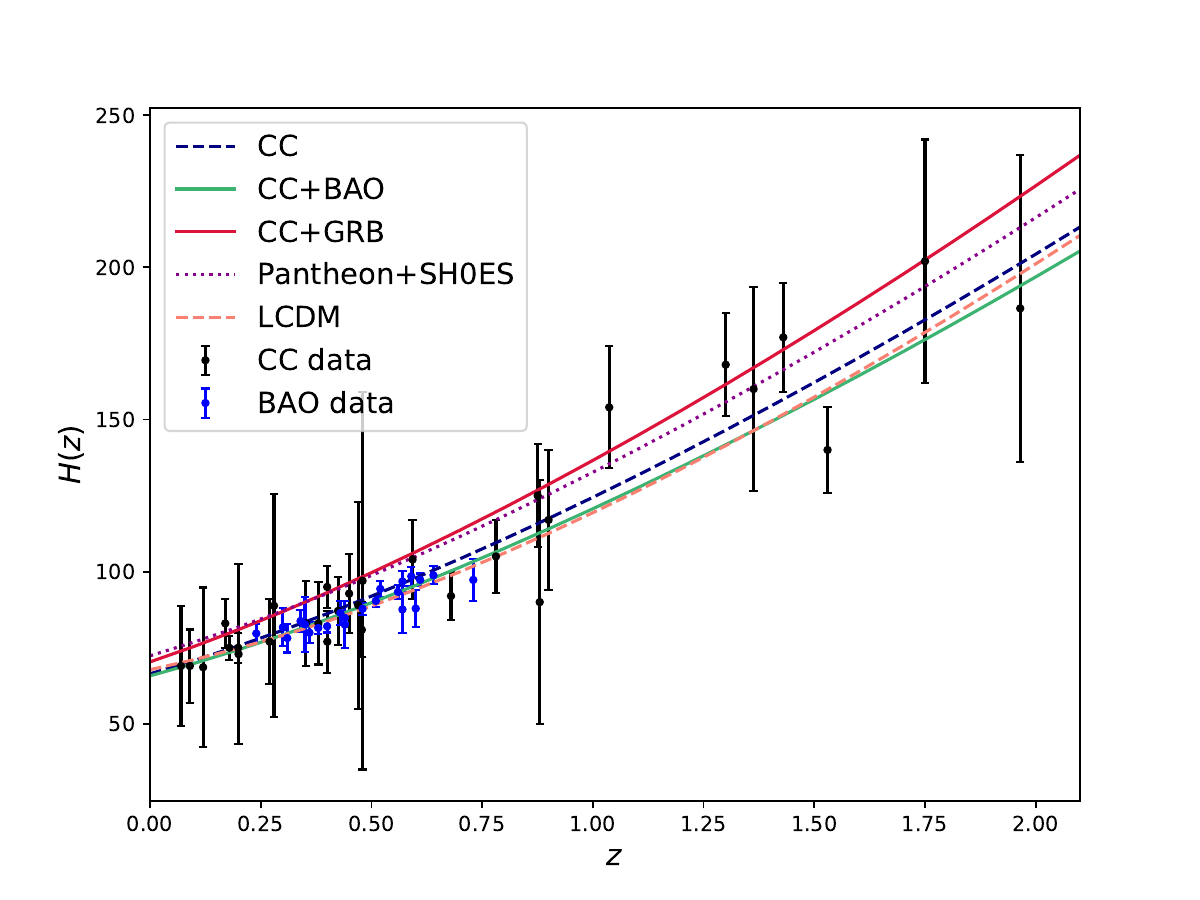}
    \caption{Hubble profile of the Hybrid model (left panel) and LSR model (right panel)  
    against the error bars of CC and BAO}
    \label{chap2.1/fig:H}
\end{figure}

\begin{figure}
    \centering
    \includegraphics[width=0.49\linewidth]{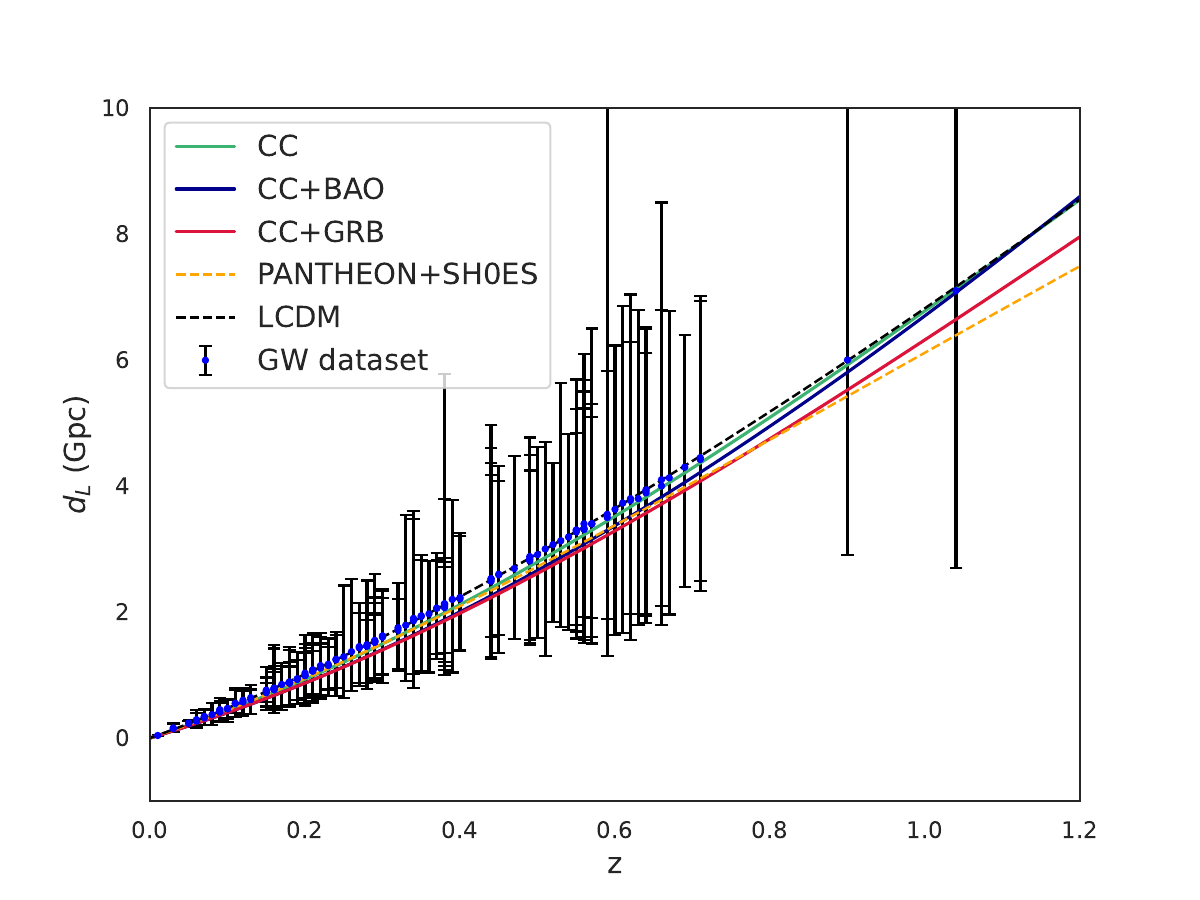}
    \includegraphics[width=0.49\linewidth]{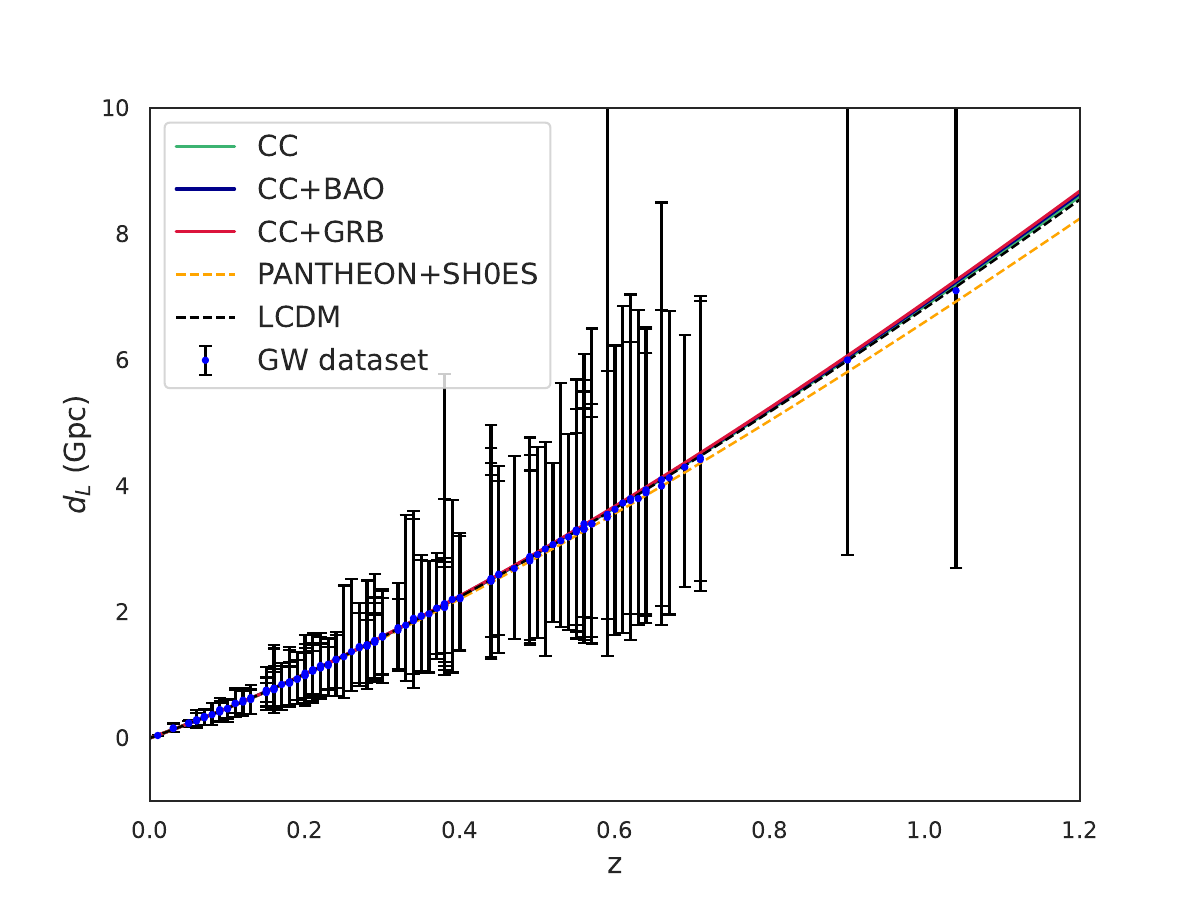}
    \caption{Luminosity distance profile of the Hybrid model (left panel) and LSR model (right panel)  
    against the error bars of VIRGO and LIGO data.}
    \label{fig:GW}
\end{figure}

Next, we verify the credibility of the constrained models by comparing them with some widely used datasets. From \autoref{chap2.1/fig:mu}, we depict that the distance modulus profile, for each combination of both the models, is in excellent agreement with 162 GRB and 1701 PANTHEON+SH0ES data points. In \autoref{chap2.1/fig:H}, the Hubble function shows a very good trend against the 31 CC and 26 BAO points for most of the cases. Furthermore, we compare the results with the recent gravitational wave (GW) data from the LIGO, VIRGO, and KAGRA collaborations through the luminosity distance. All the combinations are found to be perfectly aligned with observations (see \autoref{fig:GW}). For all the comparisons above, we consider the standard cosmological model as a role model along with the datasets. Our results appear as an excellent match to the $\Lambda$CDM model in most of the cases.

\begin{figure}
    \centering
    \includegraphics[width=0.49\linewidth]{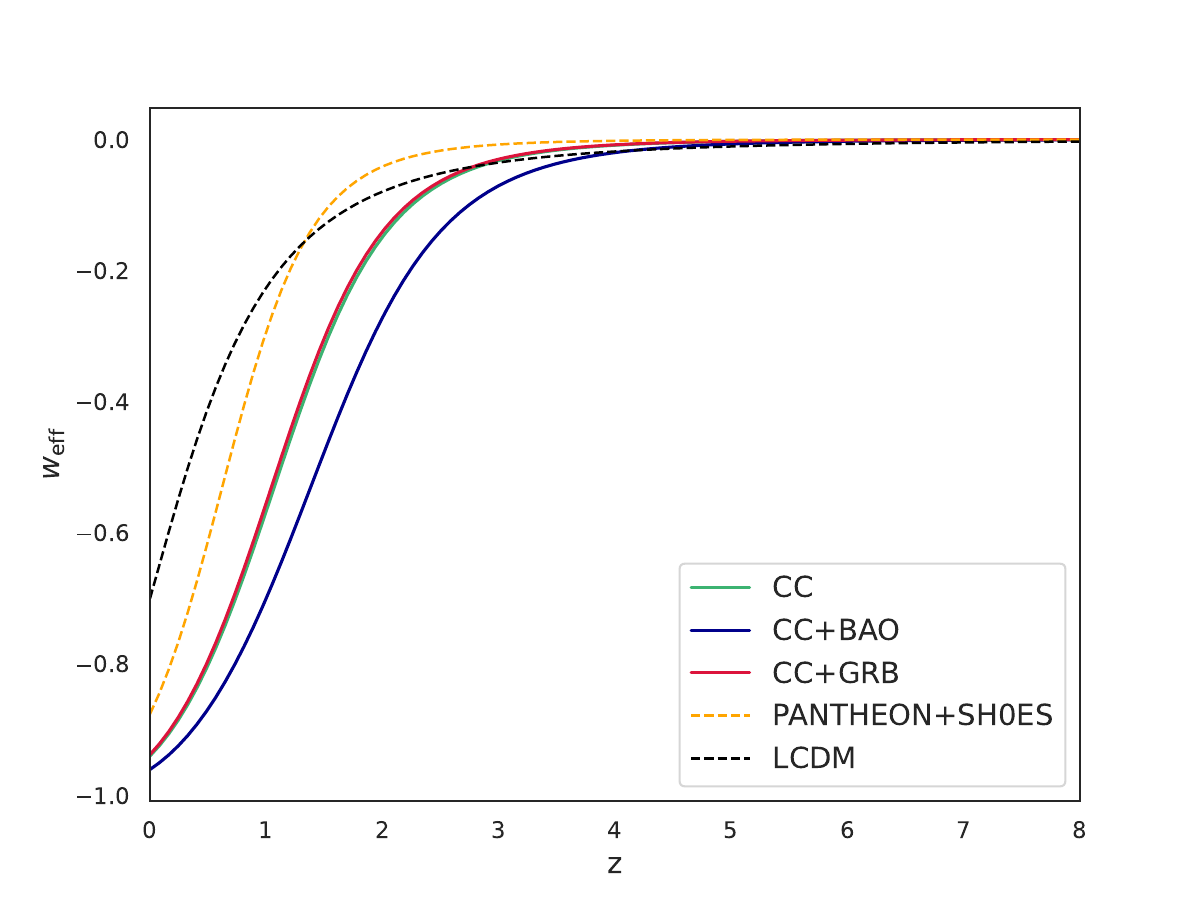}
    \includegraphics[width=0.49\linewidth]{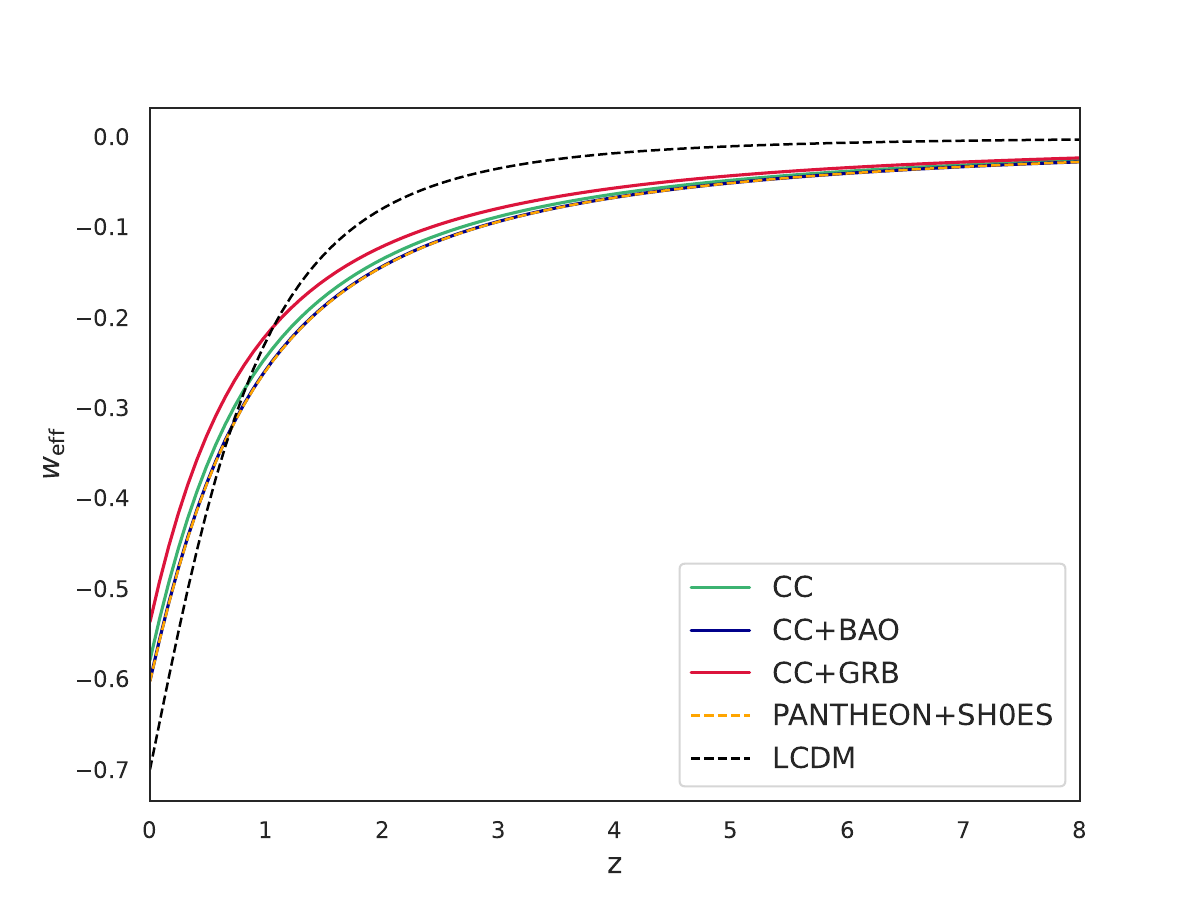}
    \caption{Evolution of effective Equation of State parameters ($\omega_{eff}$) for both Hybrid model (left panel) and LSR model (right panel). }
    \label{chap2.1/fig:eos}
\end{figure}

\begin{figure}
    \centering
    \includegraphics[width=0.49\linewidth]{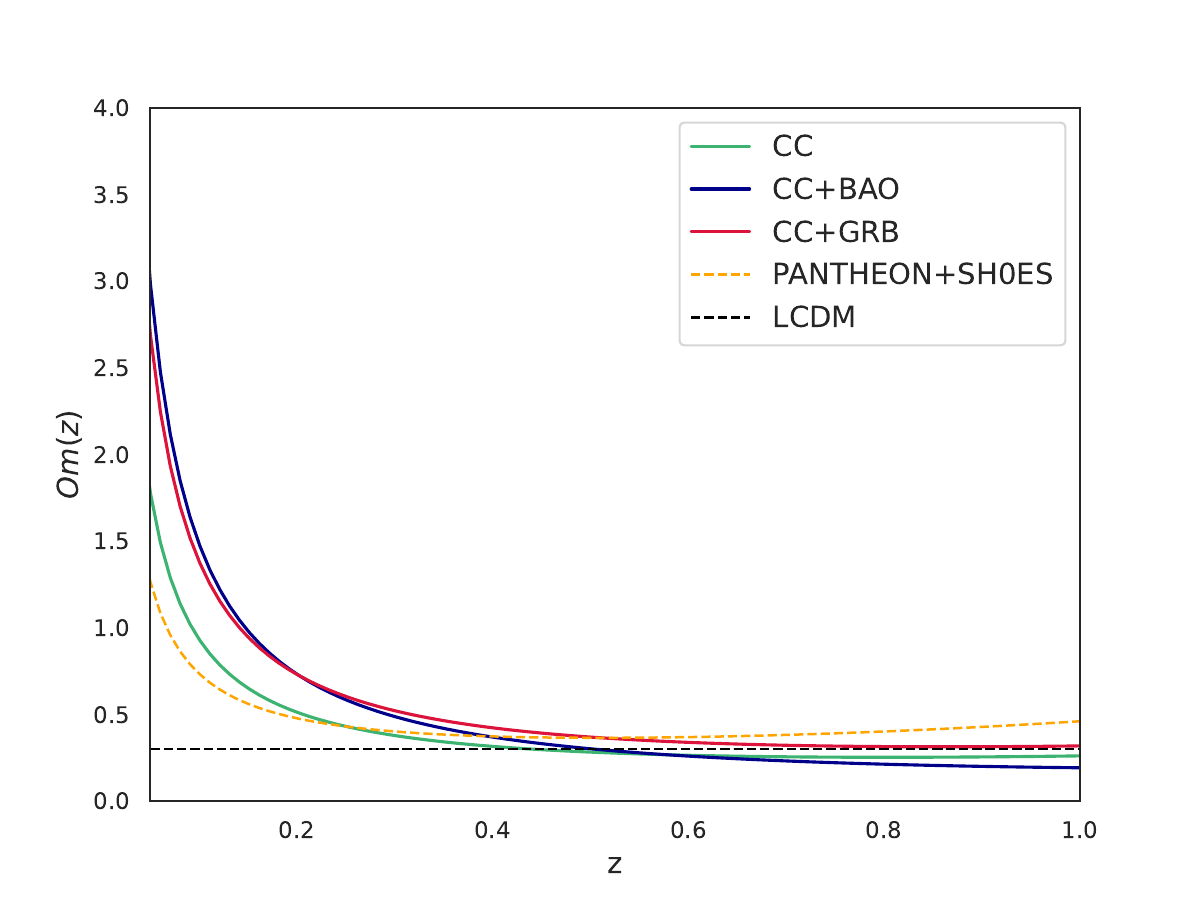}
    \includegraphics[width=0.49\linewidth]{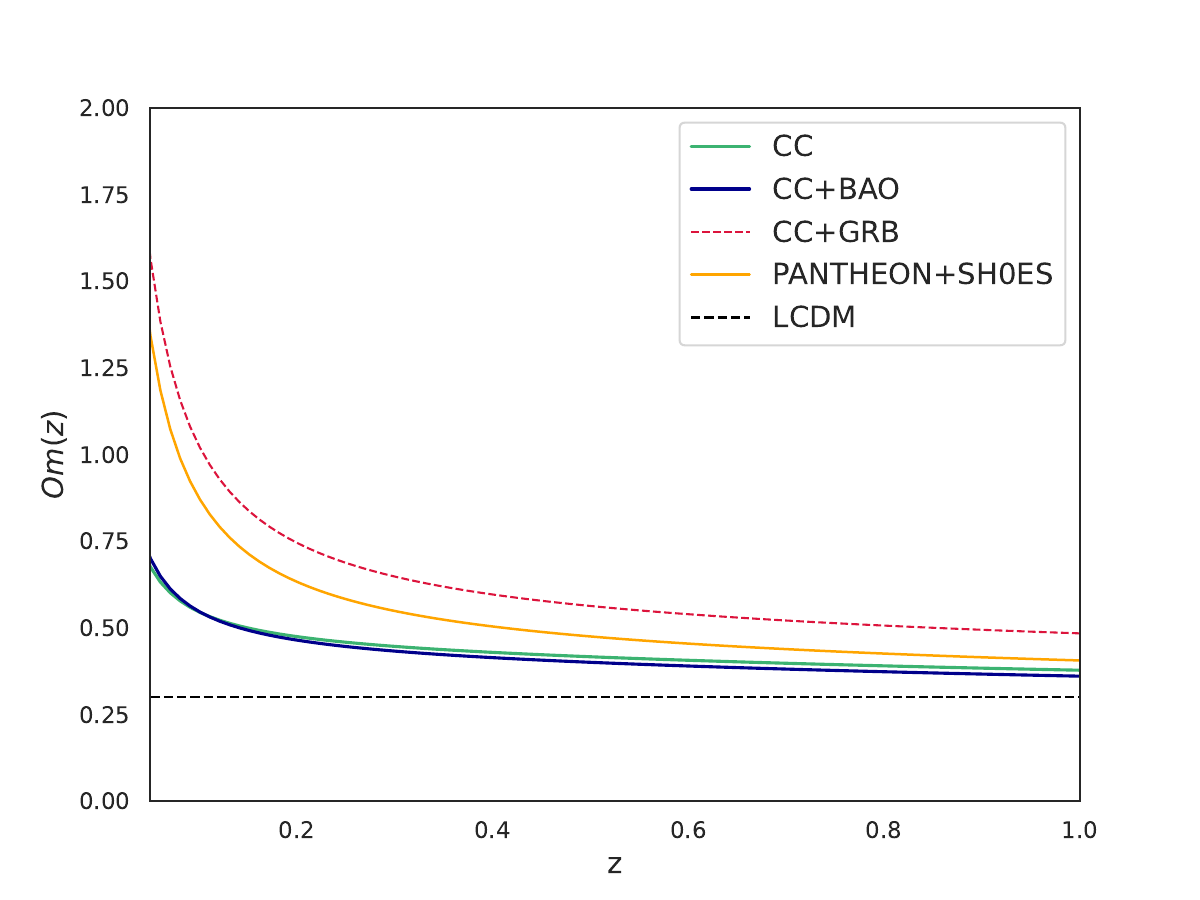}
    \caption{Evolution of $Om(z)$ for the Hybrid model (left panel) and LSR model (right panel).}
    \label{chap2.1/fig:Om}
\end{figure}

The models describe the late-time accelerated expansion of the Universe so conveniently through the distance-measuring functions and observations. Another important cosmological tool, the effective EoS parameter $(w_{eff})$, has been considered for testing their efficacy in late time. One can observe the quintessence behavior of both models in \autoref{chap2.1/fig:eos}, indicating the late time acceleration again. To probe the evolution of the Universe with additional evidence, we consider the $Om(z)$ diagnostics \footnote{For a flat Universe, the function is defined as $Om(z)=\frac{\left( \frac{H(z)}{H_{0}}\right) ^{2}-1}{(1+z)^{3}-1}$.}. This tool, proposed by \cite{Sahni:2008xx}, predicts the nature of dark energy. A constant $Om$ value confirms the behavior of dark energy like cosmological constant. In case of variation, it indicates Phantom Universe for +ve slope and Quintessence Universe for -ve slope. From \autoref{chap2.1/fig:Om}, we depict that both teleparallel models predict Quintessence behavior which supplements evidence to the previous results.

\begin{figure}
    \centering
    \includegraphics[width=0.49\linewidth]{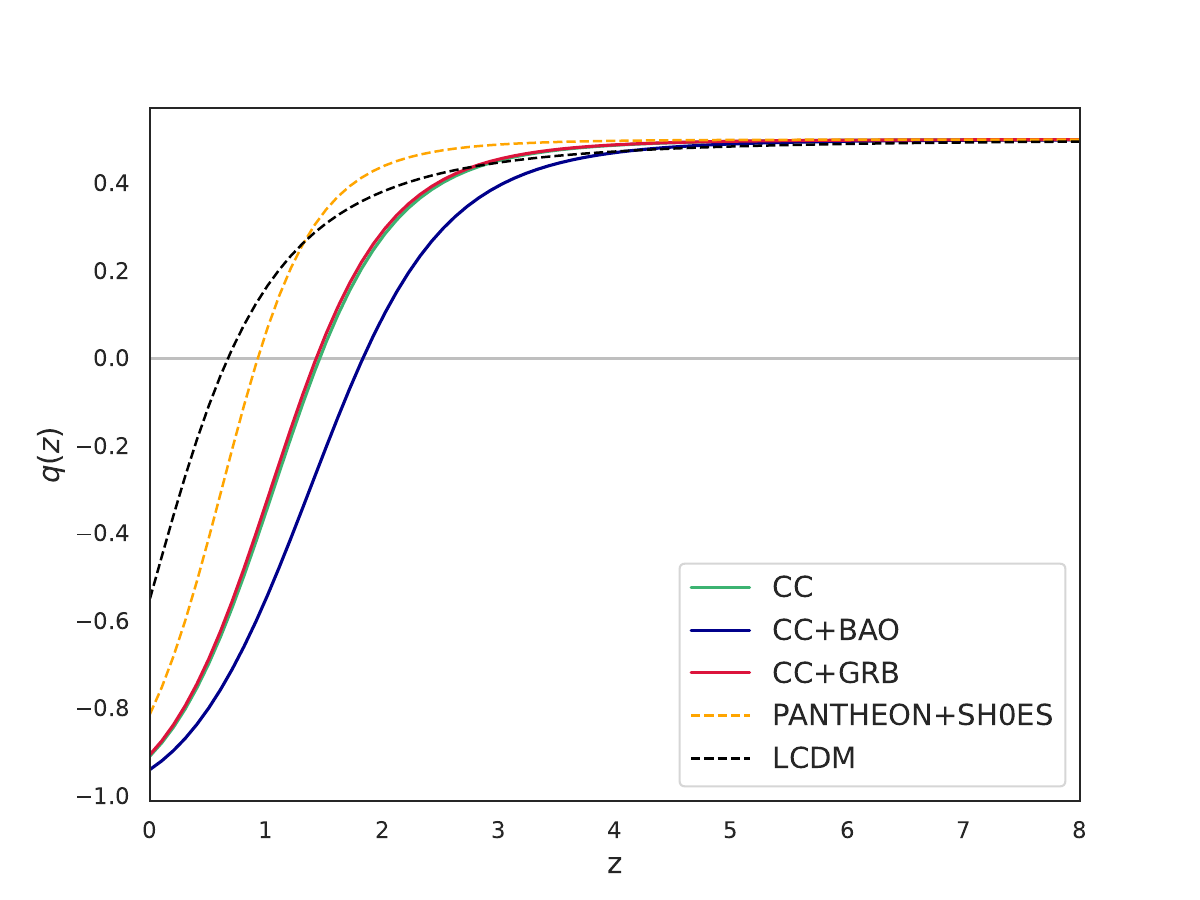}
    \includegraphics[width=0.49\linewidth]{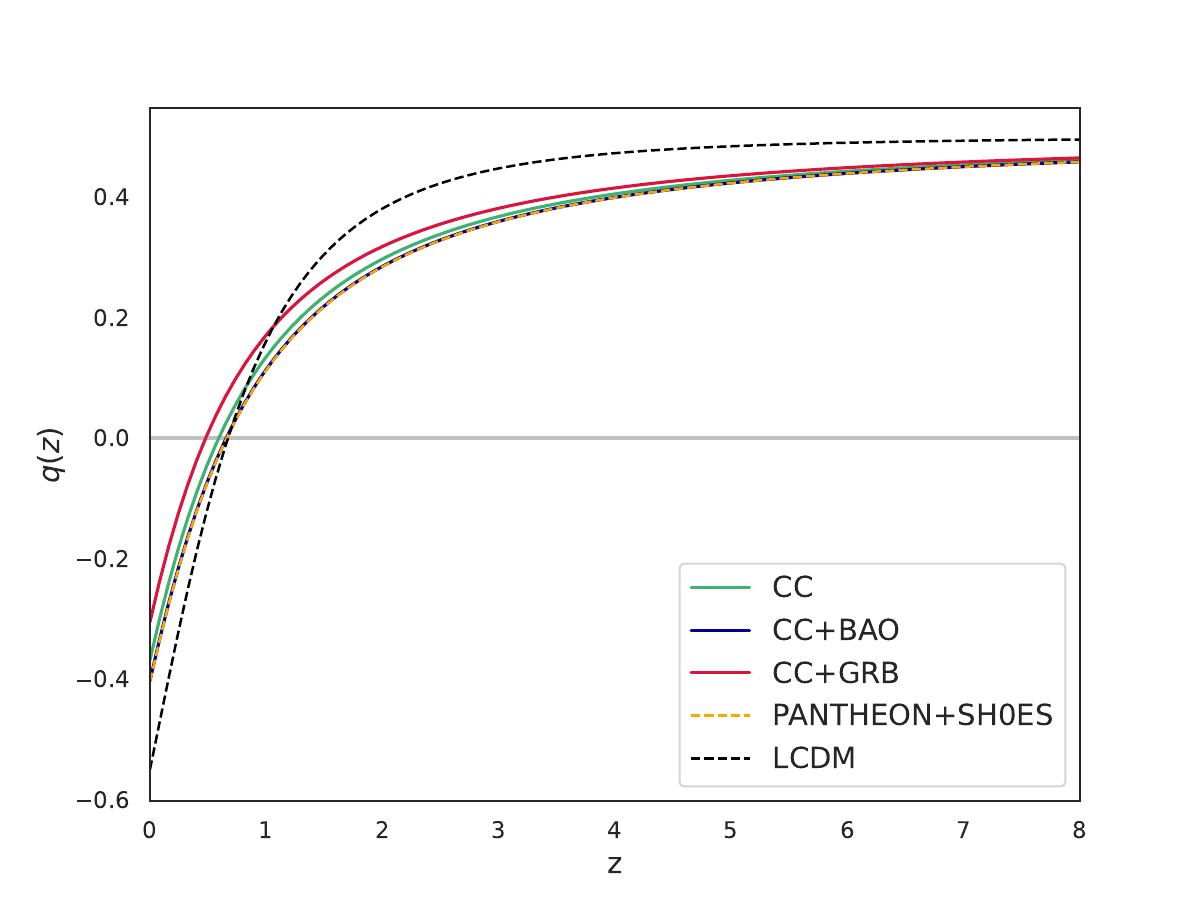}
    \caption{Evolution of deceleration parameters for both Hybrid model (left panel) and LSR model (right panel).}
    \label{chap2.1/fig:q}
\end{figure}

So far, our intention was to investigate the credibility of our results in the late time era. Now, we consider the deceleration parameter which gives insight into the intermediate phase. In \autoref{chap2.1/fig:q}, one can observe that all four cases for each model show a $\Lambda$CDM-like behavior with a transition from acceleration to deceleration. Though for some cases of the Hybrid model, the transition redshift slightly deviates from the $\Lambda$CDM, they behave exactly like the standard model for higher redshifts. Meanwhile, the LSR model has a very similar nature to the standard model in all contexts.
\begin{table}
 \centering
 
 \caption{Best fit $1\sigma$ ranges from the MCMC for the Hybrid model.}
 \label{chap2/table1}
    \begin{tabular}{c|c|c|c|c|}
    
 \cline{1-5}
\multicolumn{1}{|c|}  {\it{Dataset}} & {$H_0\,\ (km \,\ s^{-1} Mpc^{-1})$}  & $\alpha$ & $\beta$ & $\Omega_{m0}$ \\ \hline \hline
\multicolumn{1}{ |c| }{CC} & $67.5 \pm 3.2$ & $9.5 \pm 3.8$ & $0.186^{+ 0.019}_{- 0.025}$ & $0.291 \pm 0.063$     \\ 
\multicolumn{1}{ |c| }{CC + BAO} & $67.6 \pm 2.9$ & $24 \pm 4$ & $0.227^{+ 0.017}_{- 0.022}$ & $0.283 \pm 0.027$   \\ 
\multicolumn{1}{ |c| }{CC+GRB} & $68.3 \pm 2.8$ & $7.0^{+ 3.0}_{- 3.8}$ & $0.206^{+ 0.018}_{- 0.025}$ & $0.277 \pm 0.067$ \\ 
\multicolumn{1}{ |c| }{Pantheon+SH0ES} & $68.5 \pm 2.8$ & $3.6^{+ 1.5}_{- 3.1}$ & $0.148^{+ 0.010}_{- 0.014}$ & $0.357^{+ 0.098}_{- 0.12}$  \\       \hline

    \end{tabular}
\end{table}

\begin{table}
 \centering
 \caption{Best fit $1\sigma$ ranges from the MCMC for the LSR model.
 }
 \label{chap2/table2}
    \begin{tabular}{c|c|c|c|c|}
    
 \cline{1-5}
\multicolumn{1}{|c|}  {\it{Dataset}} & {$H_0\,\ (km \,\ s^{-1} Mpc^{-1})$}  & $\epsilon$ & $\eta$ & $\Omega_{m0}$ \\ \hline \hline
\multicolumn{1}{ |c| }{CC} & $65.2 \pm 2.0$ & $0.427 \pm 0.060$ & $1.30 \pm 0.27$ & $0.277^{+ 0.038}_{- 0.046}$     \\ 
\multicolumn{1}{ |c| }{CC + BAO} & $64.3 \pm 1.9$ & $0.461 \pm 0.048$ & $1.44 \pm 0.27$ & $0.260^{+ 0.019}_{- 0.021}$    \\ 
\multicolumn{1}{ |c| }{CC+GRB} & $65.3 \pm 1.9$ & $0.428 \pm 0.059$ & $1.30 \pm 0.27$ & $0.350^{+ 0.039}_{- 0.048}$ \\ 
\multicolumn{1}{ |c| }{Pantheon+SH0ES} & $67.7 \pm 1.9$ & $0.452 \pm 0.049$ & $1.27^{+0.24}_{-0.28}$ & $0.283^{+ 0.025}_{- 0.029}$  \\       \hline

    \end{tabular}
\end{table}
\begin{table}
 \centering
 \caption{Best fit $1\sigma$ ranges from the MCMC for the $\Lambda$CDM model.
 }
 \label{chap2/tab:LCDM}
    \begin{tabular}{c|c|c|c|}
    
 \cline{1-4}
\multicolumn{1}{|c|}  {\it{Dataset}} & {$H_0\,\ (km \,\ s^{-1} Mpc^{-1})$}  & $\Omega_{m0}$ & $\Omega_{DE0}$ \\ \hline \hline
\multicolumn{1}{ |c| }{CC} &  $67.4^{+2.0}_{-3.5}$ & $0.334^{+0.045}_{-0.056}$ & $0.71^{+ 0.13}_{- 0.13}$  \\ 
\multicolumn{1}{ |c| }{CC + BAO} & $68.2^{+1.5}_{-1.5}$ & $0.282^{+0.016}_{-0.018}$ & $0.770^{+ 0.052}_{- 0.052}$   \\ 
\multicolumn{1}{ |c| }{CC+GRB} & $68.6^{+1.7}_{-1.7}$ & $0.39.8^{+0.038}_{-0.043}$ & $0.73^{+ 0.11}_{- 0.11}$ \\ 
\multicolumn{1}{ |c| }{Pantheon+SH0ES} & $73.40^{+0.50}_{-0.50}$ & $0.352^{+0.021}_{-0.021}$ & $0.629^{+ 0.03}_{- 0.03}$  \\       \hline

    \end{tabular}
\end{table}

\begin{table}[H]
\centering
\small
\begin{tabular}{| c ||c | c | c | c | c|} 
 \hline
  & $\chi^2_{min}$ & AIC & BIC & $ \Delta $AIC & $ \Delta $BIC \\ [0.5ex] 
 \hline\hline
  MODEL& Hybrid \hspace{5mm} $ \Lambda $CDM & Hybrid \hspace{5mm} $ \Lambda $CDM & Hybrid \hspace{5mm} $ \Lambda $CDM & & \\ [0.5ex]
 \hline

  CC & 33.15 \hspace{5mm} 32.132 & 41.15\hspace{6mm} 40.132 & 46.88\hspace{5mm} 45.86 & 1.02&1.02  \\ 
  CC+BAO & 71.42 \hspace{6mm} 69.8 & 79.42 \hspace{6mm} 77.8 & 85.15 \hspace{5mm} 83.53 & 1.62 &1.62 \\
  CC+GRB & 458.07 \hspace{4.5mm} 460.84 &466.07 \hspace{5mm} 468.84&479.12 \hspace{5mm}  481.88& 2.76 & 2.76 \\
  Pantheon+SH0ES & 1615.26 \hspace{4.5mm} 1609.91 &  1623.26\hspace{5mm} 1617.91& 1645.01\hspace{5mm} 1639.66 & 5.35 &5.35\\
 \hline
  \hline
 MODEL & LSR \hspace{1.5mm} $ \Lambda $CDM & LSR \hspace{1.5mm} $ \Lambda $CDM & LSR \hspace{1.5mm} $ \Lambda $CDM & & \\ [0.5ex]
 \hline
  CC & 32.93 \hspace{5mm} 32.132 & 40.93\hspace{6mm} 40.132 & 46.66\hspace{6mm} 45.86 & 0.79 & 0.79 \\ 
  CC+BAO & 70.01 \hspace{7mm} 69.8 &78.01  \hspace{7mm} 77.8 & 86.18 \hspace{5mm} 85.97 & 0.21 & 0.21 \\
  CC+GRB & 459.92 \hspace{4.5mm} 460.84 &467.92  \hspace{5mm} 468.84&480.97 \hspace{5mm}  481.88& 0.92 & 0.92 \\
  Pantheon+SH0ES & 1610.71  \hspace{4.5mm} 1609.91 & 1618.71 \hspace{5mm} 1617.91& 1640.46\hspace{5mm} 1639.66 & 0.8 & 0.8\\
 \hline
\end{tabular}
\caption{Statistical comparison of our results with $ \Lambda $CDM model.}
\label{chap2/table3}
\end{table}
Before proceeding to the discussion on tensions, we would like to examine how favorable our models are to the concerned data. To this end, the AIC and BIC are employed as statistical tools for model selection. Their definitions and interpretative ranges have been introduced in detail in \autoref{sec:modelselection}. In the present analysis, we evaluate the relative evidence for or against each model using those established criteria.  It can be observed in \autoref{chap2/table3} that for the Hybrid model, the GRB and SH0ES cases, the models are moderately favored, while the other two cases are strongly favored. All cases of the LSR model are compatible with the respective data.

\subsubsection{Addressing Hubble and \texorpdfstring{$S_8$}{S8} Tensions}

\begin{figure}
    \centering
    \includegraphics[width=0.48\linewidth]{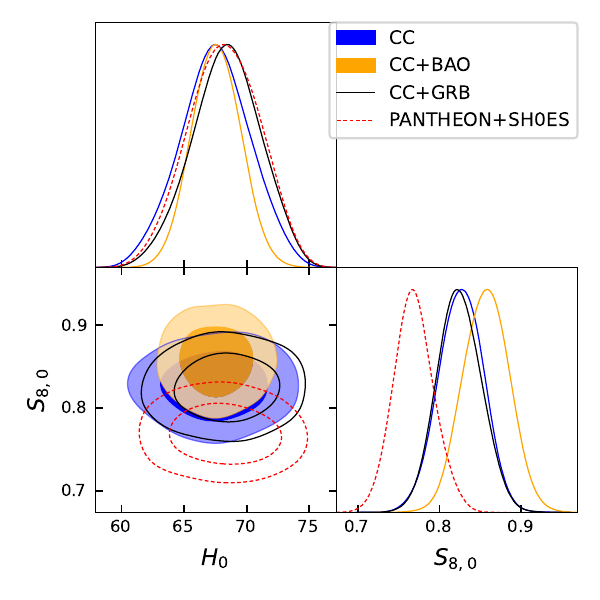}
    \includegraphics[width=0.48\linewidth]{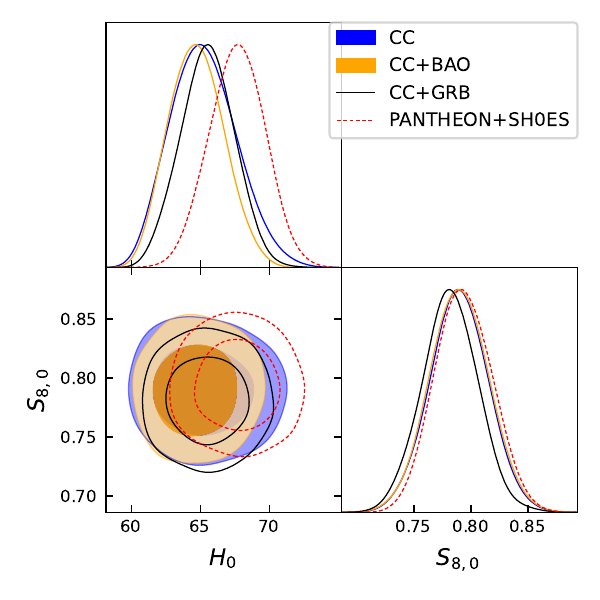}
    \caption{2D contours and Likelihood probability distributions using RSD data for Hybrid model (left panel) and LSR model (right panel).}
    \label{chap2.1/fig:contours8}
\end{figure}

\begin{figure}
    \centering
    \includegraphics[width=0.49\linewidth]{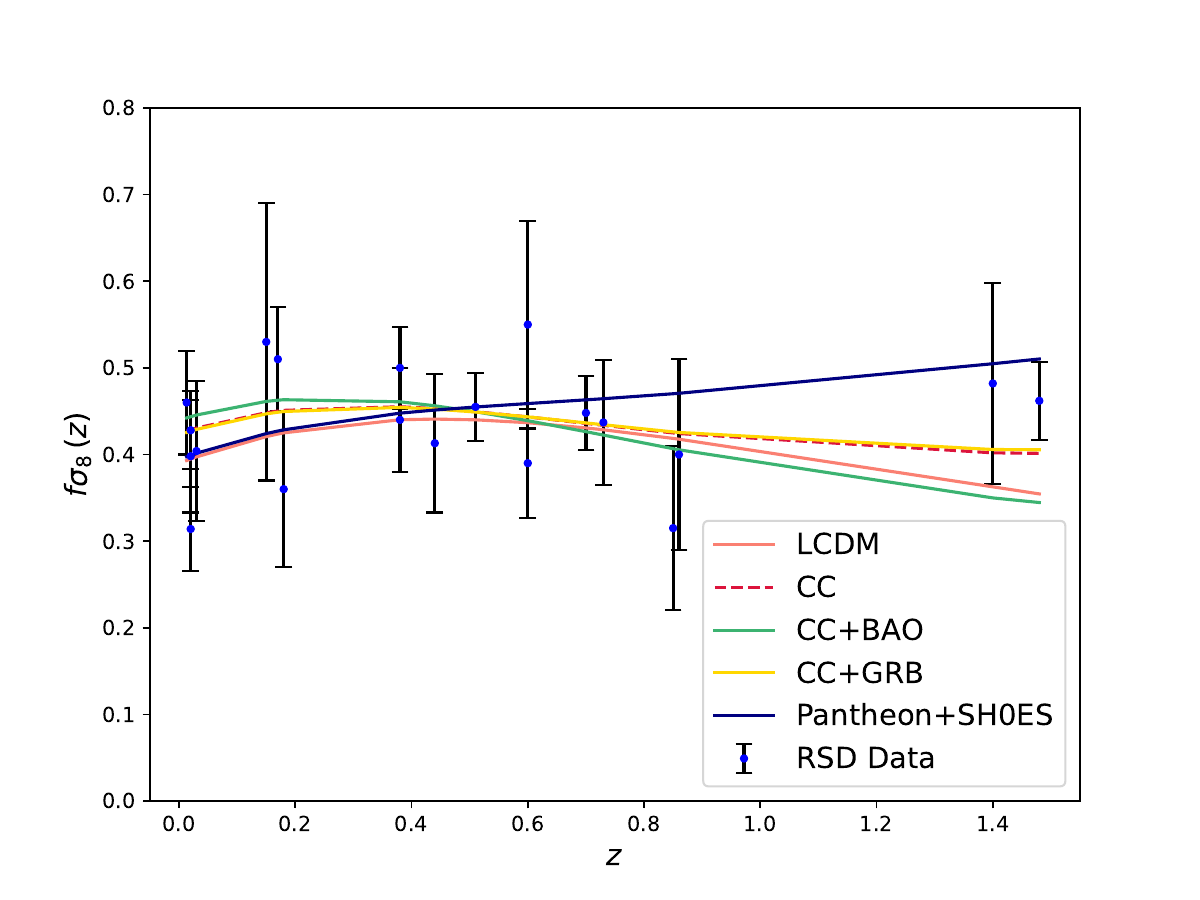}
    \includegraphics[width=0.49\linewidth]{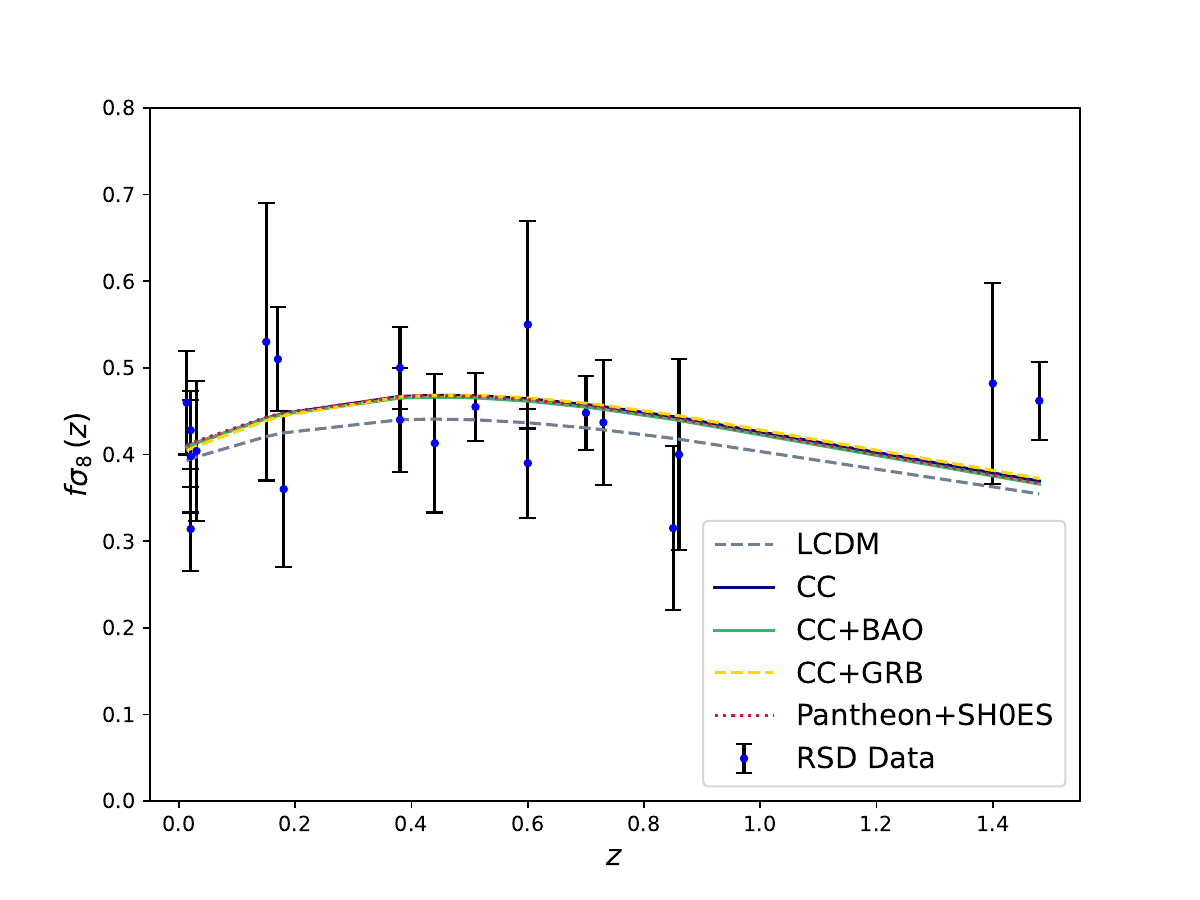}
    \caption{$f \sigma_8$ profile of the Hybrid model (left panel) and LSR model (right panel)  
    against the error bars of RSD data.}
    \label{chap2.1/fig:fs8}
\end{figure}

\begin{figure}
    \centering
    \includegraphics[width=0.85\linewidth]{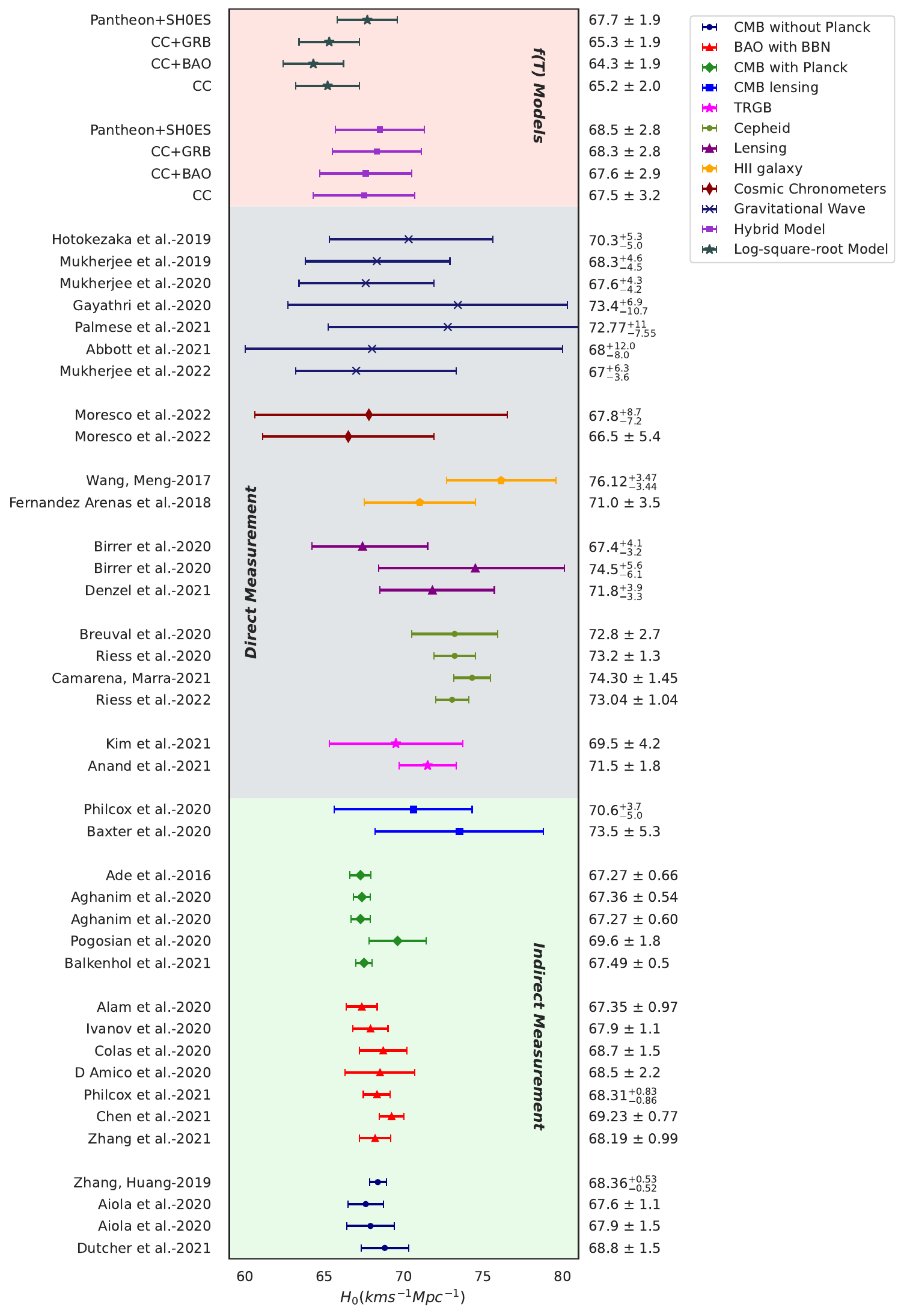}

    \caption{Comparison of the $f(T)$ models with the $H_0$ value from observational estimations}
    \label{fig:compH0}
\end{figure}

\begin{figure}
    \centering
    \includegraphics[width=1\linewidth]{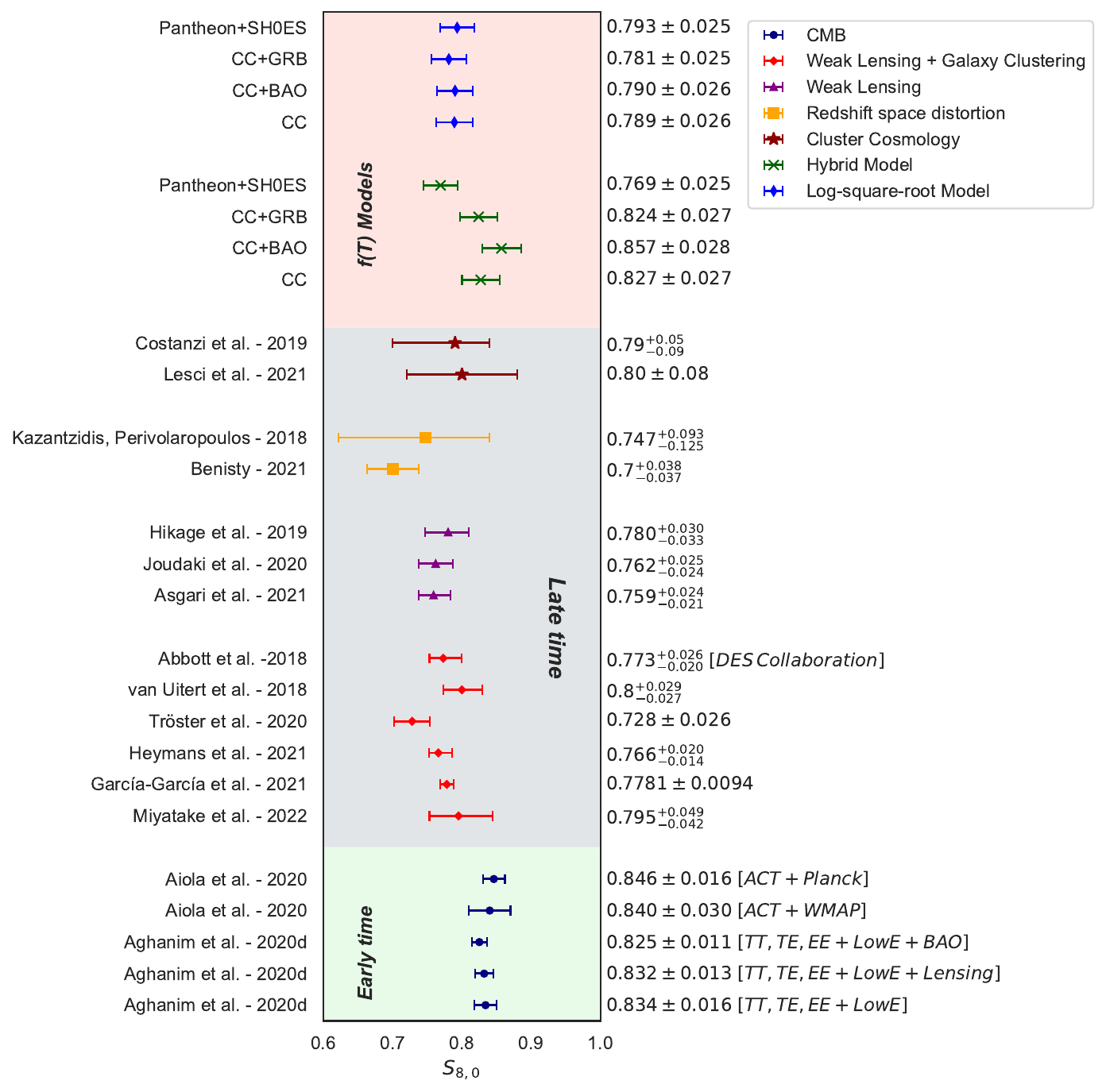}
    \caption{Comparison of the $f(T)$ models with the $S_8$ value from observational estimations}
    \label{fig:comps8}
\end{figure}

Using RSD data, we constrained the parameters \(S_{8,0}\) (representing \(S_8(z=0)\)) and \(H_0\). For this purpose, the values of the remaining model parameters were fixed based on the results of the MCMC analysis, as presented in \autoref{chap2/table1} and \ref{chap2/table2} for both models. The corresponding 2D likelihood contours are shown in \autoref{chap2.1/fig:contours8}. For the Hybrid model, the values of \(S_{8,0}\) obtained are $0.827 \pm 0.027$, $0.857 \pm 0.028$, $0.824 \pm 0.027$, and $0.769 \pm 0.025$, respectively, for the model parameter values derived from the CC, CC+BAO, CC+GRB, and Pantheon+SH0ES data combinations. Similarly, for the second model (i.e., the LSR model), the corresponding values are $0.789 \pm 0.026$, $0.790 \pm 0.026$, $0.781 \pm 0.025$, and $0.793 \pm 0.025$. For the obtained mean values the profile of $f\sigma_8(z)$ is plotted against the RSD data and is depicted in \autoref{chap2.1/fig:fs8}.
It is noticed that the LSR model behaves similarly for each combination, while for the Hybrid model, the PANTHEON+SH0ES curve slightly deviates. However, in all cases, the curves suitably match with the  RSD data.

\begin{figure}
    \centering
     \includegraphics[width=0.8\linewidth]{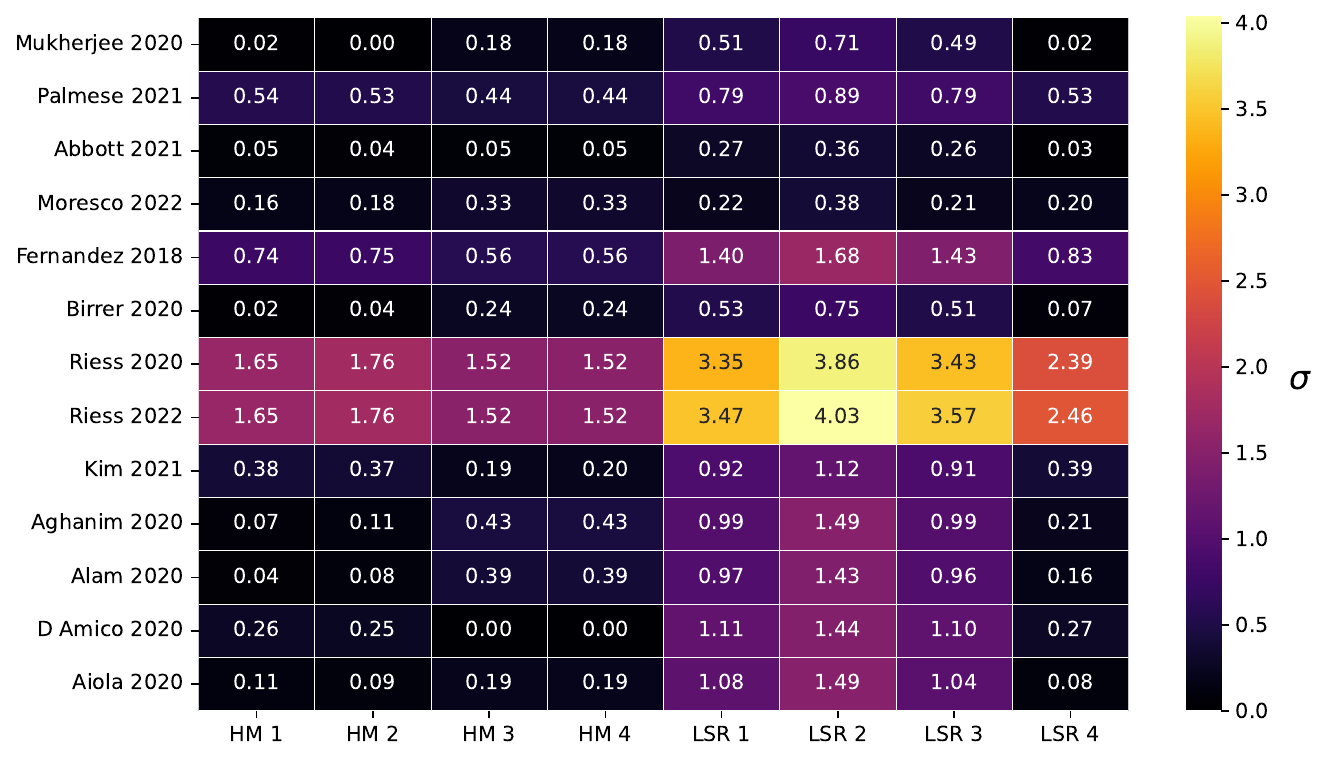}\\
      \includegraphics[width=0.8\linewidth]{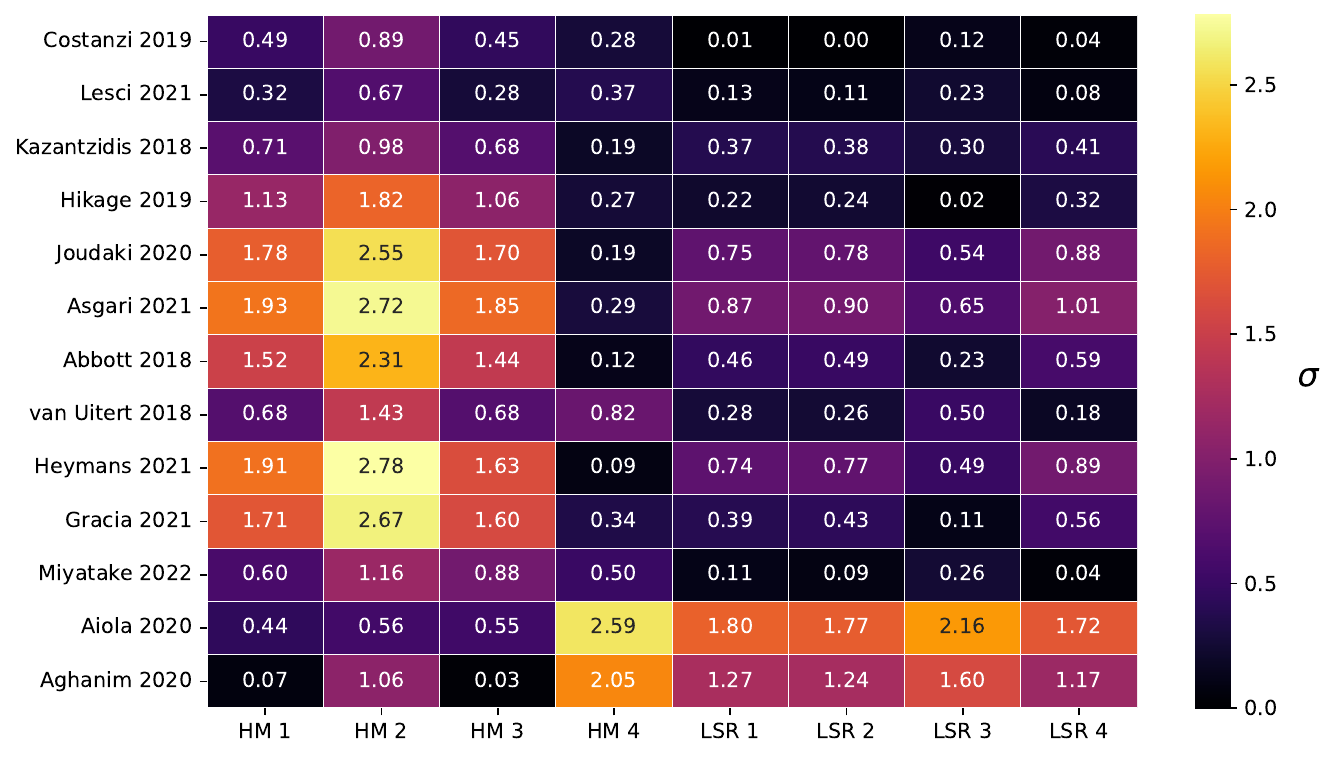}
    \caption{Heat map of tension metrics for theoretical models compared with observational surveys. Here, HM and LSR represent the Hybrid and Log-square-root models, respectively, and the indices from 1 to 4 correspond to the dataset combinations: CC, CC+BAO, CC+GRB, and Pantheon+SH0ES. The top panel represents the $H_0$ tension analysis and the bottom panel represents that of $S_8$. }
    \label{fig:heat}
\end{figure}

 The results from our theoretical models are compared with some prominent recent surveys in \autoref{fig:compH0} and \autoref{fig:comps8}. From the obtained results, it is evident that the Hybrid model significantly alleviates the $H_0$ tension. When compared with the various observational surveys reported in the literature, as illustrated in \autoref{fig:compH0}, the model yields consistency at the $0$--$2\sigma$ level across datasets. This indicates a substantial reduction in the statistical significance of the discrepancy relative to the standard $\Lambda$CDM model, primarily due to the modified expansion history and the broader allowed parameter space within the Hybrid $f(T)$ framework. On the other hand, the LSR model shows a \(3\)-\(4\sigma\) tension for the Cepheid data, but for all other surveys, it also achieves a \(0\)-\(2\sigma\) tension. Although some tension persists, the level of discrepancy is notably smaller than the current tensions observed between late time and early-time measurements. Interestingly, when addressing the \(S_8\) tension, the LSR model effectively reduces discrepancies across all the surveys considered, as shown in \autoref{fig:comps8}, with all tensions falling within the \(0\)-\(2\sigma\) range. Similarly, the Hybrid model performs well, particularly for the CC+GRB and CC data combinations, showing good agreement with all observations and alleviating tension. However, for the CC+BAO and Pantheon+SH0ES combinations, mild tension is observed with Weak Lensing and CMB data, respectively. Nevertheless, for other surveys, the tension metrics remain within the desired range. Furthermore, we have schematically represented the heat maps of the tension metric in \autoref{fig:heat}. The lighter the shade, the higher the tension. 

\section{Tensions in the \texorpdfstring{$f(T,\mathcal{T})$}{f(T,T)} framework}\label{chap2/tensionfTT}

The methodology employed to address the cosmological tensions in the previous section is now extended to a more unified and streamlined framework. 
In this part of the analysis, we combine the DESI and RSD datasets in all possible configurations, enabling a simultaneous study of the parameters $r_d$ and $S_8$ along with the other cosmological quantities. Unlike the previous analysis, the present work employs the updated CC dataset with 34 measurements, incorporating the full covariance matrix (more details can be found in the Appendices). 
The underlying theoretical framework considered here is an extension of the $f(T)$ gravity theory, where the matter sector couples to the geometry through the trace of the energy--momentum tensor. 
In particular, we focus on the Hybrid $f(T,\mathcal{T})$ model\footnote{
The Hybrid model is defined by $f(T,\mathcal{T})=\frac{a T_0^2}{T}+b \mathcal{T}$.} \eqref{mod;hybfTT}, which was introduced in Chapter~\ref{ch:intro}.

\subsection{Results}\label{chap2.2/results}
\begin{figure}
    \centering
    \includegraphics[scale=0.6]{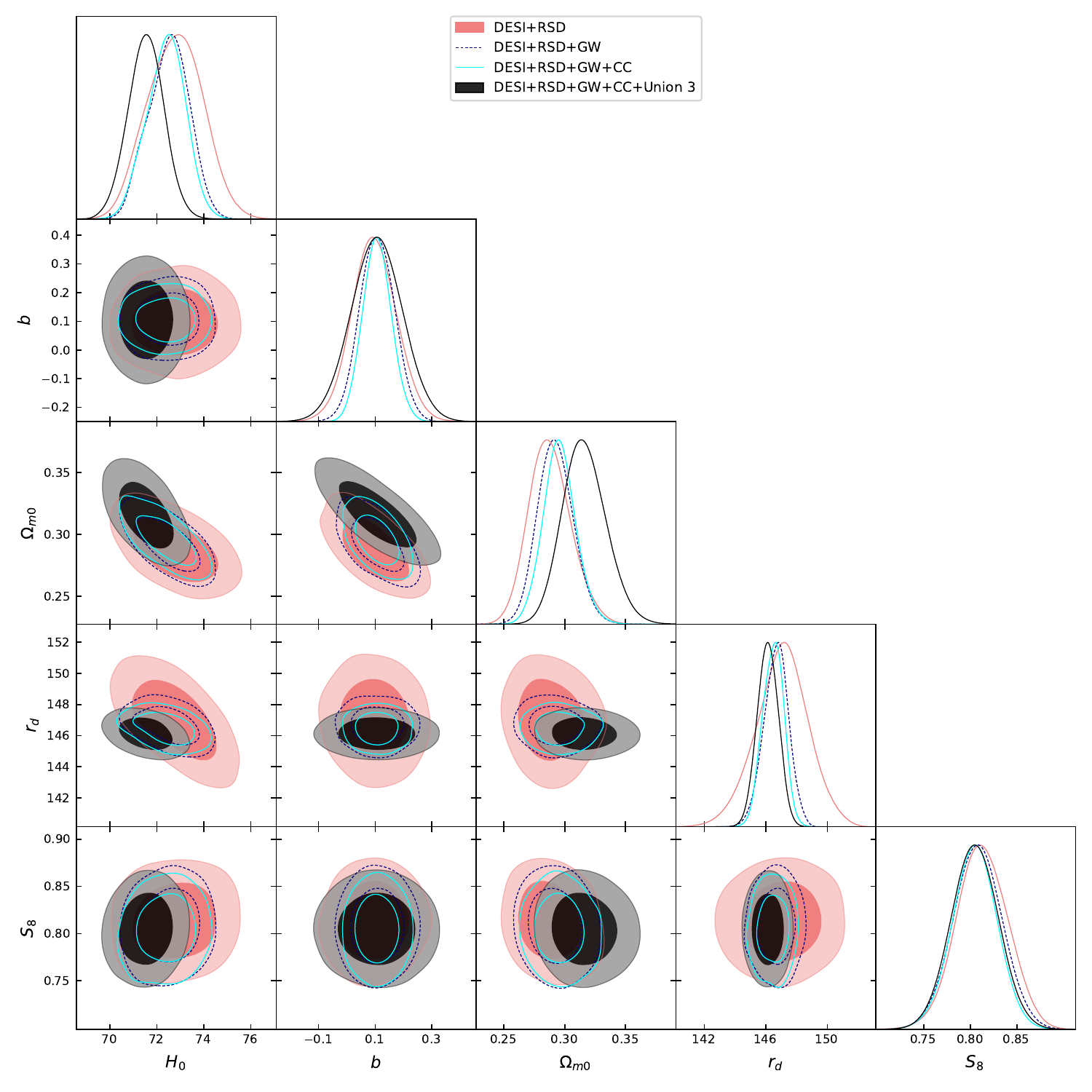}
    \caption{2D contours of the cosmological parameters upto $2 \sigma$ confidence level.}
    \label{fig:mcmc}
\end{figure}

The extended teleparallel theory has been constrained using MCMC with a different set of combinations of local and early Universe datasets. The Hubble expression \eqref{eq:hub} plays the key role in finding the theoretical predictions in each case. Though some datasets do not directly predict the observational value of Hubble, the theoretical counterpart of the cosmological quantities they predict can be easily achieved using Eq.~(\ref{eq:hub}). The two-dimensional contours up to $2 \sigma$ confidence level are presented in \autoref{fig:mcmc}. As the cosmological parameters have already been observed through various methods in the literature, it is easy to compare the results. In that context, our aim is to explore the long-standing tensions. The precise numerical values up to $1\sigma$ can be found in \autoref{chap2.2/table1}. The negative correlation between $H_0$ and $\Omega_{m0}$ indicates that as the Hubble value increases, the matter density parameter decreases. In fact, our model parameter $b$ shows a similar kind of relation with $\Omega_{m0}$.

\begin{figure}
    \centering
    \begin{minipage}{0.5\textwidth}
        \centering    \includegraphics[width=\linewidth]{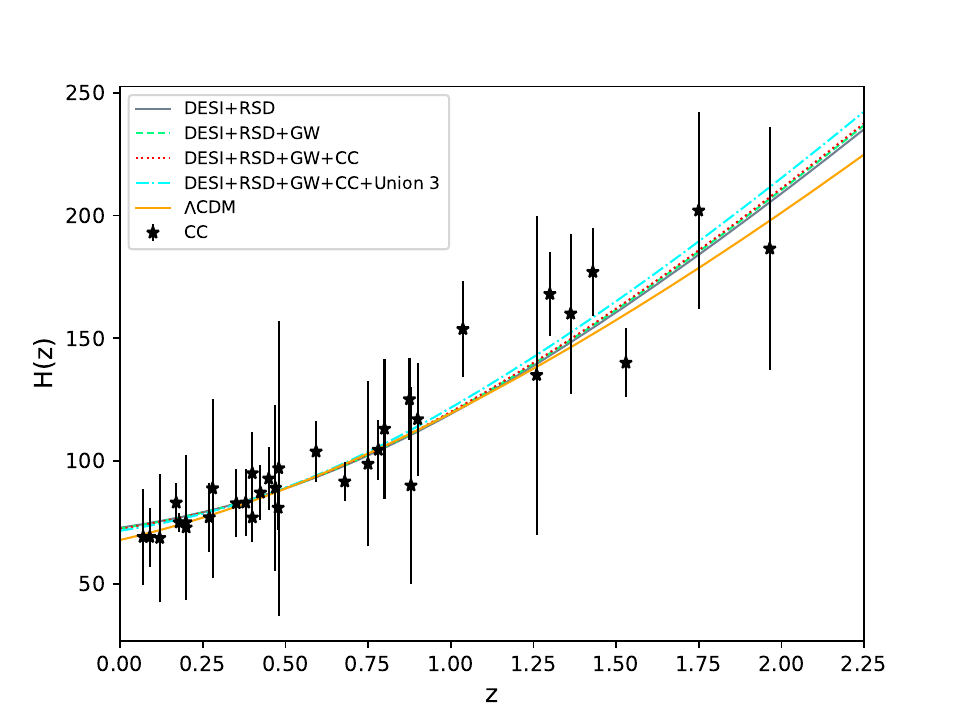}
    \caption{Hubble parameter against redshift with 34 error bars of CC.}
        \label{fig:Hub}   \end{minipage}\hfill
    \begin{minipage}{0.5\textwidth}
        \centering
        \includegraphics[width=\linewidth]{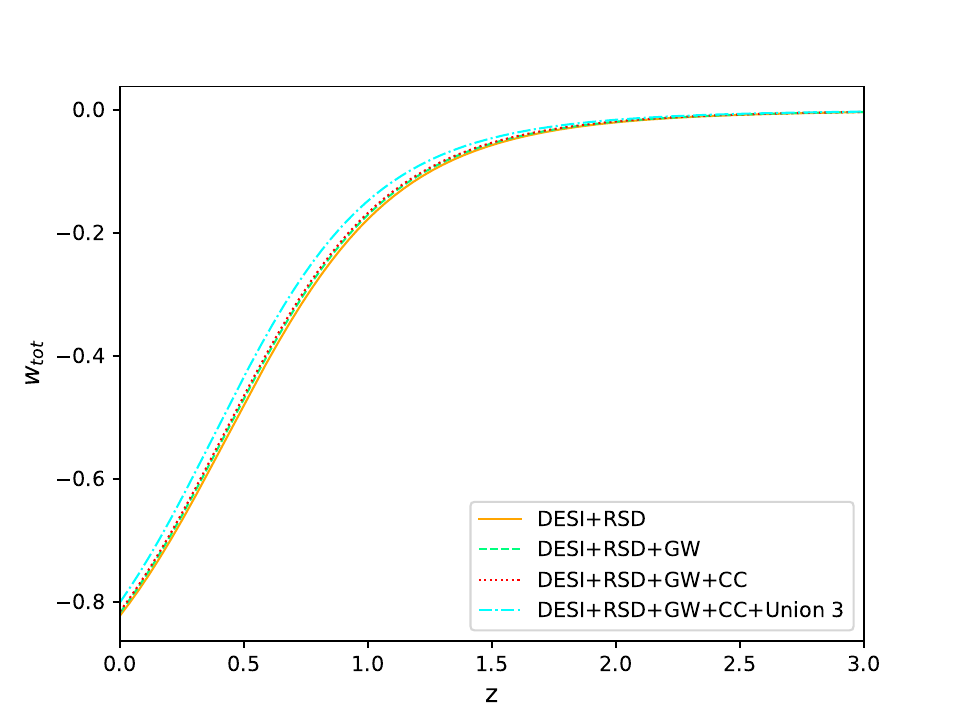}
        \caption{Total Equation of State parameter vs redshift.}
        \label{chap2.2/fig:eos}
    \end{minipage}
\end{figure}

\begin{figure}
    \centering
\includegraphics[width=0.55\textwidth]{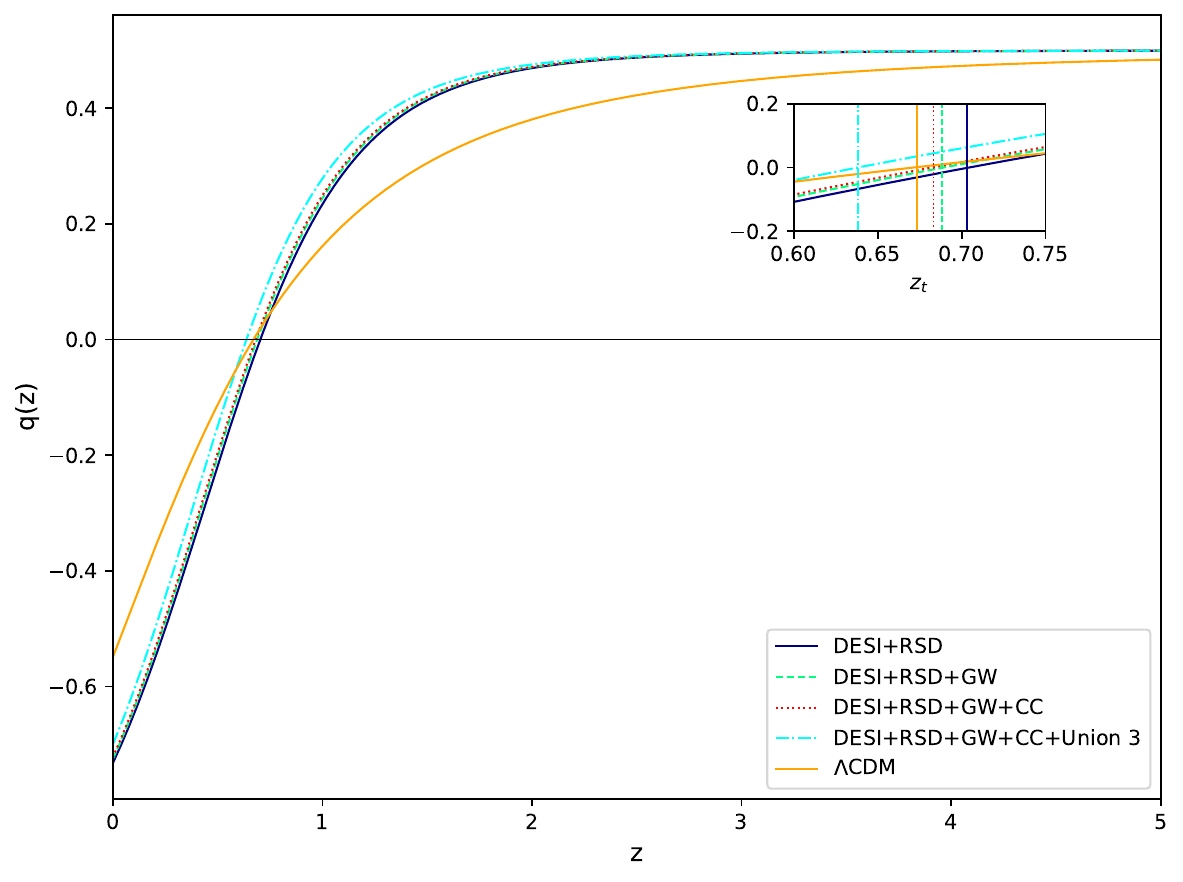}
    \caption{Behaviour of the deceleration parameter in pre- and post-transitional era with a magnified region of phase transition.}
    \label{fig:dec}
\end{figure}

Further, the resulting values are examined against the datasets by plotting the cosmological entities. In all of those analyses, we have also included the famous $\Lambda$CDM model for comparison purposes. In \autoref{fig:Hub}, one can see that the Hubble parameter for all combinations fits the 34 error bars of the CC data and the standard cosmological model. The deceleration parameters in \autoref{fig:dec} show a transition of the Universe from deceleration to the acceleration phase. The magnified region provides a clear image of the transition point ($z_t$) for each case. The present values of the deceleration parameter and the transition redshifts are summarized in \autoref{chap2.2/table2}. As for dark energy, the $w$ value is constant throughout (i.e., $-1$), we have just plotted the cases of our model in \autoref{chap2.2/fig:eos}. The value $-1<w_{tot}<-1/3$ indicates a quintessence-like behaviour, while $w_{tot}<-1$ indicates phantom-like behaviour \footnote{$w_{tot}$ is same as $w_{eff}$, $w_{tot}=\frac{p_{tot}}{\rho_{tot}}=\frac{p+p_{DE}}{\rho+\rho_{DE}}$.}. The present values of our models lie in the quintessence region (see \autoref{chap2.2/table2}), confirming the late-time acceleration. For our model, the deceleration and EoS parameters are calculated using Eq.~(\ref{eq:dec}) and Eq.~(\ref{eq:eos}), respectively.

\begin{figure}
    \centering
    \begin{minipage}{0.5\textwidth}
        \centering    \includegraphics[width=\linewidth]{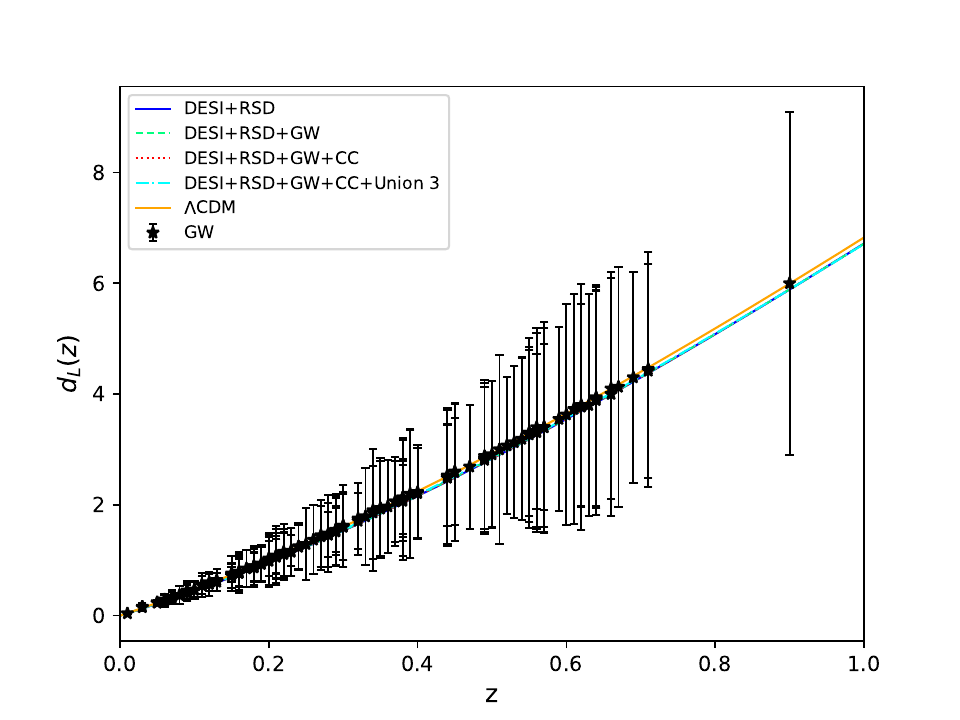}
    \caption{Luminosity distance profile against redshift with GW data.}
        \label{fig:dl}   \end{minipage}\hfill
    \begin{minipage}{0.5\textwidth}
        \centering
        \includegraphics[width=\linewidth]{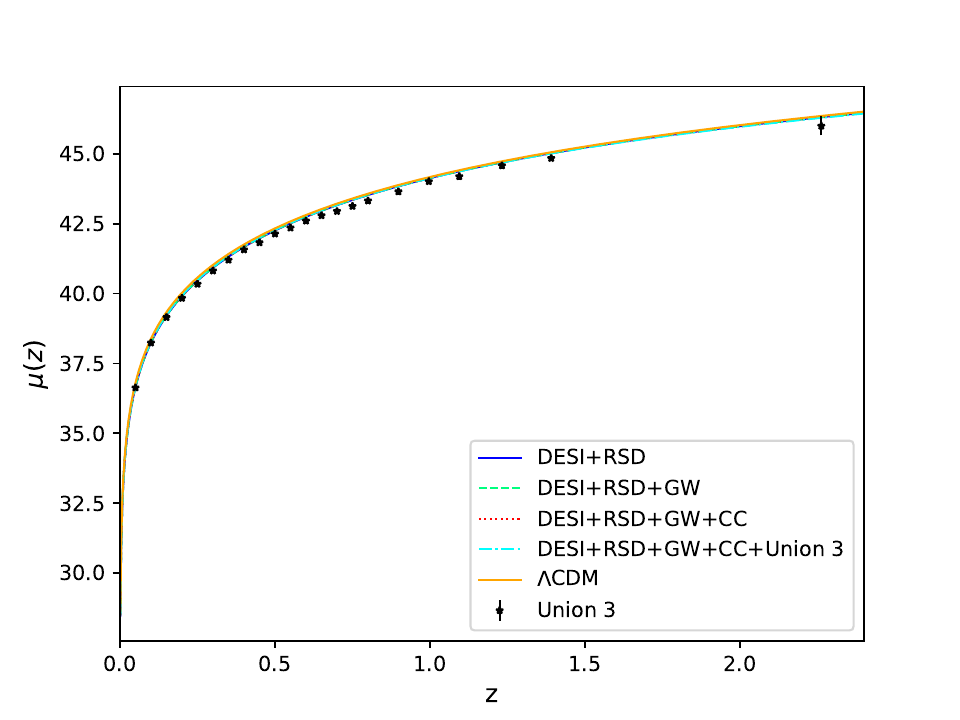}
        \caption{Distance modulus profile against redshift with Union3 data.}
        \label{chap2.2/fig:mu}
    \end{minipage}
\end{figure}

\begin{figure}
    \centering
    \includegraphics[scale=0.6]{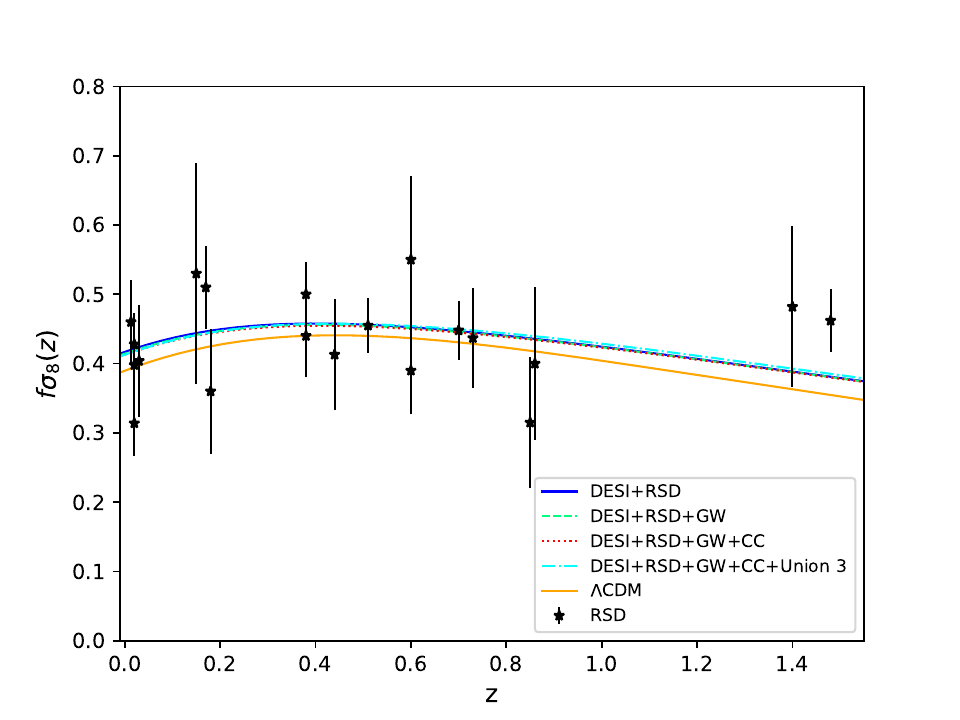}
    \caption{Behaviour of $f\sigma_8$ against redshift with error bars of RSD data.}
    \label{fig:fs8}
\end{figure}

\begin{figure}
    \centering
    \includegraphics[scale=0.48]{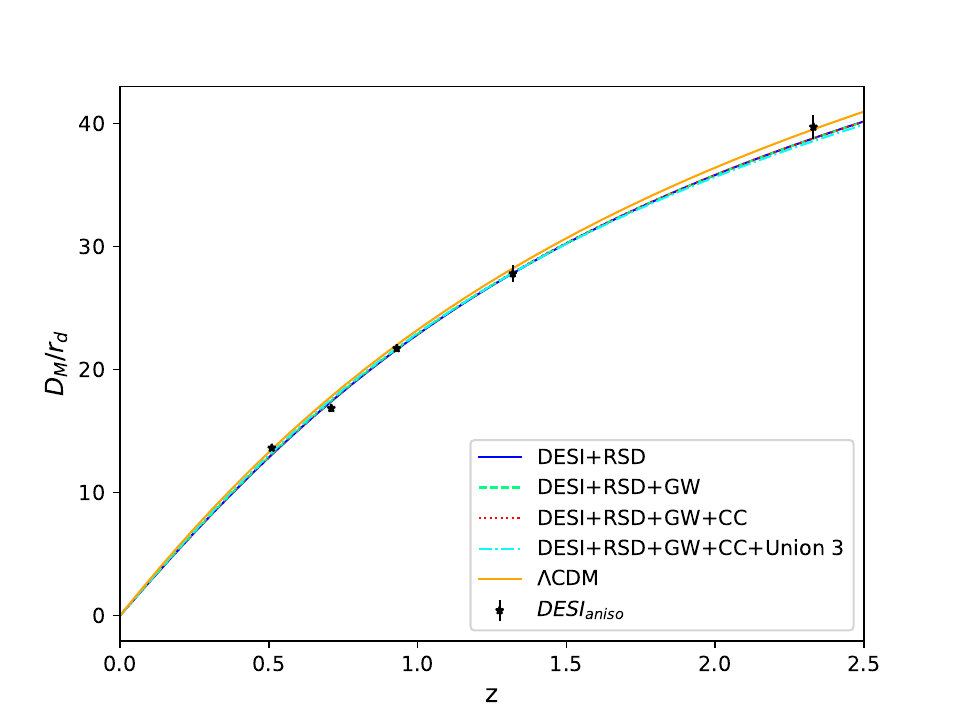}
    \includegraphics[scale=0.48]{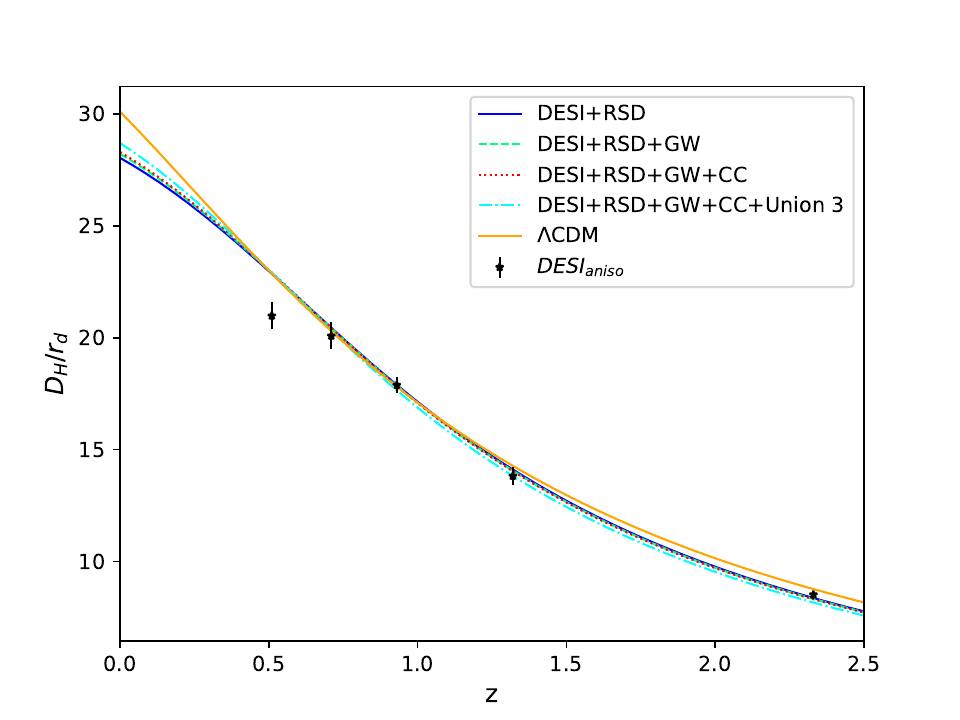}
    \includegraphics[scale=0.5]{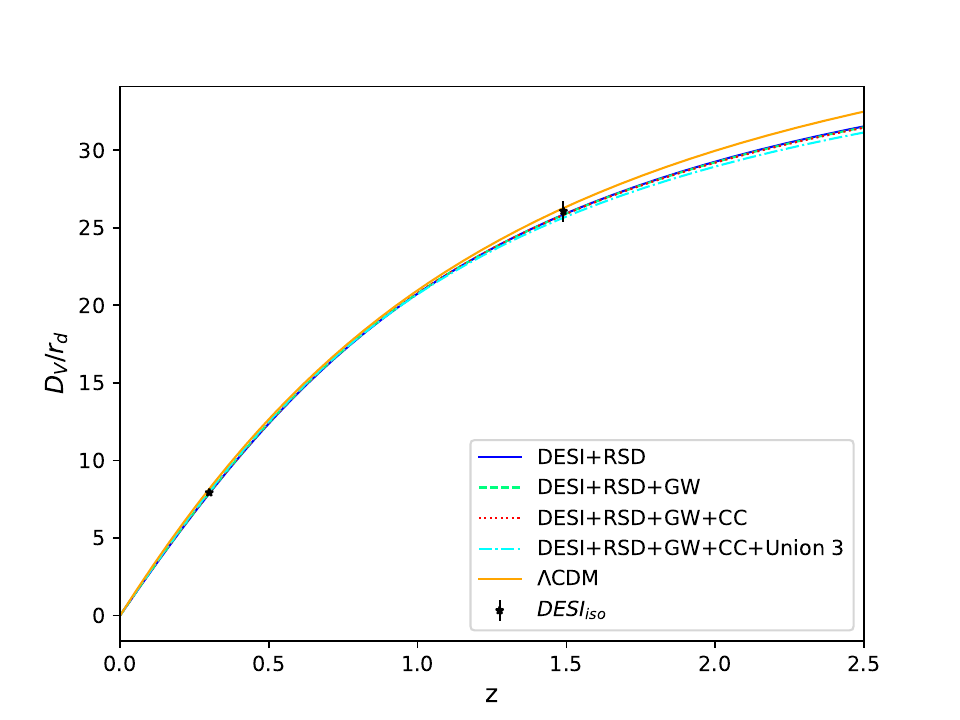}
    \caption{The distance measuring functions against redshift with DESI data.}
    \label{fig:Desi}
\end{figure}

Since GW observations provide measurements of the luminosity distance, we compare the theoretical predictions with the observational data, including error bars, as shown in \autoref{fig:dl}. The models demonstrate a good fit to the data, comparable to that of the $\Lambda$CDM model. A similar analysis is carried out for the Union3 supernova dataset, where we compare the theoretical distance modulus with the observed values, incorporating the associated error bars. The results indicate that the models considered offer excellent agreement with the Union3 data (see \autoref{chap2.2/fig:mu}), reinforcing their consistency with standard cosmological observations. In \autoref{fig:fs8}, we examine the behavior of the cases against the RSD data through $f_{\sigma_8}$. When plotted alongside the observational measurements and their uncertainties, the models show strong compatibility, effectively capturing the growth rate of cosmic structures across the redshift range. Finally, the distance measuring functions involved in DESI BAO are compared. The anisotropic and isotropic measurements are presented separately in \autoref{fig:Desi}. A similar trend to $\Lambda$CDM can be observed in all scenarios. In order to assess the performance of the model against the concordance $\Lambda$CDM scenario, we have computed the minimum $\chi^2$ values for all dataset combinations, as summarized in \autoref{chap2.2/table3}. 
The differences $\Delta\chi^2$ are found to be small (typically $\mathcal{O}(1)$), indicating that the proposed $f(T,\mathcal{T})$ extension achieves a fit to current data that is statistically comparable to $\Lambda$CDM, thereby showing that it does not deteriorate the concordance with current observations.

Having established the model's credibility through various observational comparisons in this section, we now turn our attention to the discussion of key cosmological tensions.

\begin{table}
 \centering
  \scriptsize
 \caption{Summary of the $1\sigma$ results obtained from the MCMC for the $f(T,\mathcal{T})$ model.}
 \label{chap2.2/table1}
    \begin{tabular}{c|c|c|c|c|c|}
    
 \cline{1-6}
\multicolumn{1}{|c|}  {\it{Dataset}} & {$H_0\,\ (km \,\ s^{-1} Mpc^{-1})$}  & $b$ & $\Omega_{m0}$ & $r_d$ & $S_8$ \\ \hline \hline
\multicolumn{1}{ |c| }{DESI+RSD} & $72.8 \pm 1.2$ & $0.097 \pm 0.079$ & $0.288^{+ 0.015}_{- 0.019}$ & $147 \pm 1.7$ & $0.813 \pm 0.027$   \\ 
\multicolumn{1}{ |c| }{DESI+RSD+GW} & $72.51^{+0.95}_{-0.82} $ & $0.107 \pm 0.061$ & $0.293^{+ 0.013}_{- 0.015}$ & $146.66 \pm 0.81$  & $0.808 \pm 0.026$\\ 
\multicolumn{1}{ |c| }{DESI+RSD+GW+CC} & $72.39^{+0.89}_{-0.76}$ & $0.107 \pm 0.051$ & $0.296 \pm 0.014$ & $146.47 \pm 0.68$  & $0.805 \pm 0.024$\\ 
\multicolumn{1}{ |c| }{DESI+RSD+GW+CC+Union3} & $71.54 \pm 0.76$ & $0.105 \pm 0.091$ & $0.316^{+ 0.016}_{- 0.019}$ & $146.13 \pm 0.69$  & $0.805 \pm 0.025$\\    \hline

    \end{tabular}
\end{table}

\begin{figure}
    \centering
    \includegraphics[width=0.75\linewidth]{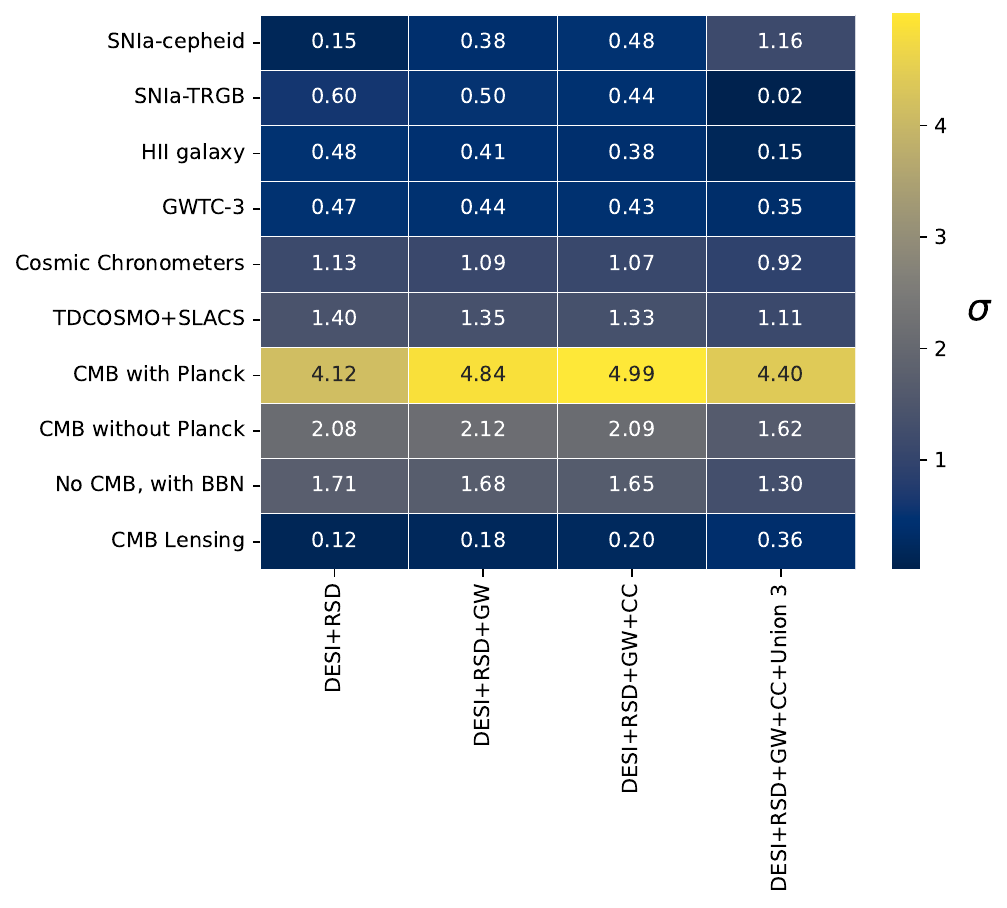}
    \caption{Heat map of the $H_0$ tension between our combinations and various measurements.}
    \label{fig:heatH0}
\end{figure}

\begin{figure}
    \centering
    \includegraphics[width=0.48\linewidth]{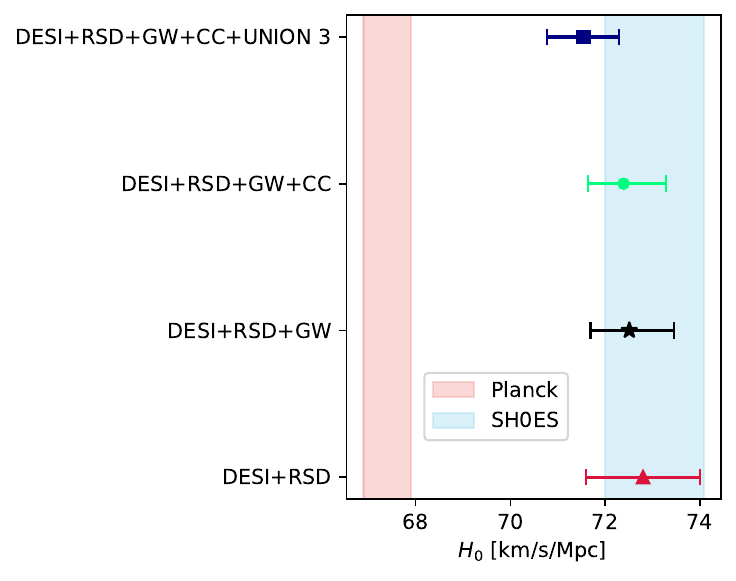}
  \includegraphics[width=0.48\linewidth]{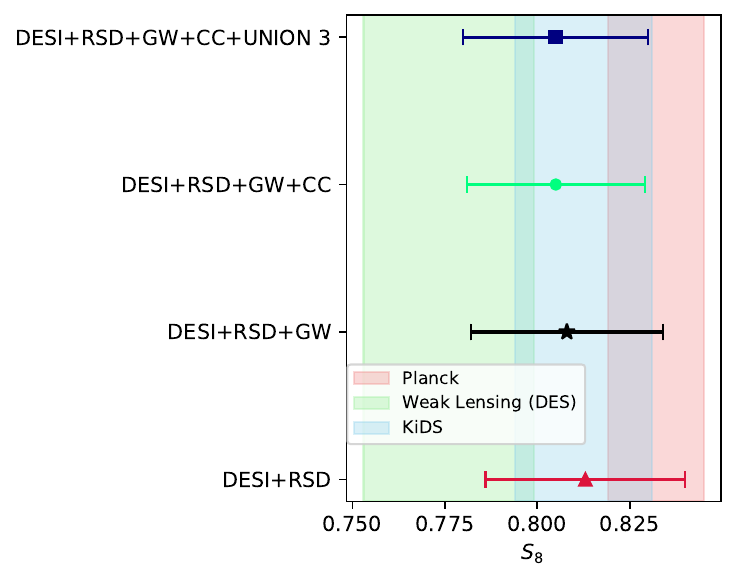}
    \caption{Comparison of our results with the landmark early and late-time surveys.}
    \label{fig:comp}
\end{figure}

\begin{table*}
 \centering
 
 \caption{Present values of the cosmological parameters.}
 \label{chap2.2/table2}
    \begin{tabular}{c|c|c|c|}
    
 \cline{1-4}
\multicolumn{1}{|c|}  {\it{Combinations}} & $q_0$  & $z_t$ & $w_{tot_0}$  \\ \hline \hline
\multicolumn{1}{ |c| }{DESI+RSD} & -0.73 & 0.70& -0.82 \\ 
\multicolumn{1}{ |c| }{DESI+RSD+GW} & -0.72&0.68 & -0.81\\ 
\multicolumn{1}{ |c| }{DESI+RSD+GW+CC} & -0.72& 0.68& -0.81\\ 
\multicolumn{1}{ |c| }{DESI+RSD+GW+CC+Union3} &-0.70 & 0.63& -0.80\\
\multicolumn{1}{ |c| }{$\Lambda$CDM} & -0.55& 0.67 & -1\\\hline

    \end{tabular}
\end{table*}
\begin{table*}
 \centering
 
 \caption{Comparison of the minimum $\chi^2$ with the reference $\Lambda$CDM model.}
 \label{chap2.2/table3}
    \begin{tabular}{c|c|c|c|}
    
 \cline{1-4}
\multicolumn{1}{|c|}  {\it{Combinations}} & $\chi^2_{min}$(Model)  & $\chi^2_{min}$($\Lambda$CDM) & $\Delta \chi^2$  \\ \hline \hline
\multicolumn{1}{ |c| }{DESI+RSD} & 33.27 & 32.06 & 1.21  \\ 
\multicolumn{1}{ |c| }{DESI+RSD+GW} & 36.24 & 35.71& 0.53 \\ 
\multicolumn{1}{ |c| }{DESI+RSD+GW+CC} & 52.86 & 50.92& 1.94\\ 
\multicolumn{1}{ |c| }{DESI+RSD+GW+CC+Union3} & 110.63 & 110.25&0.38 \\\hline

    \end{tabular}
\end{table*}
  The tension metric is used to calculate the deviation of our $H_0$ values from the prominent reviewed results in \autoref{chap2/tensions}. The \autoref{fig:heatH0} presents a heat map of tension ($\sigma$) between our model with those surveys. One can observe that for most of the observations, our model persists $0-2\sigma$ tension, which indicates strong evidence in favour of the model. Though in the CMB with Planck, the tension is a little higher, it is reduced compared to the tension between Planck and SH0ES. In the first panel of \autoref{fig:comp}, our results are compared against the Planck and SH0ES collaborations.

\begin{figure}
    \centering
    \includegraphics[width=0.7\linewidth]{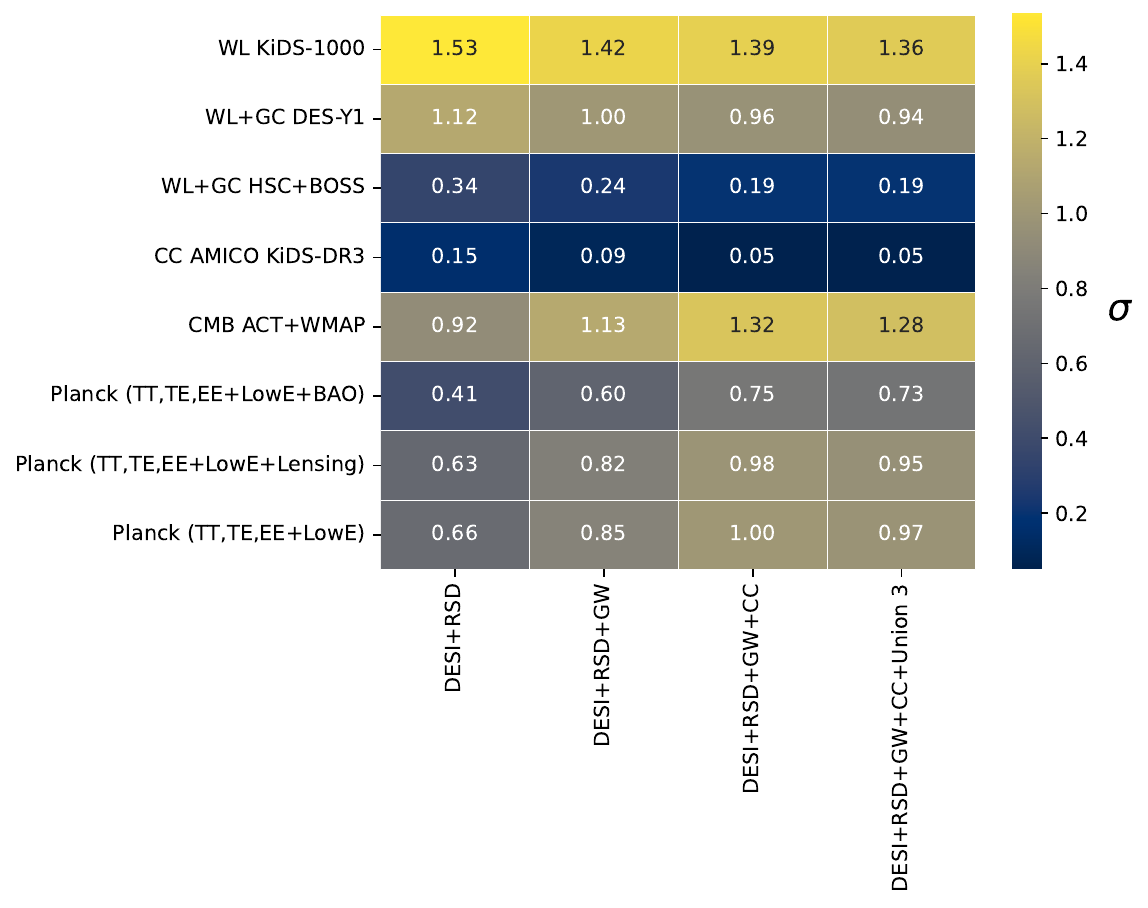}
    \caption{Heat map of the $S_8$ tension between our combinations and various measurements.}
    \label{fig:heatS8}
\end{figure}

  The \autoref{fig:heatS8} represents a heat map of $S_8$ tension of our model with the aforementioned results. It is evident that our model has excellent agreement with both the early and late-time surveys. The maximum observed tension is $1.5\sigma$, while in the majority of cases it remains within $1\sigma$.\ Also, the second panel of \autoref{fig:comp} demonstrates a clear visualization of the model's efficacy in alleviating the $S_8$ tension.

\begin{figure}
    \centering
    \includegraphics[width=0.5\linewidth]{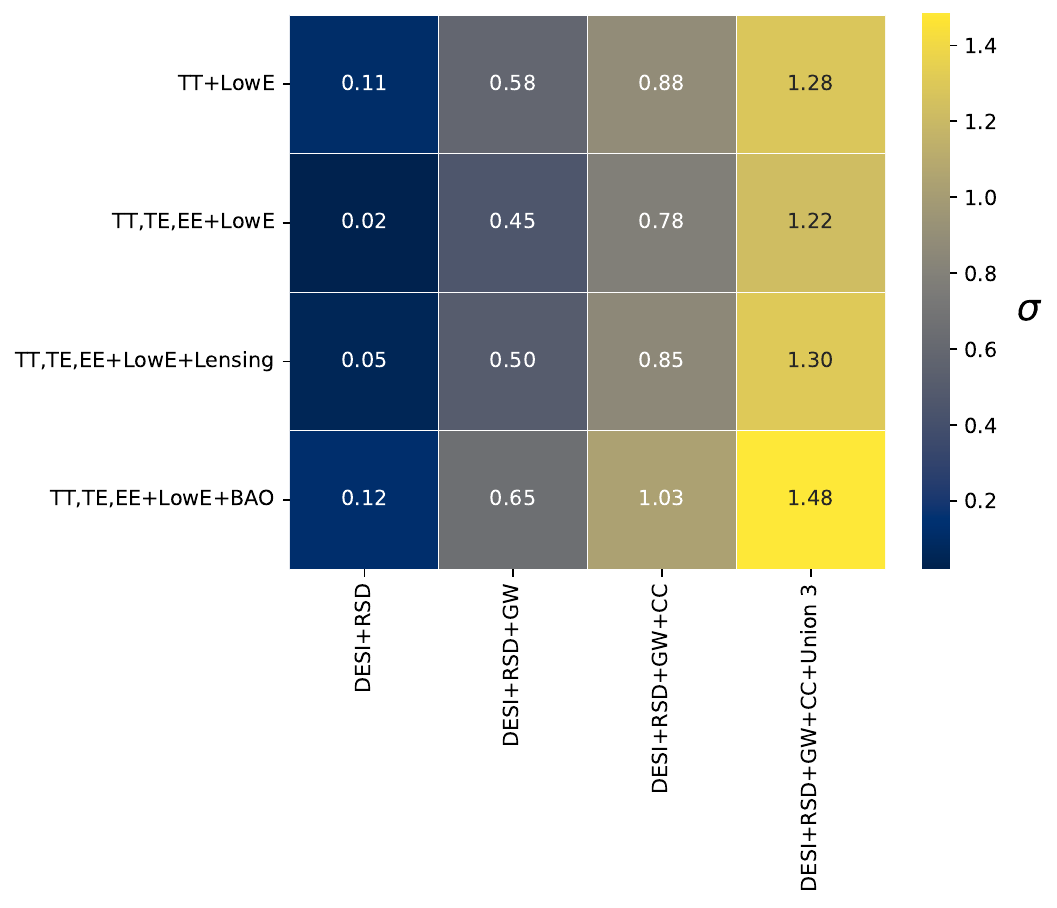}
    \caption{Heat map of the $r_d$ tension between our combinations and Planck measurements.}
    \label{fig:heatrd}
\end{figure}

  Finally, we explore the sound horizon at the drag epoch ($r_d$). It is well known that $H_0$ is negatively correlated with $r_d$. Even in our contours (\autoref{fig:mcmc}), this behaviour is observed. This is the reason, the late time surveys with high $H_0$ predict a lower $r_d$ compared to Planck. Particularly when the BAO surveys are combined with $H_0$-based measurements like H0LiCOW, SH0ES, and SNIa calibrations, they predict $r_d$ around 135-140 Mpc. However, despite having higher $H_0$ values than the Planck results, our models produce suitable sound horizon values. To get a clear overview of this, we have compared our results with the four cases of the Planck results through a heat map in \autoref{fig:heatrd}. For all the combinations, the tension is within $1.5 \sigma$, which confirms the model's excellence in resolving the sound horizon problem.

\section{Conclusion}\label{chap2/conclusion}
Since the latest inclusions of more accurate observations still have tensions with the $\Lambda$CDM cosmology, it led the researchers to probe for an ultimate theory that alleviates the tensions. Several attempts have been made in this context so far in the literature \cite{Scherer:2025esj, Gomez-Valent:2024tdb}. In the former part of this chapter, we adopted the theoretical model-dependent approach by considering two $f(T)$ models, which successfully reduced both $H_0$ and $S_8$ tensions. A similar approach can be found in \cite{Wang:2020dsc}, where the authors tried to address the tensions in Hu-Sawicki
$f(R)$ gravity, but the model fails.  The power-law, Linder, and exponential $f(T)$ models have been constrained in \cite{Briffa:2023ozo} using the early and late time datasets. Though, to some extent, their models reduce the $S_8$ tension, they are unable to resolve both tensions simultaneously. 

In the initial segment of our work, the Hubble constant and model parameters are constrained using the updated Pantheon+SH0ES data and the combinations of early-time datasets among themselves. As the concerned datasets do not contain the observational value of $f \sigma_8$, a separate analysis has been performed using the RSD data later. The results obtained from MCMC are further verified individually against the popular datasets along with the standard cosmological model, and the curves fit them perfectly.
As it is well-known, GRBs data are subject to larger observational uncertainties compared to Type Ia Supernovae, primarily due to the intrinsic scatter in empirical correlations (e.g., the Amati relation or the luminosity–variability relation) and the lack of direct calibration at lower redshifts. To mitigate these issues, we have also calibrated distance modulus using the Pantheon+ dataset, which serves viable results for low-redshift standard candles. Furthermore, the joint likelihood analysis was performed to ensure the contributions from other higher-quality datasets like CC and BAO. In addition, GRBs contribute valuable information at high redshifts (\(z > 2\)) and their inclusion provides us a platform to understand the dynamics of the Universe up to redshift around 9. Thus, using differently calibrated datasets helps constrain the model to ensure consistency with all of them.
As a new aspect, to unify the trending GW observations in cosmological studies, we verified our result against the recent GW data using the luminosity distance function. Subsequently, the models came out to be an excellent match to the data. Also, the models describe the late-time acceleration through the quintessence behavior of the EoS parameter, corroborated by the inclusion of the $Om(z)$ diagnostic. The statistical analysis also confirms that almost all the cases of our models are strongly favored by the corresponding data. We studied the phase transition of the Universe through the deceleration parameter for both models and compared their nature with the $\Lambda$CDM model.

Further, we performed MCMC analysis for the four combinations using RSD data. We yield the $H_0$ and $S_{8,0}$ values from the 2D contours, while all other parameters are fixed from the previous analysis. A thorough comparison between our models and the existing observations showcased their importance in alleviating the tensions. By considering the tension metric, we found that for most of the cases, our models have negligible tensions with the existing literature. Even in the scenarios where they have mild tensions, it is noticed that they are reduced significantly than the original tensions between observations.

Further, in the latter part, we consider a non-minimal coupling of torsion with matter to investigate the cosmological tensions. 
To constrain the parameters, we commence with the MCMC technique with a new set of combinations of datasets. The DESI BAO data is considered due to its strong constraining power. Combining RSD opens up the possibility to constrain and study the $S_8$ parameter. The latest GW observations have come up as very promising observations in the low redshift paradigm. The CC method is very popular due to its differential aging technique. Finally, we have considered the recent Supernova data, Union3. 

Furthermore, the credibility of the obtained results is substantiated by examining the cosmic evolution. The model exhibits a smooth and consistent phase transition, as demonstrated by the behavior of the deceleration parameter. Its excellent fit to the observational datasets, along with consistency with the $\Lambda$CDM model, highlights its potential as a strong candidate for describing the late-time acceleration of the Universe.

A comparative analysis of the three torsion-based models considered in this work reveals that the reduction of cosmological tensions arises from different underlying physical mechanisms. The Hybrid $f(T)$ model mainly modifies the late-time expansion through nonlinear torsional contributions, leading to a significant weakening of the $H_0$ tension by enlarging the allowed parameter space. The Log-square-root model produces a more moderate but robust modification of the expansion history, maintaining consistency across multiple observational datasets. The $f(T,\mathcal{T})$ model, on the other hand, introduces a direct torsion-matter coupling, resulting in an effective interaction between geometry and matter that impacts both the expansion rate and structure growth, thereby contributing to the alleviation of the $S_8$ tension. This comparative behavior highlights that torsion-based extensions of gravity provide multiple viable pathways for addressing current cosmological discrepancies without invoking exotic dark energy components.

Finally, we observe that the model successfully alleviates the $S_8$ tension. The $H_0$ tension with Planck results is notably reduced, while it is effectively resolved when compared with other observational surveys. The higher inferred value of $H_0$ arises from the modified gravitational dynamics induced by the torsion-matter coupling, rather than from a phantom-like dark energy component. The additional terms in the field equations lead to a dynamical effective contribution that enhances the late-time expansion rate while keeping the total EoS parameter in the non-phantom (quintessence) regime. Consequently, the model accommodates a higher $H_0$ through geometric modifications without invoking exotic phantom behavior. Additionally, the derived sound horizon values show excellent agreement with those reported by the Planck results. The joint parameter space analysis shows that the model can shift the inferred values of $H_0$, $S_8$, and $r_d$ toward alleviating the respective tensions. This is achieved through the combined action of the inverse torsion term, which influences background expansion, and the $\mathcal{T}$ coupling, which modifies the growth sector.

The future aspect of this work is quite interesting, and we can expect prominent development in the field. The combination of innovative methodologies like standard sirens and advanced observational capabilities like those offered by the Euclid mission heralds a transformative era in addressing cosmic tensions. They provide an independent measurement of the Hubble constant by leveraging GW signals, effectively bypassing the systematic uncertainties inherent in traditional methods like supernovae and Cepheid variable studies. As the sensitivity of GW detectors (e.g., LIGO, Virgo, KAGRA) continues to improve, the expected increase in observed binary neutron star and black hole merger events will refine constraints on \(H_0\). This will provide a promising area to test and potentially resolve the long-standing discrepancy between local and early-Universe measurements of the expansion rate. Since Teleparallelism provides a viable geometric framework, exploring future surveys such as DESI, Euclid, and JWST may not only reconcile existing tensions but also support the proposed modification, laying the groundwork for an era of discovery.
\newpage
\thispagestyle{empty}

\vspace*{\fill}
\begin{center}
    {\Huge \color{RedViolet} \textbf{CHAPTER 3}}\\
    \
    \\
    {\Large\color{Emerald}\textsc{\textbf{Model-independent Teleparallel Cosmography}}}\\
    \end{center}

\begin{center}  
\textbf{\color{RedViolet}PUBLICATION}\\
\textbf{``Constraining extended teleparallel gravity via cosmography: A model-independent approach"}\\
 Sai Swagat Mishra et al. 2024 \textit{The Astrophysical Journal} \textbf{970} 57.\\ \vspace{0.2 in}
\includegraphics[width=0.1\linewidth]{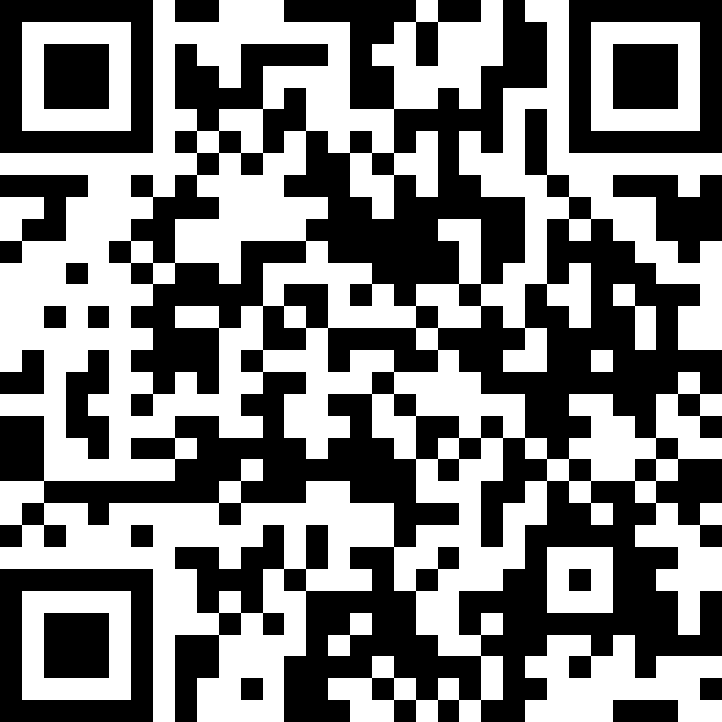} \\ DOI: \href{https://iopscience.iop.org/article/10.3847/1538-4357/ad5555/meta}{10.3847/1538-4357/ad5555}
\end{center}
\vspace*{\fill}

\pagebreak

\def\baselinestretch{1}
\chapter{\textsc{Model-independent Teleparallel Cosmography}}\label{chap3}
\def\baselinestretch{1.5}
\pagestyle{fancy}
\lhead{ \emph{Chapter 3. Model-independent Teleparallel Cosmography}}
\rhead{\thepage}
\noindent\textbf{Highlights}
\begin{itemize}
    \item[$\star$] Explores cosmography as a model-independent tool to reconstruct cosmic expansion using kinematic variables.  
    \item[$\star$] Extends the cosmographic approach to coupled gravities, specifically the $f(T,\mathcal{T})$ framework, which has been less explored in this context.  
    \item[$\star$] Employs Taylor series expansion with a minimally coupled form to constrain the cosmographic parameters.  
    \item[$\star$] Performs MCMC analysis with CC, BAO, and Pantheon+SH0ES datasets.  
    \item[$\star$] Compares the constrained results against key cosmological parameters for consistency and validation.  
    \item[$\star$] Benchmarks the findings with three well-known $f(T,\mathcal{T})$ models, highlighting similarities and deviations.  
\end{itemize}

\section{Introduction} \label{sec:intro}

In this chapter, we explore a model-independent method for reconstructing the $f(T,\mathcal{T})$ function based on cosmological observations. The reconstruction of a gravitational theory involves determining the functional form of the gravity model based on certain criteria. For instance, references \cite{Gadbail:2024syv,Vogt:2024pws,Gomes:2023xzk,Panda:2023jwe,Saez-Gomez:2016wxb,Chen:2024xkv} utilized diverse methods to reconstruct gravitational theories. Among these, some rely on a particular gravity model, while others utilize model-independent techniques such as the application of Raychaudhuri equations \cite{Gadbail:2024syv,Panda:2023jwe,Choudhury:2019zod, Mishra:2025cyq} and observational data \cite{Vogt:2024pws,Gomes:2023xzk}. In this chapter, we explore a notable model-independent approach known as \textit{Cosmography}. It presents a broad approach for handling cosmological parameters using kinematic quantities. We can utilize these quantities to examine the dynamics of the universe without assuming a specific cosmological model.

In the literature, numerous studies have explored cosmography within geometry theories such as $f(R)$, $f(T)$, and $f(Q)$ \cite{Capozziello:2008qc,Capozziello:2011hj,Sabiee:2022iyo,Mandal:2020buf}. However, there has been relatively less emphasis on reconstructing coupled gravities. Thus, in this work, we consider a generic family of minimal functional forms of $f(T,\mathcal{T})$ gravity and deduce an observationally agreeable scenario for an arbitrary function. The kinematical cosmographic series approach involves a greater number of parameters that need to be constrained by data compared to the simplest cosmological models. To streamline the analysis, we assume flat spacetime and uphold the validity of the cosmological principle. These minimal assumptions help simplify the problem, enabling us to employ kinematical cosmography as a model-independent tool to ascertain the compatibility of various $f(T,\mathcal{T})$ models with current observations.

Recent success of $f(T,\mathcal{T})$ theory as a fine alternative to the coherence model prompts us to review some prominent models of this theory. Harko et al. \cite{Harko:2014aja} considered the non-minimally coupled form $\alpha T^n \mathcal{T}+\Lambda $ and the quadratic form $\alpha \mathcal{T}+\gamma T^2$ to constrain through the energy conditions and found that the resulting models are consistent with the stability conditions. Two additive and multiplicative extensions of TEGR, $T+g(\mathcal{T})$ and $T\times g(\mathcal{T})$, have been considered by \cite{Saez-Gomez:2016wxb} to restrict by fitting the predicted distance modulus to that measured from type Ia supernovae. Further, the models are found to be good candidates to reproduce the late-time acceleration. In \cite{Mandal:2023bzo}, the authors attempted to alleviate the $H_0$ tension by performing statistical analysis on two new $f(T,\mathcal{T})$ models, $T e^{\,\alpha\, \frac{T_0}{T}}+\beta \mathcal{T}$ and $\alpha T_0(1-e^{-\beta \sqrt{T/T_0}}) + \gamma \mathcal{T}$. The deviation of their resulting $H_0$ value from Planck2018, SH0ES, and H0LiCOW experiments are analyzed, leading to partially solving the tension. The interesting results of the theory develop inquisitiveness to look for the ultimate model that can describe evolution. In this regard, in the next section, we adopt a generic family of models without relying on any particular model to constrain them through the cosmographic approach.

The outline of this chapter is as follows: In \autoref{chap3/sec:cosmo}, one can see the cosmographic analysis and governing field equations of isotropic and homogeneous universe. In \autoref{chap3/sec:result}, we present the observational constraints, including a cosmographic description of parameters and the corresponding results. In the final \autoref{chap3/sec:conclusion}, we conclude with a detailed interpretation and discussion of the results.

\section{Cosmographical Analysis}\label{chap3/sec:cosmo}
We start this section with the basic assumptions of cosmographic parameters.
The Taylor series expansion of the scale factor $a(t)$ at the present time $t_0$ \cite{Visser:2004bf,Visser:2015iua} is
\begin{equation}\label{eq:a(t)}
    a(t)= 1+ \left.\sum_{n=1}^\infty \frac{1}{n!} \frac{d^na}{dt^n}\right|_{t=t_0} (t-t_0)^n.
\end{equation}
The coefficients of $(t-t_0)^n$ in the above series are the cosmographic parameters. We are going to mention the leading cosmographic parameters which will be used further \cite{Sahni:2002fz,Poplawski:2006na, Poplawski:2006ew}.
\begin{gather}
   \label{chap3/eq:H(t)}Hubble: H(t)=\frac{1}{a} \frac{da}{dt},\\
   \label{eq:q(t)}Deceleration: q(t)=-\frac{1}{aH^2}\frac{d^2a}{dt^2},\\
   \label{eq:j(t)}Jerk: j(t)=\frac{1}{aH^3}\frac{d^3a}{dt^3},\\
   \label{eq:s(t)}Snap: s(t)=\frac{1}{aH^4}\frac{d^4a}{dt^4}.
\end{gather}
The above parameters are entirely free from any cosmological models, which makes our task easy to execute this work in a model-independent approach. By straightforward calculation on the above equations, one can find the derivatives of Hubble parameters involving the cosmographic parameters as \cite{Xu:2010hq, Luongo:2013rba},
\begin{gather}
\begin{aligned}\label{eq:derH}
    \dot{H}&=-H^2(1+q),\\
    \ddot{H}&=H^3(j+3q+2),\\
    \dddot{H}&=H^4[s-4j-3q(q+4)-6],
\end{aligned}
\end{gather}
 From now on dot $(\dot{})$ must be understood as the derivative with respect to cosmic time $(t)$. Furthermore, one can obtain the derivatives of the torsion scalar and trace of energy-momentum tensor at present time $t=t_0$ using the above expressions as
 \begin{align}
 \begin{split}
     \label{eq:derT}
     T_0=&-6H_0^2,\\
     \dot{T}_0=&12 H_0^3 (q_0+1),\\
     \ddot{T}_0=&-12 H_0^4 [j_0+q_0(q_0+5)+3],\\
     \dddot{T}_0=&12 H_0^5 \left[j_0(3 q_0+7)+3 q_0 (4 q_0+9)-s_0+12\right],
     \end{split}\\
     \begin{split}\label{eq:dertau}
    \mathcal{T}_0=& 3  H_0^2 \Omega_{m0},\\     \dot{\mathcal{T}}_0=& -9 H_0^3  \Omega _{m0}, \\   \ddot{\mathcal{T}}_0=& 9  
 H_0^4 (q_0+4) \Omega _{m0}, \\    \dddot{\mathcal{T}}_0=& -9 H_0^5 (j_0+12 q_0+20) \Omega _{m0}.
 \end{split}
     \end{align}

The ultimate motivation behind finding all these terms and their derivatives is to find the Taylor series expansion of arbitrary function of the $f(T, \mathcal{T})$ theory. Now we define the Taylor series expansion around the point $(T_0,\mathcal{T}_0)$,
\begin{multline}    f(T,\mathcal{T})=f(T_0,\mathcal{T}_0)+\frac{1}{2!}f_T(T_0,\mathcal{T}_0)(T-T_0)+\frac{1}{2!}f_{\mathcal{T}}(T_0,\mathcal{T}_0)(\mathcal{T}-\mathcal{T}_0)+\frac{1}{3!}f_{TT}(T_0,\mathcal{T}_0)(T-T_0)^2 \\ +\frac{1}{3!}f_{T \mathcal{T}}(T_0,\mathcal{T}_0)(T-T_0)(\mathcal{T}-\mathcal{T}_0)+\frac{1}{3!}f_{\mathcal{T} \mathcal{T}}(T_0,\mathcal{T}_0)(\mathcal{T}-\mathcal{T}_0)^2+(\mathcal{O}),
\end{multline}
where $(\mathcal{O})$ is the term containing higher-order derivatives. For simplification, we assume a minimally coupling form of the theory as $f(T,\mathcal{T})=g(T)+h(\mathcal{T})$. So the $f_{T \mathcal{T}}$ terms will be reduced to zero in the above series. One can also consider some other kind of generic functional types, such as $g(T)\times h(\mathcal{T})$ or $exp(g(T)+ h(\mathcal{T}))$ or any such non-minimal form. Now in terms of the new function $g(T)$ and $h(\mathcal{T})$, we rewrite the series upto $4^{th}$ degree,
\begin{multline}\label{eq:series}    f(T,\mathcal{T})=g(T_0)+h(\mathcal{T}_0)+\frac{1}{2}g_T(T_0)(T-T_0)+\frac{1}{2}h_\mathcal{T}(\mathcal{T}_0) (\mathcal{T}-\mathcal{T}_0)\\+ \frac{1}{6}g_{TT}(T_0)(T-T_0)^2+\frac{1}{6}h_{\mathcal{T} \mathcal{T}}(\mathcal{T}_0) (\mathcal{T}-\mathcal{T}_0)^2 + \frac{1}{24}g_{TTT}(T_0)(T-T_0)^3+\frac{1}{24}h_{\mathcal{T} \mathcal{T} \mathcal{T}}(\mathcal{T}_0) (\mathcal{T}-\mathcal{T}_0)^3 \\ +\frac{1}{120}g_{TTTT}(T_0)(T-T_0)^4+\frac{1}{120}h_{\mathcal{T} \mathcal{T} \mathcal{T} \mathcal{T}}(\mathcal{T}_0) (\mathcal{T}-\mathcal{T}_0)^4.
\end{multline}

 Now, inserting the assumed form $g(T)+h(\mathcal{T})$ in the motion equations \eqref{eq:fttmot1} and \eqref{eq:fttmot2} at the present time gives us
\begin{equation}
\label{eq:motion3}
    6 H_0^2 \left(2 g^{(1)}+1\right)+g+h-2 \mathcal{T}_0 \left(h^{(1)}+1\right)=0,
\end{equation}
\begin{equation}
\label{eq:motion4}
    -A H_0^2(1+q_0) + \mathcal{T}_0 \left(h^{(1)}+1\right)=0
\end{equation}
with $A=2 \left(g^{(1)}-12 H_0^2 g^{(2)}+1\right)$, $g\equiv g(T_0)$, $h\equiv h(\mathcal{T}_0)$, $g^{(1)}\equiv g_T(T_0)$, $g^{(2)}\equiv g_{TT}(T_0)$ and $h^{(1)}\equiv h_\mathcal{T}(\mathcal{T}_0)$. From now on, the derivatives will be denoted in this manner. The following equations (\ref{eq:x}, \ref{eq:y}) are obtained by taking further derivatives of the second motion equation \eqref{eq:fttmot2}.
\begin{align}
\begin{split}
\label{eq:x}
   &H_0 \Big\{2 g^{(1)} (j_0+3 q_0+2)+ 3 H_0^2 \Big[-8 g^{(2)} (j_0+3 q_0(q_0+3)+5)+96 H_0^2 (q_0+1)^2 g^{(3)} \\ &-  9 h^{(2)} \Omega _{m0}^2\Big]  -9 (h^{(1)}+1) \Omega _m{0}+2 (j_0+3 q_0+2)\Big\}=0,
\end{split}\\
\begin{split}
\label{eq:y}
     &288 H_0^4 (q_0+1) \left\{g^{(3)} [j_0+3 q_0 (2 q_0+5)+8]-12 H_0^2 (q_0+1)^2 g^{(4)}\right\}=\\
     &2 (g^{(1)}+1) \left[-4 j_0-3 q_0 (q_0+4)+s_0-6\right]+48 H_0^2 (q_0+1) (j_0+3 q_0+2) \left(g^{(2)}-12 H_0^2 g^{(3)}\right)\\&+24 H_0^2 g^{(2)} \left\{j_0 (7 q_0+11)+q_0 \left[3 q_0 (q_0+11)+56\right]-s_0+23\right\}+ 9 (q_0+4) \Omega _{m0} \\&\times \left(3 H_0^2 h^{(2)} \Omega _{m0}+h^{(1)}+1\right)+81 H_0^2 \Omega _{m0}^2 \left(3 H_0^2 h^{(3)} \Omega _{m0}+2 h^{(2)}\right).
\end{split}
\end{align}
By solving the Eqs. (\ref{eq:motion4}-\ref{eq:y}), one can find $q_0$, $j_0$ and $s_0$ in terms of the unknown parameters,
\begin{align}
\begin{split}\label{eq:q0}
    q_0=& \frac{1}{A}\left[3 (h^{(1)}+1) \Omega _{m0}\right]-1,
\end{split}\\
\begin{split}\label{eq:j0}
    j_0=& \frac{1}{A}\Big(-4 - 6 q_0 - 2 g^{(1)} (2 + 3 q_0) + 
   9 (1 + h^{(1)}) \Omega _{m0} \\
   &+3 H_0^2 \left\{-96 g^{(3)} H_0^2 (1 + q_0)^2 + 
      8 g^{(2)} [5 + 3 q_0 (3 + q_0)] + 9 h^{(2)} 
{\Omega _{m0}}^2\right\}\Big),
\end{split}\\
\begin{split}\label{eq:s0}
s_0 = & -\frac{1}{A} \Bigg[ -2 g^{(1)} \left[4 j_0 + 3 q_0 (q_0 + 4) + 6\right] 
+ 4 \big[162 H_0^2 g^{(2)} - 864 H_0^4 g^{(3)} + 864 H_0^6 g^{(4)}-3 \\
& + 9 (h^{(1)} + 1) \Omega_{m0} - 2 j_0 - 6 q_0 \big] 
- 6 \Big( -4 H_0^2 g^{(2)} \big[j_0 (9 q_0 + 13) 
+ 3 q_0 (q_0 + 2)(q_0 + 11) \big] \\
& + 144 H_0^4 g^{(3)} \big[(j_0 + 11) q_0 + j_0 + 2 q_0^3 + 9 q_0^2 \big] 
+ q_0 \big\{q_0 - 576 H_0^6 \left[q_0 (q_0 + 3) + 3\right] g^{(4)} \big\} \Big) \\
& + 27 H_0^2 (q_0 + 10) h^{(2)} \Omega_{m0}^2 
+ 243 H_0^4 h^{(3)} \Omega_{m0}^3 
+ 9 q_0 (h^{(1)} + 1) \Omega_{m0} \Bigg].
\end{split}
\end{align}

\section{Observational Constraints}\label{chap3/sec:result}

\subsection{Cosmographic Description of Parameters}

Thus far, we have examined the present-time assessment of cosmographic parameters. These evaluations serve as the basis for our analysis in understanding various cosmological parameters in terms of redshift. To begin, let us first revisit the relationship between redshift and emission time. We start with the standard relation $a(t)=\frac{1}{1+z}$, with $a(t_0)=1$. By employing the Taylor series expansion of $a(t)$ in terms of cosmographic parameters (see Eqs. (\ref{eq:a(t)} - \ref{eq:s(t)})), one can obtain the series expansion of $t(z)$. This methodology proves instrumental in deducing expansions for other parameters in terms of redshift.

The Hubble parameter is a key concept in contemporary cosmology and an essential aspect of our understanding of the dynamics and evolution of the Universe. It represents the rate of expansion of the Universe, illustrating the relationship between the velocity at which distant galaxies are receding from us and their respective distances. Furthermore, it plays a crucial role in the development of the cosmic distance ladder, which is a hierarchical set of techniques used for calculating distances to astronomical objects through cosmological observations. Using the cosmographic technique to derive the series expansion of emission time and employing the definition of $H$ (Eq.~(\ref{chap3/eq:H(t)})), along with (Eq.~(\ref{eq:s(t)})), yields \cite{Gao:2024}
\begin{equation}\label{eq:H(z)}
    H(z) = H_0 \left(1+\alpha_1 z+\alpha_2 z^2+\alpha_3 z^3+\xi_H\right),
\end{equation}
where $\xi_H$ represents the higher order terms of $H(z)$ and $\alpha_i$'s are coefficients in terms of cosmographic parameter given by 
\begin{align*}
    \alpha_1&=(1+q_0),\\
    \alpha_2&=\frac{1}{2!}(j_0-q_0^2),\\
    \alpha_3&=-\frac{1}{3!}(s_0+3j_0+4j_0q_0-3q_0^2-3q_0^3).
\end{align*}
An important quantity that quantifies the relationship between the angular size of an object and its physical size when observed from Earth is the angular diameter distance $d_A$ (already introduced in Eq.~(\ref{eq:da})). It enables us to determine the actual size of distant objects from their apparent angular size and is essential for measuring the separations between galaxies, galaxy clusters, and other cosmic formations \cite{Tonghua:2023hdz}.
It shows how the apparent brightness of distant celestial objects is affected by the expansion of the Universe. The quantity, $d_L(z)$, represents the luminosity distance expressed in megaparsecs, which accounts for the redshift caused by the Universe's expansion and denotes the distance at which an object with a known intrinsic luminosity would appear \cite{Busti:2015xqa, Demianski:2012ra, Piedipalumbo:2015jya}. The mathematical description is presented in Eq.~(\ref{eq:dl}).
Using the definition, the series expansion of $d_L(z)$ is given by
\begin{equation}\label{eq:d_L(z)}
    d_L(z) = \frac{cz}{H_0} \left(1+\beta_1 z+\beta_2 z^2+\beta_3 z^3+\xi_{d_L}\right),
\end{equation}
where $\xi_{d_L}$ denotes the higher-order terms of $d_L(z)$, while the coefficients $\beta_i$ are expressed in terms of cosmographic parameters, as follows
\begin{align*}
    \beta_1&=\frac{1}{2!}(1-q_0),\\
    \beta_2&=-\frac{1}{3!}(1-q_0-3q_0^2+j_0),\\
    \beta_3&=\frac{1}{4!}(2-2q_0-15q_0^2-15q_0^3+5j_0+10q_0j_0+s_0).
\end{align*}
In light of Eq.~(\ref{eq:d_L(z)}) and the Taylor series expansion of $(1+z)^{-2}$, the angular diameter distance \eqref{eq:da} can be reformulated as
\begin{equation}
    d_A(z) = \frac{cz}{H_0} \left(1+\gamma_1 z+\gamma_2 z^2+\gamma_3 z^3+\xi_{d_A}\right).
\end{equation}
Here, $\xi_{d_A}$ represents the higher-order terms of $d_A(z)$, with the coefficients $\gamma_i$ specified in terms of cosmographic parameters, as outlined below
\begin{align*}
    \gamma_1&=-\frac{1}{2!}(3+q_0),\\
    \gamma_2&=\frac{1}{3!}(11+7q_0+3q_0-j_0),\\
    \gamma_3&=-\frac{1}{4!}(50+46q_0+39q_0^2+15q_0^3-13j_0-10q_0j_0-s_0).
\end{align*}
Eq.~(\ref{eq:d_L(z)}) enables us to compute the distance modulus $\mu$, which represents the difference in magnitude between the apparent magnitude $(m)$ and the absolute magnitude $(M)$. The distance modulus proves useful in estimating the distances of objects based on observed brightness when direct distance measurements are not feasible. The theoretical expression for distance modulus is given by
\begin{equation}\label{eq:mu}
    \mu(z) = 5\log_{10}\left(\frac{d_L(z)}{1\ \text{Mpc}} \right) + 25.
\end{equation}
Thus, the series expansion of the above function can be provided as
\begin{equation}\label{eq:mu(z)}
    \mu(z)=\frac{5}{\log 10} \left[\log z+\delta_1 z+\delta_2 z^2+\delta_3 z^3+\xi_{\mu}\right].
\end{equation}
In this expression, $\xi_\mu$ denotes the higher-order terms of $\mu(z)$, while the coefficients $\delta_i$ represent parameters within the cosmographic framework, defined as
\begin{align*}
    \delta_1&=-\frac{1}{2}(-1+q_0),\\
    \delta_2&=-\frac{1}{24}(7-10q_0-9q_0^2+4j_0),\\
    \delta_3&=\frac{1}{24}(5-9q_0-16q_0^2-10q_0^3+7j_0+8q_0j_0+s_0).
\end{align*}
\subsection{Methodology}
To reconstruct $f(T,\mathcal{T})$ gravity and accurately reflect the dynamics of our Universe, it is essential to make use of observational data and employ suitable methodologies for parameter estimation. In this section, we provide an overview of the methodology adopted to constrain the model parameters. 

In our study, we have considered the general minimal form of $f(T,\mathcal{T})$ coupling in the form of $g$ and $h$, where the latter is a pure function of $\mathcal{T}$ and the former is a pure function of $T$ alone. The Taylor series expansion of the same is provided in Eq.~(\ref{eq:series}). Since we are relying on a model-independent approach, $g$ and $h$ are not assigned specific functional forms. Thus, the present values of these functions and their derivatives are unknown. To this end, we consider $g^{(1)}, g^{(2)}, g^{(3)}, g^{(4)}, h^{(1)}, h^{(2)}$, and $h^{(3)}$ as free parameters whose values need to be calibrated for the present epoch. These parameters along with $H_0$ and $\Omega_{m0}$ are constrained using the statistical Bayesian approach.

To determine the best-fit values for the parameters, we minimize the $\chi^2$ function. Given the relationship between $\chi^2$ and likelihood as $\mathcal{L} \propto e^{-\frac{\chi^2}{2}}$, minimizing $\chi^2$ is equivalent to maximizing the likelihood, which is further equivalent to minimizing the negative log-likelihood. To obtain constraints on the free parameters, we utilize the MCMC sampling method. In order to achieve this, we employ the following data sets: CC, BAO, and Pantheon+SH0ES. A detailed description of the methodology and datasets is already discussed in the chapter~\ref{ch:intro}
.
\begin{table}
 \centering
 \caption{ Best fit range of the parameters with $1-\sigma$ confidence level for constrained parameters.
 }
 
 \label{table1}
 \begin{footnotesize}
    \begin{tabular}{|c||c|c|c|}
    \hline
    
          & $CC$ & $BAO$ & $Pantheon+SH0ES$ \\
    \hline
    \hline

    $H_0$ & $(72.424,72.450)$ & $(72.387,72.405)$ & $(72.392,72.418)$ \\
    \hline
    
     $g^{(1)}$ & $(2371730.05,2371730.85)$ & $(2369798.02,2369798.82)$ & $(2369819.41,2369820.39)$  \\
    \hline
    
     $g^{(2)}$ & $(-2250802.13,-2250801.35)$ & $(-2465604.79,-2465603.99)$ & $(-1508338.51,-1508337.51)$  \\
    \hline
    
     $g^{(3)}$ & $(-9457280.63,-9457279.83)$ & $(-9613894.84,-9613894.04)$ &  $(-9624514.53,-9624513.53)$\\
    \hline
    
     $g^{(4)}$ & $(1131.8,1132.56)$ & $(1151.77,1152.35)$ & $(1153.09,1153.89)$ \\
    \hline
    
     $h^{(1)}$ & $(191000553,191000553.80)$ & $(190784544.07,190784544.87)$ & $(190782097.84,190782098.82)$ \\
    \hline
    
     $h^{(2)}$ & $(56396393.89,56396394.69)$ & $(57087663.57,57087664.37)$ & $(57127315.97,57127316.93)$ \\
    \hline
    
     $h^{(3)}$ & $(20639685.58,20639686.38)$ & $(20639602.75,20639603.55)$ & $(20639605.24,20639606.20)$ \\
    \hline
    
     $\Omega_{m0}$ & $(0.2815,0.2867)$ & $(0.28225,0.28421)$ & $(0.2853,0.2883)$ \\
    \hline
   
    \end{tabular}
    \end{footnotesize}
\end{table}
\begin{table}
\caption{Best fit range of the parameters with $1-\sigma$ confidence level for the combined dataset.}
    \centering
    \begin{tabular}{|c||c|}
    \hline
        Parameter & Range \\
        \hline\hline
         $H_0$ & $(72.98,73.42)$ \\\hline
         $q_0$ & $(-0.509,-0.44)$ \\\hline
         $j_0$ & $(0.63,1.01)$ \\\hline
         $s_0$ & $(0.48,0.12)$ \\\hline
    \end{tabular}
    
    \label{tab:combined}
\end{table}

\subsection{Results}
\begin{figure}
    \centering
    \includegraphics[width=0.8\linewidth]{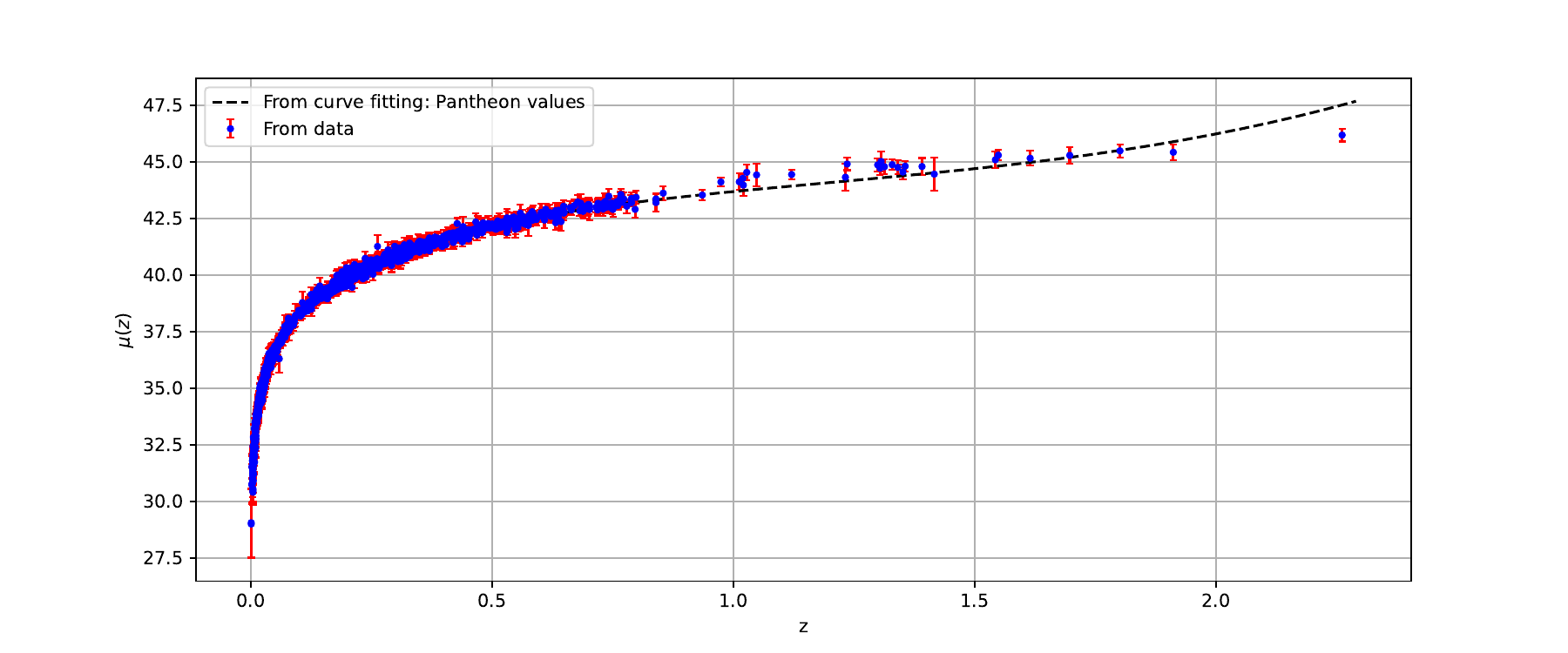}
    \caption{Distance modulus function: Plot illustrating the $\mu(z)$ profile resulting from the MCMC analysis of parameters $(H_0,g^{(1)},g^{(2)},g^{(3)},g^{(4)},h^{(1)},h^{(2)},h^{(3)},\Omega_{m0})$ using  Pantheon+SH0ES dataset. The error bars represent 1701 data points of the Pantheon+SH0ES sample.}
    \label{fig:muz}
\end{figure}

\begin{figure}
    \centering
    \includegraphics[width=\linewidth]{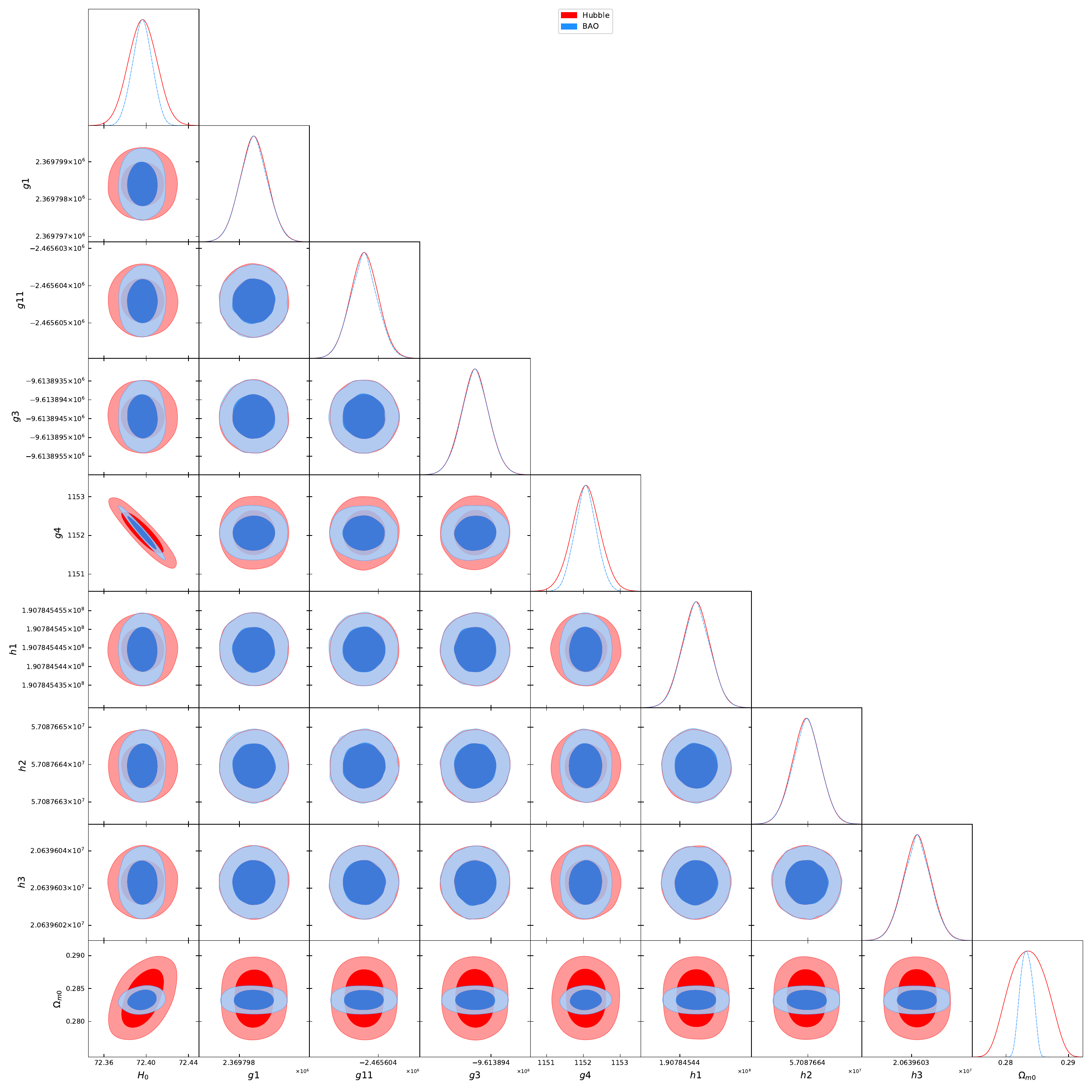}
    \caption{2D likelihood contours obtained from MCMC analysis of CC and BAO datasets.}
    \label{fig:hubble_bao}
\end{figure}

\begin{figure}
    \centering
    \includegraphics[width=\linewidth]{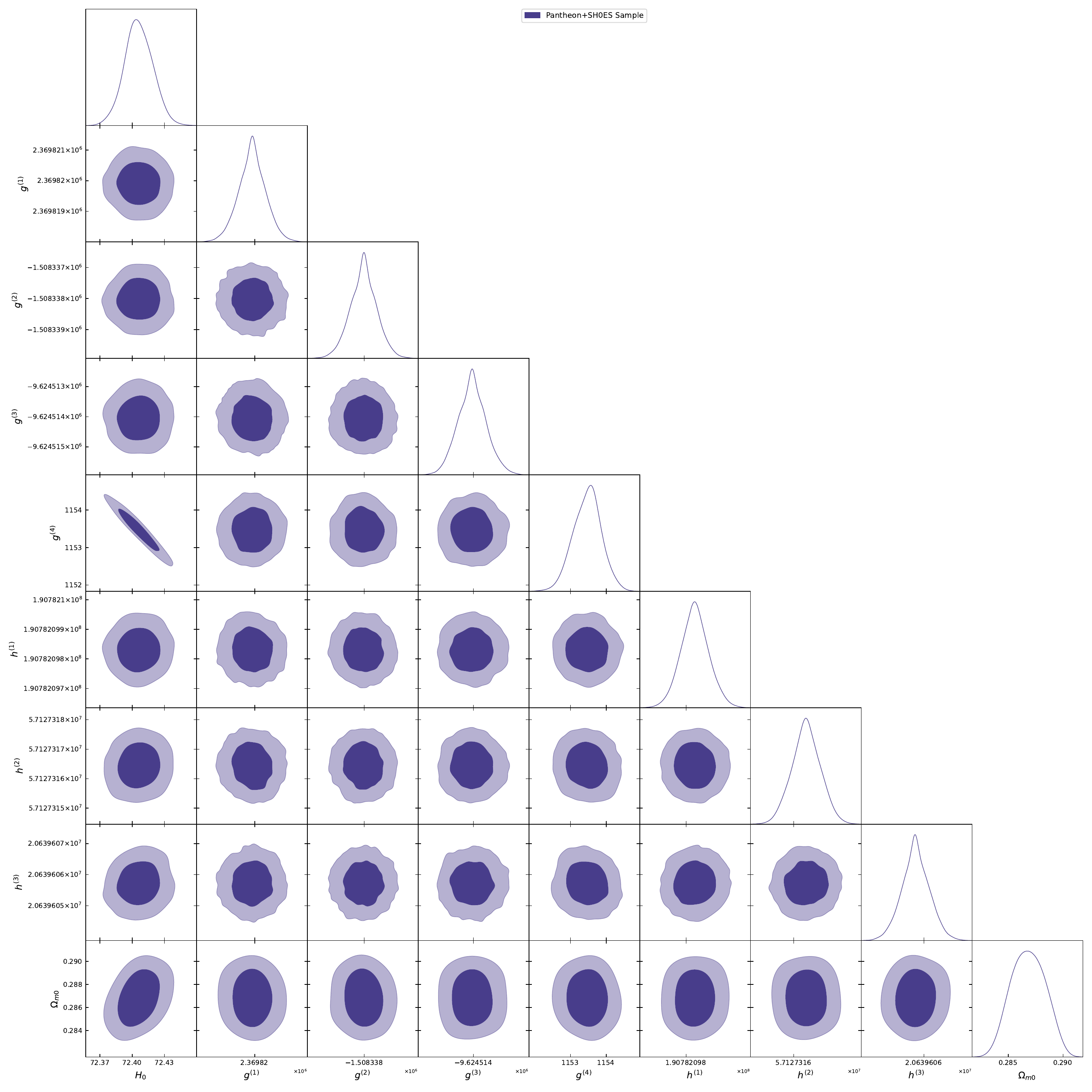}
    \caption{2D likelihood contours obtained from MCMC analysis of Pantheon+SH0ES dataset.}
    \label{fig:pantheon}
\end{figure}

\begin{figure}
    \centering
    \includegraphics[width=0.6\linewidth]{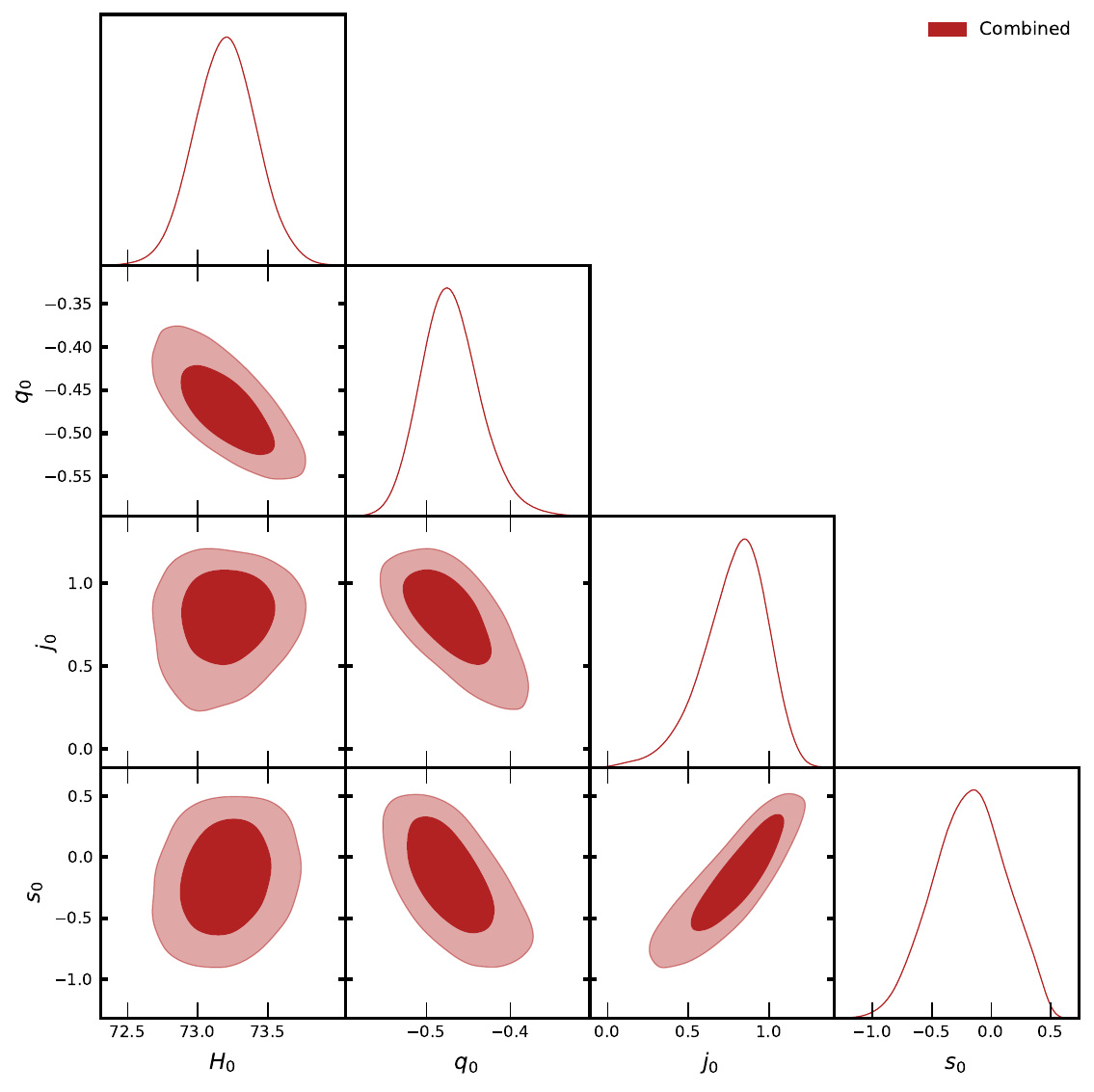}
    \caption{2D likelihood contours obtained from MCMC analysis of combined CC, BAO, Pantheon+SH0ES dataset.}
    \label{fig:qjs}
\end{figure}

This section is dedicated to discussing the outcomes obtained from curve fitting and MCMC using various datasets. We verify the credibility of those results by plotting cosmological parameters. From \autoref{fig:muz}, we depict that the distance modulus function is in excellent agreement with 1701 data points of the Pantheon+SH0ES sample. The 2D contours for all the datasets up to $2-\sigma$ CL are presented in Figures \ref{fig:hubble_bao} and \ref{fig:pantheon}.  We summarize the best-fit $1-\sigma$ range in \autoref{table1}. It is worth to note that a considerable deviation in ranges for $1-\sigma$ CL can be observed from MCMC analysis of all three datasets, while for $2-\sigma$ CL, parameters lie in almost similar ranges. In addition, we conducted MCMC analysis with $H_0$, $q_0$, $j_0$, and $s_0$ as free variables. To accomplish this, we utilized the combined dataset (CC+BAO+Pantheon+SH0ES), and the corresponding 2D likelihood contour up to $2-\sigma$ is depicted in \autoref{fig:qjs}. These values are summarized in \autoref{tab:combined}.  


\begin{figure}
    \centering
    \includegraphics[width=0.47\linewidth]{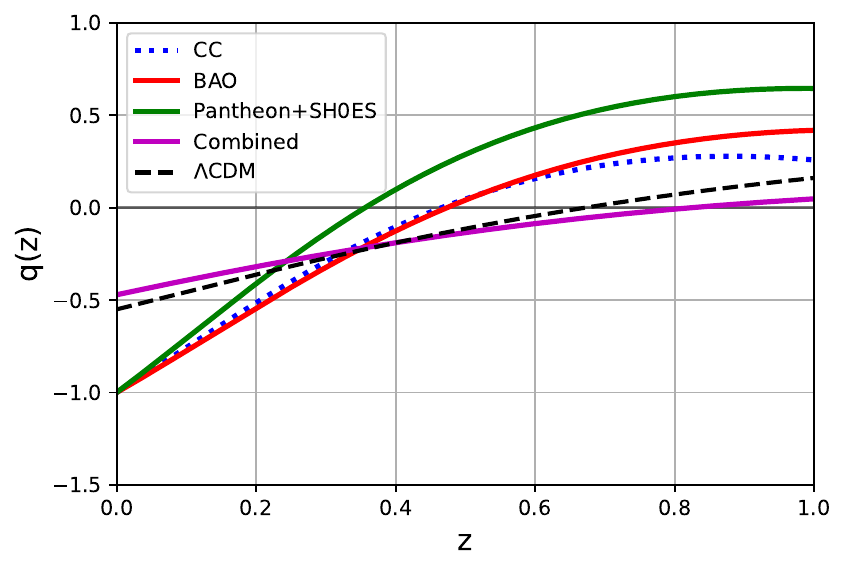}
    \caption{Deceleration Parameter: Plot illustrating the $q(z)$ profile resulting from the MCMC analysis of parameters $(H_0,g^{(1)},g^{(2)},g^{(3)},g^{(4)},h^{(1)},h^{(2)},h^{(3)},\Omega_{m0})$ using (\mycirc[blue]) CC, (\mycirc[red]) BAO, and (\mycirc[ForestGreen]) Pantheon+SH0ES datasets, and using parameters $(H_0, q_0, j_0, s_0)$ for (\mycirc[magenta]) the combined dataset. For (\mycirc) $\Lambda$CDM, the curve is plotted with $\Omega_{m_0}=0.3$, $\Omega_{\Lambda_0}=0.7$ and $H_0=67.8~\text{km}~\text{s}^{-1}~\text{Mpc}^{-1}$}
    \label{chap3/fig:q}
\end{figure}
\begin{figure}
    \centering
    \includegraphics[width=0.47\linewidth]{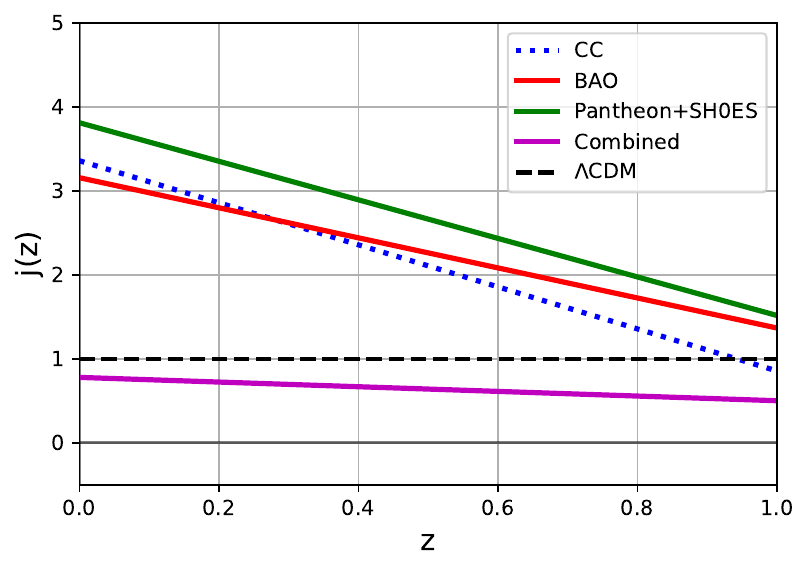}
    \caption{Jerk Parameter:  Plot illustrating the $j(z)$ profile resulting from the MCMC analysis of parameters $(H_0,g^{(1)},g^{(2)},g^{(3)},g^{(4)},h^{(1)},h^{(2)},h^{(3)},\Omega_{m0})$ using (\mycirc[blue]) CC, (\mycirc[red]) BAO, and (\mycirc[ForestGreen]) Pantheon+SH0ES datasets, and using parameters $(H_0, q_0, j_0, s_0)$ for (\mycirc[magenta]) the combined dataset. For (\mycirc) $\Lambda$CDM, the curve is plotted with $\Omega_{m_0}=0.3$, $\Omega_{\Lambda_0}=0.7$ and $H_0=67.8~\text{km}~\text{s}^{-1}~\text{Mpc}^{-1}$}
    \label{chap3/fig:j}
\end{figure}

For combined analysis, the obtained ranges of $q_0$, $j_0$, and $s_0$ are comparable to $\Lambda$CDM. Through the deceleration parameter in \autoref{chap3/fig:q}, one can observe the dynamic phase transition from deceleration to acceleration at certain redshifts, which are explicitly mentioned in \autoref{table2}. To describe the current accelerated expansion $q_0$ has to be in $-1 < q_0 \leq 0$, while $q_0=-1$ explains the de Sitter Universe, completely dominated by dark energy. The transition redshift ($z_t$) for $\Lambda$CDM is around $z_t \sim 0.7$. The redshifts we obtained from \autoref{chap3/fig:q} are well within the best-fit $1-\sigma$ ranges found from MCMC analysis in \cite{Muthukrishna:2016evq}. 

\begin{figure}
    \centering
    \includegraphics[width=0.47\linewidth]{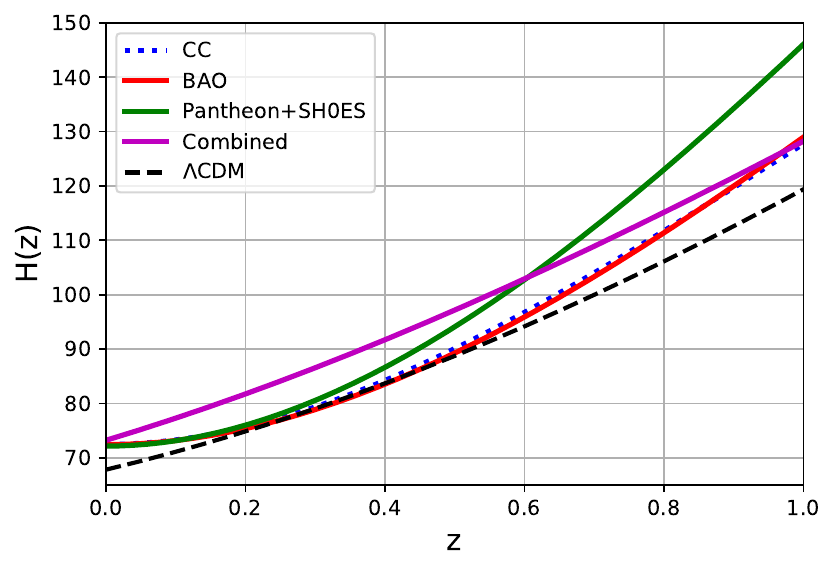}
    \caption{Hubble Parameter:  Plot illustrating the $H(z)$ profile resulting from the MCMC analysis of parameters $(H_0,g^{(1)},g^{(2)},g^{(3)},g^{(4)},h^{(1)},h^{(2)},h^{(3)},\Omega_{m0})$ using (\mycirc[blue]) CC, (\mycirc[red]) BAO, and (\mycirc[ForestGreen]) Pantheon+SH0ES datasets, and using parameters $(H_0, q_0, j_0, s_0)$ for (\mycirc[magenta]) the combined dataset. For (\mycirc) $\Lambda$CDM, the curve is plotted with $\Omega_{m_0}=0.3$, $\Omega_{\Lambda_0}=0.7$ and $H_0=67.8~\text{km}~\text{s}^{-1}~\text{Mpc}^{-1}$}
    \label{chap3/fig:H}
\end{figure}

\begin{figure}
    \centering
    \includegraphics[width=0.47\linewidth]{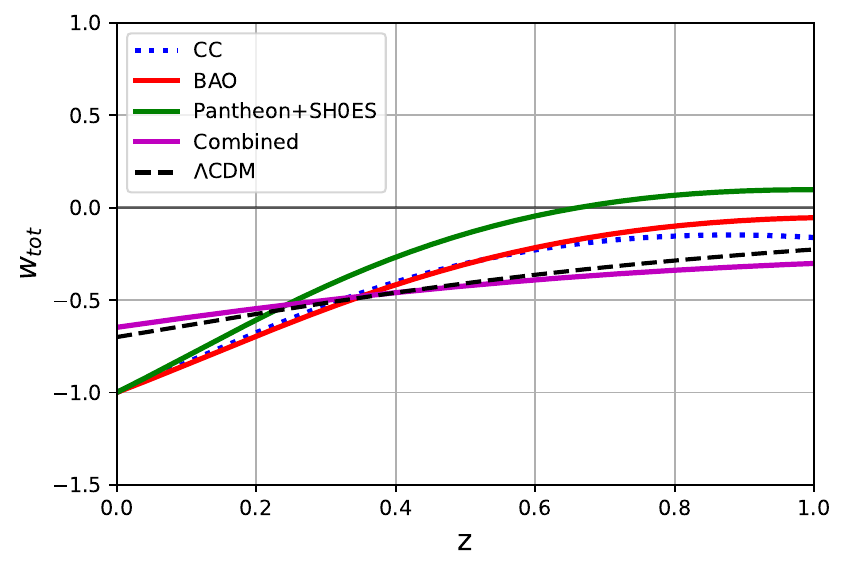}
    \caption{Total EoS Parameter:  Plot illustrating the $w_{\text{tot}}$ profile resulting from the MCMC analysis of parameters $(H_0,g^{(1)},g^{(2)},g^{(3)},g^{(4)},h^{(1)},h^{(2)},h^{(3)},\Omega_{m0})$ using (\mycirc[blue]) CC, (\mycirc[red]) BAO, and (\mycirc[ForestGreen]) Pantheon+SH0ES datasets, and using parameters $(H_0, q_0, j_0, s_0)$ for (\mycirc[magenta]) the combined dataset. For (\mycirc) $\Lambda$CDM, the curve is plotted with $\Omega_{m_0}=0.3$, $\Omega_{\Lambda_0}=0.7$ and $H_0=67.8~\text{km}~\text{s}^{-1}~\text{Mpc}^{-1}$}
    \label{chap3/fig:w}
\end{figure}

The jerk parameter obtained from the second-order derivative of the scale factor series should be positive at the present time.
  Though the behavior of the jerk parameter (see \autoref{chap3/fig:j}) obtained from the result of MCMC is slightly deviating from $\Lambda$CDM, for which it is always $1$, positivity at the present redshift makes it a viable candidate to explain the acceleration. From \autoref{chap3/fig:H}, we found that for all individual samples along with the combined one, the Hubble parameter shows very similar behavior to the standard cosmological model. Furthermore, the total equation of state parameter (denoted as $w_{tot}$) is plotted against redshift in \autoref{chap3/fig:w}. From \autoref{table2} it can be seen that the present values of $w_{tot}$ are not subceeding $-1$, which confirms the quintessence behavior of the Universe.

\begin{table}
 \centering
 \caption{ Present values of the cosmographic parameters for the three datasets.
 }
 
 \label{table2}
 
    \begin{tabular}{|c||c|c|c|}
    \hline
    
          & $CC$ & $BAO$ & $Pantheon+SH0ES$ \\
    \hline
    \hline

    $H_0$ & 72.4361776
 & 72.3961722
 & 72.138
 \\
    \hline
    
     $q_0$ & -0.999426 & -0.999477 &  -0.999371 \\
    \hline
    
     $j_0$ & 3.359057 & 3.157094 & 3.811645  \\
    \hline
    
     $s_0$ & 5.855289 & 4.941443 & 6.098992 \\
    \hline
    
     $w_0$ & -0.999617 & -0.999651 & -0.999581 \\
    \hline
    
     $z_t$ & 0.465654 & 0.473282 & 0.355881 \\
    \hline

    \end{tabular}
\end{table}
\section{Conclusion}\label{chap3/sec:conclusion}
 To reconstruct the theory, we started with the definition of the Taylor series function of $f(T,\mathcal{T})$. We have constrained the independent free parameters involved in the series, using which one can find the remaining ones such as $g(T_0)$, $h(\mathcal{T}_0)$, and $h^{(4)}(\mathcal{T}_0)$. One can directly obtain $g(T_0)+h(\mathcal{T}_0)$ from the motion equation (Eq.~(\ref{eq:motion3})), by substituting the resulting values from MCMC. Also, the fourth derivative of $h$ can be computed by differentiating the nonpresent form of Eq.~(\ref{eq:motion3}) three times.

Cosmography is used as an essential tool to reconstruct a theory that already has a plethora of successful models describing various phases of the evolution of the Universe. All Taylor expansions related to cosmography are valid within the observable domain where $z \ll 1$, allowing one to establish constraints on the present-day universe. The $\mu(z)$ and $H(z)$ functions involving cosmographic parameters are used to constrain the $f(T,\mathcal{T})$ theory by using the Markov Chain Monte Carlo method for the SNeIa, BAO, and Hubble datasets. Also, we obtained an independent range for the parameters $H_0$, $q_0$, $j_0$, and $s_0$ using the combined dataset (CC+BAO+SNeIa). All ranges that the MCMC analysis yielded are summarized in tabular form in \autoref{table1}. Further, the constrained functions from all datasets are examined and compared through various cosmological parameters against low redshifts. Since $\Lambda$CDM is in fine agreement with empirical data, we set it as a benchmark for comparison in each analysis. The common redshift range is considered $z \in [0,1]$, where we observe that the models agree with the cosmographic parameters as well as the transition redshifts. 
Since cosmography confines only cosmological quantities that are not strictly dependent on a model, it alleviates degeneracy. The utilization of Taylor series expansion highly permits this alleviation in the lower redshift range. However, for a higher redshift range, this may lead to some extent of degeneracy \cite{Capozziello:2017uam}.

\begin{figure}
    \centering
    \includegraphics[width=0.7\linewidth]{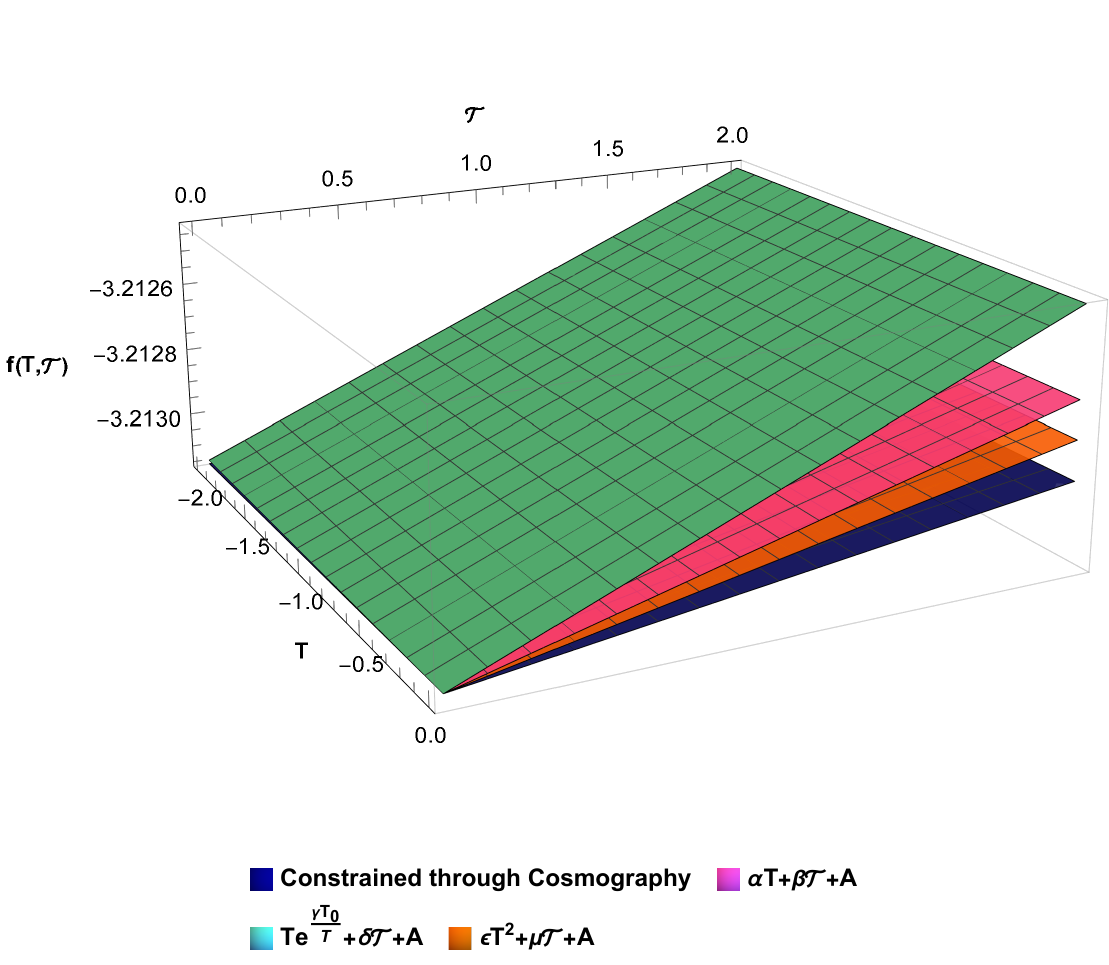}
    \caption{Comparision of three $f(T,\mathcal{T})$ models with the constrained function. The considered values of model parameters are $\alpha=3.5,\,\ \beta=2.25 \times 10^{14},\,\ \gamma=-0.00336,\,\ \delta =3.55 \times 10^{14}, \,\ A =-3.21314\times 10^{18}, \,\ \mu=1.7 \times 10^{14} \,\ \text{and} \,\ \epsilon=0.75.$}
    \label{fun}
\end{figure}

As the work has been executed in a model-independent manner, we interpret that any minimally coupled class of model from $f(T,\mathcal{T})$ theory should show a similar kind of behavior to the reconstructed functional form. For instance, in \autoref{fun} we consider the well-known exponential model $f(T,\mathcal{T})=T e^{\frac{\gamma T_0}{T}}+\delta \mathcal{T}+A$, the linear model $f(T,\mathcal{T})=\alpha T +\beta \mathcal{T}+A$ and the quadratic correction $f(T,\mathcal{T})=\epsilon T^2 +\mu \mathcal{T}+A$ to compare with the constrained model. From the literature review, the above particular models have been taken into account due to their efficiency in explaining the evolution of the Universe. For certain fixed values of the free parameters, we find that all three models make an excellent match with our model (free parameter values mentioned in the Figure Caption). While explaining the dark sector responsible for the evolution of the universe, scalar-tensor theories provide promising results, as one can see in \cite{Brans:1961sx, Fujii:2003pa, Faraoni:2004pi}. Since the value of $\alpha$ shows a deviation from the usual value steaming from the gravitational constant when matching with the model-independent Taylor series form, we believe there is potential in this approach to connect the results of scalar-tensor theories. However, to make a definitive claim, we need to investigate its behavior in the early stages of the universe. Given that the Taylor series works well in regions with lower redshifts, we could obtain interesting results if the study is conducted using different orthogonal polynomials or series expansions with the intention of exploring the scenarios at higher redshifts. If this approach sheds light on scalar-tensor theories through cosmography, it could lead to revolutionary results. Since this is a promising idea, one can aim to reconnect scalar-tensor theory with cosmography using different series expansions as a future perspective.

One can also consider the non-minimally coupled form to constrain the theory using this method, such as $g(T) \times h(\mathcal{T})$, $exp(g(T)+ h(\mathcal{T}))$, etc. In order to cosmographically connect the early Universe with its late-time dynamics, the analysis can be extended to higher redshift using the other series expansions, such as Pad\'e polynomials and Chebyshev polynomials. If one can successfully establish this connection, then the dark energy candidate can be fully explained solely in terms of cosmographic parameters, without necessarily applying prior modifications to Einstein's equations. This could potentially open up new dimensions of research.\\
\clearpage
\thispagestyle{empty}

\vspace*{\fill}
\begin{center}
    {\Huge \color{RedViolet} \textbf{CHAPTER 4}}\\[2ex]
    {\Large \color{Emerald} \textbf{Pad\'e Cosmography and its insight into Torsional theories}}
\end{center}

\begin{center}
\textbf{\color{RedViolet}PUBLICATION}\\
\textbf{``Pad\'e Cosmography and its insight into Teleparallel Gravity"}\\
    Sai Swagat Mishra et al. 2025 \textit{Monthly Notices of the Royal Astronomical Society} \textbf{543} 2816–2835.\\\vspace{0.2 in}
\includegraphics[width=0.1\linewidth]{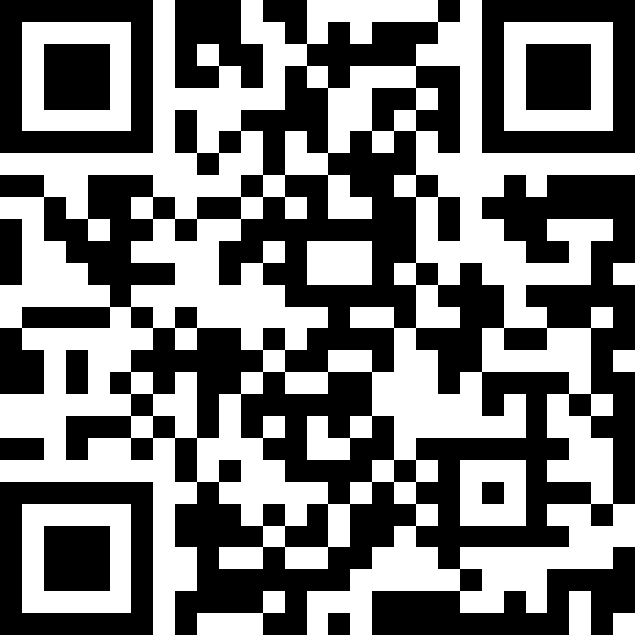} \\
    DOI: \href{https://doi.org/10.1093/mnras/staf1492}{10.1093/mnras/staf1492}
\end{center}
\vspace*{\fill}

\clearpage
\chapter[Pad\'e Cosmography and its insight into Torsional theories]%
{Pad\'e Cosmography and its insight into Torsional theories}\label{chap4}

\pagestyle{fancy}
\lhead{\emph{Chapter 4. Pad\'e Cosmography and its insight into Torsional theories}}
\rhead{\thepage}
\noindent\textbf{Highlights}
\begin{itemize}
    \item[$\star$] Implements two complementary approaches for cosmological parameter estimation:  
        \begin{itemize}
            \item Padé(2,1) expansion of the luminosity distance $d_L(z)$ for stable reconstruction across redshift.  
            \item Direct solution of the modified Friedmann equation for dynamical consistency.  
        \end{itemize}
    \item[$\star$] Performs a comprehensive MCMC analysis with the latest cosmological datasets: CC, GW standard sirens, DESI BAO DR2, Pantheon+SH0ES, and Union3.  
    \item[$\star$] Demonstrates that when tested against observational data points and derived cosmological parameters, both approaches yield reliable and concordant results.  
    \item[$\star$] Finds strong consistency between Padé-based and dynamical approaches, reinforcing the robustness of the $f(T)$ model.  
    \item[$\star$] Both methods provide a statistically better fit than $\Lambda$CDM when confronted with DESI DR2 and Union3 data.   
\end{itemize}

\section{Introduction}
In light of the issues of GR, there is a growing need for model-independent approaches that rely solely on kinematics rather than assuming a specific cosmological model. Cosmography, introduced by Weinberg (1972) \cite{Weinberg:1972kfs}, is one such approach. It describes the expansion history of the Universe using only the cosmological principle, homogeneity and isotropy, without invoking Einstein's equations. Cosmography typically expands observables like the luminosity distance $d_L(z)$ in a Taylor series around the current epoch ($z = 0$). However, this traditional method performs poorly at high redshifts ($z \gtrsim 1$), where the Taylor series ceases to converge, significantly limiting its application to recent cosmological data.

However, recent observational advances have significantly expanded the redshift range that is available to cosmological studies. The Pantheon+ compilation \cite{Riess:2021jrx} now includes Type Ia supernovae up to redshifts $z \approx 2.26$, providing a much deeper probe into the expansion history than the previous compilations. In addition to this, the Union3 dataset \cite{Rubin:2023ovl} covers the range $z \sim 0.05$ to $z \sim 2.26$, still providing valuable legacy constraints. Meanwhile, the Dark Energy Spectroscopic Instrument extends the reach even further, mapping large-scale structure and the Hubble flow through galaxies and quasars up to $z \sim 3.5$ \cite{DESI:2016fyo}. Additionally, early detections of GW standard sirens, while currently limited to $z \lesssim 1$, open an independent, calibration-free window into cosmic expansion, with future facilities expected to push this limit to much higher redshifts.

Together, these diverse datasets form a robust and multiscale framework for investigating the expansion history of the universe. They provide an unprecedented opportunity to scrutinize the $\Lambda$CDM paradigm across a wide redshift range and to search for signs of deviations that could hint at new physics beyond the standard cosmological model.

In light of these, this work lies at the intersection of cosmography and geometrically extended theories of gravity, adopting a novel approach that constrains gravitational models through a model-independent cosmographic framework. Specifically, we have considered the extension of TEGR to integrate with cosmography.

To overcome the convergence issues that arise at high redshifts in traditional Taylor expansions, we employ Pad\'e cosmography, which is particularly well-suited for this purpose. Pad\'e approximants provide improved convergence properties and enable robust comparisons between observational data and theoretical predictions beyond $z \sim 1$, a regime where signatures of new physics are most likely to emerge.

In this study, we present and compare results obtained from both the traditional approach (referred to as the direct method) and the Pad\'e cosmographic method. The novelty of our work lies in: (i) for the first time in the literature, we implement a two-track validation of the same $f(T)$ model, using both Pad\'e${(2,1)}$ cosmography of $d_L(z)$ and direct integration of the modified Friedmann equation, and show their consistency; (ii) performing the analysis with the latest datasets (DESI DR2, Union3, Pantheon+SH0ES, CC, and GW sirens) in a unified Bayesian framework, which has not been jointly applied before; and (iii) demonstrating that Pad\'e cosmography provides a stable, unbiased tool for testing modified gravity. In this chapter, the Pantheon+SH0ES compilation is referred to as SN 22 for brevity.

 This chapter is organized as follows: In \autoref{sec:directapp}, we discuss the background $f(T)$ model utilized in the work. Next, the Pad\'e Cosmography is explored in \autoref{sec:padecosmo}. Following that, a detailed discussion on the results are presented in \autoref{sec:results}. Finally, we conclude with a summary of our findings in \autoref{sec:padeconc}.

\section{Direct Approach}\label{sec:directapp}
In this chapter, we revisit the LSR model introduced in Eq.~(\ref{model;log}) of Chapter~\ref{ch:intro} with a different choice of notations\footnote{
The parameters $\epsilon$ and $\eta$ used in Chapter~\ref{ch:intro} are denoted by $\lambda$ and $\zeta$ here, respectively, for consistency with the conventions of this chapter. Hence, the functional form reads $f(T)=\lambda T_0 \sqrt{\frac{T}{\zeta \, T_0}} \ln\left(\frac{\zeta \, T_0}{T}\right)$.}.
From the structural form of the model, it is evident that the parameter $\zeta$ must remain positive to ensure physical viability. Moreover, an additional constraint on $\zeta$ arises by expressing it in terms of the other model parameter $\lambda$, derived from the present-time evaluation of the Hubble expression (Eq.~(\ref{eq;hubblelsr})). This leads to the following relation
\begin{equation}
    \zeta = \frac{4 \lambda^2}{(1 - \Omega_{m0})^2}.
\end{equation}
Substitution of the above relation makes the Hubble expression (Eq.~(\ref{eq;hubblelsr})) independent of model parameters, and the equation reads
\begin{equation}
    H(z) = \frac{H_0}{\sqrt{2}} \, (1 - \Omega_{m0}) \left[ (1 + 2X) + \sqrt{1 + 4X} \right]^{1/2},
\end{equation}
where $X = \frac{\Omega_{m0}(1 + z)^3}{(1 - \Omega_{m0})^2}
$. As a result, the analysis can be effectively carried out within the reduced parameter space $\{H_0, \Omega_{m0}\}$, which forms the basis for the subsequent cosmological investigation. Given that \( \Omega_{m0} \) is restricted to the prior \([0.2, 0.4]\), it directly follows that \( H(0) = H_0 \) within this domain. We refer to this method as the \textit{Direct Approach} hereafter.

\section{Pad\'e\texorpdfstring{${(2,1)}$}{(2,1)} Cosmography}
\label{sec:padecosmo}
Cosmography offers a kinematic description of the Universe's expansion history by expanding cosmological observables in terms of the redshift $z$, without assuming any specific underlying gravitational theory \cite{Visser:2004bf, Cattoen:2007sk, Luongo:2013rba, Capozziello:2019cav, Capozziello:2017nbu, Luongo:2012dv,Luongo:2011zz,Mishra:2024oln,Kavya:2024bpj,Mishra:2024shg,Aviles:2012ay,Luongo:2012dv,Bamba:2012cp, Capozziello:2008qc,Capozziello:2011hj, Capozziello:2020awd, Sabiee:2022iyo,Mandal:2020buf}.
However, the traditional Taylor series expansion of quantities like the luminosity distance $d_L(z)$ or Hubble parameter $H(z)$ around $z = 0$ suffers from convergence issues at higher redshifts ($z \gtrsim 1$), limiting its applicability to low-$z$ data. To overcome this, rational approximations such as Pad\'e approximants have been proposed \cite{Gruber:2013wua,Capozziello:2018jya}.

The Pad\'e polynomials, originally developed by Steven and Harold \cite{Steven1992}, are constructed from the standard Taylor expansion but offer improved convergence behavior, especially at higher redshifts ($z \gtrsim 1$). Unlike truncated Taylor series, which may diverge or lose accuracy beyond low-redshift regimes, Pad\'e approximants often yield more reliable functional representations and remain effective even in cases where the Taylor series fails to converge \cite{Wei:2013jya}. Owing to these advantageous properties, Pad\'e expansions have been increasingly employed in high-redshift cosmography and have demonstrated superior performance in reconstructing cosmological observables \cite{Wei:2013jya, Demianski:2016zxi, Capozziello:2020ctn}.

The construction of Pad\'e approximation comes from the usual Taylor series expansion of the function $f(z)=\sum_{i=0}^\infty c_iz^i$, where $c_i$ is the set of coefficients. The function is approximated by means of a $(n,m)$ Pad\'e approximation, given by the rational polynomial,
\begin{equation}
P_{n, m}(z)=\frac{\displaystyle{\sum_{i=0}^{n}a_{i} z^{i}}}{1+\displaystyle{\sum_{j=1}^{m}b_j z^{j}}}\,,
\label{eq:defPade}
\end{equation}
which has $n+m+1$ independent coefficients. The Taylor series expansion reproduces the coefficients of the above polynomial up to the maximum possible order as
\begin{align}\label{am}
&P_{n,m}(0)=f(0)\,,\\
&P_{n,m}'(0)=f'(0)\,,\\
&\vdots	\\	
&P_{n,m}^{(n+m)}(0)=f^{(n+m)}(0)\,.
\end{align}
For a more detailed explanation of the Pad\'e approximants, refer to \cite{Capozziello:2020ctn}. Here, we shift our attention to a specific Pad\'e approximation, i.e., Pad\'e${(2,1)}$. The luminosity distance for Pad\'e${(2,1)}$ can be defined as \cite{Hu:2022udt}
\begin{equation}\label{chap4/eq:dlpade}
    d_L(z)=\frac{c}{H_0}\left(\frac{ z \left\{[-2 j_0+q_0 (3 q_0+8)-5]z+6 (q_0-1)\right\}}{ 2 q_0 (3 q_0 z+z+3)-2 (j_0 z+z+3)}\right).
\end{equation}
Further, we note down the Hubble expression calculated from the above $d_L$
\begin{equation}\label{chap4/eq:hubpade}
H(z) = \frac{2 H_0 (1+z)^2 \left[\mathcal{N}(z)\right]^2}{\mathcal{D}(z)},
\end{equation}
where
\begin{align}
\mathcal{N}(z) &= j_0 z - q_0 (3 q_0 z + z + 3) + z + 3, \\
\mathcal{D}(z) &= z^2 \big[2 j_0^2 + j_0 (7 - 9 q_0^2 - 10 q_0) + q_0 (9 q_0^3 + 18 q_0^2 + 17 q_0 - 40) + 14 \big] \notag \\
&\quad + 6 (q_0 - 1) z \left[-2 j_0 + q_0 (3 q_0 + 8) - 5 \right] 
+ 18 (q_0 - 1)^2.
\end{align}
In what follows, we continue to employ the set of Eqs.~(\ref{chap3/eq:H(t)})--(\ref{eq:derT}) introduced in the previous chapter.
\subsection{\texorpdfstring{$f(T)$}{f(T)} Cosmography}
Performing up to the third derivative of the first motion equation (Eq.~(\ref{eq:ftmot1})), and substituting the expressions for the Hubble parameter derivatives (Eq.~(\ref{eq:derH})) and the torsion scalar derivatives (Eq.~(\ref{eq:derT})), yields a system of three equations involving the three unknowns $\{q_0,j_o,s_0\}$. Solving the system, one can achieve

\begin{align}
\begin{split}
q_0 =& \left[3 - \Omega_{m0} \left(4 + \Omega_{m0} + 3 \epsilon \right)\right]
\Big/ \left[(-3 + \Omega_{m0})(1 + \Omega_{m0})\right],
\end{split}\\
\begin{split}
    j_0 =& \Big[
-27 + \Omega_{m0} \Big(
-54 + 18 \Omega_{m0} \left(6 + 7 \epsilon \right)+ \Omega_{m0}\big
\{
-4 \left(56 + 27 \epsilon \right)   \\&
  + \Omega_{m0} \left[165 + 2 \Omega_{m0} (-21 + 5 \Omega_{m0}) + 54 \epsilon \right] 
\big\} \Big) \Big] \Big/ \left[(-3 + \Omega_{m0})^3 (1 + \Omega_{m0})^3\right],
\end{split}\\
\begin{split}
   s_0 =& \Bigg[
-243 + \Omega_{m0} \Big(
81 \left(8 + 9 \epsilon \right) 
- 54 \Omega_{m0} \left(38 + 69 \epsilon \right) + \Omega_{m0} \Big\{
14715 \epsilon+22986 \\&
  - 6 \Omega_{m0} \left(6631 + 4362 \epsilon \right) + \Omega_{m0} \Big[
44350 + 21699 \epsilon
+ \Omega_{m0} \big(
-35200 - 11070 \epsilon \\&   + \Omega_{m0} \left\{
53 (302 + 81 \epsilon) 
+ \Omega_{m0} \left[
-3827 - 504 \epsilon + 2 \Omega_{m0} \left(373 - 22 \Omega_{m0} 
+ 18 \epsilon \right)
\right] \right\} \big) \Big] \Big\} \Big) \Bigg] \\&
\Big/ \left[(-3 + \Omega_{m0})^5 (1 + \Omega_{m0})^5\right], 
\end{split}
\end{align}
where $-1 + \Omega_{m0}\ =\epsilon$.

\begin{figure*}
    \centering
    \includegraphics[width=0.4\linewidth]{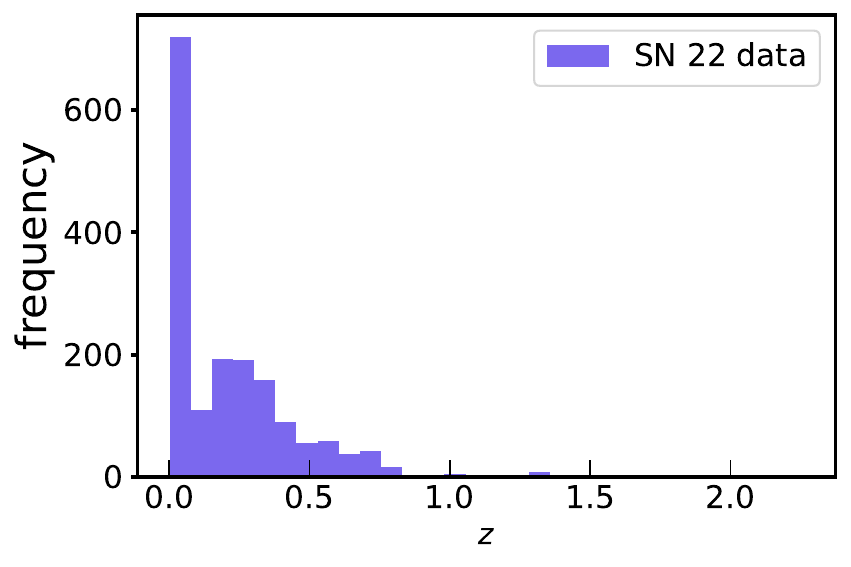}
    \includegraphics[width=0.39\linewidth]{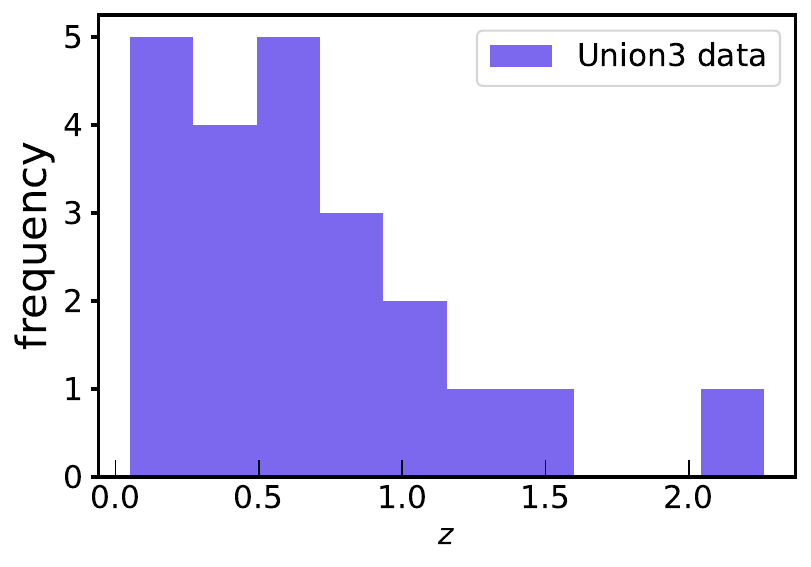}\\
    \includegraphics[width=0.4\linewidth]{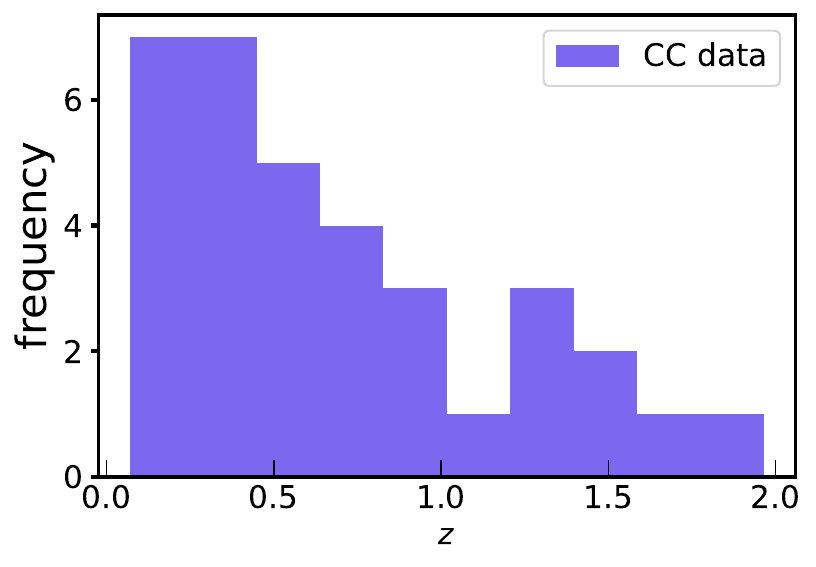}
    \includegraphics[width=0.4\linewidth]{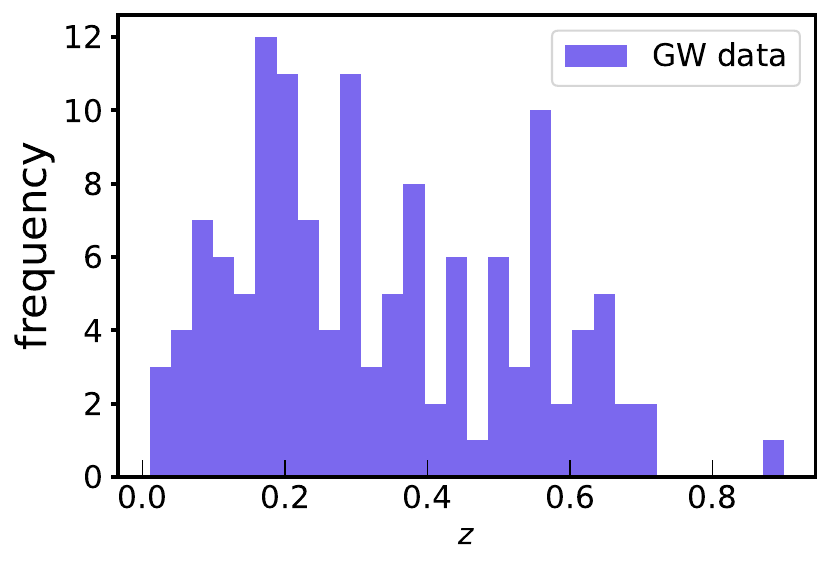}
    \caption{Figure representing the distribution of data used in MCMC over the redshift $z$.}
    \label{fig:hist}
\end{figure*}

The above solution set can be substituted into Eq.~(\ref{chap4/eq:dlpade}) and Eq.~(\ref{chap4/eq:hubpade}) to derive the theoretical predictions for $d_L(z)$ and $H(z)$. This facilitates a convenient implementation of parameter estimation for the model within the framework of the Pad\'e cosmographic approach. 
\autoref{fig:hist} illustrates the redshift distribution of the data points for all the datasets utilized in this analysis. In the Appendices, a brief overview of the sources and methodologies associated with each dataset is discussed.

\begin{figure}
    \centering
    \includegraphics[width=0.5\linewidth]{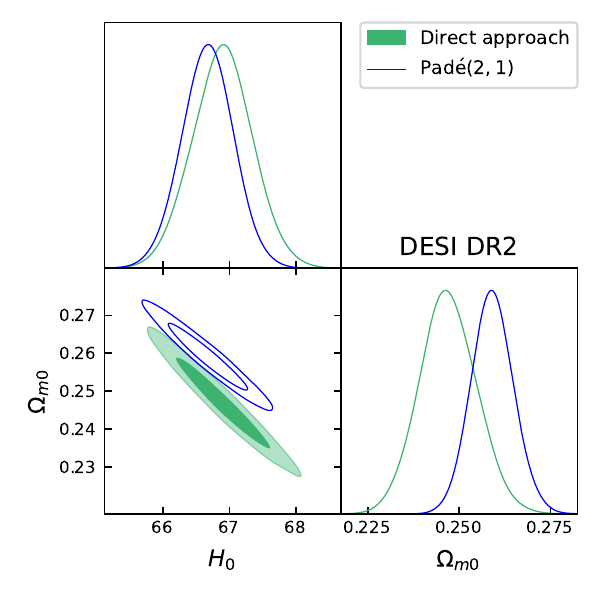}
    \caption{Two-dimensional contours for the parameter space $\{H_0,\Omega_{m0}\}$ using DESI DR2 dataset.}
    \label{fig:desicon}
\end{figure}
\begin{figure}
    \centering
    \includegraphics[width=0.5\linewidth]{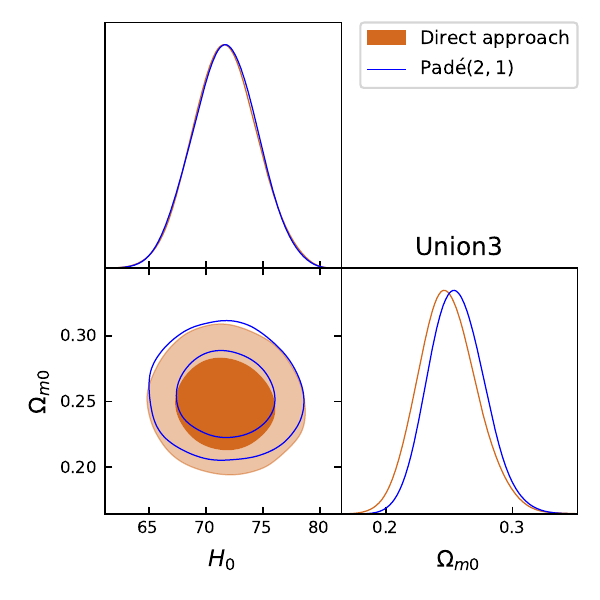}
    \caption{Two-dimensional contours for the parameter space $\{H_0,\Omega_{m0}\}$ using Union3 dataset.}
    \label{fig:unicon}
\end{figure}
\begin{figure}
    \centering
    \includegraphics[width=0.5\linewidth]{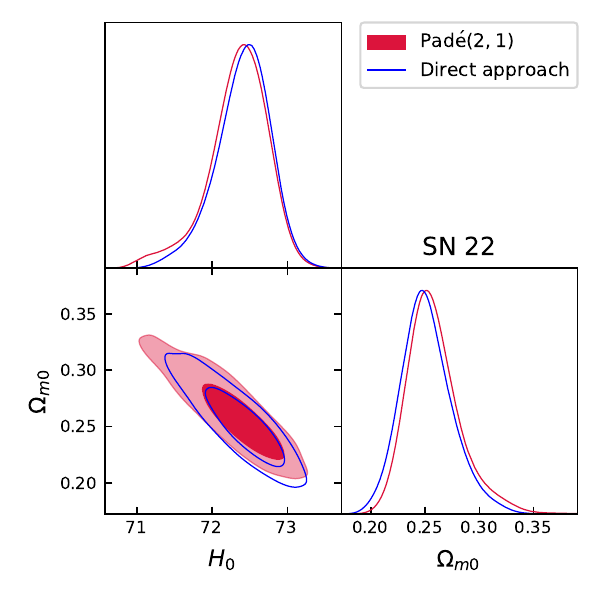}
    \caption{Two-dimensional contours for the parameter space $\{H_0,\Omega_{m0}\}$ using SN 22 dataset.}
    \label{fig:sncon}
\end{figure}
\begin{figure}
    \centering
    \includegraphics[width=0.5\linewidth]{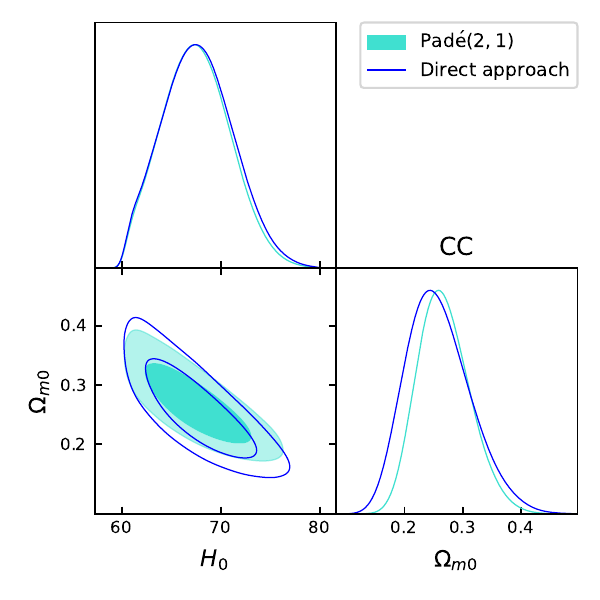}
    \caption{Two-dimensional contours for the parameter space $\{H_0,\Omega_{m0}\}$ using CC dataset.}
    \label{fig:cccon}
\end{figure}
\begin{figure}
    \centering
    \includegraphics[width=0.5\linewidth]{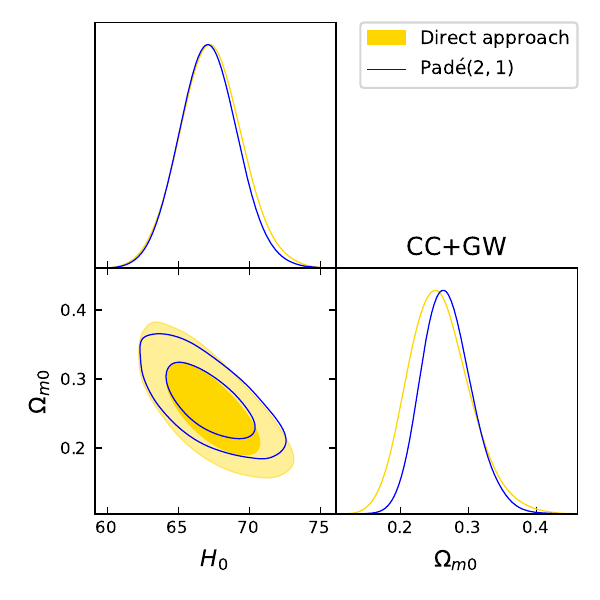}
    \caption{Two-dimensional contours for the parameter space $\{H_0,\Omega_{m0}\}$ using CC+GW dataset.}
    \label{fig:ccgwcon}
\end{figure}
\begin{figure}
    \centering
    \includegraphics[width=0.6\linewidth]{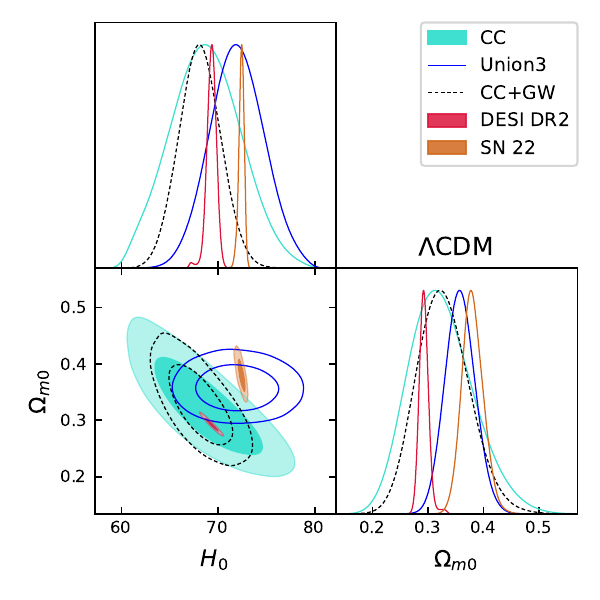}
    \caption{The contour plot of $\Lambda$CDM model for all datasets.}
    \label{fig:lcdmcon}
\end{figure}

\section{Results}\label{sec:results}
In this section, we present the results of our analysis along with their cosmological implications. The contours in Figures~\ref{fig:desicon} -- \ref{fig:ccgwcon} represent a comparative plot between the two approaches: the first is obtained from the Pad\'e approximation, and the second from the first motion equation (Direct approach). While the second approach is commonly adopted in the literature, constraining a model using Pad\'e cosmography offers a novel perspective. It is evident that, although the best-fit mean values are quite similar, the associated uncertainties differ. Furthermore, some discrepancies are also observed in the resulting values of the cosmological parameters. In this context, we intend to investigate the further implications of these differences in detail. The obtained numerical $1\sigma$ values of the involved parameters for both methods are summarized in \autoref{tab:padeparams} and \autoref{tab:directparams}, respectively.

\begin{figure}
    \centering
    \includegraphics[width=0.49\linewidth]{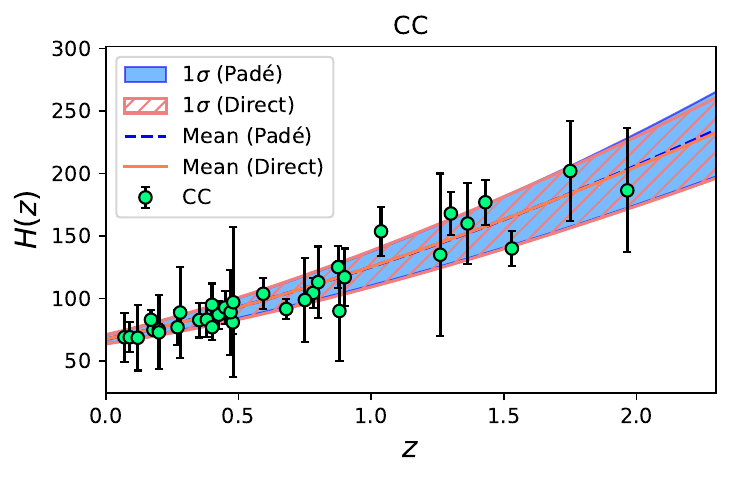}
    \includegraphics[width=0.49\linewidth]{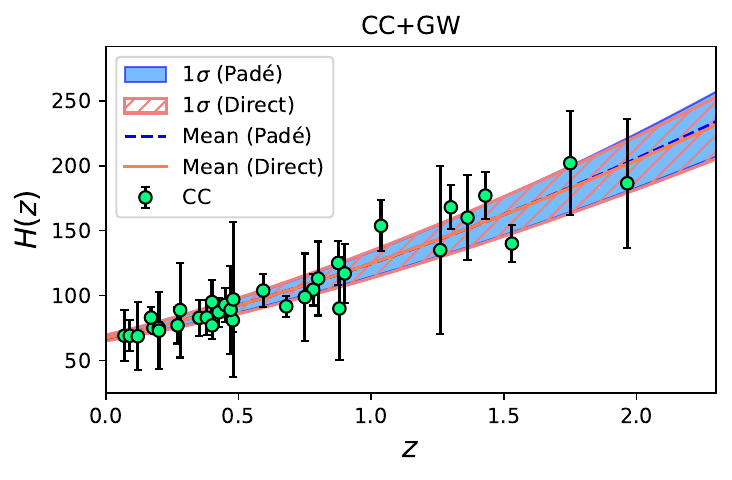}\\
    \includegraphics[width=0.49\linewidth]{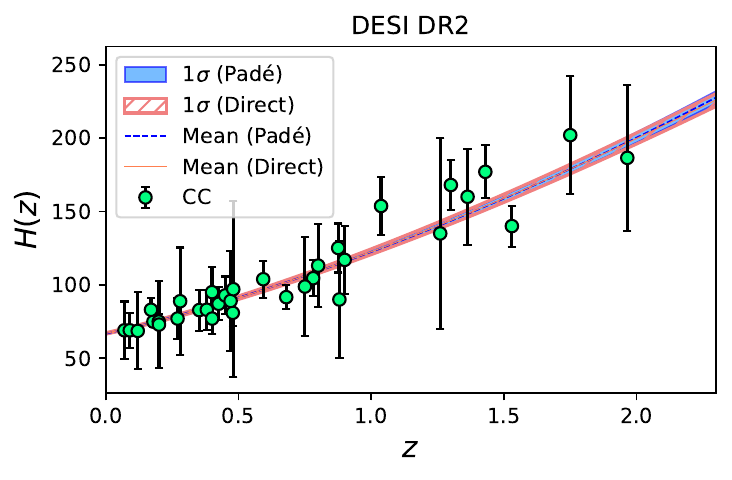}
    \includegraphics[width=0.49\linewidth]{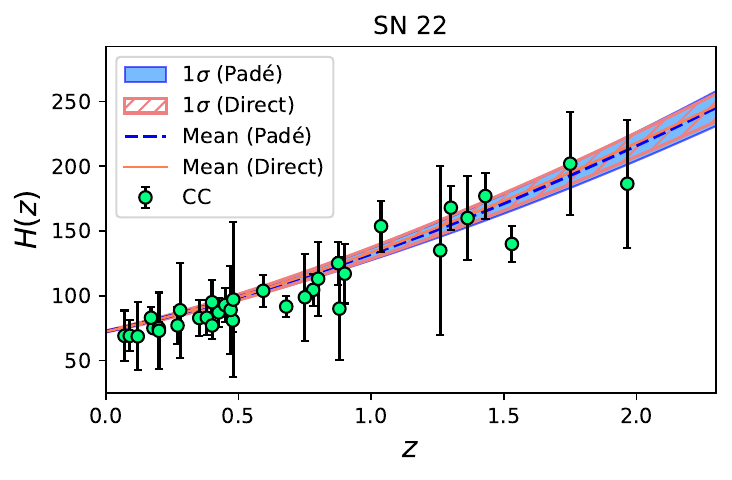}\\
    \includegraphics[width=0.5\linewidth]{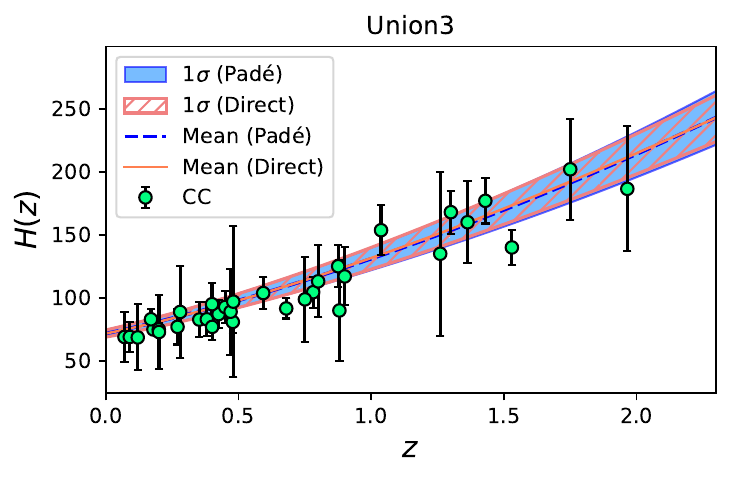}
    \caption{The Hubble profile for the resulting values of each case against the 34 error bars of CC data}
    \label{fig:herr}
\end{figure}

\begin{figure}
    \centering
    \includegraphics[width=0.49\linewidth]{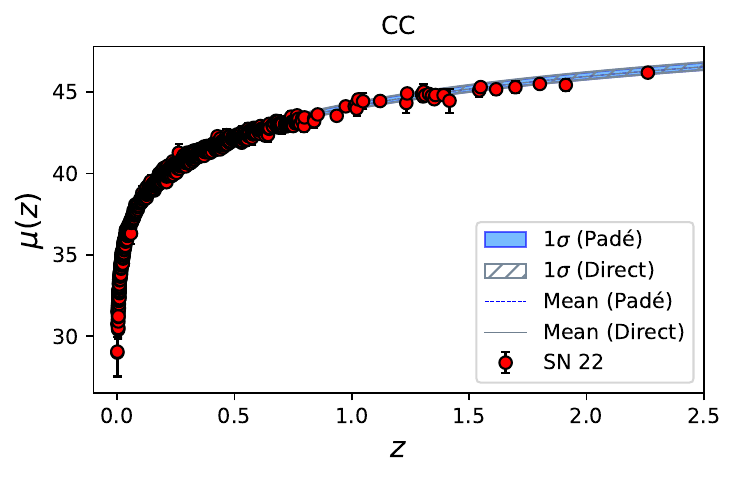}
    \includegraphics[width=0.49\linewidth]{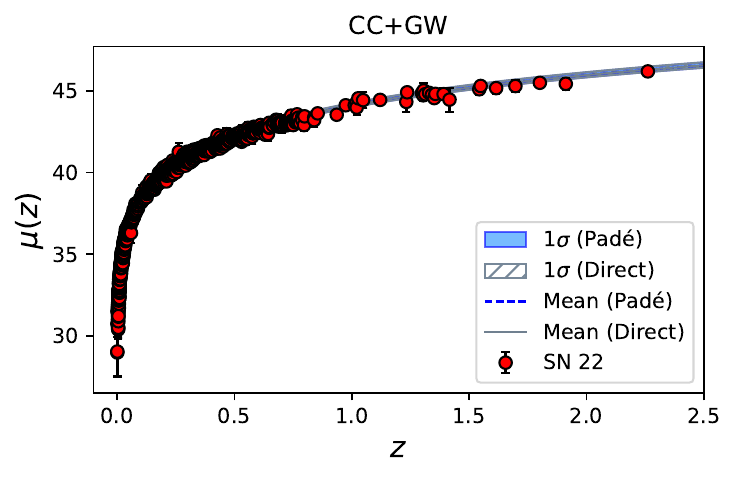}\\
    \includegraphics[width=0.49\linewidth]{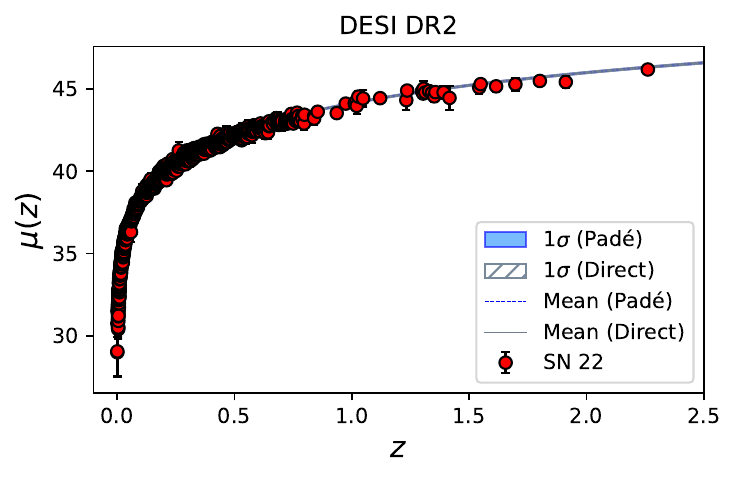}
    \includegraphics[width=0.49\linewidth]{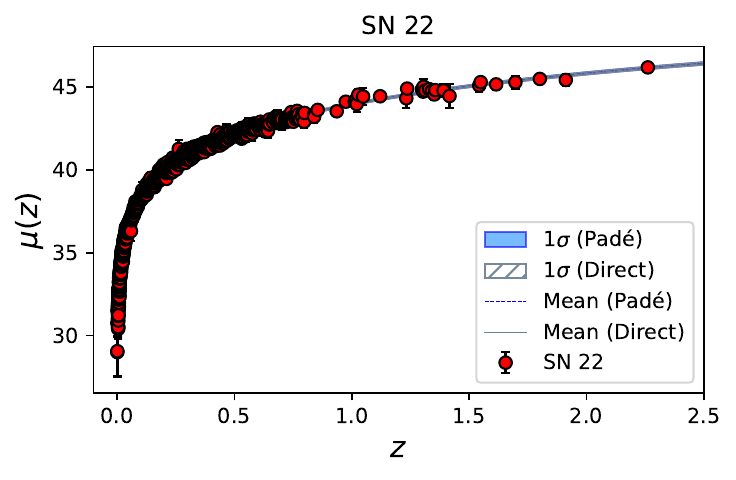}\\
    \includegraphics[width=0.49\linewidth]{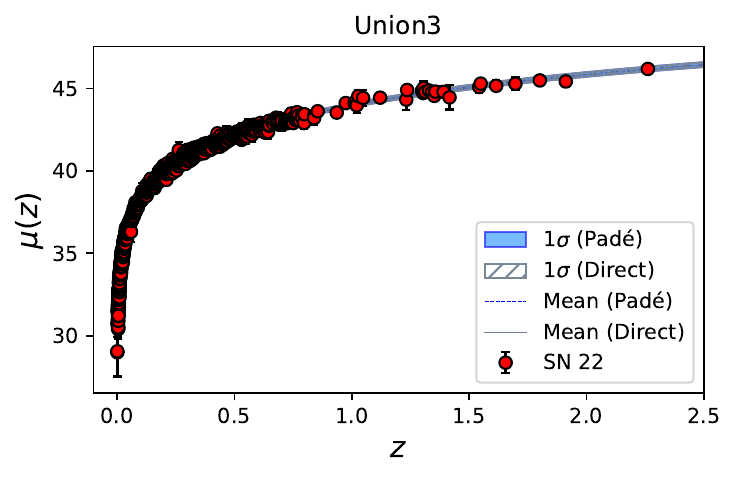}
    \caption{The Distance modulus profile for the resulting values of each case against the 1701 error bars of Pantheon+SH0ES data}
    \label{fig:muerrsn}
\end{figure}

\begin{figure}
    \centering
    \includegraphics[width=0.49\linewidth]{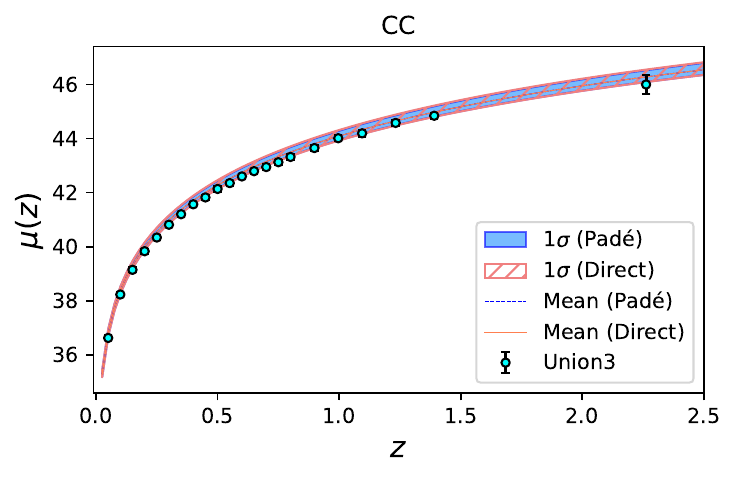}
    \includegraphics[width=0.49\linewidth]{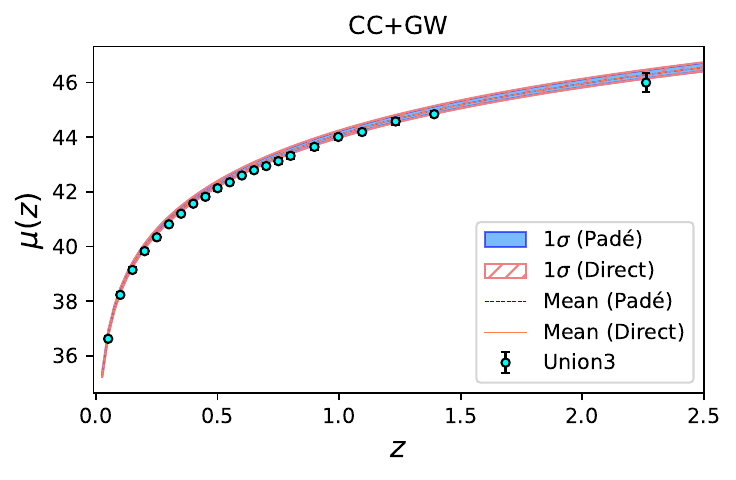}\\
    \includegraphics[width=0.49\linewidth]{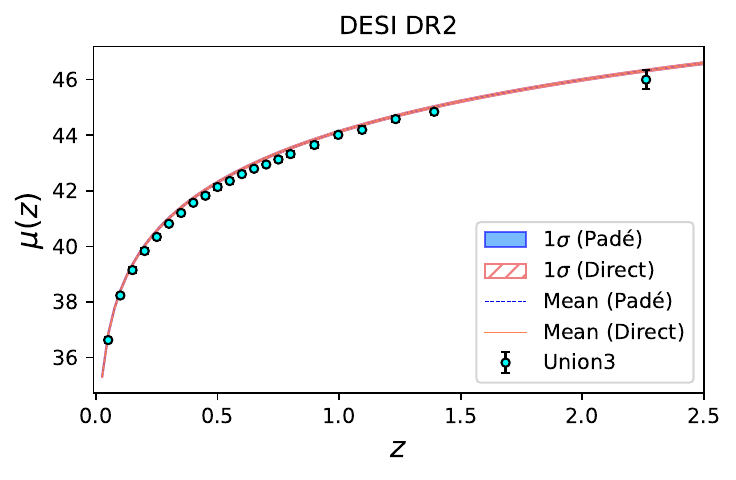}
    \includegraphics[width=0.49\linewidth]{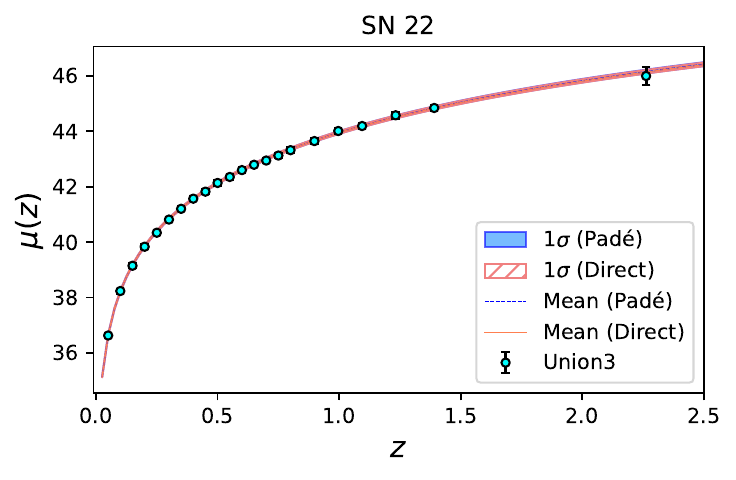}\\
    \includegraphics[width=0.49\linewidth]{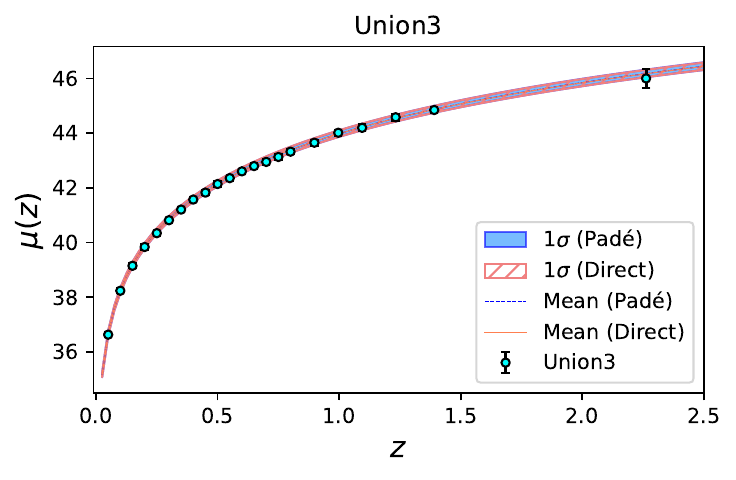}
    \caption{The Distance modulus profile for the resulting values of each case against the 22 error bars of Union3 data}
    \label{fig:muerruni}
\end{figure}

\begin{figure}
    \centering
    \includegraphics[width=0.49\linewidth]{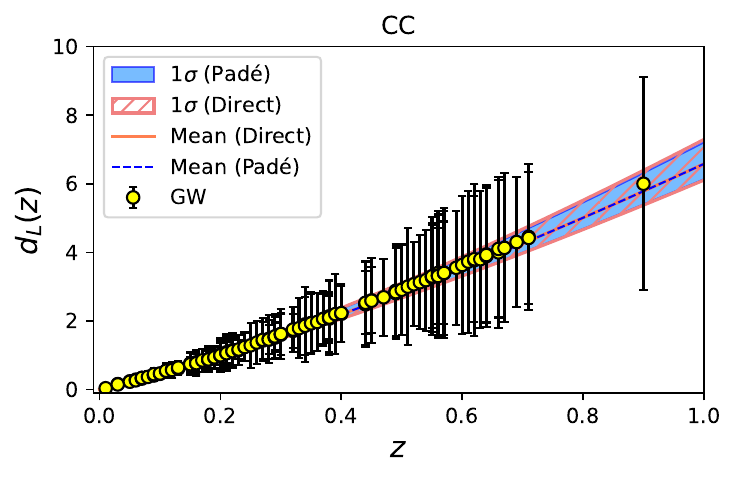}
    \includegraphics[width=0.49\linewidth]{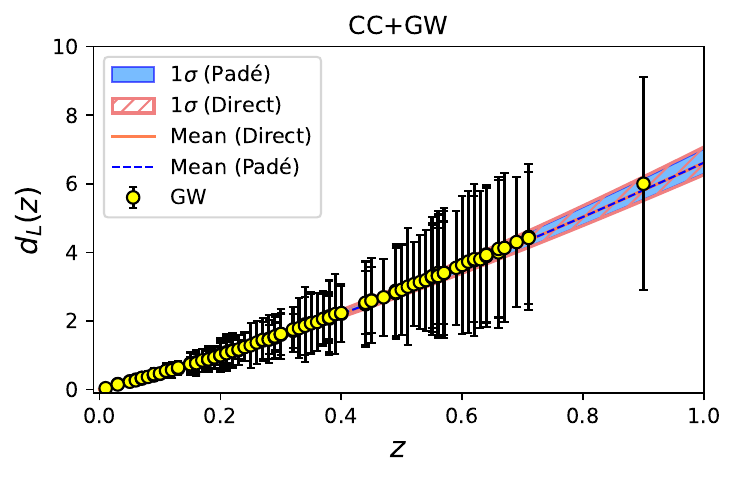}\\
    \includegraphics[width=0.49\linewidth]{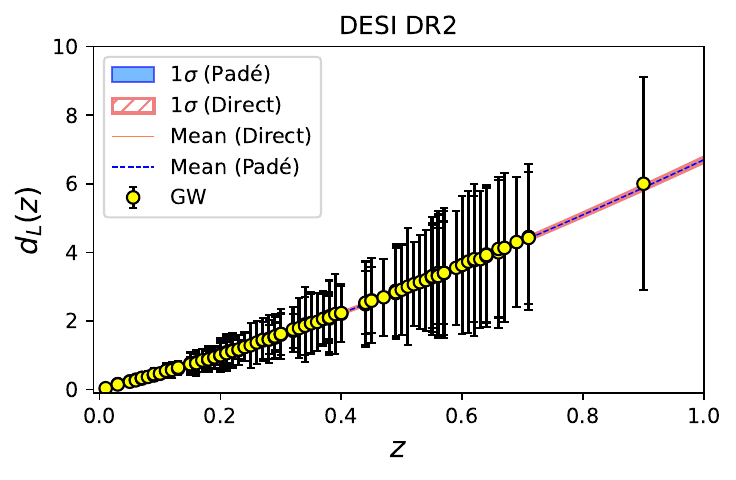}
    \includegraphics[width=0.49\linewidth]{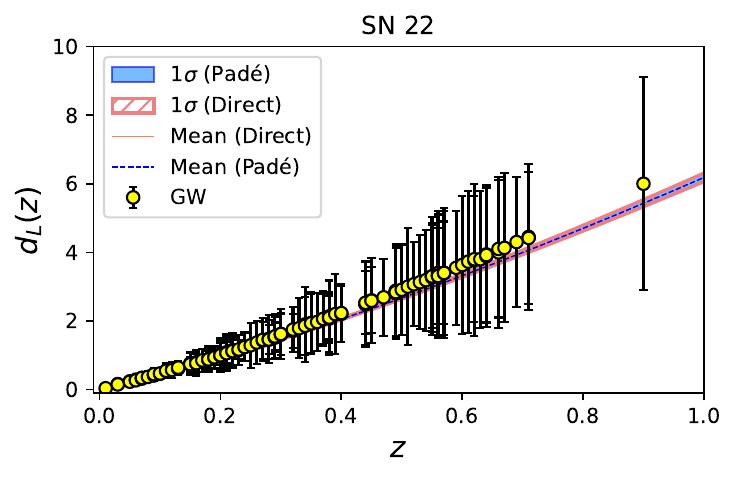}\\
    \includegraphics[width=0.49\linewidth]{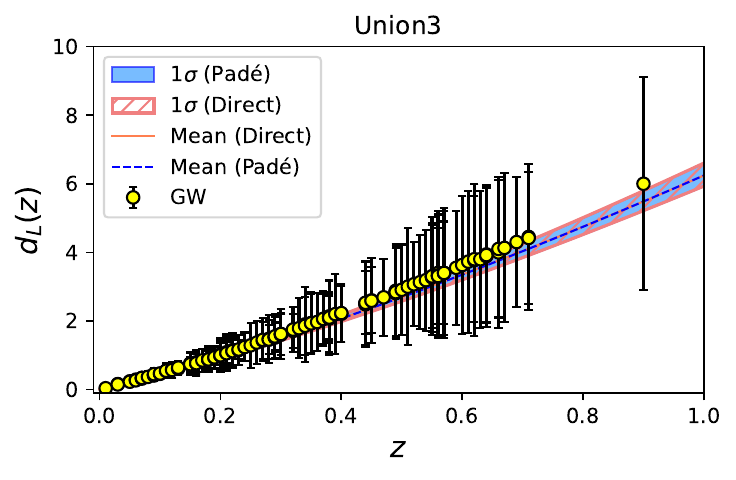}
    \caption{The Luminosity distance profile for the resulting values of each case against the error bars of Gravitational Wave data.}
    \label{fig:dlerr}
\end{figure}

\begin{figure}
    \centering
    \includegraphics[width=0.49\linewidth]{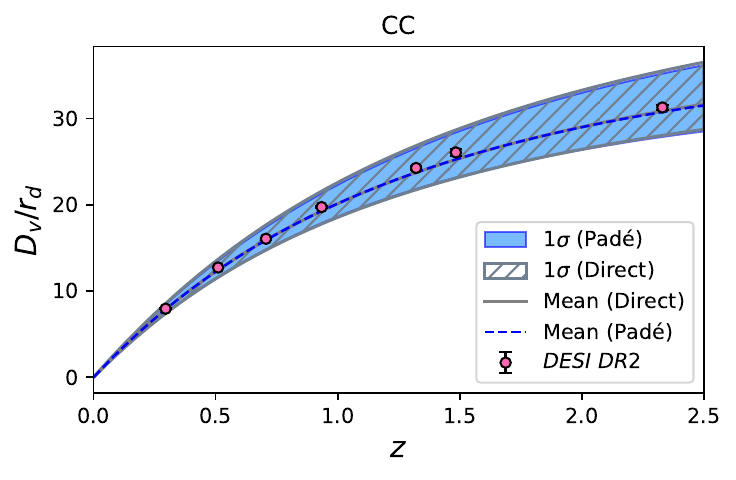}
    \includegraphics[width=0.49\linewidth]{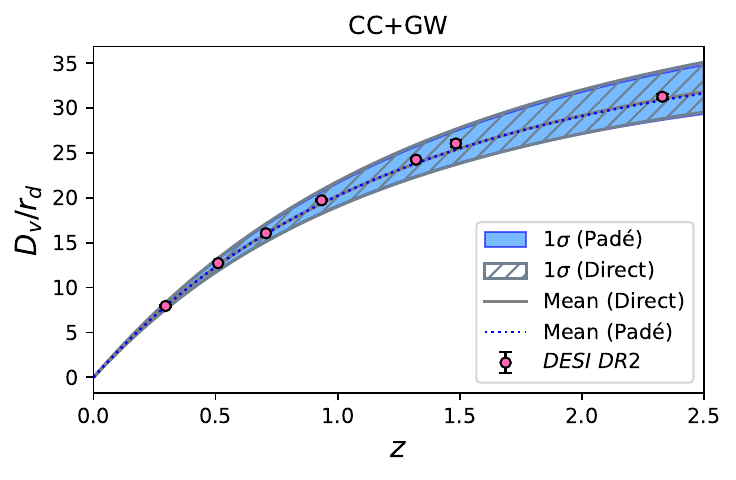}\\
    \includegraphics[width=0.49\linewidth]{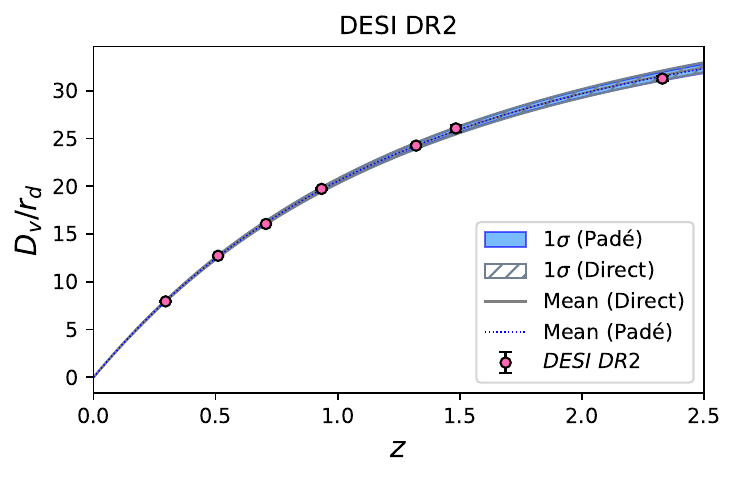}
    \includegraphics[width=0.49\linewidth]{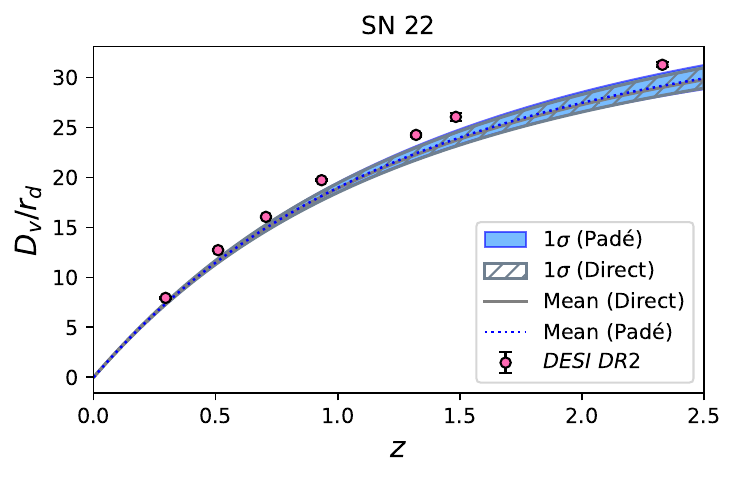}\\
    \includegraphics[width=0.49\linewidth]{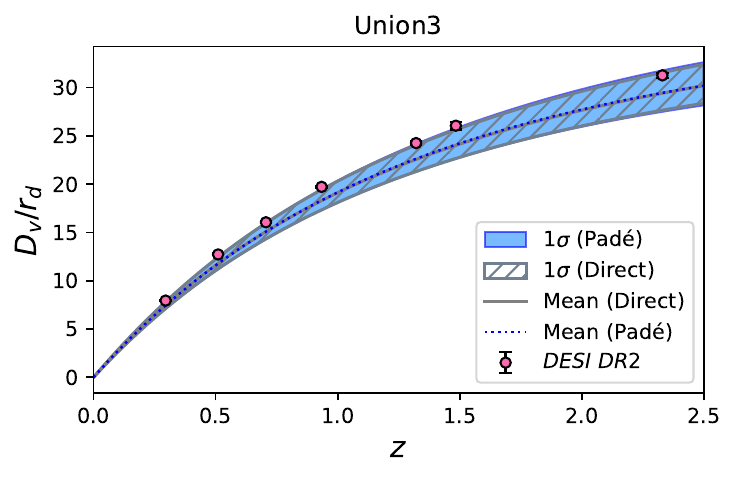}
    \caption{The $D_v/r_d$ profile against DESI BAO measurements.}
    \label{fig:dv}
\end{figure}

\begin{figure}
    \centering
    \includegraphics[width=0.49\linewidth]{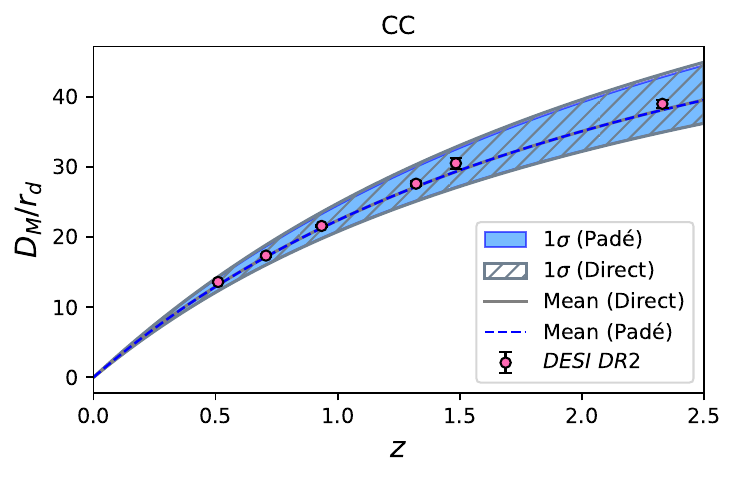}
    \includegraphics[width=0.49\linewidth]{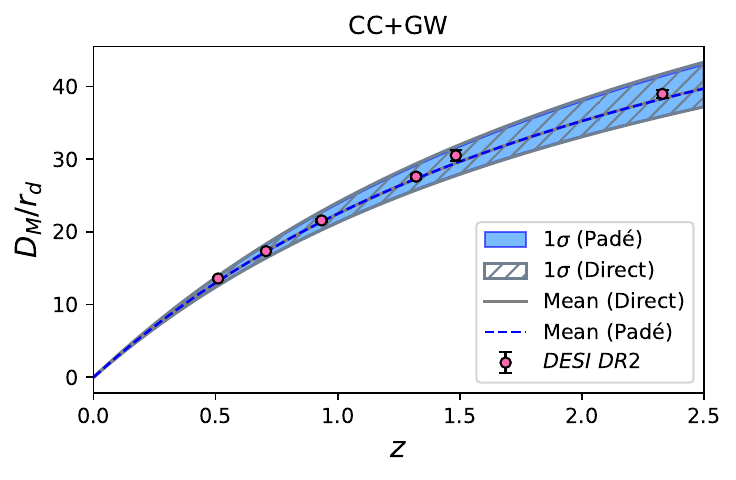}\\
    \includegraphics[width=0.49\linewidth]{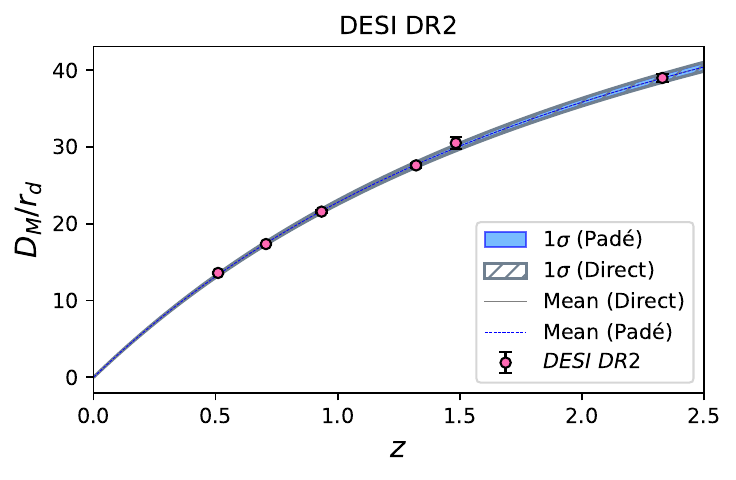}
    \includegraphics[width=0.49\linewidth]{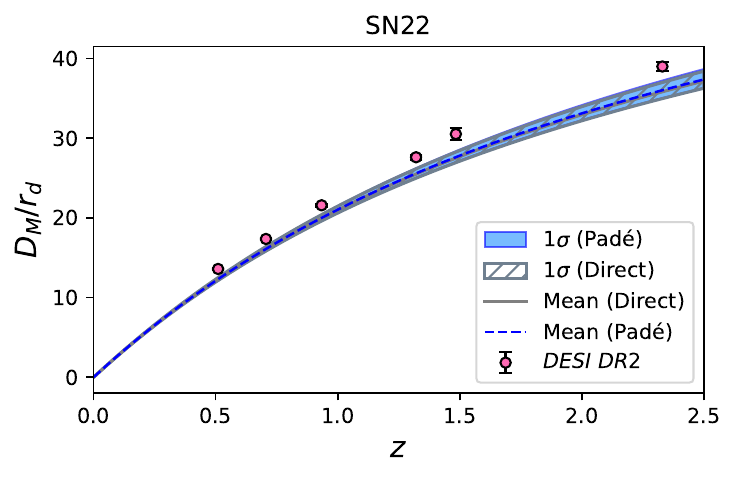}\\
    \includegraphics[width=0.49\linewidth]{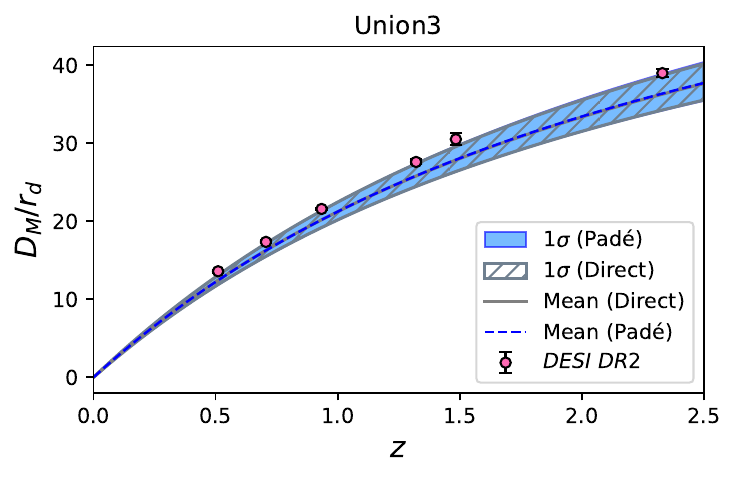}
    \caption{The $D_M/r_d$ profile against DESI BAO measurements.}
    \label{fig:dm}
\end{figure}

\begin{figure}
    \centering
    \includegraphics[width=0.49\linewidth]{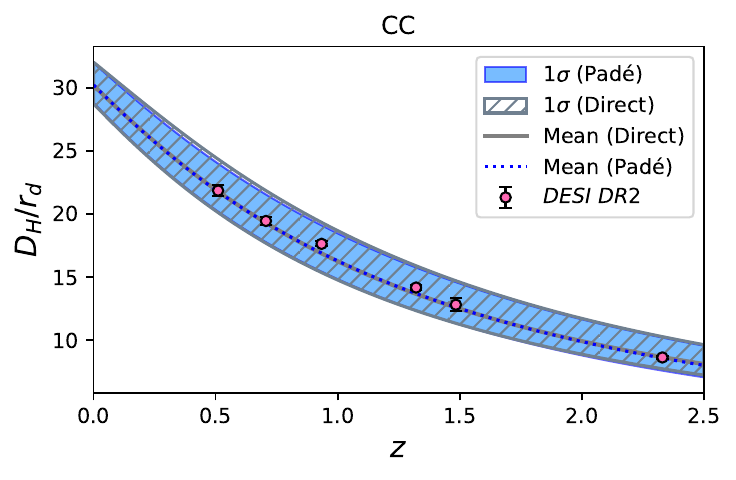}
    \includegraphics[width=0.49\linewidth]{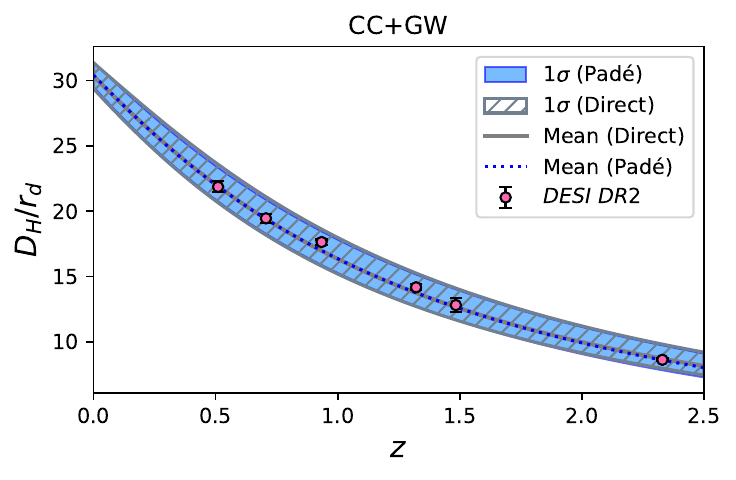}\\
    \includegraphics[width=0.49\linewidth]{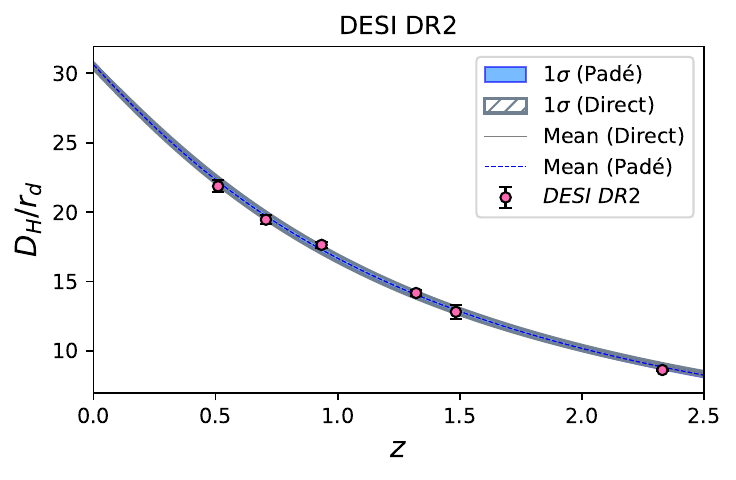}
    \includegraphics[width=0.49\linewidth]{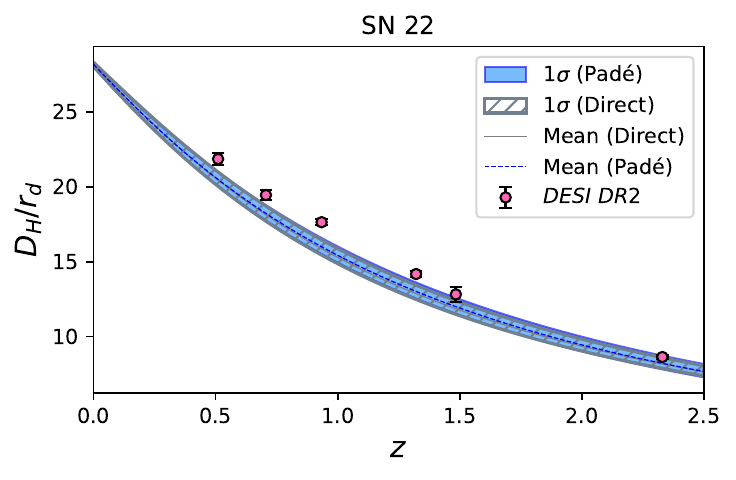}\\
    \includegraphics[width=0.49\linewidth]{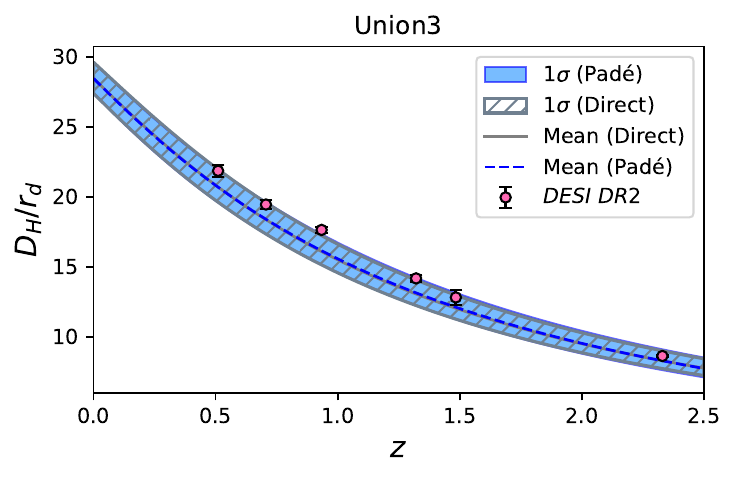}
    \caption{The $D_H/r_d$ profile against DESI BAO measurements.}
    \label{fig:dh}
\end{figure}

For each dataset, the obtained mean values and corresponding $1\sigma$ uncertainties are validated by comparing them with the respective error bars of certain cosmological parameters. Throughout the figures, the two methods are distinguished by different visual styles, one depicted using solid colors and the other with hatched patterns. In \autoref{fig:herr}, we specifically present the Hubble parameter as a function of redshift, along with the 34 error bars corresponding to the CC data. Further, the distance modulus is plotted with the Pantheon+SH0ES dataset in \autoref{fig:muerrsn} and with the Union3 dataset in \autoref{fig:muerruni}. Since GWs provide a direct measurement of the luminosity distance, we plot it as a function of redshift, including the associated error bars from the GW data, in \autoref{fig:dlerr}. This provides an independent probe of the cosmic expansion history, complementing the constraints from electromagnetic observations. Finally, the three distance measures, namely the volume-averaged distance $D_V/r_d$, the comoving angular diameter distance $D_M/r_d$, and the Hubble distance $D_H/r_d$, are plotted against the DESI DR2 data in Figures~\ref{fig:dv} -- \ref{fig:dh}, respectively.

\begin{figure}
    \centering
    \includegraphics[width=0.49\linewidth]{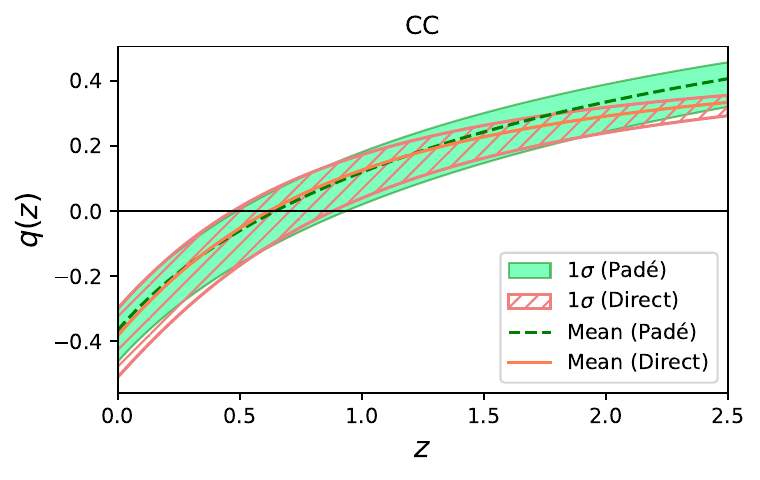}
    \includegraphics[width=0.49\linewidth]{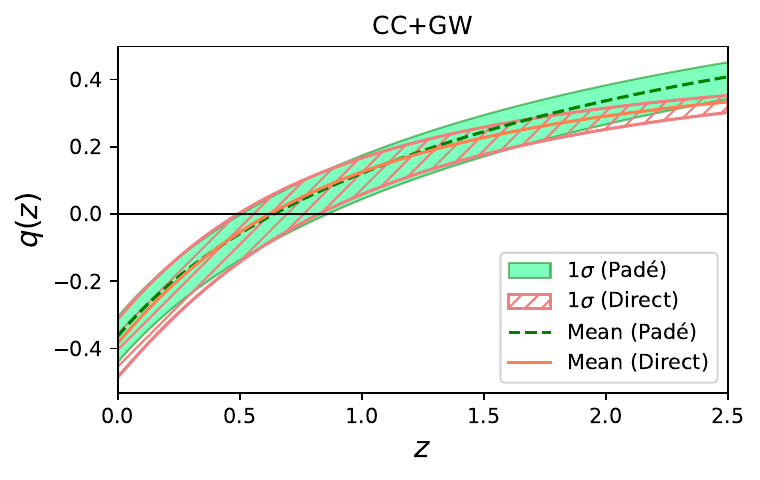}\\
    \includegraphics[width=0.49\linewidth]{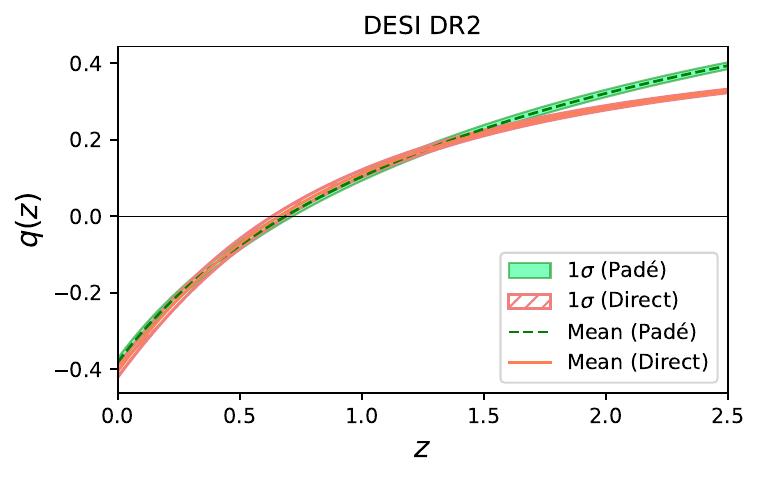}
    \includegraphics[width=0.49\linewidth]{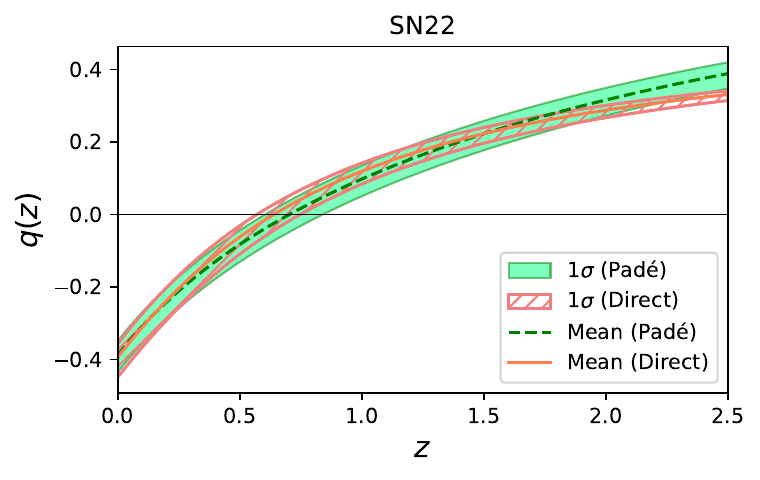}\\
    \includegraphics[width=0.49\linewidth]{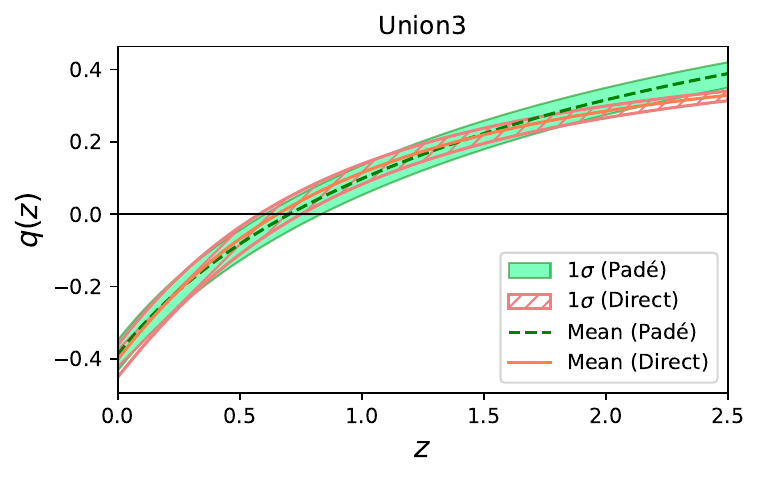}
    \caption{The evolution of the Universe from decelerated to accelerated phase.}
    \label{fig:q}
\end{figure}

Next, we consider two key cosmological parameters that characterize the evolution of the Universe over a range of redshifts: the deceleration parameter $q(z)$ and the effective equation of state parameter $\omega_{\mathrm{eff}}(z)$. From \autoref{fig:q}, it can be observed that all cases exhibit a transition of the Universe from a decelerating phase to an accelerating phase. It is well known that the transition redshift $(z_t)$ for the $\Lambda$CDM model is $z_t \sim 0.67$ \cite{Mishra:2025msl}. The $1\sigma$ estimates of $z_t$ obtained from the Pad\'e method and the direct method, as listed in \autoref{tab:padezt} and \ref{tab:directzt}, respectively, show excellent agreement with the $\Lambda$CDM value. The effective equation of state parameter, \( \omega_{\mathrm{eff}} \), serves as a comprehensive diagnostic for understanding the expansion history of the Universe. It characterizes the net pressure-to-density ratio of the cosmic fluid, encapsulating the combined dynamical effects of all energy components. A value of \( \omega_{\mathrm{eff}} = 0 \) corresponds to a pressureless matter-dominated era, typically associated with high redshifts. For cosmic acceleration to occur, the condition \( \omega_{\mathrm{eff}} < -\tfrac{1}{3} \) must be satisfied. In the standard \( \Lambda \)CDM model, \( \omega_{\mathrm{eff}} \) smoothly transitions from \( 0 \) in the early Universe to approximately \( -0.7 \) at the present epoch, reflecting the growing influence of the cosmological constant.

\begin{figure}
    \centering
    \includegraphics[width=0.49\linewidth]{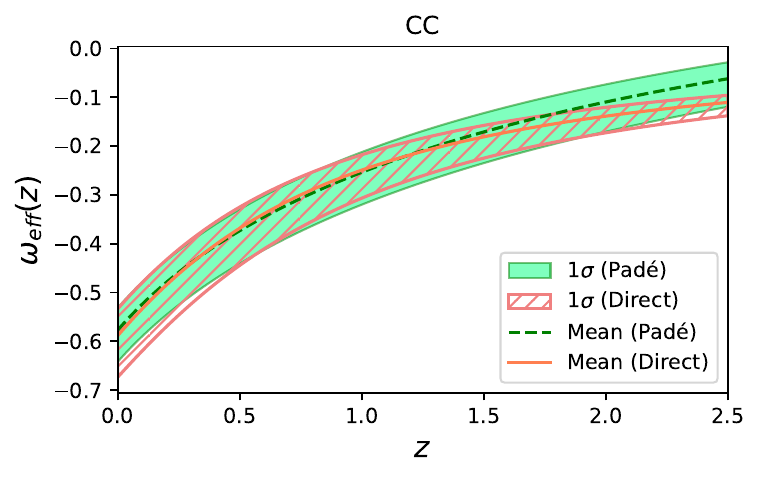}
    \includegraphics[width=0.49\linewidth]{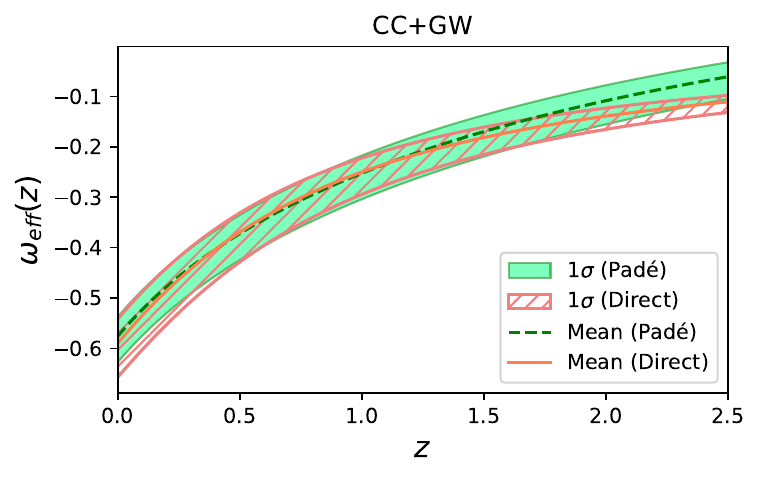}\\
    \includegraphics[width=0.49\linewidth]{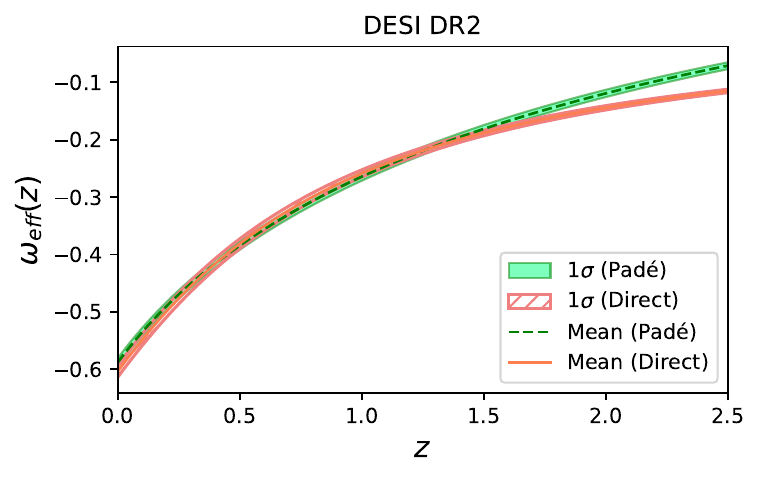}
    \includegraphics[width=0.49\linewidth]{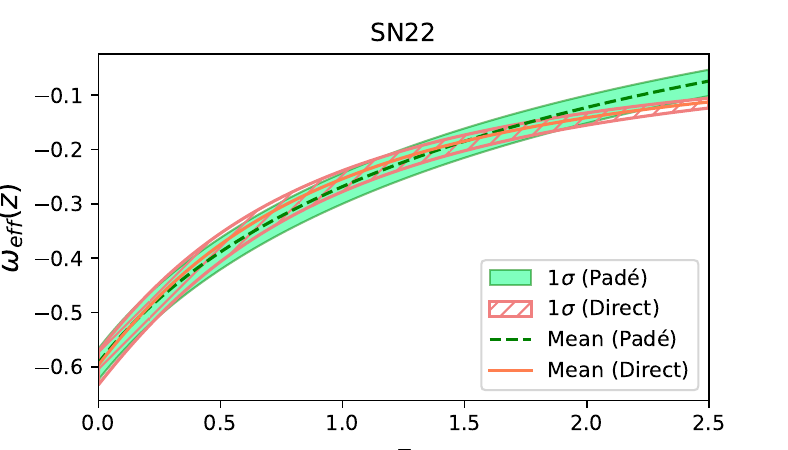}\\
    \includegraphics[width=0.49\linewidth]{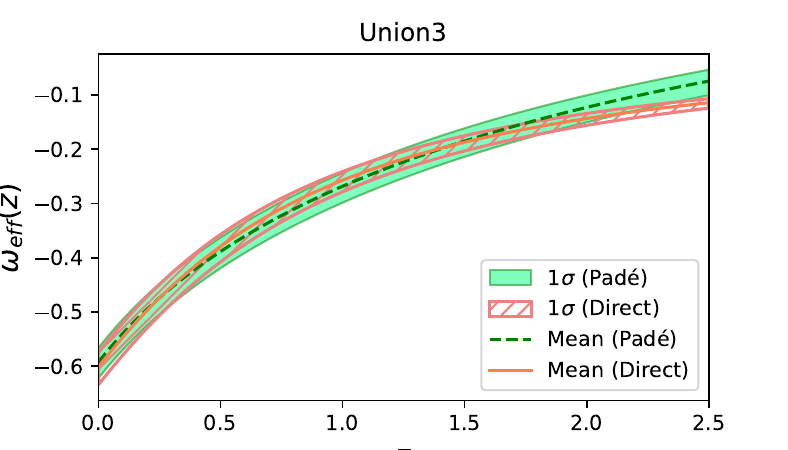}
    \caption{The effective EoS parameter indicating a quintessence behavior.}
    \label{fig:w}
\end{figure}

From \autoref{fig:w}, it is evident that both the Padé and Direct models successfully capture this expected evolution of \( \omega_{\mathrm{eff}} \). At early times, the Universe undergoes a matter-dominated phase with \( \omega_{\mathrm{eff}} \approx 0 \), followed by a transition to an accelerated phase where \( \omega_{\mathrm{eff}} \) drops below \( -\tfrac{1}{3} \). These results reinforce the capability of our models to replicate the key dynamical features of the standard cosmological scenario while offering flexibility in capturing possible deviations at different redshifts.

\begin{figure}
    \centering
    \includegraphics[width=0.45\linewidth]{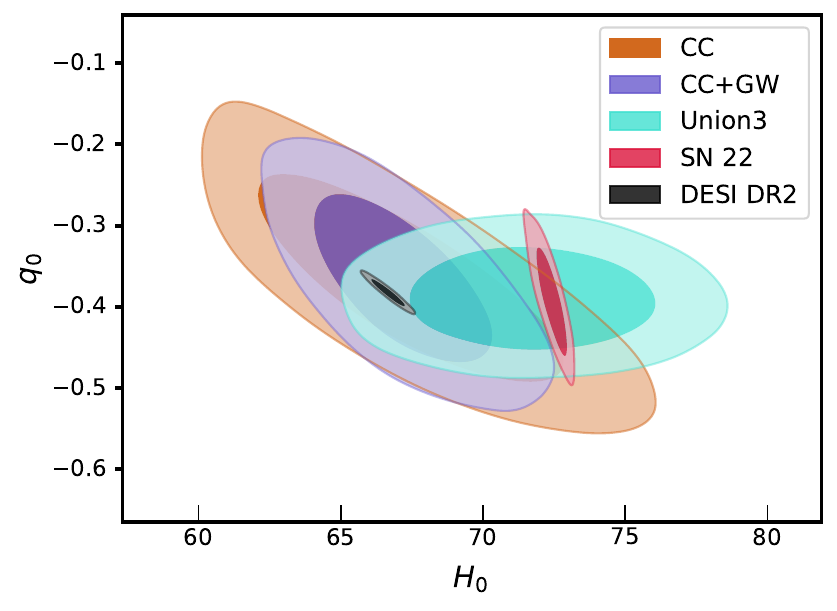}
    \includegraphics[width=0.45\linewidth]{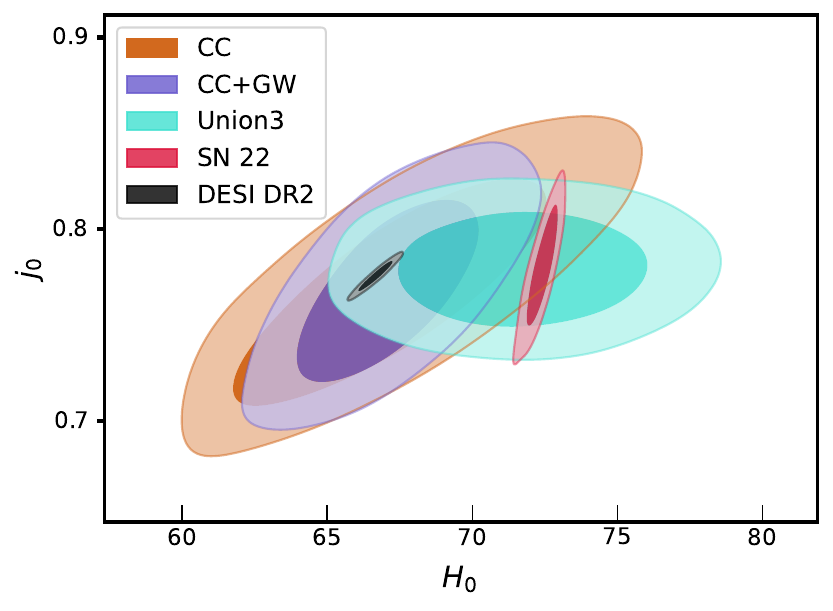}\\
    \includegraphics[width=0.45\linewidth]{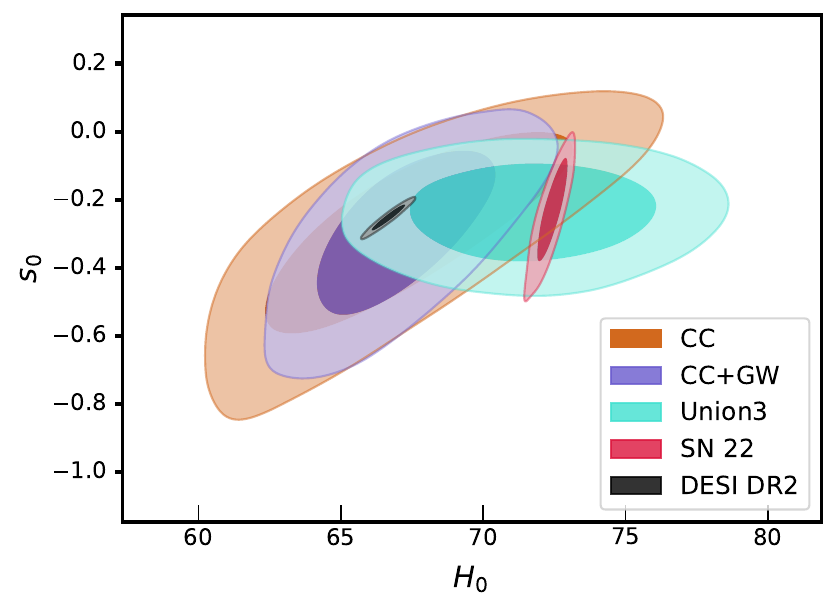}
    \includegraphics[width=0.45\linewidth]{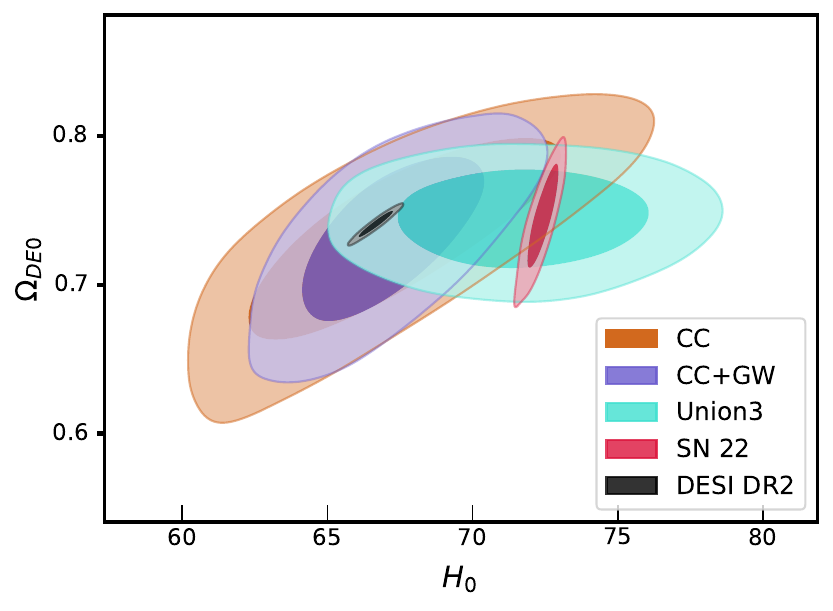}
    \caption{Plot depicting the 2D likelihood contour for cosmological parameters for Pad\'e polynomial}
    \label{fig:pade_parameters}
\end{figure}

\begin{figure}
    \centering
    \includegraphics[width=0.45\linewidth]{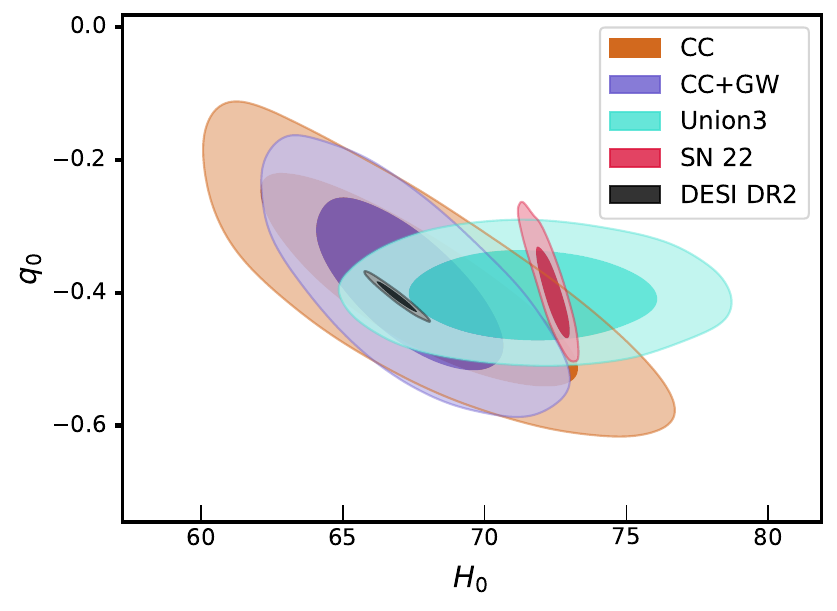}
    \includegraphics[width=0.45\linewidth]{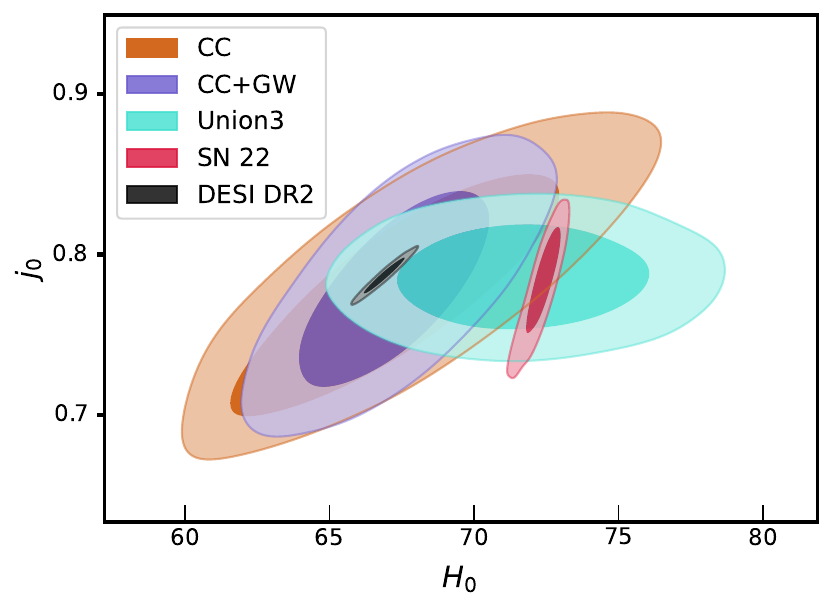}\\
    \includegraphics[width=0.45\linewidth]{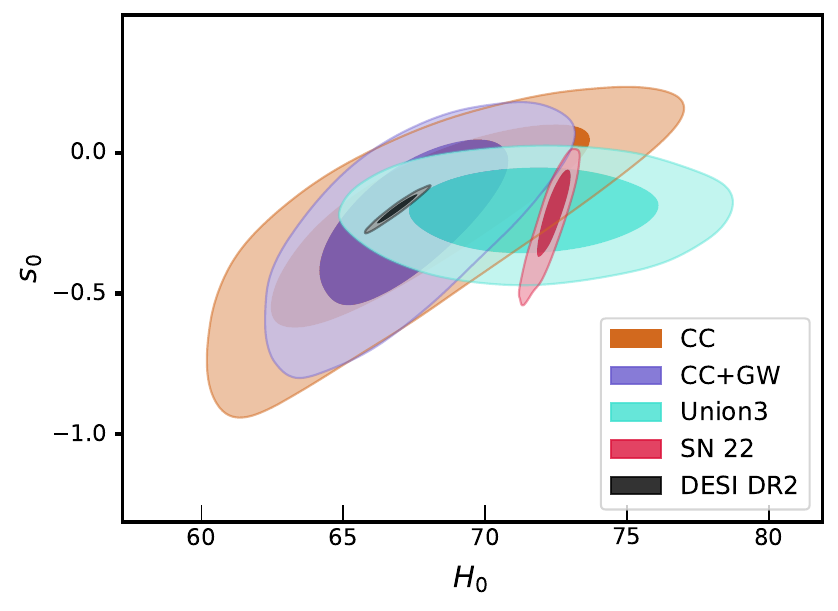}
    \includegraphics[width=0.45\linewidth]{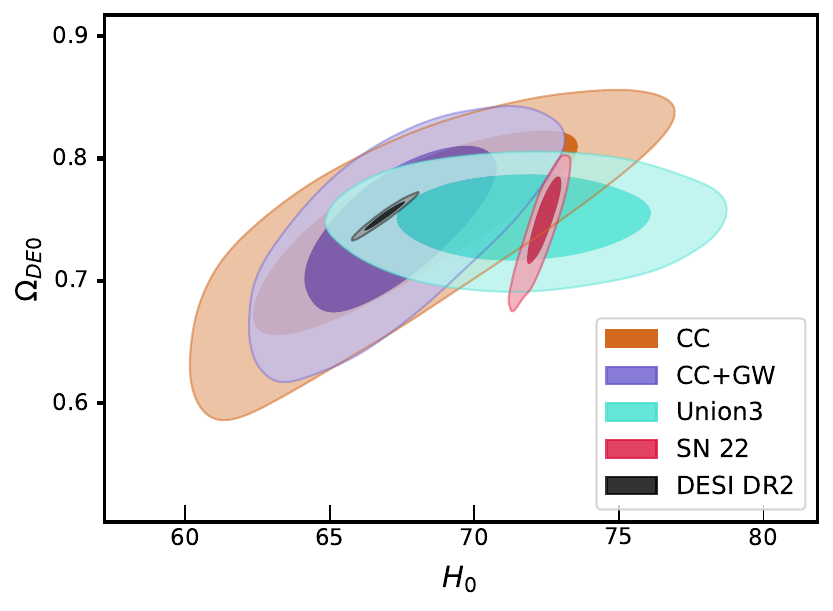}
    \caption{Plot depicting the 2D likelihood contour for cosmological parameters calibrated for the direct approach}
    \label{fig:model_parameters}
\end{figure}

Finally, a direct interpretation of the results in terms of the cosmographic parameters is presented in \autoref{fig:pade_parameters} and \autoref{fig:model_parameters}. Additionally, we have analyzed the behavior of the parameter $\Omega_{\mathrm{DE}0}$ against $H_0$. In this context, the contribution of dark energy is effectively captured by the modified gravity terms in the model, and therefore, $\Omega_{\mathrm{DE}0}$ can be computed directly from Eq.~(\ref{eq:densityparam}) and Eq.~(\ref{model;log}). The values summarized in \autoref{tab:padeqjs} and \autoref{tab:directqjs} show excellent agreement with the results reported in \cite{Pourojaghi:2024bxa}.

\subsection*{Comparison with the $\Lambda$CDM Model: AIC and BIC Analysis}

To further assess the statistical performance of our proposed methods in comparison to the $\Lambda$CDM framework, we employ two widely used information criteria: AIC and BIC. These quantities are already introduced and defined in \autoref{ch:intro}.

Table~\ref{tab:aicbic} summarizes the $\Delta$AIC and $\Delta$BIC values for both the Pad\'e and the Direct approach with respect to the $\Lambda$CDM baseline. In the DESI DR2 and Union3 datasets, both alternative models yield lower AIC and BIC values, indicating a statistically better fit than $\Lambda$CDM. Notably, the Pad\'e approach shows a stronger preference due to its comparatively lower minimum $\chi^2$ values (see \autoref{tab:chi2}). Even in datasets where $\Lambda$CDM marginally outperforms in terms of $\chi^2$, the differences in AIC/BIC remain small, implying that the performance of our models is statistically competitive.

The reason why the $f(T)$ model can outperform $\Lambda$CDM in certain datasets lies in its underlying physical mechanism. Unlike the cosmological constant of $\Lambda$CDM, which has a fixed equation of state, the effective dark energy emerging from $f(T)$ gravity is dynamical, with an equation of state that can evolve with redshift. This additional degree of freedom allows the expansion history to deviate slightly from $\Lambda$CDM, thereby offering greater flexibility to accommodate the latest BAO and SNe data (such as DESI DR2 and Union3), while still remaining consistent with other cosmological probes.


\begin{table}
    \centering
    \caption{Comparison of the minimum $\chi^2$ of our models and $\Lambda$CDM.}
    \label{tab:chi2}
    \begin{tabular}{|c|c|c|c|}
    \hline
    Dataset & $\chi^2_{min}(\text{Pad\'e})$ & $\chi^2_{min}(\text{Direct})$ & $\chi^2_{min}(\text{$\Lambda$CDM})$\\
    \hline
    \hline
    DESI DR2 & $12.141$ & $13.801$ &  $15.536$ \\
    \hline
    Union3 & $22.510$ & $22.739$ & $23.958$  \\
    \hline
    SN 22 & $813.429$ & $813.666$ &  $812.612$ \\
    \hline
    CC & $14.780$ & $14.685$ & $14.535$  \\
    \hline
    CC+GW & $15.901$ & $15.805$ & $15.642$  \\
    \hline    
    \end{tabular}
\end{table}

\begin{table}
    \centering
    \caption{$1\sigma$ values of the independent parameters $\{H_0,\Omega_{m0}\}$ for the Pad\'e method.}
    \label{tab:padeparams}
    \begin{tabular}{|c|c|c|}
    \hline
    Dataset & $H_0$ & $\Omega_{m0}$\\
    \hline
    \hline
    DESI DR2 & $66.67\pm0.39$  &  $0.2592\pm0.0058$ \\
    \hline
    Union3 & $71.7\pm2.8$ & $0.256^{+0.021}_{-0.023}$  \\
    \hline
    SN 22 & $72.40^{+0.39}_{-0.32}$ &   $0.256^{+0.021}_{-0.024}$ \\
    \hline
    CC & $67.5^{+3.3}_{-3.7}$ &  $0.268^{+0.038}_{-0.050}$  \\
    \hline
    CC+GW & $67.1\pm2.1$ & $0.269^{+0.032}_{-0.040}$  \\
    \hline    
    \end{tabular}
\end{table}

\begin{table}
    \centering
    \caption{$1\sigma$ values of the independent parameters $\{H_0,\Omega_{m0}\}$ for the Direct approach.}
    \label{tab:directparams}
    \begin{tabular}{|c|c|c|}
    \hline
    Dataset & $H_0$ & $\Omega_{m0}$\\
    \hline
    \hline
    DESI DR2 & $66.89\pm0.46$ & $0.247\pm0.0078$ \\
    \hline
    Union3 & $71.7\pm2.8$ &  $0.249^{+0.021}_{-0.025}$\\
    \hline
    SN 22 & $72.37^{+0.44}_{-0.33}$  & $0.253^{+0.020}_{-0.028}$ \\
    \hline
    CC & $67.6^{+3.4}_{-3.9}$  & $0.259^{+0.045}_{-0.064}$ \\
    \hline
    CC+GW & $67.3\pm2.2$  & $0.258^{+0.039}_{-0.051}$ \\ 
    \hline
    \end{tabular}
\end{table}

\begin{table}
    \centering
    \caption{$1\sigma$ values of the independent parameters $\{H_0,\Omega_{m0}\}$ for the $\Lambda$CDM model.}
    \label{tab:lcdmparams}
    \begin{tabular}{|c|c|c|}
    \hline
    Dataset & $H_0$ & $\Omega_{m0}$\\
    \hline
    \hline
    DESI DR2 & $69.30^{+0.52}_{-0.44}$ & $0.293^{+0.007}_{-0.009}$ \\
    \hline
    Union3 & $71.9\pm2.8$ &  $0.358\pm0.027$\\
    \hline
    SN 22 & $72.40^{+0.30}_{-0.24}$  & $0.380^{+0.017}_{-0.021}$ \\
    \hline
    CC & $68.7\pm3.6$  & $0.325^{+0.049}_{-0.065}$ \\
    \hline
    CC+GW & $68.1\pm2.2$  & $0.328^{+0.043}_{-0.051}$ \\
    \hline    
    \end{tabular}
\end{table}
\begin{table}
    \centering
    \caption{$1\sigma$ values of $z_t$ and $\omega_{eff_0}$ for the Pad\'e method.}
    \label{tab:padezt}
    \begin{tabular}{|c|c|c|}
    \hline
    Dataset & $z_t$ & $\omega_{eff_0}$ \\
    \hline
    \hline
    DESI DR2 & $0.68\pm0.03$ & $-0.58\pm0.01$ \\
    \hline
    Union3 & $0.70^{+0.13}_{-0.10}$ & $-0.59\pm0.03$ \\
    \hline
    SN 22 & $0.70^{+0.14}_{-0.10}$ & $-0.59\pm0.03$ \\
    \hline
    CC & $0.64^{+0.29}_{-0.16}$ &  $-0.57^{+0.04}_{-0.07}$\\
    \hline
    CC+GW & $0.63^{+0.22}_{-0.13}$ & $-0.57^{+0.04}_{-0.05}$ \\
    \hline    
    \end{tabular}
\end{table}

\begin{table}
    \centering
    \caption{$1\sigma$ values of $z_t$ and $\omega_{eff_0}$ for the Direct approach.}
    \label{tab:directzt}
    \begin{tabular}{|c|c|c|}
    \hline
    Dataset & $z_t$ & $\omega_{eff_0}$ \\
    \hline
    \hline
    DESI DR2 & $0.66\pm0.03$ & $-0.60\pm0.01$ \\
    \hline
    Union3 & $0.65^{+0.10}_{-0.07}$ & $-0.60\pm0.03$ \\
    \hline
    SN 22 & $0.64^{+0.10}_{-0.07}$ & $-0.59^{+0.02}_{-0.04}$ \\
    \hline
    CC & $0.61^{+0.27}_{-0.14}$ & $-0.58^{+0.05}_{-0.09}$ \\
    \hline
    CC+GW & $0.62^{+0.20}_{-0.13}$ & $-0.58^{+0.04}_{-0.07}$ \\
    \hline    
    \end{tabular}
\end{table}
\begin{table*}
    \centering
    \caption{$1\sigma$ values of the cosmographic parameters and $\Omega_{DE0}$ for the Pad\'e method.}
    \label{tab:padeqjs}
    \begin{tabular}{|c|c|c|c|c|}
    \hline
    Dataset & $q_0$ & $j_0$ & $s_0$ & $\Omega_{DE0}$\\
    \hline
    \hline
    DESI DR2 & $-0.382\pm0.011$ & $0.7757\pm0.0051$ & $-0.253\pm0.025$ & $0.7408\pm0.0058$\\
    \hline
    Union3 & $-0.389\pm0.042$ & $0.779\pm0.02$ & $-0.239^{+0.10}_{-0.088}$ & $0.744^{+0.023}_{-0.021}$\\
    \hline
    SN 22 & $-0.390^{+0.043}_{-0.043}$ & $0.78^{+0.020}_{-0.020}$ & $-0.237^{+0.11}_{-0.088}$ & $0.744^{+0.024}_{-0.021}$\\
    \hline
    CC & $-0.369^{+0.078}_{-0.089}  $ & $0.771\pm{0.037}$& $-0.29^{+0.22}_{-0.16}$& $0.732^{+0.05}_{-0.038}$\\
    \hline
    CC+GW & $-0.367^{+0.064}_{-0.072}$ & $0.770^{+0.031}_{-0.031}$ & $-0.30^{+0.18}_{-0.14}$ & $0.731^{+0.040}_{-0.032}$\\
    \hline    
    \end{tabular}
\end{table*}

\begin{table*}
    \centering
    \caption{$1\sigma$ values for the cosmographic parameters and $\Omega_{DE0}$ for the Direct approach.}
    \label{tab:directqjs}
    \begin{tabular}{|c|c|c|c|c|}
    \hline
    Dataset & $q_0$ & $j_0$ & $s_0$ & $\Omega_{DE0}$\\
    \hline
    \hline
    DESI DR2 & $-0.406\pm0.015$ & $0.7868\pm0.0073$ & $-0.2\pm0.034$ & $0.753\pm0.0078$\\
    \hline
    Union3 & $-0.403\pm0.045$ & $0.786\pm0.021$ & $-0.209^{+0.11}_{-0.09}$ & $0.751^{+0.025}_{-0.021}$\\
    \hline
    SN 22 & $-0.395^{+0.039}_{-0.053}$ & $0.782^{+0.025}_{-0.020}$ & $-0.227^{+0.12}_{-0.083}$ & $0.747^{+0.028}_{-0.02}$\\
    \hline
    CC & $-0.387^{+0.096}_{-0.11}  $ & $0.780\pm{0.047}$& $-0.26^{+0.28}_{-0.19}$& $0.741^{+0.064}_{-0.045}$\\
    \hline
    CC+GW & $-0.388^{+0.080}_{-0.091}$ & $0.780^{+0.039}_{-0.039}$ & $-0.25^{+0.22}_{-0.16}$ & $0.742^{+0.051}_{-0.039}$\\
    \hline    
    \end{tabular}
\end{table*}
\begin{table}
    \centering
    \caption{Difference in the Akaike Information Criterion ($\Delta$AIC) and Bayesian Information Criterion ($\Delta$BIC) between the Padé and Direct models relative to the $\Lambda$CDM model. Negative values indicate a statistical preference over $\Lambda$CDM.}
    \label{tab:aicbic}
    \begin{tabular}{|c|c|c|c|c|}
    \hline
    Dataset & $\Delta AIC_{\text{Pad\'e}}$ & $\Delta BIC_{\text{Pad\'e}}$ & $\Delta AIC_{\text{Direct}}$ & $\Delta BIC_{\text{Direct}}$\\
    \hline
    \hline
    DESI DR2 & $-3.395$ & $-3.395$ & $-1.735$ & $-1.735$ \\
    \hline
    Union3 & $-1.448$ & $-1.448$ & $-1.219$ & $-1.219$ \\
    \hline
    SN 22 & $0.817$ & $0.817$ & $1.054$ & $1.054$ \\
    \hline
    CC & $0.245$ & $0.245$ & $0.150$ & $0.150$ \\
    \hline
    CC+GW & $0.259$ & $0.259$ & $0.163$ & $0.163$ \\
    \hline    
    \end{tabular}
\end{table}

\section{Conclusion}
\label{sec:padeconc}
Pad\'e cosmography is highly regarded among cosmographic techniques due to its improved convergence properties at high redshifts. Pad\'e approximants employ rational functions that can capture the behavior of cosmological observables more accurately over a wider redshift range. This makes it particularly suitable for analyzing observational data that extends to moderate and high redshifts. In the literature, cosmographic techniques are extensively used as model-independent tools for reconstructing the kinematics of the Universe. However, the perspective of these techniques on specific gravitational Lagrangians has not been widely explored so far. Motivated by this, here we compared the traditional approach (constraining models using the motion equations) and the Pad\'e method for a very successful model from the literature. Although introduced phenomenologically, the adopted Lagrangian (Eq.~(\ref{model;log})) is not arbitrary, as it can reproduce late-time acceleration without a cosmological constant, mimic effective dark energy, and allow smooth cosmic transitions. Preliminary checks suggest that the model may also avoid the Dolgov--Kawasaki instability and admit a stable de Sitter solution under suitable parameters. For completeness, relevant discussions of stability conditions in modified gravity can be found in the original work of Dolgov and Kawasaki \cite{Dolgov:2003px} and in reviews of torsion-based gravity theories such as \cite{Cai:2015emx,Bamba:2013jqa}. A full theoretical consistency study, however, lies beyond the scope of this work and will be pursued in future research. These promising features nevertheless justify the present model as a useful framework for confronting modified gravity with observations.

Our analysis demonstrates that both methods yield nearly identical results, as expected. This consistency highlights the reliability of the Pad\'e approach in constraining cosmological models. The choice of datasets in this study, DESI DR2, GW data, Union3, Pantheon+SH0ES, and CC, reflects a comprehensive and up-to-date strategy for cosmological parameter estimation. Each dataset brings unique strengths that complement one another, offering a robust foundation for constraining the model. DESI DR2 provides high-precision measurements of large-scale structure and BAO, essential for tracking the expansion history. GW standard sirens serve as independent distance indicators, free from traditional cosmic distance ladder systematics. The inclusion of both Union3 and the more recent Pantheon+SH0ES supernova compilations allows for cross-validation and improved constraints on late-time cosmic acceleration. Meanwhile, CC data offer a model-independent approach to estimating the Hubble parameter, especially at intermediate redshifts. Together, these datasets span a wide redshift range and different observational channels, making them particularly powerful for reliable and precise MCMC parameter estimation. This diverse combination ensures that the model is tested against a rich, multi-probe dataset, enhancing both the accuracy and credibility of the inferred cosmological parameters. 

The results obtained from the MCMC analysis for both the Pad\'e and direct approaches have been examined from multiple perspectives. For a detailed comparison with the standard cosmological model, we also performed MCMC for $\Lambda$CDM and the results are presented in \autoref{fig:lcdmcon} and \autoref{tab:lcdmparams}. Further, we compared the $\chi^2_{min}$, AIC, and BIC values of our model with the standard model. This comparative analysis highlights the strength of the Padé cosmographic technique, which consistently yields lower minimum $\chi^2$ values across most datasets. The Direct approach, while simpler in its formulation and interpretation, performs comparably well in many cases. However, the Padé method offers slightly better performance overall, particularly due to its improved flexibility in capturing the detailed features of the cosmic expansion history. Both approaches remain competitive with the standard $\Lambda$CDM model and, in several datasets, even outperform it in terms of statistical criteria such as AIC and BIC. Overall, our analysis shows that the $f(T)$ model remains consistent with current observations while providing improved agreement over $\Lambda$CDM in the DESI DR2 and Union3 datasets.

Recent studies reveal that dynamical dark energy models yield better fits to DESI DR2 compared to the standard $\Lambda$CDM scenario, thereby directing the scientific community toward new avenues of investigation \cite{DESI:2025zgx, Wang:2025bkk, Chaudhary:2025pcc}. The $\Lambda$CDM model has long served as the benchmark cosmological framework, consistently providing excellent agreement with a wide range of observational datasets. However, the DESI DR2 results indicate that dynamical dark energy offers a statistically better fit than $\Lambda$CDM, challenging the conventional expectation of its universality and opening new perspectives on the nature of cosmic acceleration. Our model also exhibits this trend. Nevertheless, to make a definitive assessment of which model ultimately provides the best description of the Universe, the results of forthcoming surveys will be crucial. For now, based on DESI DR2 results, these alternative theories, including our model, emerge as a serious contender to $\Lambda$CDM.

 The figures and tables for the cosmological parameters clearly demonstrate that both models provide excellent agreement with observational data, maintaining consistency within acceptable confidence intervals. Moreover, the reconstructed expansion history, capturing features such as the transition from deceleration to acceleration and late-time cosmic dynamics, aligns well with independent cosmological probes. These findings reinforce the potential of cosmography-based frameworks as robust tools to describe the Universe's evolution and to test for deviations from the standard cosmological paradigm.

Future work will aim to extend the Pad\'e approximation methodology to constrain more complex and physically motivated cosmological models. With the continual release of high-precision data, including upcoming gravitational wave observations and next-generation galaxy surveys, an updated and expanded dataset collection will be employed to improve parameter constraints. Additionally, other rational approximation techniques, such as continued fractions and Chebyshev approximants, will be explored to assess their effectiveness in modeling cosmological quantities and capturing the underlying dynamics more accurately. These efforts are expected to enhance the robustness and versatility of analytical approaches in cosmological modeling.\\
\newpage
\thispagestyle{empty}

\vspace*{\fill}
\begin{center}
    {\Huge \color{RedViolet} \textbf{CHAPTER 5}}\\
    \
    \\
    {\Large\color{Emerald}\textsc{\textbf{Gravitational Baryogenesis in $f(T)$ gravity}}}\\
    \end{center}

\begin{center} 
\textbf{\color{RedViolet}PUBLICATION}\\
\textbf{``Effects of DESI and GW observations on $f(T)$ gravitational baryogenesis"}\\
 Sai Swagat Mishra et al. 2025, \textit{Physics Letters B} \textbf{872}, 140036.\\\vspace{0.2 in}
\includegraphics[width=0.1\linewidth]{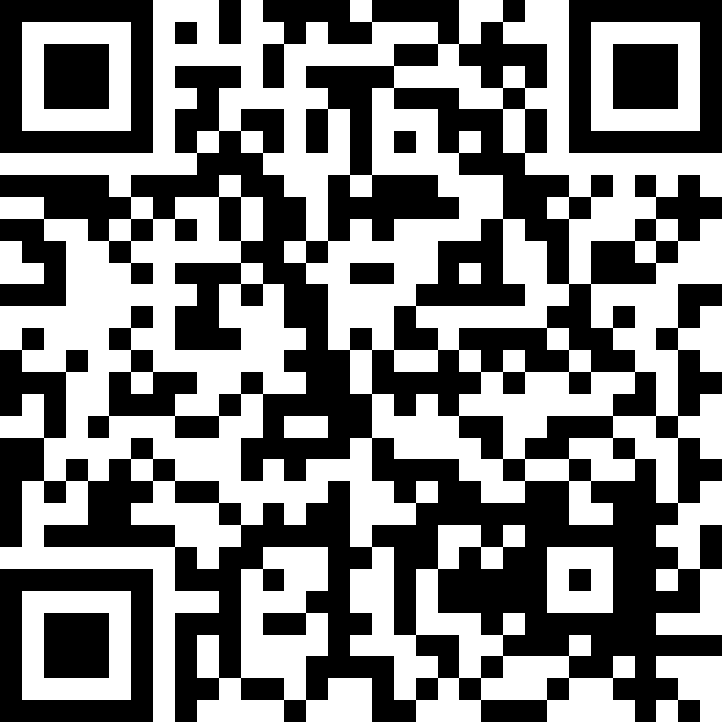} \\
    DOI: \href{https://doi.org/10.1016/j.physletb.2025.140036}{10.1016/j.physletb.2025.140036}

\end{center}
\vspace*{\fill}

\pagebreak

\def\baselinestretch{1}
\chapter{\textsc{Gravitational Baryogenesis in \texorpdfstring{$f(T)$}{f(T)} gravity}}\label{chap5}
\def\baselinestretch{1.5}
\pagestyle{fancy}
\lhead{\emph{Chapter 5. Gravitational Baryogenesis in $f(T)$ gravity}}
\rhead{\thepage}
\noindent\textbf{Highlights}
\begin{itemize}
    \item[$\star$] Explores baryogenesis as the mechanism behind the observed matter–antimatter asymmetry in the Universe.  
    \item[$\star$] Proposes a novel teleparallel gravity-based framework linking late-time observations to early-Universe baryogenesis.  
    \item[$\star$] Incorporates DESI and gravitational wave datasets, ensuring strong observational relevance.  
    \item[$\star$] Investigates the role of intermediate cosmic epochs using the deceleration parameter.  
    \item[$\star$] Demonstrates that the baryon-to-entropy ratio predicted by the model is in excellent agreement with current observational estimates.  
\end{itemize}

\section{Introduction}
Grand Unified Theories (GUTs) seek to combine the strong, weak, and electromagnetic interactions, along with quarks and leptons, into a single theoretical framework. These theories are built on non-Abelian symmetry groups such as \( SU(5) \), \( SO(10) \), or \( E_6 \). A key consequence of GUTs is the emergence of new interactions that violate the baryon number (\( B \)) and the lepton number (\( L \)), leading to proton decay. Another prediction is the existence of stable, supermassive magnetic monopoles, which arise from topologically stable configurations of gauge and Higgs fields. Both of these phenomena carry important implications for cosmology.  

Currently, baryon-number-violating interactions appear to be extremely weak, as suggested by the remarkably long lifetime of the proton. These interactions can be described by a coupling constant similar to the Fermi coupling constant but suppressed by at least 25 orders of magnitude
\begin{equation*}
    G_{AB} \sim M^{-2} \lesssim 10^{-30} \text{ GeV}^{-2}.
\end{equation*}
Here, \( M \) represents the unification energy scale, estimated to be around \( 10^{14} \) GeV or higher. However, at temperatures near or above \( M \), these interactions-if they exist-could be as strong as other fundamental forces. These interactions could enable the Universe to transition from a baryon-symmetric state to one with a baryon asymmetry sufficient to account for the observed baryon-to-photon ratio. The process of baryogenesis stands as one of the central challenges in particle cosmology, providing a theoretical framework that connects particle physics with the observed baryon asymmetry, even though the exact mechanism remains unknown. Even in the absence of direct evidence for proton decay, baryogenesis offers compelling, albeit indirect, support for the idea that quarks and leptons are unified at a fundamental level.

The validity of this discovery is confirmed through various lines of evidence, including predictions of the Big Bang Nucleosynthesis, the precise observations of the cosmic microwave background, and the lack of strong radiation signals from matter-antimatter annihilation. Various baryogenesis models attempt to explain why there is more matter than antimatter in the Universe, with these processes potentially taking place during either the matter-dominated or radiation-dominated eras. The presence of Charge (C) and Charge-Parity (CP) violation indicates a fundamental asymmetry between matter and antimatter, which plays a crucial role in the generation of this imbalance. Some interesting approaches to exploring gravitational baryogenesis can be found in the literature (see \cite{Arbuzova:2017vdj, Mishra:2023khd, Bhattacharjee:2020jfk, Mishra:2024rut}). 

In this chapter, we explore baryogenesis within the framework of modified gravity, motivated by the limitations of the standard cosmological model. Given the well-documented challenges associated with GR in explaining certain cosmological phenomena, investigating possible modifications or extensions of GR is a natural and necessary approach (see \autoref{ch:intro}). Since the observed late-time cosmic acceleration, as reported in \cite{SupernovaSearchTeam:1998fmf}, is a key feature that any viable alternative theory must account for, an effective framework should be capable of explaining both early- and late-time cosmic evolution. 
In the present chapter, we adopt the teleparallel gravity framework, where the geometric structure is characterized solely by torsion, with both nonmetricity and curvature explicitly set to zero.

The structure of this chapter is as follows. The \autoref{sec3} explores gravitational baryogenesis in the teleparallel framework. In \autoref{sec5}, we constrain the model against the observational datasets and examine its influence in both the early and late Universe. Finally, we summarize our key findings and present our conclusions in \autoref{sec6}.

\section{Baryogenesis in teleparallel framework}\label{sec3}
The precise origin of the baryonic asymmetry is still under investigation. So far, the three Sakharov conditions \cite{Sakharov:1967dj} are believed to be the reasons behind the matter-antimatter asymmetry. Namely, the conditions are as follows: 
\begin{itemize} 
\item Baryon number Violation,
\item Charge and Charge Parity Violation, 
\item Interactions out of thermal equilibrium.
\end{itemize}
In 2004, Davoudiasl et al. \cite{Davoudiasl:2004gf} proposed a CP-violating interaction term that dynamically violates Charge Parity during the expansion of the Universe. Following the second criterion of Sakharov, this form, an interaction between the derivative of the Ricci scalar and the baryonic matter current $(J^\mu)$, could lead to asymmetry. The form is
\begin{equation}
\label{eq:cpr}
    \frac{1}{{M_*}^2} \int \sqrt{-g} \,\ d^4x (\partial_\mu R)J^\mu,
\end{equation}
where $M_*$ refers to an effective cutoff scale or suppression scale of the higher-dimensional operator responsible for baryon number violation. In a similar line, for the teleparallel framework, the $CP$-Violating interaction term takes the form
\begin{equation}
\label{eq:cp}
    \frac{1}{{M_*}^2} \int \sqrt{-g} \,\ d^4x \left(\partial_\mu(-T)\right)J^\mu.
\end{equation}
Proceeding with an assumption of the existence of thermal equilibrium, the energy density in terms of decoupling temperature (${\mathcal{T}_D}^*$) can be defined as 
\begin{equation}
\label{eq:rhobaryo}
    \rho = \frac{\pi^2\, g_*}{30}  \left({\mathcal{T}_D}^*\right)^4 ,
\end{equation}
where $g_*$ represents the number of degrees of freedom of the effectively massless particles that contribute to the global entropy of the Universe and is defined as $g_*= 45 \,s/2 \pi^2 ({\mathcal{T}_D}^*)^3$. Before we move on, it is necessary to understand how the interaction affects the baryonic asymmetry. The interaction in Eq.~\eqref{eq:cp} produces opposite-sign energy contributions that vary for particle and antiparticle in an expanding Universe with $T\sim-H^2$ and $\dot{T}$ nonzero, dynamically violating CPT.  Similar to an induced chemical potential $\mu\sim \pm \frac{\dot{T}}{{M_*}^2}$, this affects the thermal equilibrium distributions, leading the Universe towards nonzero equilibrium baryonic asymmetry. Further, the asymmetry gets frozen when the Universe cools down, and the temperature decreases to below the decoupling temperature. Hence, the remaining net baryon-to-entropy ratio can be defined as \cite{Oikonomou:2016jjh}
\begin{equation}
\label{eq:nbs}
    \frac{n_B}{s} \simeq \frac{15g_b}{4\pi^2 g_*}\left(\frac{1}{{M_*}^2 \mathcal{T}^*}\dot{T} \right)_{\mathcal{T}^*={\mathcal{T}_D}^*}. 
\end{equation}
Here, $g_b$ represents the total number of intrinsic degrees of freedom of baryons. Using the theoretical expression above, one can constrain the teleparallel theory against observations. 

The existing literature mostly relies on a specific choice of a scale factor in terms of cosmic time. The novelty of this work is bypassing the dependence on the scale factor and directly investigating the effect of late-time constraints on gravitational baryogenesis. To do so, we incorporate Eq.~\eqref{eq:rhobaryo} into both the motion equations \eqref{eq:ftmot1} and \eqref{eq:ftmot2}. As the only matter of concern now is obtaining the term $\dot{T}$, we can use the relation $\dot{T}=-12 H \dot{H}$ to perform the straightforward calculation. The LSR model introduced in Chapter \ref{ch:intro} serves as the background framework in this work, albeit expressed in an alternative notation\footnote{In this notation, the model takes the form $f(T)=A T_0 \sqrt{\frac{T}{B T_0}} ln\left(\frac{B T_0}{T}\right)$.}. For this form, the theoretical expression of $n_b/s$ reads
\begin{equation}\label{eq:nbsmodel}
   \frac{n_b}{s}\simeq\frac{B \, g_b \sqrt{\frac{-T}{B {H_0}^2}} \sqrt{30 \sqrt{6} A {H_0}^2 \sqrt{\frac{-T}{B {H_0}^2}}+\pi ^2 g_* {\mathcal{T}_D}^*}}{3 \sqrt{10} {M_*}^2 \left(B \sqrt{\frac{-T}{B {H_0}^2}}-\sqrt{6} A\right)}.
\end{equation}

We intend to investigate the above teleparallel baryon-to-entropy ratio using the results obtained from the late-time epoch. Regarding this, a detailed discussion has been presented in the upcoming section.
\section{Results}\label{sec5}
\begin{figure}
    \centering
    \includegraphics[scale=0.7]{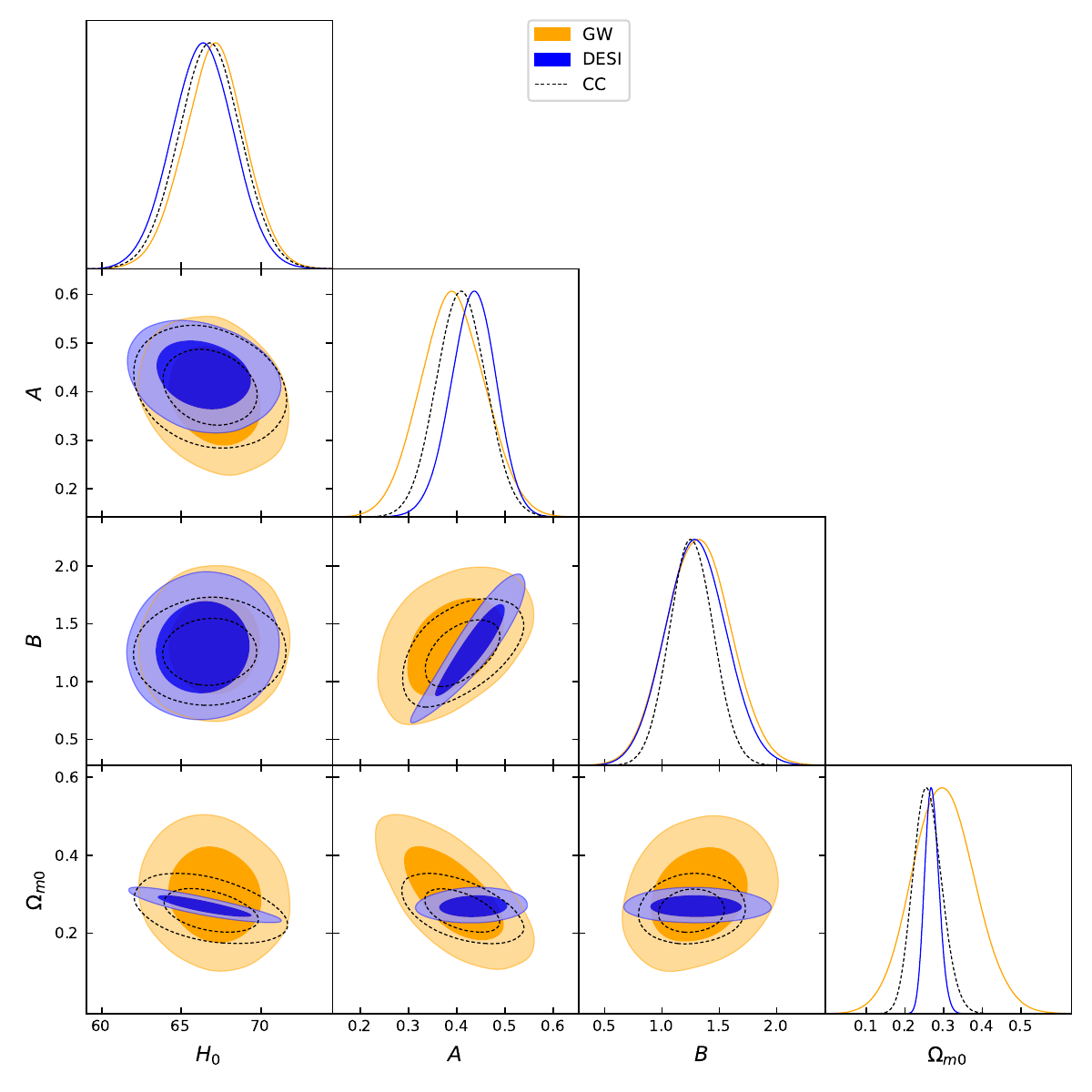}
    \caption{Contour plot of the parameters upto $2 \sigma$ confidence level for different datasets.}
    \label{chap5/mcmc}
\end{figure}

In this section, we discuss the perspective of our theory in various epochs. To begin with the late-time era, we use a challenging statistical technique, MCMC, which draws samples from a probability distribution. This simulation constrains our model with the utilization of very promising recent datasets. The methodologies, as discussed in the first chapter, are employed, and the 2D likelihoods are obtained (see \autoref{chap5/mcmc}). One can clearly observe the overlapping regions for all parameters, including the controversial ``$H_0$." It shows the efficiency of this model in describing the low-redshift paradigm.

\begin{figure}
\includegraphics[width=0.48\linewidth]{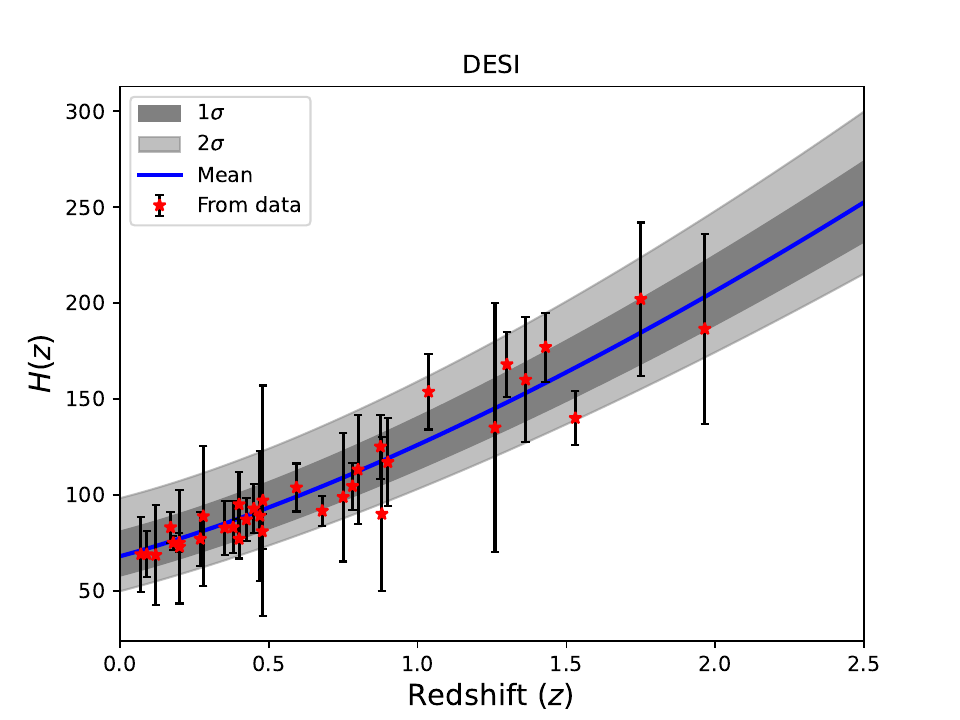}
\includegraphics[width=0.48\linewidth]{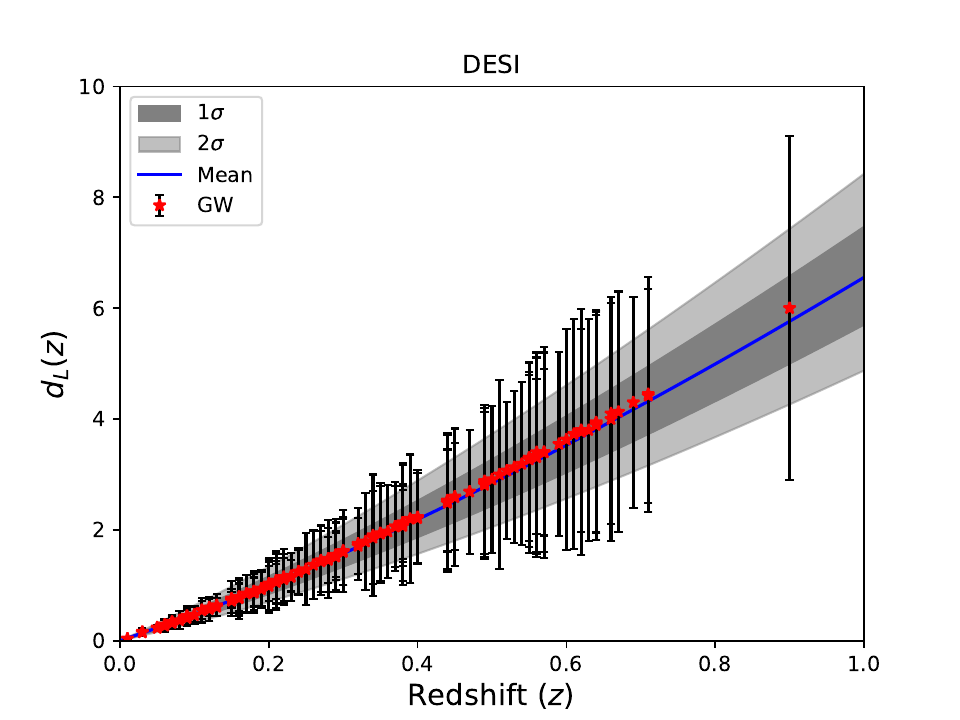}\\
\includegraphics[width=0.48\linewidth]{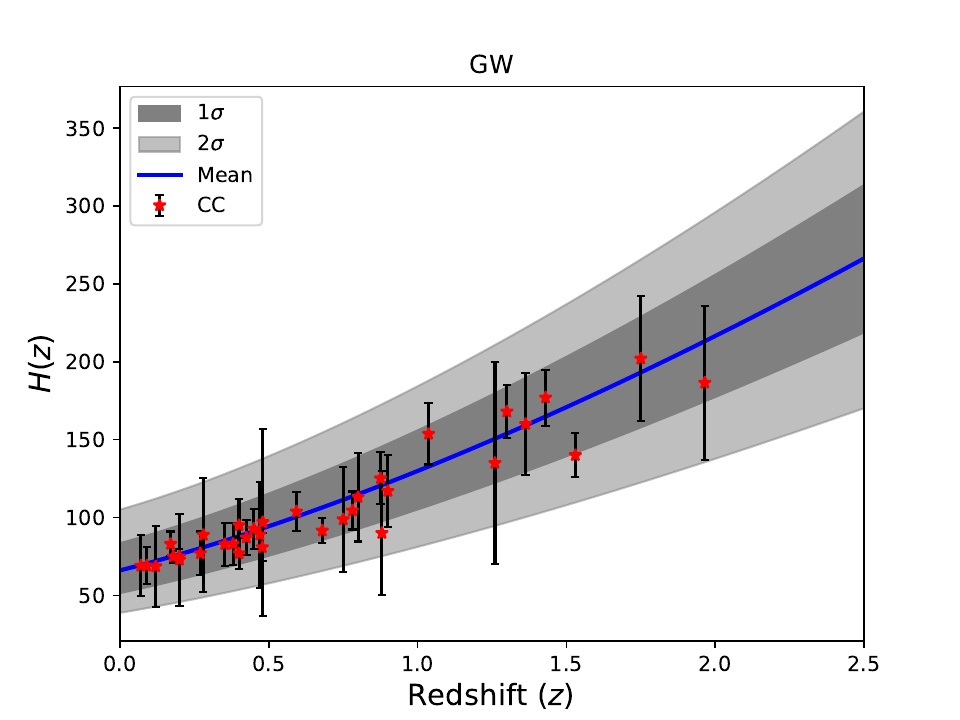}
\includegraphics[width=0.48\linewidth]{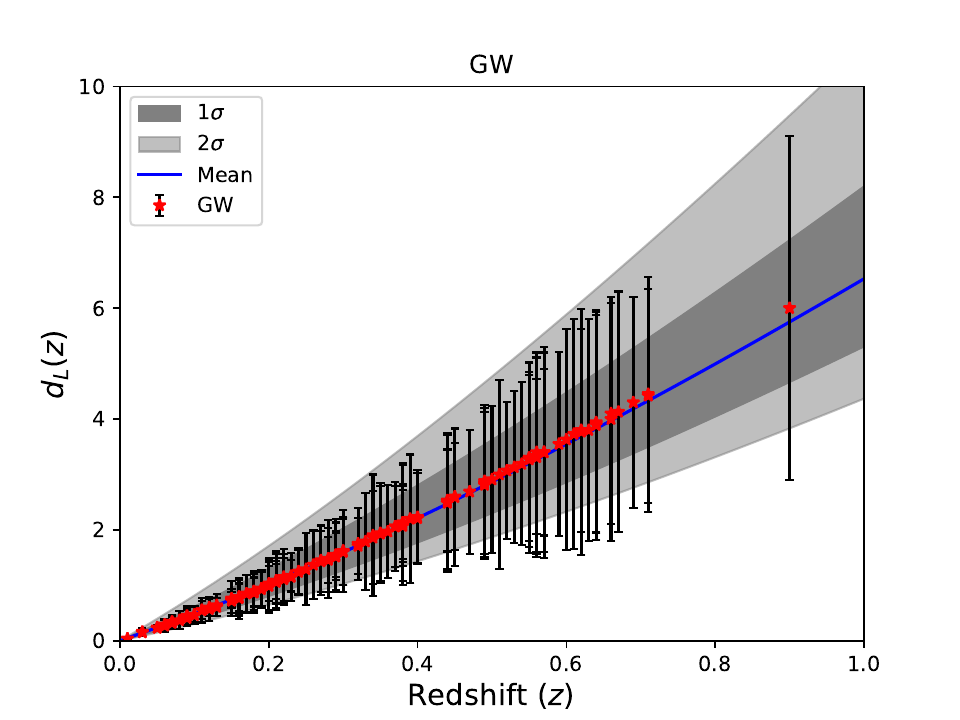}\\
\includegraphics[width=0.48\linewidth]{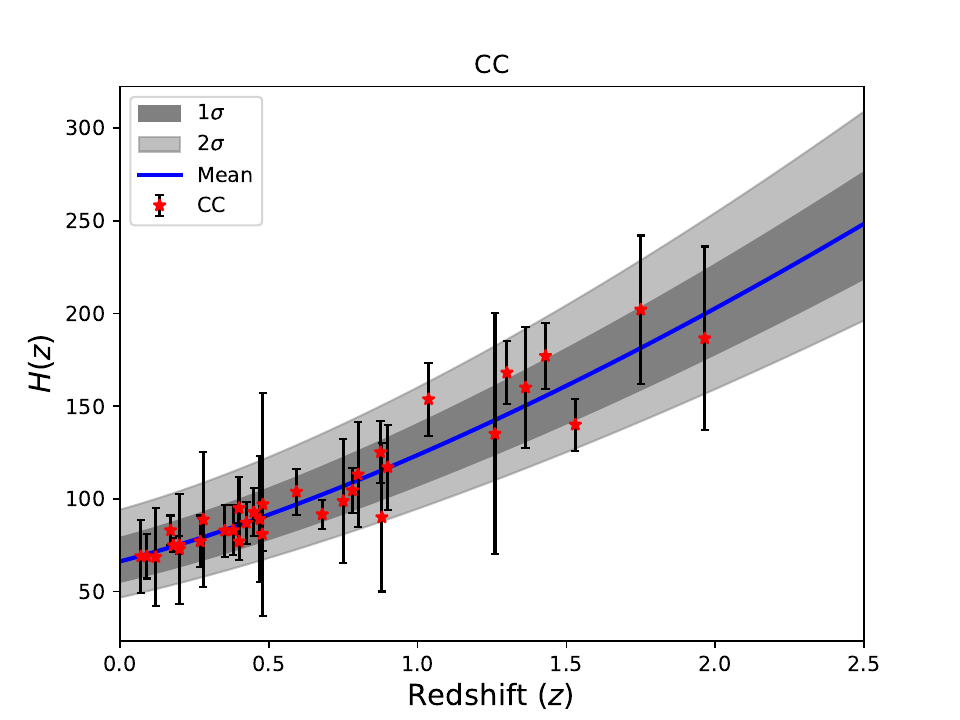}
\includegraphics[width=0.48\linewidth]{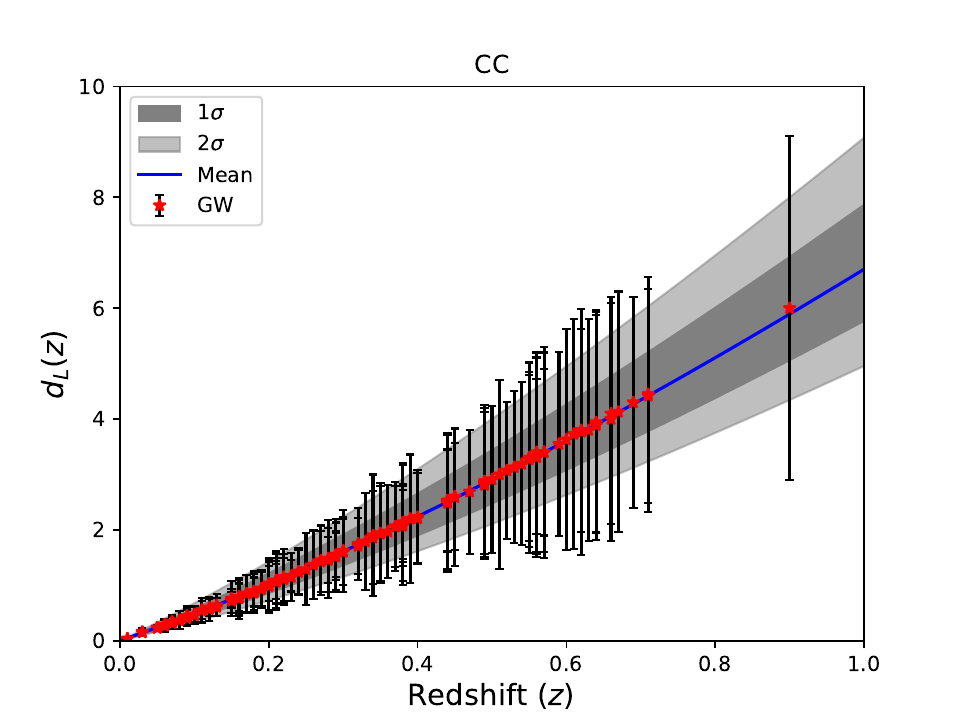}
    \caption{Hubble parameter (left panel) and luminosity distance (right panel) for each dataset}
    \label{chap5/HDL}
\end{figure}

\begin{figure}
\includegraphics[width=0.32\linewidth]{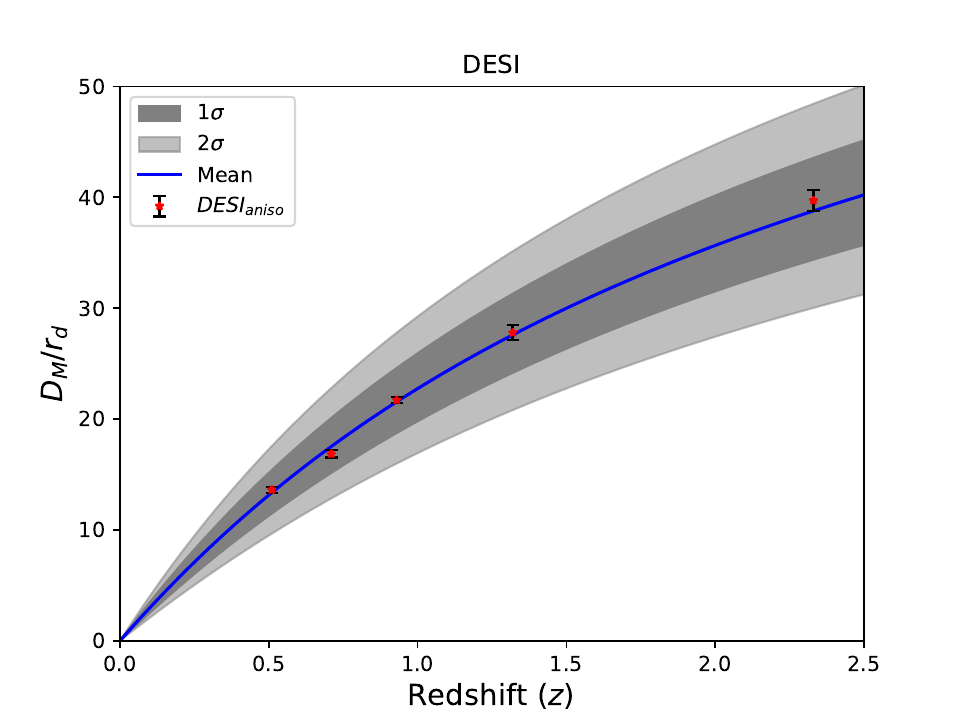}
\includegraphics[width=0.32\linewidth]{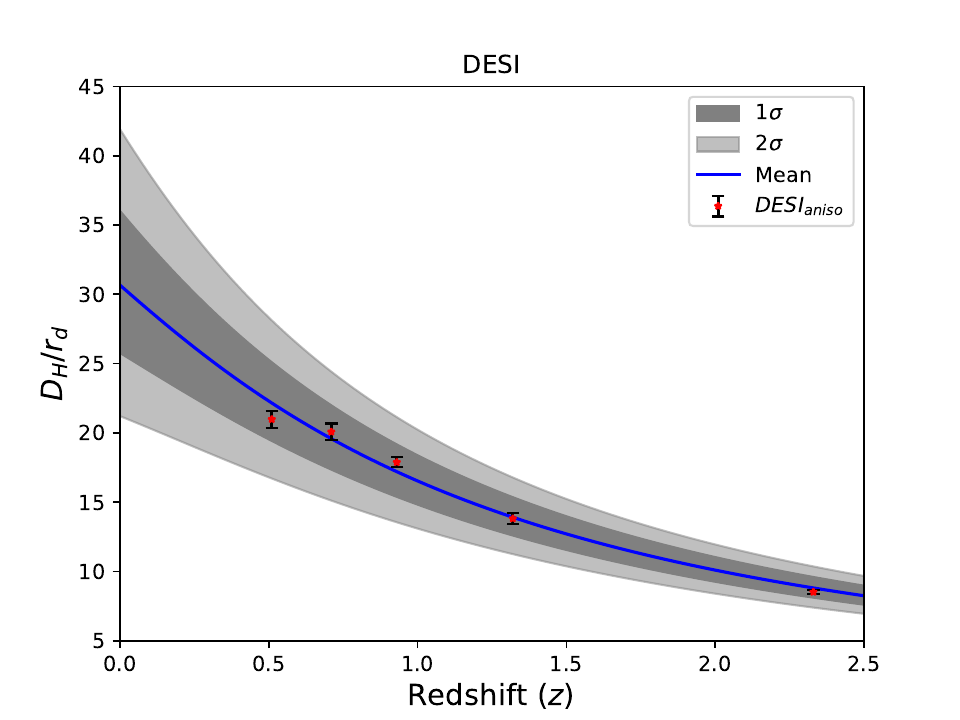}
\includegraphics[width=0.32\linewidth]{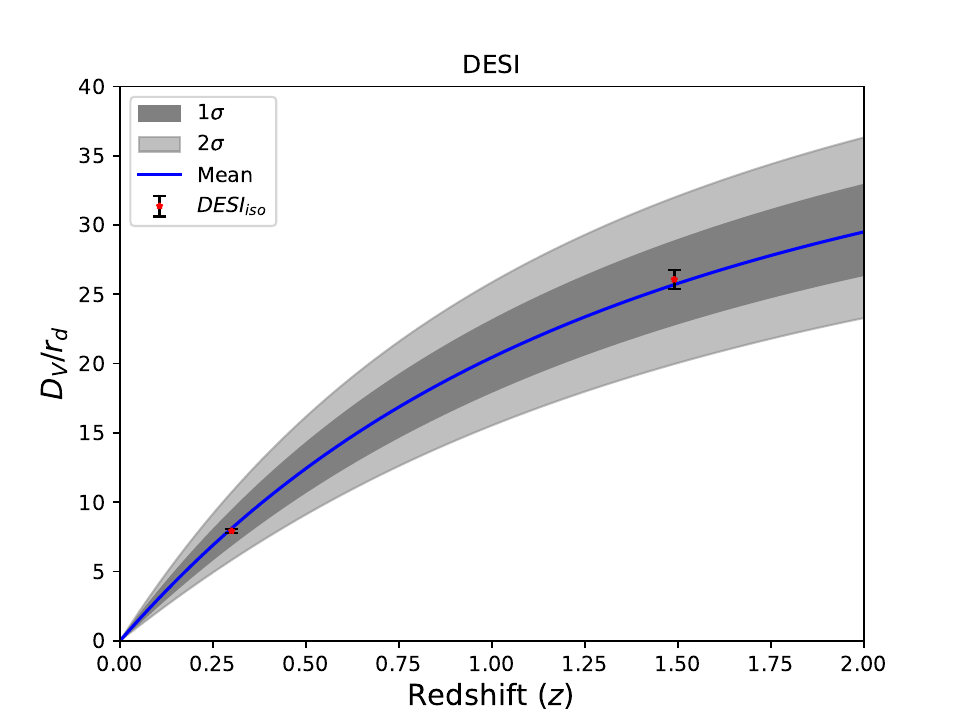}\\
\includegraphics[width=0.32\linewidth]{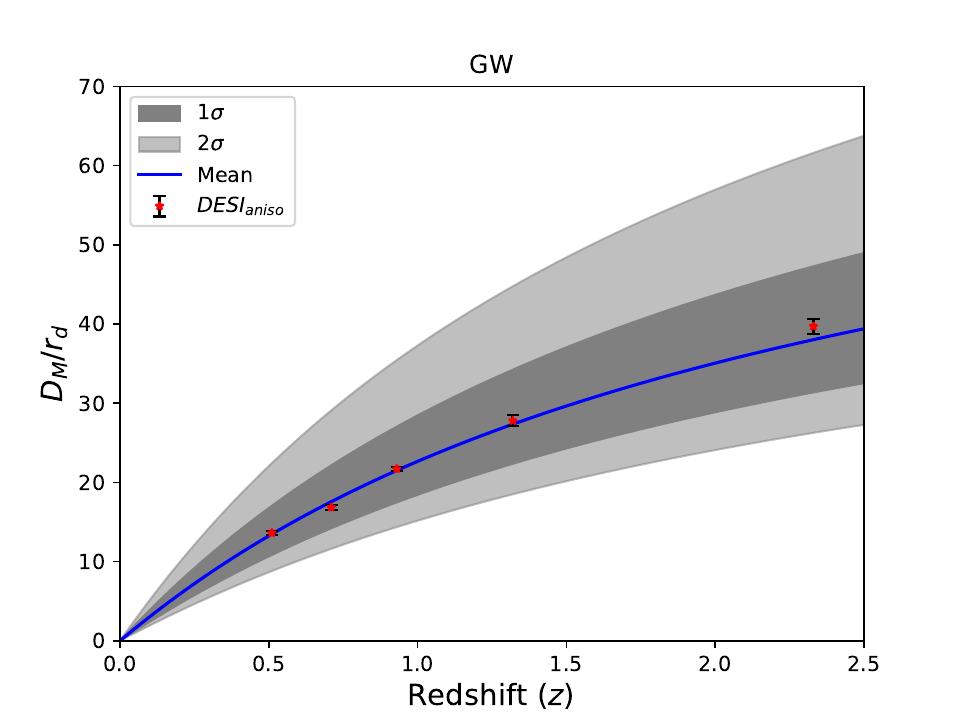}
\includegraphics[width=0.32\linewidth]{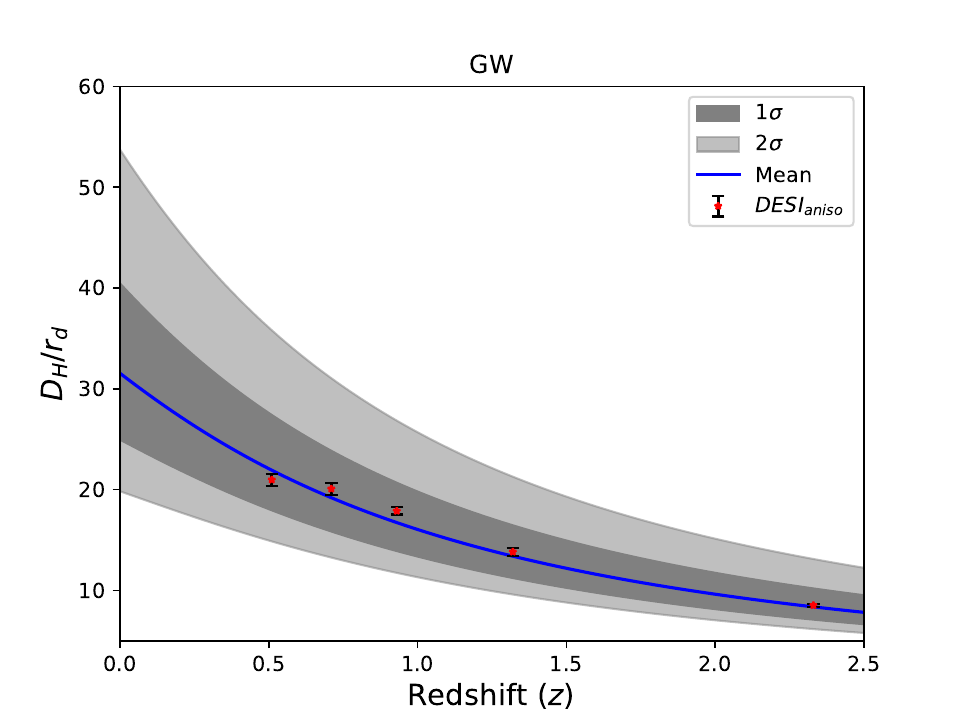}
\includegraphics[width=0.32\linewidth]{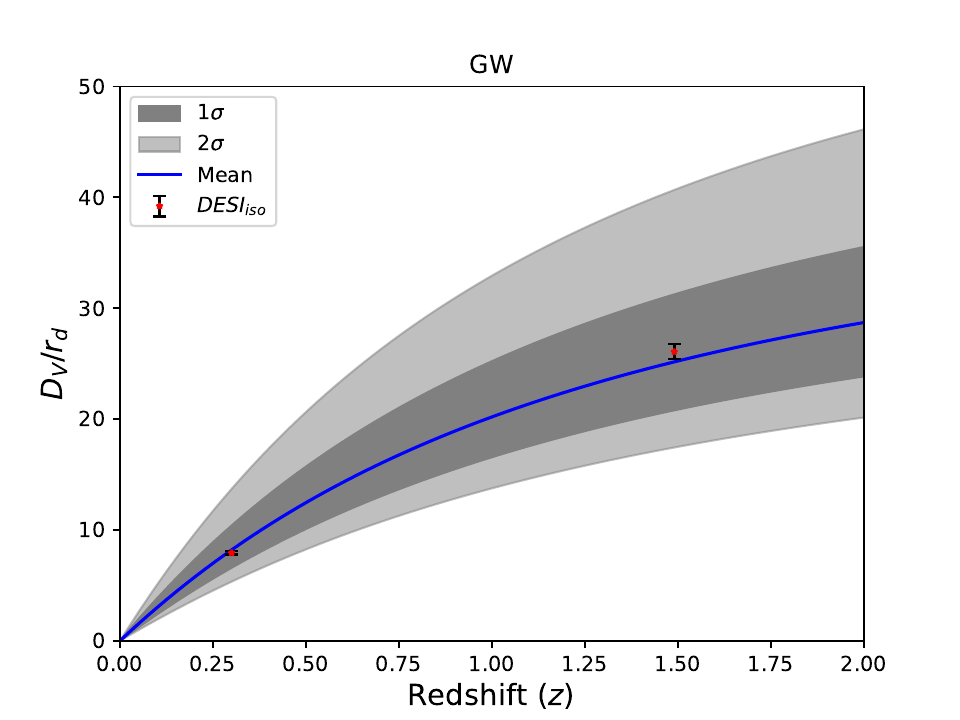}\\
\includegraphics[width=0.32\linewidth]{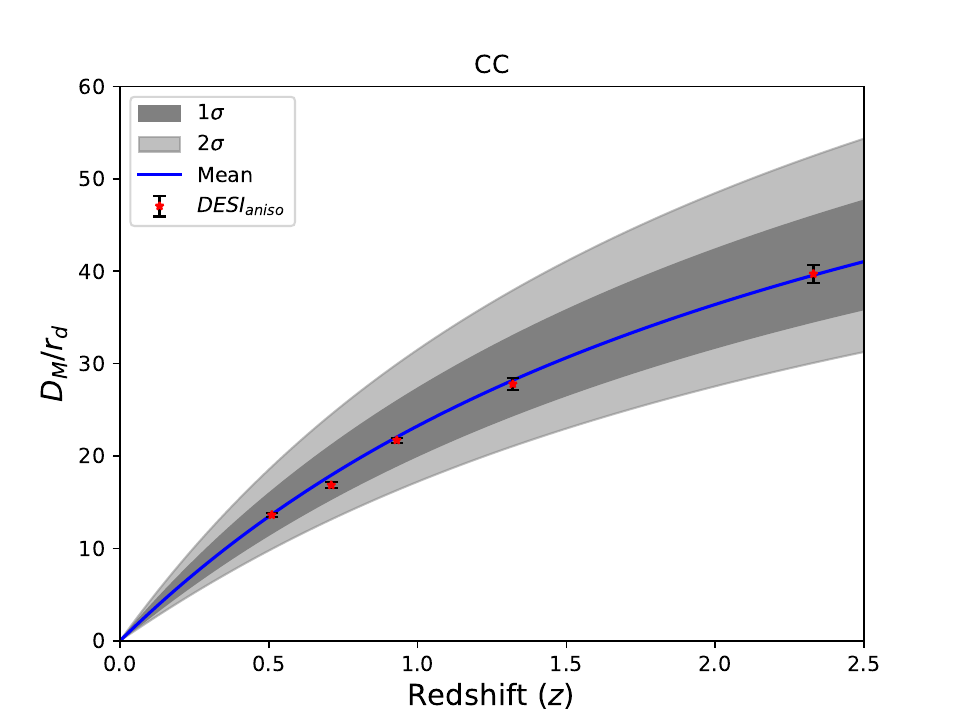}
\includegraphics[width=0.32\linewidth]{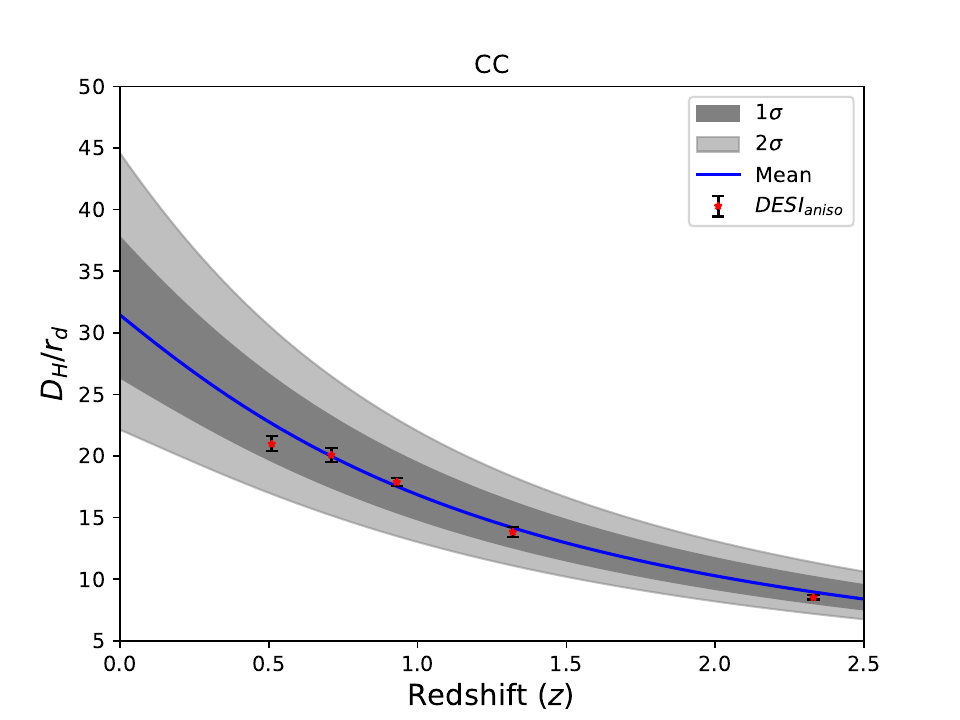}
\includegraphics[width=0.32\linewidth]{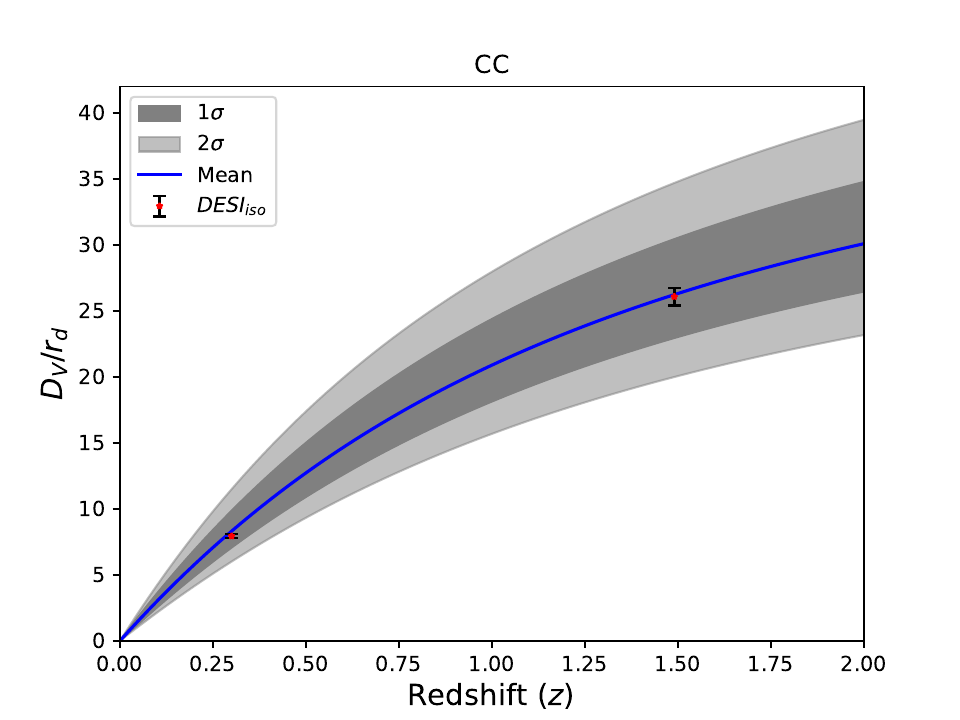}\\
    \caption{$\frac{D_M}{r_d}$ and $\frac{D_H}{r_d}$ with anisotropic  (first two columns) and $\frac{D_V}{r_d}$ with isotropic (last column) DESI BAO measurements.}
    \label{chap5/desi}
\end{figure}

The results are further plotted against the parameters whose observational predictions are available in the considered datasets. In each analysis, we plot the mean curve along with the $1\sigma$ and $2\sigma$ bands obtained from the results. It provides a very clear illustration of the fit of our model to the data with the margin of uncertainty. The first column of \autoref{chap5/HDL} represents the Hubble parameter, and the second column represents the luminosity distance for each dataset. The three distance measuring parameters of the DESI BAO are then plotted in \autoref{chap5/desi}. The anisotropic BAO measurements are presented in the first two columns, while the isotropic ones are presented in the last column.

\begin{figure}
\includegraphics[width=0.48\linewidth]{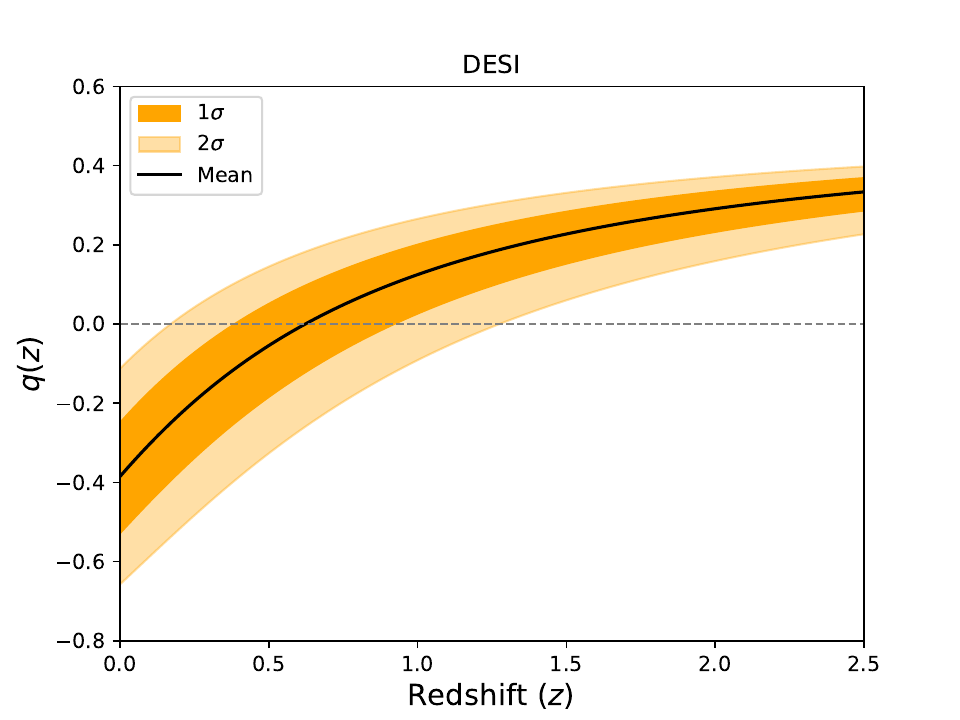}
\includegraphics[width=0.48\linewidth]{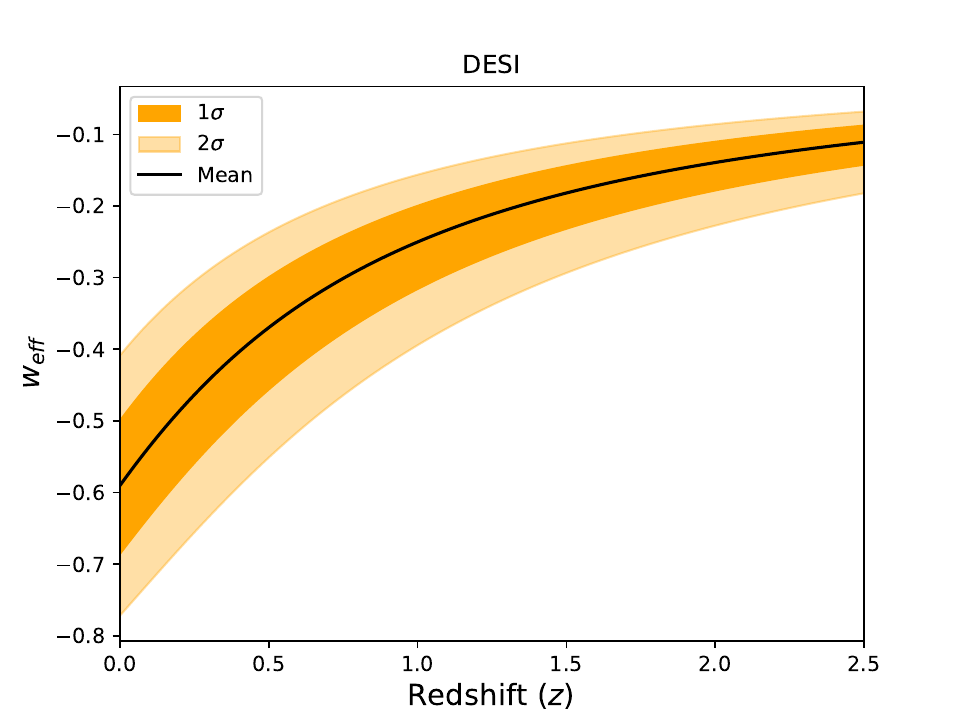}\\
\includegraphics[width=0.48\linewidth]{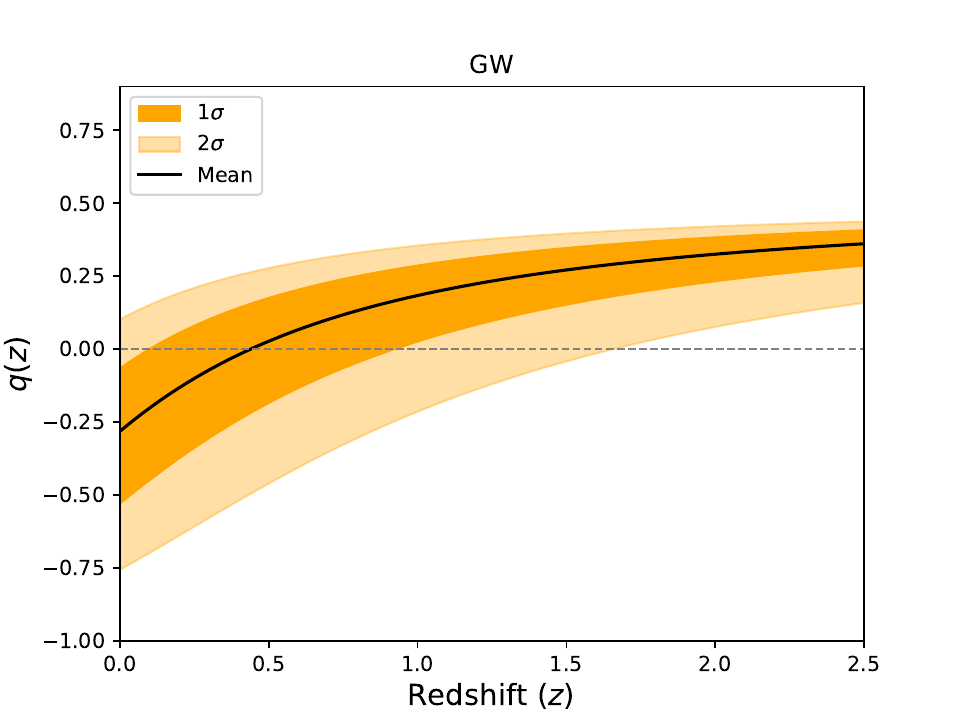}
\includegraphics[width=0.48\linewidth]{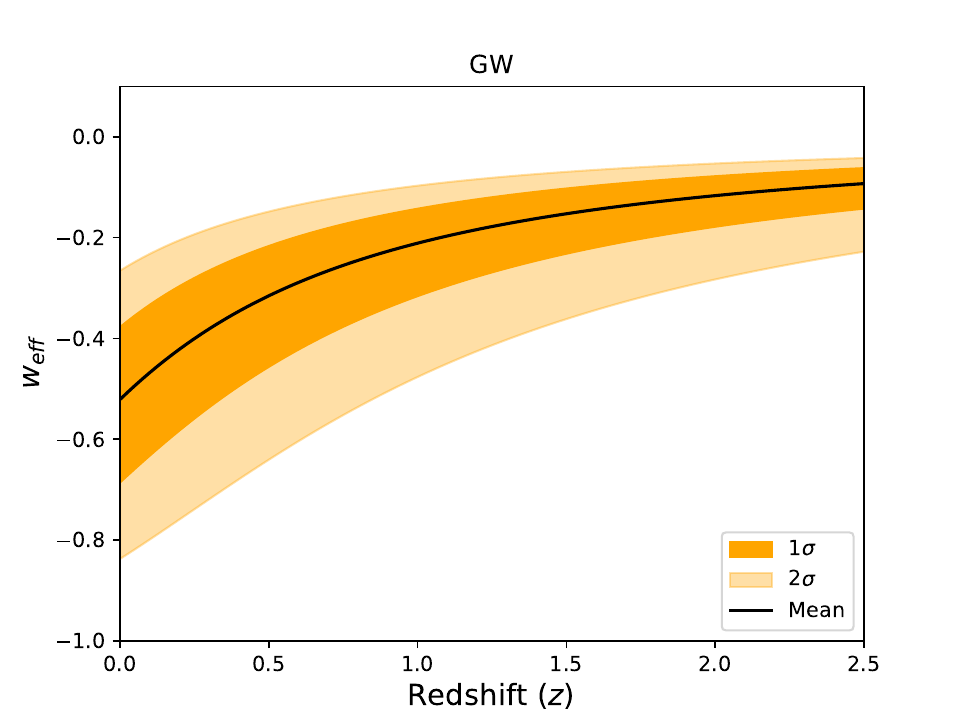}\\
\includegraphics[width=0.48\linewidth]{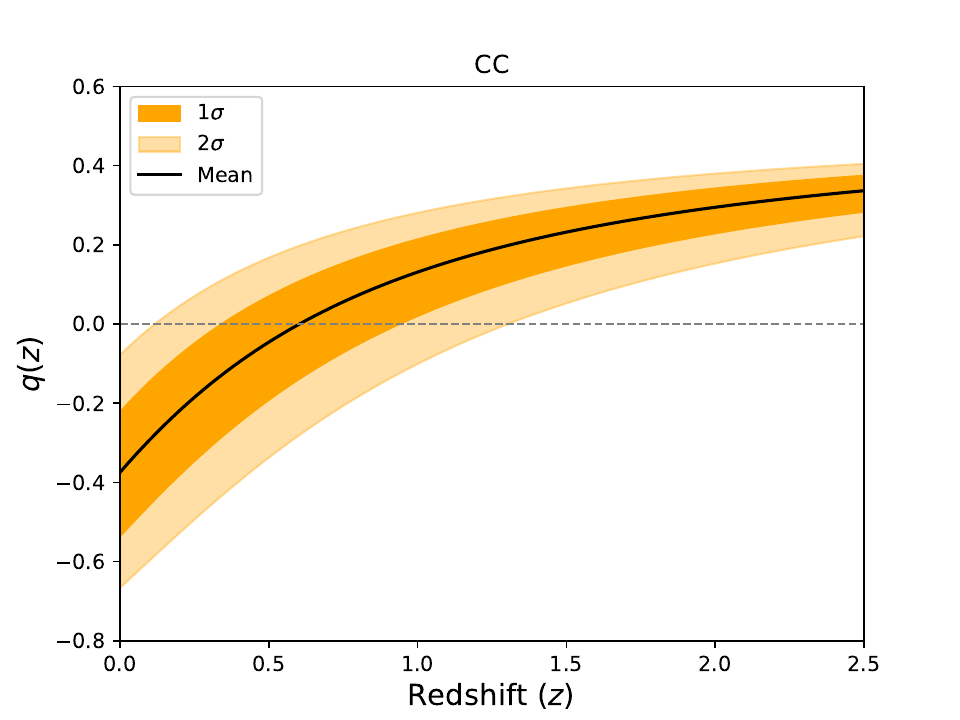}
\includegraphics[width=0.48\linewidth]{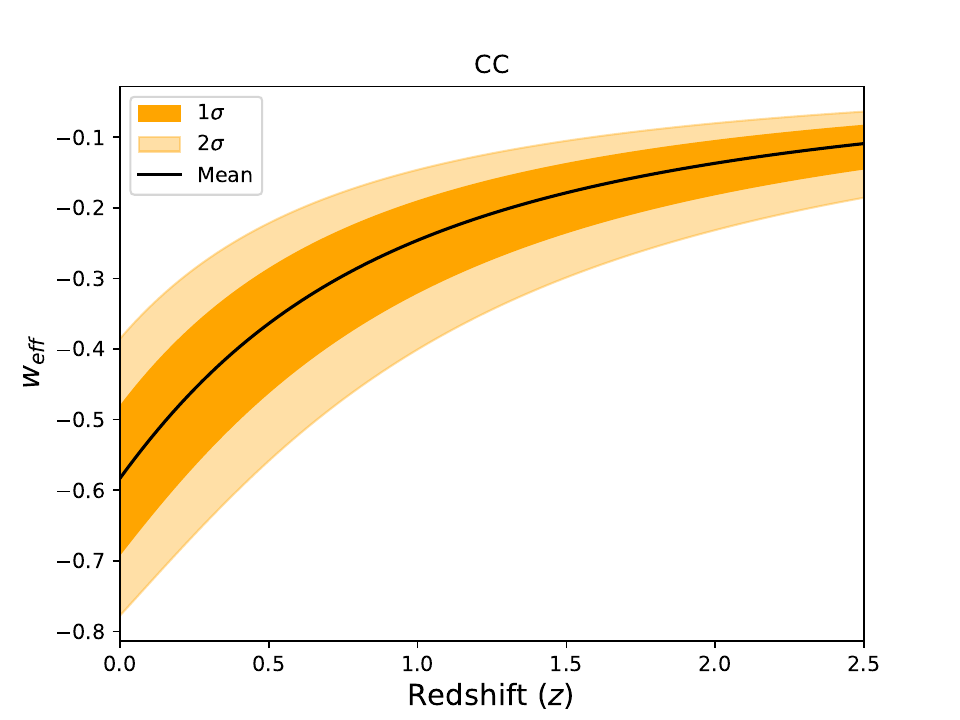}
    \caption{Deceleration parameter (left panel) and effective EoS parameter (right panel) for each dataset}
    \label{chap5/qw}
\end{figure}

Further, we analyze the transitional behavior of the Universe using the deceleration parameter. In the first column of \autoref{chap5/qw}, one can observe that, for all cases, the curve evolves from deceleration to acceleration at a certain redshift. The numerical figures of the transition redshift ($z_t$) up to $1 \sigma$ are summarized in \autoref{tab:cp}. In the same table, one can also notice that the present values obtained corroborate the observational values. The second column of \autoref{chap5/qw} corresponds to the effective EoS ($w_{eff}$) parameter. A quintessence-like behavior is observed for all the cases, as the values lie within the range $-1<w_{eff}<-1/3$ for late time. The present values of $w_{eff}$ are also summarized in \autoref{tab:cp}. The evidence confirms the accelerated Universe, and the present values suitably match the observations \cite{Planck:2018vyg}.

\begin{figure}
    \centering
    \includegraphics[width=0.6\linewidth]{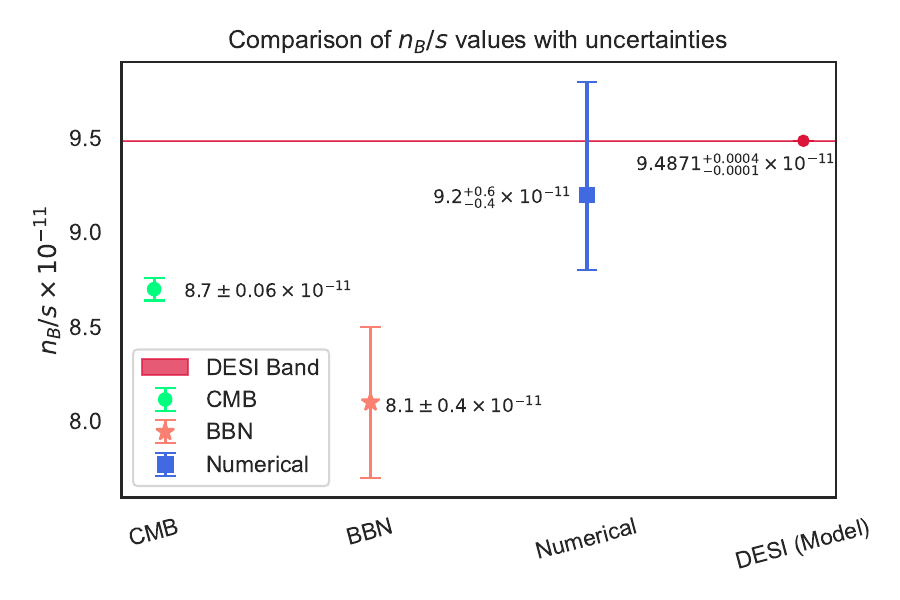}
    \includegraphics[width=0.6\linewidth]{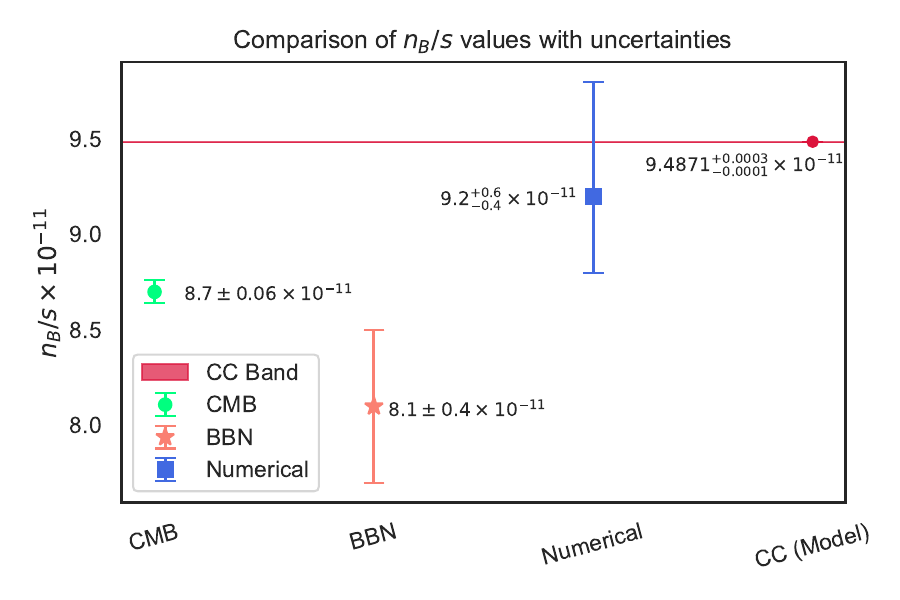}
    \includegraphics[width=0.6\linewidth]{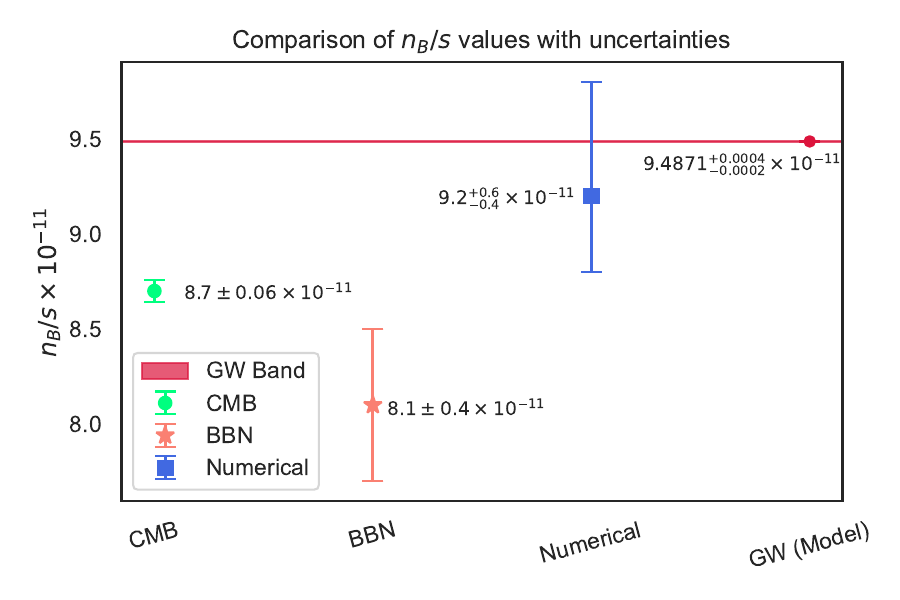}
    \caption{Observed baryon-to-entropy ratio ($n_B/s$) with $1\sigma$ uncertainties from CMB, BBN, numerical estimates, and model predictions (DESI, CC, GW).}
    \label{chap5/baryo}
\end{figure}

Finally, we move to the outcomes from the early-time scenario in the gravitational baryogenesis context. As the theoretical expression (Eq.~(\ref{eq:nbsmodel})) involves the model parameters $A$, $B$, and the cosmological parameter $H_0$, we have utilized their respective values obtained from MCMC. Furthermore, to constrain the quantity $n_B/s$, the values of other parameters are considered as $g_b\approx \mathcal{O}(1)$, $T_D^*=8\times10^{10} GeV$, $M_*=1.01\times10^{8} GeV$ \cite{Daido:2015gqa,Dev:2017wwc,Bodeker:2020ghk} and $g_*=106.75$ \cite{Kolb:1990vq,Saikawa:2018rcs}. The \autoref{chap5/baryo} illustrates the value of $n_B/s$ obtained for our model for each dataset, along with a comparison with the observational predictions. For each case, we also present the associated $2\sigma$ uncertainty in the form of a shaded red band. The observed baryon-to-entropy ratio, \( n_B/s \), from BBN is \( (8.1 \pm 0.4) \times 10^{-11} \) \cite{Pitrou2018, Burles:2000ju}, while the CMB prediction is \( (8.9 \pm 0.06) \times 10^{-11} \) \cite{Planck:2018vyg, WMAP:2003ivt} at a 95\% confidence level. Numerically, the observed value of the baryon-to-entropy ratio is \( n_B/s = 9.2^{+0.6}_{-0.4} \times 10^{-11} \) \cite{Davoudiasl:2004gf}, which is greater than the BBN and CMB values. In \autoref{chap5/baryo}, it can be seen that the values obtained from our model are in close agreement with the numerical predictions and lie well within the observationally inferred range. This consistency supports the reliability of the emerging datasets.

\begin{table}
 \centering
 \caption{Best fit $1\sigma$ ranges obtained from the MCMC.
 }
 \label{tab:params}
    \begin{tabular}{c|c|c|c|c|}  
 \cline{1-5}
\multicolumn{1}{|c|}  {\it{Dataset}} & {$H_0\,\ (km \,\ s^{-1} Mpc^{-1})$}  & $A$ & $B$ & $\Omega_{m0}$ \\ \hline \hline
\multicolumn{1}{ |c| }{DESI} &  $66.4^{+1.9}_{-1.9}$ & $0.433^{+0.047}_{-0.047}$ & $1.3^{+ 0.26}_{- 0.26}$ & $0.27^{+ 0.017}_{- 0.020}$ \\ 
\multicolumn{1}{ |c| }{GW} & $67.1^{+1.9}_{-1.9}$ & $0.389^{+0.067}_{-0.067}$ & $1.33^{+ 0.28}_{- 0.28}$& $0.305^{+0.082}_{-0.082}$  \\ 
\multicolumn{1}{ |c| }{CC} & $66.8^{+2}_{-2}$ & $0.410^{+0.052}_{-0.052}$ & $1.26^{+ 0.19}_{- 0.19}$ & $0.259^{+0.033}_{-0.040}$
\\  \hline
    \end{tabular}
\end{table}

\begin{table}
 \centering
 \caption{$1\sigma$ ranges of the cosmological parameters.
 }
 \label{tab:cp}
    \begin{tabular}{c|c|c|c|}  
 \cline{1-4}
\multicolumn{1}{|c|}  {\it{Dataset}} & {$q_0$}  & $z_t$ & ${w_{eff}}_0$ \\ \hline \hline
\multicolumn{1}{ |c| }{DESI} &  $-0.38^{+0.14}_{-0.14}$ & $0.62^{+0.30}_{-0.24}$ & $-0.59^{+ 0.10}_{- 0.09}$  \\ 
\multicolumn{1}{ |c| }{GW} & $-0.28^{+0.22}_{-0.24}$ & $0.43^{+0.49}_{-0.33}$ & $-0.52^{+ 0.15}_{- 0.16}$  \\ 
\multicolumn{1}{ |c| }{CC} & $-0.37^{+0.15}_{-0.16}$ & $0.60^{+ 0.33}_{- 0.26}$ & $-0.58^{+ 0.10}_{- 0.11}$ 
\\  \hline
    \end{tabular}
\end{table}


\section{Conclusion}\label{sec6}
There is a plethora of research on the reasons for baryonic asymmetry and the net remaining baryon-to-entropy ratio. However, none of them has come to definitive conclusions. In light of this, following the methodology of Davoudiasl et al. \cite{Davoudiasl:2004gf}, we verified the effect of a model that is extremely efficient in describing the late-time evolution. Specifically, this work proposes a novel approach to connecting late-time observational data to primordial events. The recent observations from the gravitational wave data, along with the emerging DESI BAO and CC data, have been used to constrain the Log Square Root Model.

 Interestingly, the results of the parameters corroborate the analysis performed in \cite{Mishra:2025rhi} for different data choices. This indicates that the model might serve as a fine alternative to the existing GR models. The resulting values of the cosmological entities are verified in several contexts. The confrontation against error bars appeared as a suitable match in all cases. Furthermore, in the context of the cosmological parameters $q$ and $w$, the model was in good alignment with the observations. 
 
 The numerically observed baryon asymmetry for our model is as follows: 
 \begin{itemize}[label=$\star$]
    \item ${\frac{n_b}{s}}_{DESI} = \left( 9.4871^{+0.0004}_{-0.0001} \right) \times 10^{-11}
$
     \item ${\frac{n_b}{s}}_{GW}=\left( 9.4871^{+0.0004}_{-0.0002} \right) \times 10^{-11}$
     \item ${\frac{n_b}{s}}_{CC} = \left( 9.4871^{+0.0003}_{-0.0001} \right) \times 10^{-11}.
$
 \end{itemize}
 In this chapter, a direct consequence of the results from late time is observed in the baryogenesis context. The analysis is performed without any prior assumptions about the cosmic phenomenon. The positivity of the net remaining asymmetry predicts the dominance of matter over antimatter. In particular, we find that the predicted baryon-to-entropy ratio lies in good agreement with the numerical estimates and close to the bounds inferred from CMB and BBN observations. This consistency highlights that the late-time datasets from DESI, CC, and GW not only provide strong constraints on the underlying cosmological parameters but also support a viable framework for baryogenesis. The remarkable precision of our model further demonstrates that the mechanism is stable against the inclusion of different observational inputs.

The results, therefore, indicate that the framework employed here offers a coherent picture of baryogenesis, linking early-universe physics with late-time cosmological observations. This strengthens the case for modified gravity scenarios as a consistent approach to addressing both the cosmic acceleration and the origin of the matter-antimatter asymmetry. Future work incorporating complementary probes, such as CMB polarization or high-redshift structure formation, will be important in further testing and refining the connection between baryogenesis and the evolving cosmic background. The latest observations from the upcoming detectors can be utilized to check the consequences. The contribution of the generalized gravitational baryogenesis term can also be investigated.


\chapter{Conclusions and Future Perspectives} 

\label{Chapter6} 

\lhead{\emph{Chapter 6. Conclusions}} 


The central theme of this thesis has been to explore the extent to which teleparallel gravity and its extensions can address the fundamental challenges of modern cosmology. Motivated by the limitations of GR and the persisting tensions within the standard $\Lambda$CDM model, we have developed and analyzed a series of frameworks based on torsional gravity, examining both their theoretical structure and their confrontation with observational data. The work presented here provides a coherent narrative that begins with cosmological tensions, progresses through model-independent reconstruction techniques, incorporates complementary approaches to parameter estimation, and culminates in a novel connection between late-time cosmology and the physics of the early Universe. 

\medskip
\noindent
In \autoref{chap2}, we focused on the pressing issues of the Hubble constant and matter clustering tensions, which represent two of the most significant discrepancies in precision cosmology. By implementing two distinct classes of $f(T)$ and $f(T,\mathcal{T})$ gravity models, we demonstrated how torsional modifications can shift late-time expansion rates and structure growth in ways that naturally reduce the severity of these tensions. Importantly, our analysis incorporated multiple and complementary datasets, including cosmic chronometers, baryon acoustic oscillations, Type Ia supernovae, and standard siren gravitational wave data. This comprehensive observational strategy established the robustness of our results and showed that modified teleparallel models can statistically compete with $\Lambda$CDM while offering fresh physical insights into the origin of these tensions.

The \autoref{chap3} extended the investigation by employing cosmography as a model-independent tool to reconstruct and constrain $f(T,\mathcal{T})$ theory. While cosmography has been successfully applied to curvature-based and other modified gravity theories, its application to coupled torsional gravities remains relatively unexplored. Our analysis showed that cosmographic expansions, including Taylor-based approaches, provide an efficient pathway for constraining higher-order extensions without committing to a specific functional form. Through detailed MCMC analyses using CC, BAO, and supernova data, we obtained tight bounds on the cosmographic parameters and translated these constraints into viable ranges for extended $f(T,\mathcal{T})$ models. This approach not only validated the consistency of coupled teleparallel theories with observations but also emphasized the utility of cosmography as a diagnostic tool in the context of non-standard gravitational frameworks.

In \autoref{chap4}, we presented two complementary strategies for parameter estimation within the $f(T)$ framework: a Pad\'e-based reconstruction of the luminosity distance and a direct dynamical approach derived from the modified Friedmann equations. The comparison between these methods revealed a striking internal consistency. Both approaches produced parameter constraints that were mutually compatible, and when tested against DESI BAO DR2 and Union3 datasets, they outperformed the predictions of $\Lambda$CDM in certain regimes. The Padé method, in particular, proved to be an exceptionally powerful tool for extending the convergence of cosmological expansions to higher redshifts, thereby enabling stable reconstructions across the entire expansion history. At the same time, the dynamical approach provided a direct link to the fundamental field equations of torsional gravity. Together, these results highlighted not only the viability of the $f(T)$ framework but also the methodological richness that emerges when combining kinematic and dynamical reconstructions.

The \autoref{chap5} established a novel connection between late-time cosmology and early-Universe physics through the lens of gravitational baryogenesis. By embedding baryogenesis within the teleparallel framework, we showed how the baryon-to-entropy ratio can be traced to late-time cosmic acceleration and tested against contemporary datasets such as DESI and gravitational wave observations. The analysis revealed that the resulting baryon asymmetry is in excellent agreement with observationally inferred values. This outcome is particularly significant because it links two seemingly disconnected domains--the origin of matter in the early Universe and the accelerated expansion observed today--within a unified torsional paradigm. The study thus demonstrates that teleparallel gravity is not only a framework for addressing late-time observational puzzles but also a viable candidate for probing fundamental questions about the Universe’s earliest moments.

\medskip
\noindent
Taken together, the findings of this thesis support the view that torsion-based modifications of gravity, particularly $f(T)$ and $f(T,\mathcal{T})$ models, represent a fertile ground for addressing key open questions in cosmology. By consistently alleviating cosmological tensions, offering competitive fits to data, and linking early- and late-time physics, these theories go beyond being mere alternatives to $\Lambda$CDM. Instead, they stand as promising frameworks that could form the foundation of a more complete understanding of gravitation and cosmic evolution. The interplay between different methodologies, cosmography, Padé expansions, dynamical reconstructions, and Bayesian parameter estimation, has further underscored the importance of employing diverse and complementary tools when testing theories of gravity against high-precision data.

\medskip
\noindent
Looking forward, several avenues naturally arise from the present work. One promising direction is the extension of the analysis to more general torsional theories, such as non-local $f(T)$ models or higher-order torsion invariants, which could further refine the description of late-time acceleration and structure formation. Upcoming cosmological surveys, including Euclid, LSST, and SKA, will provide unprecedented datasets with which to test these frameworks at both background and perturbative levels. In particular, redshift-space distortions, weak lensing, and gravitational wave standard sirens will offer new consistency checks on growth dynamics and propagation of perturbations in teleparallel models. Another important frontier lies in the strong-field regime, where gravitational wave observations from future detectors such as LISA may probe signatures unique to torsional gravity. Furthermore, the link between baryogenesis and cosmic acceleration established here suggests that extensions involving $f(Q)$ or $f(Q,T)$ models could reveal deeper connections between modified gravity and high-energy particle physics. Finally, the quest for a consistent quantum theory of torsional gravity remains an open challenge, with potential bridges to be explored between teleparallelism and approaches such as loop quantum gravity or string-inspired models.

\medskip
\noindent
In conclusion, this thesis demonstrates that teleparallel gravity and its extensions provide not only phenomenologically viable models for the late-time Universe but also conceptual frameworks that unify disparate aspects of cosmology. By addressing current observational tensions, offering new reconstruction methodologies, and connecting the physics of the early and late Universe, these theories contribute meaningfully to the ongoing effort to move beyond the standard paradigm. The results presented here are a step toward the broader goal of uncovering the fundamental nature of gravity and its role in shaping cosmic evolution.




\addtocontents{toc}{\vspace{2em}} 
\backmatter


\cleardoublepage
\pagestyle{fancy}
\label{References}
\pagestyle{fancy}
\lhead{\emph{References}} 
\rhead{\thepage}

\bibliographystyle{apsrev4-2-titles.bst}
\bibliography{ref.bib}
\cleardoublepage
\phantomsection
\addcontentsline{toc}{chapter}{Appendices} 

\chapter*{Appendices}
\label{ch:appn}
\lhead{\emph{Appendices}} 

\renewcommand{\theequation}{A.\arabic{equation}}
\renewcommand{\thetable}{A.\arabic{table}}

\setcounter{equation}{0}
\setcounter{table}{0}

\section*{Observational Datasets}
\subsection*{DESI BAO}
The methodology of the DESI Data Release 1 \& 2 (DR1 \& DR2) baryon acoustic oscillation (BAO) \cite{DESI:2024mwx,DESI:2025zgx} analysis reflects a comprehensive strategy that combines wide-field spectroscopy, precision clustering statistics, and Bayesian inference to constrain cosmological models. Over a span of three years, DESI collected spectra of more than 14 million galaxies and quasars using its state-of-the-art 5000-fiber spectrograph mounted on the Mayall 4-meter telescope at Kitt Peak National Observatory. The survey targeted four main tracer populations that span different redshift ranges: Bright Galaxy Sample (BGS, $z < 0.4$) \cite{Hahn:2022dnf}, Luminous Red Galaxies (LRGs, $0.4 < z < 1.0$) \cite{DESI:2022gle}, Emission Line Galaxies (ELGs, $0.6 < z < 1.6$) \cite{Raichoor:2022jab}, and Quasars (QSOs, $z > 0.8$) \cite{Chaussidon:2022pqg}, including those used to probe the intergalactic medium via the Lyman-$\alpha$ forest at $z \sim 2.3$.

Once the raw spectroscopic data were obtained, DESI applied an automated redshift determination pipeline using algorithms like Redrock, supplemented by Spectro-Perfectionism methods to ensure high-accuracy fits. After applying strict quality flags and completeness cuts, the resulting galaxy and quasar catalogs were used to compute two-point correlation functions $\xi(s)$ and power spectra $P(k)$, key statistical tools that capture the spatial distribution of tracers. The BAO feature, appearing as a peak at \~150 Mpc/h, was extracted by fitting theoretical templates that account for non-linear gravitational effects, galaxy bias, and redshift-space distortions. The signal was further sharpened using density field reconstruction techniques, which mitigate the smoothing of the BAO peak caused by large-scale motions.

Clustering was analyzed anisotropically, enabling the extraction of cosmological distance measures: the transverse comoving distance $D_M(z)$, the Hubble parameter $H(z)$, and their combination in the volume-averaged distance. These distances were compared to predictions from a fiducial cosmology using a dilation factor $\alpha = D_V^{\text{obs}} / D_V^{\text{fid}}$. Robust estimation of uncertainties was achieved using large ensembles of mock catalogs that replicate survey conditions and tracer bias, allowing for accurate covariance matrices and error propagation.

\begin{table*}
\centering
\caption{DESI BAO (DR2) measurements used in MCMC compilation \protect\cite{DESI:2025zgx}.}
\begin{tabular}{lccccc}
\hline
\textbf{Tracer} & $\mathbf{z_\mathrm{eff}}$ & $\mathbf{D_V/r_d}$ & $\mathbf{D_M/D_H}$ & $\mathbf{D_M/r_d}$ & $\mathbf{D_H/r_d}$ \\
\hline
BGS & 0.295 & $7.942 \pm 0.075$ & — & — & — \\
LRG1 & 0.510 & $12.720 \pm 0.099$ & $0.622 \pm 0.017$ & $13.588 \pm 0.167$ & $21.863 \pm 0.425$ \\
LRG2 & 0.706 & $16.050 \pm 0.110$ & $0.892 \pm 0.021$ & $17.351 \pm 0.177$ & $19.455 \pm 0.330$ \\
LRG3+ELG1 & 0.934 & $19.721 \pm 0.091$ & $1.223 \pm 0.019$ & $21.576 \pm 0.152$ & $17.641 \pm 0.193$ \\
ELG2 & 1.321 & $24.252 \pm 0.174$ & $1.948 \pm 0.045$ & $27.601 \pm 0.318$ & $14.176 \pm 0.221$ \\
QSO & 1.484 & $26.055 \pm 0.398$ & $2.386 \pm 0.136$ & $30.512 \pm 0.760$ & $12.817 \pm 0.516$ \\
Ly$\alpha$ & 2.330 & $31.267 \pm 0.256$ & $4.518 \pm 0.097$ & $38.988 \pm 0.531$ & $8.632 \pm 0.101$ \\
LRG3 & 0.922 & $19.656 \pm 0.105$ & $1.232 \pm 0.021$ & $21.648 \pm 0.178$ & $17.577 \pm 0.213$ \\
ELG1 & 0.955 & $20.008 \pm 0.183$ & $1.220 \pm 0.033$ & $21.707 \pm 0.335$ & $17.803 \pm 0.297$ \\
\hline
\end{tabular}
\label{tab:desi}
\end{table*}
\subsection*{Pantheon+SH0ES}
The Pantheon+SH0ES analysis builds upon and enhances the original Pantheon analysis. It includes a more extensive dataset of SNeIa, featuring 1701 light curves from 1550 SNeIa, collected from 18 distinct studies. These supernovae have redshifts ranging from 0.001 to 2.2613. A significant aspect of this compilation is the addition of 77 light curves from galaxies with known Cepheid distances. The Pantheon+SH0ES analysis introduces several key improvements over the original Pantheon compilation by \cite{Pan-STARRS1:2017jku}. It includes a larger sample size, particularly at lower redshifts (below 0.01), and expands the redshift range. Additionally, the analysis tackles various systematic uncertainties, such as those related to redshifts, peculiar velocities, photometric calibration, and intrinsic scatter models of SNeIa. In previous studies, the EoS of dark energy and the expansion rate ($H_0$) of the Universe were analyzed separately, even though both depend on nearly the same SNeIa. This separation arose because the determination of these parameters involves comparing SNeIa from different redshift ranges. For $H_0$, supernovae in nearby galaxies with redshifts below 0.01 are compared to those in the ``Hubble flow" with redshifts between 0.023 and 0.15, excluding higher redshifts. Conversely, EoS measurements typically include supernovae with redshifts up to around 2 but exclude those below 0.01. As a result, SNeIa within the 0.023 to 0.15 redshift range is commonly used for both parameters.

The latest advancements in the scale and calibration of SNeIa catalogs have improved the ability to constrain the properties and evolution of dark energy by analyzing its influence on the expansion history of the Universe. Researchers, including \cite{Lovick:2023tnv} and \cite{Dainotti:2024gca}, have extended Bayesian methods to compare scattering models and investigate potential non-Gaussianity in the Pantheon+ Hubble residuals.

To constrain cosmological models, the statistical and systematic covariance matrices are combined as follows
\begin{equation}
   C_{stat} + C_{syst} = C_{stat+syst}. 
\end{equation}
By minimizing the chi-square function, the parameters of the model can be derived using the equation
\begin{equation}
    \chi^2_{SN} = \Delta D (C_{stat+syst})^{-1} \Delta D^T,
\end{equation}
where the vector \( D \) represents 1701 supernova distance-modulus residuals, calculated as
\begin{equation}
    \Delta D_i = \mu_i - \mu_{model}(z_i),
\end{equation}
with \(\mu_i\) being the distance modulus of the \(i^{th}\) supernova, \(\mu_{model}(z_i)\) the theoretical distance modulus at redshift \(z_i\), and \(\mu_i = m_i - M\), where \(m_i\) is the apparent magnitude and \(M\) the fiducial magnitude of Type Ia supernovae. The theoretical distance modulus is given by
\begin{equation}
    \mu_{model}(z, \Phi) = 25 + 5\log_{10}\left(\frac{d_L(z, \Phi)}{1 \text{ Mpc}}\right),
\end{equation}
the luminosity distance \(d_L(z, \Phi)\) is already introduced in \autoref{sec:cosmologicalparams}.

While the parameters \(M\) and \(H_0\) are degenerate in SNeIa analysis, the recent SH0ES results relax these constraints. In this analysis, \(M = -19.253\) is adopted, determined from SH0ES Cepheid host distances, providing strong constraints on \(H_0\). The distance residual is then calculated as \cite{Mishra:2024shg}
\[
\Delta D'_i = 
\begin{cases} 
\mu_i - \mu_i^{Cep}, & \text{if } i \text{ belongs to Cepheid hosts} \\
\mu_i - \mu_{model}(z_i), & \text{otherwise}
\end{cases}
\]
where \(\mu_i^{Cep}\) is the Cepheid host-galaxy distance from SH0ES. The covariance matrix for Cepheid host galaxies can be combined with the SNeIa covariance matrix as
\begin{equation}
   C^{SN}_{stat+syst} + C^{cep}_{stat+syst} 
\end{equation}
This combined covariance matrix, which includes statistical and systematic uncertainties from the Pantheon+SH0ES dataset, is used to constrain cosmological models via
\begin{equation}
    \chi^2_{SH0ES} = \Delta D' (C^{SN}_{stat+syst} + C^{cep}_{stat+syst})^{-1} \Delta D'^T.
\end{equation}
\subsection*{Redshift-Space Distortions}
To study the $S_8$ tension, the growth rate data, also known as Redshift Space Distortion (RSD) data, plays a very significant role. The compilation contains the observation value of $f\sigma_8(a)$ from different surveys like WiggleZ, eBOSS, etc. The complete table of the compilation can be found in Table II of \cite{Alestas:2022gcg}. These data are commonly referred to in the literature as RSD data, due to a specific phenomenon observed at both large and small scales during galaxy observations. In essence, the peculiar velocities of galaxies cause overdense regions to occur compressed along the line of sight at large scales, while at small scales, these regions become elongated along the line of sight. This distortion affects the two-point correlation function, leading to an anisotropic distribution in the power spectrum. 

Additionally, at large scales, part of the observed anisotropy in the power spectrum may also stem from using an incorrect fiducial cosmology for $H(z)$, an effect known as the Alcock-Paczynski (AP) effect. This factor must be considered when analyzing growth rate data. The AP effect causes a correction \cite{Kazantzidis:2018rnb} which can be approximated as \cite{Nesseris:2017vor}
\begin{equation}
    f\sigma_8(a) \approx \frac{H(a)D_A(a)}{\tilde{H}(a)\tilde{D}_A(a)} \tilde{f}\sigma_8(a).
\end{equation}
Here, \( H(a) \) and \( D_A(a) \) represent the Hubble parameter and the angular diameter distance for the cosmological model under investigation, while \( \tilde{H}(a) \) and \( \tilde{D}_A(a) \) denote the corresponding quantities in the fiducial cosmology used for data analysis. The quantity \( f\sigma_8(a) \) represents the reference value provided by the data, which requires correction to account for the differences between the fiducial and target cosmological models.
\subsection*{Gamma-Ray Bursts}
We consider the relationship between the observed photon energy at the peak spectral flux, $E_{p,i}$, which corresponds to the peak in the $\nu F_{\nu}$ spectra, and the isotropic equivalent radiated energy, $E_{iso}$. This is expressed as
\begin{equation}
    \log \left( \frac{E_{\text{iso}}}{1 \, \text{erg}} \right) = b + a \log \left( \frac{E_{p,i}}{300 \, \text{keV}} \right).
\end{equation}

The constants $a$ and $b$ are defined in this relation, where $E_{p,i}$ represents the spectral peak energy in the GRB rest frame. This quantity can be obtained from the observer-frame energy, $E_p$, using the equation $E_{p,i} = E_p(1+z)$. This correlation not only helps to constrain models for the prompt emission of GRBs but also implies that GRBs can serve as reliable distance indicators. The isotropic equivalent energy, $E_{\text{iso}}$, can be derived from the bolometric fluence, $S_{\text{bolo}}$, using the following expression
\begin{equation}
    E_{\text{iso}} = 4\pi d_L^2(z, c_p) S_{\text{bolo}} (1+z)^{-1}.
\end{equation}
Here, $c_p$ refers to the set of parameters that define the underlying cosmological model. To effectively utilize GRBs as distance indicators, it is essential to ensure that this correlation is consistently calibrated. We analyze a sample of 162 long GRBs \cite{Demianski:2016zxi, Sudharani:2024mnras}, based on the updated compilation of spectral and intensity parameters by \cite{Demianski:2016dsa}. The redshift distribution of this sample spans a wide range, from $0.03$ to $9.3$, significantly exceeding the range of SNIa $(z \le 1.7)$. These data correspond to long GRBs with reliably measured redshifts and rest-frame peak energy, $E_{p,i}$. Most of the data come from joint detections by Swift/BAT and Fermi/GBM or Konus-WIND, except for a few cases where Swift/BAT directly provides $E_{p,i}$ within the $15-150$ keV range. For events detected by multiple instruments, the uncertainties in $E_{p,i}$ and $E_{\text{iso}}$ account for the measurements and uncertainties from each individual detector.
\subsection*{Cosmic Chronometers}
The Cosmic Chronometers (CC) approach offers a model-independent method to constrain the expansion history of the universe by directly measuring the Hubble parameter at different redshifts. This technique relies on identifying a class of galaxies that evolve passively, i.e., massive, early-type galaxies with minimal ongoing star formation and homogeneously old stellar populations. These galaxies are considered “cosmic chronometers” because their differential aging encodes information about the time evolution of the universe. The method is built on the Differential Age (DA) technique, which estimates the expansion rate from the relative age difference between galaxy populations observed at slightly different redshifts but formed in similar conditions.

At redshifts around $z \sim 2-3$, such passively evolving galaxies form rapidly over short cosmic timescales (on the order of $\sim 0.3$ Gyr). The Hubble parameter at a given redshift is then determined using the relation
\begin{equation}
    H(z) = -\frac{1}{1+z} \frac{\Delta z}{\Delta t},
\end{equation}
where $\Delta z$ is the redshift separation between two galaxy populations and $\Delta t$ is their differential age. The accuracy of this derivative depends on reliable age-dating of galaxies, which is performed using Stellar Population Synthesis (SPS) models applied to their observed spectra. By comparing two galaxy clusters with similar metallicity and formation histories but different redshifts, the age difference can be calculated, and thus $H(z)$ can be inferred.

In our analyses, we employ a compilation of 34 CC data points taken from \cite{Moresco:2012jh, Moresco:2015cya, Moresco:2016mzx, Jimenez:2003iv}, which provide measurements of the Hubble parameter over a redshift range $z \sim 0.1$ to $z \sim 2$. These data are particularly valuable due to their cosmological model independence, enabling robust comparisons and complementary constraints when used alongside probes such as Type Ia supernovae, BAO, and gravitational-wave standard sirens.

To incorporate the CC data into our cosmological analysis, we define the following chi-square statistic
\begin{equation}
   \chi^2_{\mathrm{CC}} = \sum_{i=1}^{34} \frac{[H_{\mathrm{th}}(z_i) - H_{\mathrm{obs}}(z_i)]^2}{\sigma_H^2(z_i)}, 
\end{equation}
where $H_{\mathrm{obs}}(z_i)$ and $\sigma_H(z_i)$ are the observed Hubble parameter and its associated error at redshift $z_i$, and $H_{\mathrm{th}}(z_i)$ is the theoretical prediction for a given cosmological model. The observed values are based on SPS-inferred galaxy age differences, while the theoretical values are calculated from the cosmological model.

The CC dataset plays a critical role in reconstructing the late-time expansion history without relying on priors from early-universe physics. Its independence from the distance ladder and minimal assumptions about the cosmological model make it an excellent cross-validation tool when combined with gravitational-wave sirens and standard candles.
\subsection*{Baryon Acoustic Oscillations}
BAOs serve as a vital tool in cosmology for investigating the LSS of the Universe. These oscillations originate from acoustic waves in the early Universe that compress baryonic matter and radiation within the photon-baryon fluid, creating a characteristic peak in the correlation function of galaxies or quasars. This peak functions as a standard ruler for distance measurements in the Universe. The comoving size of the BAO peak is set by the sound horizon at the time of recombination, which is influenced by the baryon density and the temperature of the cosmic microwave background.

At a given redshift $z$, the position of the BAO peak can be used to determine angular separation in the transverse direction, $\Delta \theta = r_d / ((1+z) D_A(z))$, and redshift separation in the radial direction, $\Delta z = r_d / D_H(z)$. Here, $D_A$ represents the angular diameter distance, $D_H$ is the Hubble distance, and $r_d$ refers to the sound horizon at the drag epoch. These quantities are already introduced in \autoref{sec:cosmologicalparams}. By measuring the BAO peak at various redshifts, constraints can be placed on cosmological parameters that govern $D_H / r_d$ and $D_A / r_d$, allowing for an estimation of $H(z)$ by choosing an appropriate value for $r_d$. This analysis uses a dataset comprising 26 uncorrelated data points obtained from line-of-sight BAO measurements \cite{BOSS:2014hwf,
BOSS:2013rlg,BOSS:2013igd, BOSS:2016zkm}.
\subsection*{Gravitational Wave}
GWs, emitted by the inspiral and merger of compact objects such as black holes and neutron stars, provide a direct means to measure the luminosity distance to the source through the amplitude and phase of the observed waveform. Unlike electromagnetic distance indicators such as Cepheid variables or Type Ia supernovae, GW-based measurements do not require external calibration and are thus referred to as “standard sirens”. This makes them valuable tools for independently constraining cosmological parameters, especially the Hubble constant $H_0$ and the matter density parameter $\Omega_m$.

In our works, we employ a compilation of GW events observed by the LIGO, Virgo, and KAGRA collaborations over successive observing runs. Specifically, our dataset includes 11 events from the GWTC-1 catalog \cite{LIGOScientific:2018mvr}, 39 from GWTC-2 \cite{LIGOScientific:2020ibl}, 36 from GWTC-3 \cite{KAGRA:2021vkt}, and 46 from GWTC-2.1 \cite{LIGOScientific:2021usb}, resulting in a total of 132 binary merger events. These include Binary Black Hole (BBH) mergers and Neutron Star–Black Hole (NSBH) systems for which high signal-to-noise ratio posterior samples of the luminosity distance have been made publicly available. These samples are obtained using Bayesian parameter estimation techniques with waveform models that take into account spin, inclination, and detector noise properties.

Each GW event provides an estimate of the luminosity distance but lacks a precise measurement of redshift unless an electromagnetic counterpart is observed. Since most BBH events do not have detected counterparts, we rely on a statistical redshift assignment approach using galaxy redshift catalogs. For each event, the sky localization map provided in HEALPix format is overlapped with large-area photometric and spectroscopic galaxy surveys such as GLADE+, 2MASS, or the DESI Legacy Imaging Survey. The galaxies within the three-dimensional localization volume are treated as potential hosts, and their redshifts are used to build a probability-weighted redshift distribution for each GW event. This statistical approach allows us to infer cosmological information by treating the redshift as a latent variable marginalized over potential host galaxies. This methodology allows us to derive constraints on the late-time expansion history of the universe using GW observations alone. \autoref{tab:gw1} - \ref{tab:gw2.1} present the complete set of GW observations incorporated in our analysis.
\begin{table}
\centering
\caption{GW data from  GWTC-1 used in MCMC compilation \cite{LIGOScientific:2018mvr}}
\begin{tabular}{lll}
\hline
Event & $z$ & $d_L$ (Mpc) \\
\hline
GW150914 & $0.09^{+0.03}_{-0.03}$ & $440^{+150}_{-170}$ \\
GW151012 & $0.21^{+0.09}_{-0.09}$ & $1080^{+550}_{-490}$ \\
GW151226 & $0.09^{+0.04}_{-0.04}$ & $450^{+180}_{-190}$ \\
GW170104 & $0.20^{+0.08}_{-0.08}$ & $990^{+440}_{-430}$ \\
GW170608 & $0.07^{+0.02}_{-0.02}$ & $320^{+120}_{-110}$ \\
GW170729 & $0.49^{+0.19}_{-0.21}$ & $2840^{+1400}_{-1360}$ \\
GW170809 & $0.20^{+0.05}_{-0.07}$ & $1030^{+320}_{-390}$ \\
GW170814 & $0.12^{+0.03}_{-0.04}$ & $600^{+150}_{-220}$ \\
GW170817 & $0.01^{+0.00}_{-0.00}$ & $40^{+7}_{-15}$ \\
GW170818 & $0.21^{+0.07}_{-0.07}$ & $1060^{+420}_{-380}$ \\
GW170823 & $0.35^{+0.15}_{-0.15}$ & $1940^{+970}_{-900}$ \\
\hline
\end{tabular}
\label{tab:gw1}
\end{table}

\begin{table}
\centering
\caption{GW data from  GWTC-2 used in MCMC compilation \cite{LIGOScientific:2020ibl}}
\label{tab:gw2}
\begin{tabular}{lcc}
\hline
Event & $z$ & $D_L$ (Gpc) \\
\hline
GW190408 & $0.29^{+0.06}_{-0.10}$ & $1.55^{+0.40}_{-0.60}$ \\
GW190412 & $0.15^{+0.03}_{-0.03}$ & $0.74^{+0.14}_{-0.17}$ \\
GW190413 & $0.59^{+0.29}_{-0.24}$ & $3.55^{+2.27}_{-1.66}$ \\
GW190413 & $0.71^{+0.31}_{-0.30}$ & $4.45^{+2.48}_{-2.12}$ \\
GW190421 & $0.49^{+0.19}_{-0.21}$ & $2.88^{+1.37}_{-1.38}$ \\
GW190424 & $0.39^{+0.23}_{-0.19}$ & $2.20^{+1.58}_{-1.16}$ \\
GW190425 & $0.03^{+0.01}_{-0.02}$ & $0.16^{+0.07}_{-0.07}$ \\
GW190426 & $0.08^{+0.04}_{-0.03}$ & $0.37^{+0.18}_{-0.16}$ \\
GW190503 & $0.27^{+0.11}_{-0.11}$ & $1.45^{+0.69}_{-0.63}$ \\
GW190512 & $0.27^{+0.09}_{-0.10}$ & $1.43^{+0.55}_{-0.55}$ \\
GW190513 & $0.37^{+0.13}_{-0.13}$ & $2.06^{+0.88}_{-0.80}$ \\
GW190514 & $0.67^{+0.33}_{-0.31}$ & $4.13^{+2.65}_{-2.17}$ \\
GW190517 & $0.34^{+0.24}_{-0.14}$ & $1.86^{+1.62}_{-0.84}$ \\
GW190519 & $0.44^{+0.25}_{-0.14}$ & $2.53^{+1.83}_{-0.92}$ \\
GW190521 & $0.64^{+0.28}_{-0.28}$ & $3.92^{+2.19}_{-1.95}$ \\
GW190521 & $0.24^{+0.07}_{-0.10}$ & $1.24^{+0.40}_{-0.57}$ \\
GW190527 & $0.44^{+0.34}_{-0.20}$ & $2.49^{+2.48}_{-1.24}$ \\
GW190602 & $0.47^{+0.25}_{-0.17}$ & $2.69^{+1.79}_{-1.12}$ \\
GW190620 & $0.49^{+0.23}_{-0.20}$ & $2.81^{+1.68}_{-1.31}$ \\
GW190630 & $0.18^{+0.10}_{-0.07}$ & $0.89^{+0.56}_{-0.37}$ \\
GW190701 & $0.37^{+0.11}_{-0.12}$ & $2.06^{+0.76}_{-0.73}$ \\
GW190706 & $0.71^{+0.32}_{-0.27}$ & $4.42^{+2.59}_{-1.93}$ \\
GW190707 & $0.16^{+0.07}_{-0.07}$ & $0.77^{+0.38}_{-0.37}$ \\
GW190708 & $0.18^{+0.06}_{-0.07}$ & $0.88^{+0.33}_{-0.39}$ \\
GW190719 & $0.64^{+0.33}_{-0.29}$ & $3.94^{+2.59}_{-2.00}$ \\
GW190720 & $0.16^{+0.12}_{-0.06}$ & $0.79^{+0.69}_{-0.32}$ \\
GW190727 & $0.55^{+0.21}_{-0.22}$ & $3.30^{+1.54}_{-1.50}$ \\
GW190728 & $0.18^{+0.05}_{-0.07}$ & $0.87^{+0.26}_{-0.37}$ \\
GW190731 & $0.55^{+0.31}_{-0.26}$ & $3.30^{+2.39}_{-1.72}$ \\
GW190803 & $0.55^{+0.26}_{-0.24}$ & $3.27^{+1.95}_{-1.58}$ \\
GW190814 & $0.05^{+0.009}_{-0.010}$ & $0.24^{+0.04}_{-0.05}$ \\
GW190828 & $0.38^{+0.10}_{-0.15}$ & $2.13^{+0.66}_{-0.93}$ \\
GW190828 & $0.30^{+0.10}_{-0.10}$ & $1.60^{+0.62}_{-0.60}$ \\
GW190909 & $0.62^{+0.41}_{-0.33}$ & $3.77^{+3.27}_{-2.22}$ \\
GW190910 & $0.28^{+0.16}_{-0.10}$ & $1.46^{+1.03}_{-0.58}$ \\
GW190915 & $0.30^{+0.11}_{-0.10}$ & $1.62^{+0.71}_{-0.61}$ \\
GW190924 & $0.12^{+0.04}_{-0.04}$ & $0.57^{+0.22}_{-0.22}$ \\
GW190929 & $0.38^{+0.49}_{-0.17}$ & $2.13^{+3.65}_{-1.05}$ \\
GW190930 & $0.15^{+0.06}_{-0.06}$ & $0.76^{+0.36}_{-0.32}$ \\
\hline
\end{tabular}
\end{table}

\begin{table}
\centering
\caption{GW data from  GWTC-3 used in MCMC compilation \cite{KAGRA:2021vkt}}
\label{tab:gw3}
\begin{tabular}{lcc}
\hline
Event & $z$ & $D_L$ (Gpc) \\
\hline
GW191103 & $0.20^{+0.09}_{-0.09}$ & $0.99^{+0.50}_{-0.47}$ \\
GW191105 & $0.23^{+0.07}_{-0.09}$ & $1.15^{+0.43}_{-0.48}$ \\
GW191109 & $0.25^{+0.18}_{-0.12}$ & $1.29^{+1.13}_{-0.65}$ \\
GW191113 & $0.26^{+0.18}_{-0.11}$ & $1.37^{+1.15}_{-0.62}$ \\
GW191126 & $0.30^{+0.12}_{-0.13}$ & $1.62^{+0.74}_{-0.74}$ \\
GW191127 & $0.57^{+0.40}_{-0.29}$ & $3.4^{+3.1}_{-1.9}$ \\
GW191129 & $0.16^{+0.05}_{-0.06}$ & $0.79^{+0.26}_{-0.33}$ \\
GW191204 & $0.34^{+0.25}_{-0.18}$ & $1.9^{+1.7}_{-1.1}$ \\
GW191204 & $0.13^{+0.04}_{-0.05}$ & $0.64^{+0.20}_{-0.26}$ \\
GW191215 & $0.35^{+0.13}_{-0.14}$ & $1.93^{+0.89}_{-0.86}$ \\
GW191216 & $0.07^{+0.02}_{-0.03}$ & $0.34^{+0.12}_{-0.13}$ \\
GW191219 & $0.11^{+0.05}_{-0.03}$ & $0.55^{+0.24}_{-0.16}$ \\
GW191222 & $0.51^{+0.23}_{-0.26}$ & $3.0^{+1.7}_{-1.7}$ \\
GW191230 & $0.69^{+0.26}_{-0.27}$ & $4.3^{+2.1}_{-1.9}$ \\
GW200105 & $0.06^{+0.02}_{-0.02}$ & $0.27^{+0.12}_{-0.11}$ \\
GW200112 & $0.24^{+0.07}_{-0.08}$ & $1.25^{+0.43}_{-0.46}$ \\
GW200115 & $0.06^{+0.03}_{-0.02}$ & $0.29^{+0.15}_{-0.10}$ \\
GW200128 & $0.56^{+0.28}_{-0.28}$ & $3.4^{+2.1}_{-1.8}$ \\
GW200129 & $0.18^{+0.05}_{-0.07}$ & $0.89^{+0.26}_{-0.37}$ \\
GW200202 & $0.09^{+0.03}_{-0.03}$ & $0.41^{+0.15}_{-0.16}$ \\
GW200208 & $0.40^{+0.15}_{-0.14}$ & $2.23^{+1.02}_{-0.85}$ \\
GW200208 & $0.66^{+0.53}_{-0.29}$ & $4.1^{+4.4}_{-2.0}$ \\
GW200209 & $0.57^{+0.25}_{-0.26}$ & $3.4^{+1.9}_{-1.8}$ \\
GW200210 & $0.19^{+0.08}_{-0.06}$ & $0.94^{+0.43}_{-0.34}$ \\
GW200216 & $0.63^{+0.37}_{-0.29}$ & $3.8^{+3.0}_{-2.0}$ \\
GW200219 & $0.57^{+0.22}_{-0.22}$ & $3.4^{+1.7}_{-1.5}$ \\
GW200220 & $0.90^{+0.55}_{-0.40}$ & $6.0^{+4.8}_{-3.1}$ \\
GW200220 & $0.66^{+0.36}_{-0.31}$ & $4.0^{+2.8}_{-2.2}$ \\
GW200224 & $0.32^{+0.08}_{-0.11}$ & $1.71^{+0.50}_{-0.65}$ \\
GW200225 & $0.22^{+0.09}_{-0.10}$ & $1.15^{+0.51}_{-0.53}$ \\
GW200302 & $0.28^{+0.16}_{-0.12}$ & $1.48^{+1.02}_{-0.70}$ \\
GW200306 & $0.38^{+0.24}_{-0.18}$ & $2.1^{+1.7}_{-1.1}$ \\
GW200308 & $1.04^{+1.47}_{-0.57}$ & $7.1^{+13.9}_{-4.4}$ \\
GW200311 & $0.23^{+0.05}_{-0.07}$ & $1.17^{+0.28}_{-0.40}$ \\
GW200316 & $0.22^{+0.08}_{-0.08}$ & $1.12^{+0.48}_{-0.44}$ \\
GW200322 & $0.59^{+1.43}_{-0.32}$ & $3.5^{+12.5}_{-2.2}$ \\
\hline
\end{tabular}
\end{table}

\begin{table}
\centering
\caption{GW data from  GWTC-2.1 used in MCMC compilation \cite{LIGOScientific:2021usb}}
\label{tab:gw2.1}
\begin{tabular}{lcc}
\hline
Event & $z$ & $D_L$ (Gpc) \\
\hline
GW190408 & $0.29^{+0.07}_{-0.11}$ & $1.54^{+0.44}_{-0.62}$ \\
GW190412 & $0.15^{+0.04}_{-0.04}$ & $0.72^{+0.24}_{-0.22}$ \\
GW190413 & $0.56^{+0.25}_{-0.21}$ & $3.32^{+1.91}_{-1.40}$ \\
GW190413 & $0.62^{+0.32}_{-0.26}$ & $3.80^{+2.48}_{-1.83}$ \\
GW190421 & $0.45^{+0.21}_{-0.19}$ & $2.59^{+1.49}_{-1.24}$ \\
GW190425 & $0.03^{+0.02}_{-0.01}$ & $0.15^{+0.08}_{-0.06}$ \\
GW190503 & $0.29^{+0.10}_{-0.10}$ & $1.52^{+0.63}_{-0.60}$ \\
GW190512 & $0.28^{+0.08}_{-0.10}$ & $1.46^{+0.51}_{-0.59}$ \\
GW190513 & $0.40^{+0.14}_{-0.13}$ & $2.21^{+0.99}_{-0.81}$ \\
GW190514 & $0.64^{+0.33}_{-0.30}$ & $3.89^{+2.61}_{-2.07}$ \\
GW190517 & $0.33^{+0.26}_{-0.15}$ & $1.79^{+1.75}_{-0.88}$ \\
GW190519 & $0.45^{+0.24}_{-0.15}$ & $2.60^{+1.72}_{-0.96}$ \\
GW190521 & $0.56^{+0.36}_{-0.27}$ & $3.31^{+2.79}_{-1.80}$ \\
GW190521 & $0.21^{+0.10}_{-0.10}$ & $1.08^{+0.58}_{-0.53}$ \\
GW190527 & $0.44^{+0.29}_{-0.19}$ & $2.52^{+2.08}_{-1.23}$ \\
GW190602 & $0.49^{+0.26}_{-0.20}$ & $2.84^{+1.93}_{-1.28}$ \\
GW190620 & $0.50^{+0.23}_{-0.20}$ & $2.91^{+1.71}_{-1.32}$ \\
GW190630 & $0.18^{+0.09}_{-0.07}$ & $0.87^{+0.53}_{-0.36}$ \\
GW190701 & $0.38^{+0.11}_{-0.12}$ & $2.09^{+0.77}_{-0.74}$ \\
GW190706 & $0.60^{+0.33}_{-0.29}$ & $3.63^{+2.60}_{-2.00}$ \\
GW190707 & $0.17^{+0.06}_{-0.08}$ & $0.85^{+0.34}_{-0.40}$ \\
GW190708 & $0.19^{+0.06}_{-0.07}$ & $0.93^{+0.31}_{-0.39}$ \\
GW190719 & $0.61^{+0.39}_{-0.30}$ & $3.73^{+3.12}_{-2.07}$ \\
GW190720 & $0.16^{+0.11}_{-0.05}$ & $0.77^{+0.65}_{-0.26}$ \\
GW190727 & $0.52^{+0.18}_{-0.18}$ & $3.07^{+1.30}_{-1.23}$ \\
GW190728 & $0.18^{+0.05}_{-0.07}$ & $0.88^{+0.26}_{-0.38}$ \\
GW190731 & $0.56^{+0.31}_{-0.26}$ & $3.33^{+2.35}_{-1.77}$ \\
GW190803 & $0.54^{+0.22}_{-0.22}$ & $3.19^{+1.63}_{-1.47}$ \\
GW190814 & $0.05^{+0.01}_{-0.01}$ & $0.23^{+0.04}_{-0.05}$ \\
GW190828 & $0.38^{+0.10}_{-0.15}$ & $2.07^{+0.65}_{-0.92}$ \\
GW190828 & $0.29^{+0.11}_{-0.11}$ & $1.54^{+0.69}_{-0.65}$ \\
GW190910 & $0.29^{+0.17}_{-0.11}$ & $1.52^{+1.09}_{-0.63}$ \\
GW190915 & $0.32^{+0.11}_{-0.11}$ & $1.75^{+0.71}_{-0.65}$ \\
GW190924 & $0.11^{+0.04}_{-0.04}$ & $0.55^{+0.22}_{-0.22}$ \\
GW190929 & $0.53^{+0.33}_{-0.20}$ & $3.13^{+2.51}_{-1.37}$ \\
GW190930 & $0.16^{+0.06}_{-0.06}$ & $0.77^{+0.32}_{-0.32}$ \\
\hline
\end{tabular}

\begin{tabular}{lcc}
\hline
Event & $z$ & $D_L$ (Gpc) \\
\hline
GW190403 & $1.18^{+0.73}_{-0.53}$ & $8.28^{+6.72}_{-4.29}$ \\
GW190426 & $0.73^{+0.41}_{-0.32}$ & $4.58^{+3.40}_{-2.28}$ \\
GW190725 & $0.20^{+0.09}_{-0.08}$ & $1.03^{+0.52}_{-0.43}$ \\
GW190805 & $0.92^{+0.43}_{-0.40}$ & $6.13^{+3.72}_{-3.08}$ \\
GW190916 & $0.77^{+0.45}_{-0.32}$ & $4.94^{+3.71}_{-2.38}$ \\
GW190917 & $0.15^{+0.05}_{-0.06}$ & $0.72^{+0.30}_{-0.31}$ \\
GW190925 & $0.19^{+0.08}_{-0.07}$ & $0.93^{+0.46}_{-0.35}$ \\
GW190926 & $0.55^{+0.44}_{-0.26}$ & $3.28^{+3.40}_{-1.73}$ \\
\hline
\end{tabular}
\end{table}
\subsection*{Union3}
The Union3 supernovae dataset is a homogeneous, high-quality compilation of Type Ia supernova observations developed by the Supernova Cosmology Project (SCP) \cite{Rubin:2023ovl}. It extends the earlier Union and Union2 compilations, incorporating improved photometric calibrations, systematic corrections, and a larger, uniformly processed sample. Union3 comprises over 1400 spectroscopically confirmed SNe Ia, covering a redshift range from $z \sim 0.01$ to $z \sim 1.4$, and provides key cosmological information for constraining the expansion history of the Universe, especially the properties of dark energy.

The dataset brings together supernova observations from multiple surveys, including SNLS, SDSS, Pan-STARRS, CSP, and low-redshift compilations, as well as space-based data from the Hubble Space Telescope (HST). To ensure internal consistency, all supernova light curves were reprocessed using a common pipeline based on the SALT2 light-curve fitter. The standardization of SNe Ia as distance indicators was performed using the Tripp formula
\begin{equation}\label{eq:trip}
    \mu = m_B - M + \alpha x_1 - \beta c,
\end{equation}
where $\mu$ is the distance modulus, $m_B$ is the observed peak B-band magnitude, $x_1$ is the light-curve stretch (or width), and $c$ is the SN color. The nuisance parameters $\alpha$ and $\beta$ are determined globally from the data through a likelihood maximization process, while $M$, the absolute magnitude, can be treated as a free parameter or externally calibrated using Cepheid data.

One of the critical improvements in Union3 is its rigorous treatment of systematic uncertainties. Unlike earlier datasets that relied predominantly on statistical errors, Union3 provides a full covariance matrix $C$, which includes both diagonal statistical uncertainties and off-diagonal terms from identified systematics
\begin{equation}
 C = D_{\text{stat}} + C_{\text{sys}}.   
\end{equation}
The statistical component $D_{\text{stat}}$ includes measurement noise, photometric errors, and intrinsic dispersion. The systematic component $C_{\text{sys}}$ accounts for uncertainties in photometric calibration (zero-points and filter functions), host galaxy corrections, dust extinction laws, light-curve model uncertainties, and peculiar velocity effects. These systematic contributions were propagated through the SALT2 fitting and cosmological analysis pipeline using Monte Carlo simulations.

The cosmological constraints are derived by comparing the observed distance moduli $\mu_{\text{obs}}$ with theoretical predictions $\mu_{\text{model}}(z;\theta)$ based on a cosmological model with parameters $\theta$ (e.g., $H_0$, $\Omega_m$, $w_0$, $w_a$). The goodness-of-fit is quantified using the chi-squared statistic
\begin{equation}
    \chi^2 = (\mu_{\text{obs}} - \mu_{\text{model}})^T C_{\text{union}}^{-1} (\mu_{\text{obs}} - \mu_{\text{model}}).
\end{equation}
The data we have used in our work is given in \autoref{tab:union}, and $C_{\text{union}}$ is given by \cite{Rubin:2023ovl}

\begin{table}
\centering 
\caption{Union Data (Union3) used in MCMC compilation \protect\cite{Rubin:2023ovl}}
\begin{tabular}{ccc}
\hline
\textbf{$z$} & $\mu(z)$ & $\mu_{err}$ \\
\hline
0.050000 & 36.630361 & 0.0086044 \\
0.100000 & 38.235499 & 0.0078677 \\
0.150000 & 39.147691 & 0.0085239 \\
0.200000 & 39.834880 & 0.0082393 \\
0.250000 & 40.340319 & 0.0083980 \\
0.300000 & 40.813875 & 0.0083796 \\
0.350000 & 41.202487 & 0.0084498 \\
0.400000 & 41.568946 & 0.0085916 \\
0.450000 & 41.822650 & 0.0087093 \\
0.500000 & 42.136311 & 0.0087613 \\
0.550000 & 42.352181 & 0.0089000 \\
0.600000 & 42.599505 & 0.0090432 \\
0.650000 & 42.793121 & 0.0093150 \\
0.700000 & 42.945322 & 0.0095681 \\
0.750000 & 43.125351 & 0.0095079 \\
0.800000 & 43.319655 & 0.0096321 \\
0.898138 & 43.647594 & 0.0109719 \\
0.996276 & 44.011256 & 0.0110380 \\
1.094415 & 44.191092 & 0.0150685 \\
1.232244 & 44.577985 & 0.0138742 \\
1.390961 & 44.842358 & 0.0125267 \\
2.262260 & 45.997159 & 0.1155183 \\
\hline
\end{tabular}
\label{tab:union}
\end{table}

\begin{equation}
    C_{\text{union}} = \begin{bmatrix}
0.0086044 & 0.0078396 & \cdots & 0.0081695 \\
0.0078396 & 0.0078677 & \cdots & 0.0079259 \\
\vdots    & \vdots    & \ddots & \vdots   \\
0.0081695 & \cdots  & \cdots & 0.1155183
\end{bmatrix}_{22\times22}.
\end{equation}
\pagestyle{fancy}
\label{Publications}
\lhead{\emph{List of Publications}}

\chapter{List of Publications}
\section*{Thesis Publications}
\begin{enumerate}

\item \textbf{S. S. Mishra}, N. S. Kavya, P. K. Sahoo, V. Venkatesha, \textit{Impact of teleparallelism on addressing current tensions and exploring the GW cosmology,} \textcolor{orange}{The Astrophysical Journal} \textbf{962}, 125 (2025).

\item \textbf{S. S. Mishra}, P. K. Sahoo, \textit{Hubble Constant, $S_8$, and Sound Horizon Tensions: A Study within the Teleparallel Framework,} \textcolor{orange}{Progress of Theoretical and Experimental Physics} \textbf{2025}, 103E03 (2025).

\item \textbf{S. S. Mishra}, N. S. Kavya, P. K. Sahoo, V. Venkatesha, \textit{Constraining extended teleparallel gravity via cosmography: A model-independent approach,} \textcolor{orange}{The Astrophysical Journal} \textbf{970}, 57 (2024). 

\item \textbf{S. S. Mishra}, N. S. Kavya, P. K. Sahoo, T. Harko, \textit{Padé Cosmography and its insight into Teleparallel Gravity,} \textcolor{orange}{Monthly Notices of the Royal Astronomical Society} \textbf{543}, 2816--2835 (2025).

\item \textbf{S. S. Mishra}, N. S. Kavya, P. K. Sahoo, \textit{Effects of DESI and GW observations on $f(T)$ gravitational baryogenesis,} \textcolor{orange}{Physics Letters B} \textbf{872}, 140036 (2025).
\end{enumerate}

\section*{Other Publications}
\begin{enumerate}
\item \textbf{S. S. Mishra}, N. S. Kavya, P. K. Sahoo, Kazuharu Bamba, \textit{DESI DR2 Meets Cosmography: A Comparative Study of Padé, Chebyshev, and Taylor Expansions,} \textcolor{orange}{Monthly Notices of the Royal Astronomical Society} 
\textbf{546}, stag197 (2026).

\item \textbf{S. S. Mishra}, Suchita Patel, P. K. Sahoo, \textit{BBN to late-time acceleration in $f (T, L_m)$ gravity,} \textcolor{orange}{Physics Letters B} 
\textbf{872}, 140098 (2026).

\item \textbf{S. S. Mishra}, Soumyakanta Bhoi, P. K. Sahoo, \textit{Exploring late-time cosmic acceleration in VCDM cosmology,} \textcolor{orange}{The European Physical Journal C}, 
In Press (2026).

\item N. S. Kavya, \textbf{S. S. Mishra}, P. K. Sahoo, \textit{Model‐Independent Cosmography with Logarithmic Polynomial using Recent Observational Data}, \textcolor{orange}{Fortschritte der Physik} \textbf{74}, e70078 (2026).
\item N. S. Kavya, \textbf{S. S. Mishra}, P. K. Sahoo, \textit{$f(Q)$ gravity as a possible resolution of the $H_0$ and $S_8$ tensions with DESI DR2}, \textcolor{orange}{Scientific Reports} \textbf{15}, 36504 (2025).

\item \textbf{S. S. Mishra}, J. A. S. Fortunato, P. H. R. S. Moraes, P. K. Sahoo, \textit{Cosmography of $f(R,L,T)$ gravity,} \textcolor{orange}{Physics of the Dark Universe} \textbf{45}, 101551 (2025).

\item \textbf{S. S. Mishra}, P. K. Sahoo, \textit{Cosmological aspects in the constrained $f(T,T)$ theory using Raychaudhuri equations,} \textcolor{orange}{Physics of the Dark Universe} \textbf{43}, 101481 (2025).

\item A. Kolhatkar, \textbf{S. S. Mishra}, P. K. Sahoo, \textit{Implications of cosmological perturbations of $\sqrt{Q}$ in STEGR,} \textcolor{orange}{European Physical Journal C} \textbf{85}, 656 (2025).

\item D. S. Rana, \textbf{S. S. Mishra}, A. Bhat, P. K. Sahoo, \textit{Viscous Cosmology in $f(Q,L_m)$ Gravity: Insights from CC, BAO, and GRB Data,} \textcolor{orange}{Universe} \textbf{11}, 355 (2025).

\item N. S. Kavya, \textbf{S. S. Mishra}, P. K. Sahoo, V. Venkatesha, \textit{Can teleparallel $f(T)$ models play a bridge between early and late time Universe?}, \textcolor{orange}{Monthly Notices of the Royal Astronomical Society} \textbf{532}, 3126--3133 (2024).

\item \textbf{S. S. Mishra}, N. S. Kavya, P. K. Sahoo, V. Venkatesha, \textit{Chebyshev cosmography in the framework of extended symmetric teleparallel theory,} \textcolor{orange}{Physics of the Dark Universe} \textbf{43}, 101473 (2024).

\item A. Kolhatkar, \textbf{S. S. Mishra}, P. K. Sahoo, \textit{Investigating early and late-time epochs in $f(Q)$ gravity,} \textcolor{orange}{European Physical Journal C} \textbf{84}, 1161 (2024).

\item C. A. S. Almeida, F. C. E. Lima, \textbf{S. S. Mishra}, G. J. Olmo, P. K. Sahoo, \textit{Thick brane in mimetic-like gravity,} \textcolor{orange}{Nuclear Physics B} \textbf{1008}, 116648 (2024).

\item \textbf{S. S. Mishra}, A. Kolhatkar, P. K. Sahoo, \textit{Big Bang Nucleosynthesis constraints on $f(T,T)$ gravity,} \textcolor{orange}{Physics Letters B} \textbf{853}, 138391 (2024).

\item \textbf{S. S. Mishra}, A. Bhat, P. K. Sahoo, \textit{Probing baryogenesis in $f(Q)$ gravity,} \textcolor{orange}{Europhysics Letters} \textbf{146}, 49001 (2024).

\item \textbf{S. S. Mishra}, S. Mandal, P. K. Sahoo, \textit{Constraining $f(T,T)$ Gravity with Gravitational Baryogenesis,} \textcolor{orange}{Physics Letters B} \textbf{842}, 137978 (2023).

\item S. Mandal, O. Sokoliuk, \textbf{S. S. Mishra}, P. K. Sahoo, \textit{$H_0$ tension in torsion-based modified gravity,} \textcolor{orange}{Nuclear Physics B} \textbf{993}, 116277 (2023).

\end{enumerate}
\cleardoublepage
\pagestyle{fancy}
\lhead{\emph{Conference and Workshop}}
\chapter*{Conference and Workshop Participation}
\addcontentsline{toc}{chapter}{Conference and Workshop Participation}

\section*{A. Conferences and Workshops Presented}

\begin{enumerate}
 \item Delivered a lecture on \textit{``Can Teleparallel $f(T)$ Models Play a Bridge Between Early and Late-time Universe?''} at the \textbf{International Conference on Neutrinos and Dark Matter (NuDM–2024)}, Cairo, Egypt.
 \item Presented on \textit{``Pad\'e Cosmography and its insight into Teleparallel Gravity"} at the \textbf{`` Particles, Gravitation and the Universe: from Quantum Mechanics to Quantum Gravity"}, IOP, VAST, Vietnam.
    \item Presented on \textit{``Constraining $f(T,T)$ Gravity with Gravitational Baryogenesis''} at the \textbf{International Conference of Differential Geometry and Relativity (ICDGR–2023)}, Department of Mathematics and Statistics, SSJ University, Almora, Uttarakhand, India.

    \item Presented a poster (with flash talk) on \textit{``Big Bang Nucleosynthesis Constraints on $f(T,T)$ Gravity''} at \textbf{COSMOGRAVITAS–2024}, Mahidol University, Nakhon Sawan Campus, Thailand.

    \item Presented a poster (with flash talk) on \textit{``Constraining Extended Teleparallel Gravity via Cosmography: A Model-independent Approach''} at the \textbf{3rd IAGRG Summer School}, International Centre for Theoretical Sciences (ICTS), Bengaluru, India.

\end{enumerate}

\vspace{1em}

\section*{B. Conferences and Workshops Attended}

\subsection*{International Events}

\begin{enumerate}
\item \textbf{Particles, Gravitation and the Universe: from Quantum Mechanics to Quantum Gravity (PGU-3)}, IOP, VAST, Hanoi, Vietnam (Nov $26$--$28\,2025$).
    \item \textbf{Theoretical Aspects of Astroparticle Physics, Cosmology, and Gravitation – 2025}, Galileo Galilei Institute, Florence, Italy (March 3–14, 2025).

    \item \textbf{Quantum Groups, Tensor Categories, and Quantum Field Theory – 2025}, University of Oslo, Norway (January 13–17, 2025).

    \item \textbf{Neutrinos and Dark Matter (NuDM–2024)}, Cairo, Egypt (December 11–14, 2024).

    \item \textbf{CosmoGravitas–2024}, Centre for Theoretical Physics and Natural Science, Mahidol University, Nakhon Sawan Campus, Thailand (June 10–14, 2024).
\end{enumerate}

\vspace{1em}

\subsection*{National and Other Events}

\begin{enumerate}
    \item \textbf{National Workshop on Contemporary Issues in Astronomy and Astrophysics – 2024}, jointly organized by Shivaji University, Kolhapur and IUCAA, Pune.

    \item \textbf{89th Annual Conference of the Indian Mathematical Society}, organized by Department of Mathematics, BITS PILANI Hyderabad Campus.

    \item \textbf{International Workshop on Galaxy Formation and Evolution Around the Cosmic Time - 2022}, organized by Department of Physics, Visva-Bharati, Santiniketan.

    \item \textbf{National Workshop on MATLAB and LaTeX}, organized by Department of Mathematics, National Institute of Technology, Tiruchirappali.

\end{enumerate}

\cleardoublepage
\pagestyle{fancy}
\lhead{\emph{Biography}}

\chapter{Biography}

\section*{Brief Biography of the Candidate:}
\textbf{Mr. Sai Swagat Mishra} is a Ph.D. Research Scholar in Theoretical Cosmology in the Department of Mathematics of Birla Institute of Technology and Science, Pilani, Hyderabad Campus, under the supervision of Prof.~Pradyumn Kumar Sahoo. His research focuses on torsion-based modified gravity theories and their implications for the cosmological evolution of the Universe, with an emphasis on addressing observational tensions and exploring novel gravitational frameworks.  

He completed his Integrated M.Sc. in Mathematics and Computing from the College of Engineering and Technology (now Odisha University of Technology and Research), Bhubaneswar, Odisha, in 2019. He secured an All-India rank of 144 in the CSIR-UGC NET JRF examination in mathematics and also qualified GATE 2022 in mathematics.  

Sai Swagat has been the recipient of several prestigious international travel grants, including support from SERB, CSIR, BITS Pilani, and the University of Oslo, which enabled him to present and participate in international conferences and workshops in Thailand, Egypt, Italy, Norway, and Vietnam. He has delivered talks and presented posters at prominent meetings such as \textit{COSMOGRAVITAS} (Thailand), the \textit{International Conference on Neutrinos and Dark Matter} (Cairo), the \textit{Galileo Galilei Institute} (Florence), and the \textit{University of Oslo}, among others.  

To date, he has authored or co-authored \textbf{22 peer-reviewed publications}, including \textbf{21 in Q1 journals and 1 in a Q2 journal}, covering a wide range of topics such as Big Bang Nucleosynthesis constraints, cosmography in extended teleparallel theories, baryogenesis in modified gravity, and the resolution of cosmological tensions. His work reflects a strong commitment to exploring alternatives to General Relativity and deepening our understanding of the Universe’s structure and dynamics.  

Beyond research, he is recognized for his active participation in academic communities, his strong presentation skills, and his passion for advancing the field of cosmology through both independent and collaborative research.

\section*{Brief Biography of the Supervisor:}

\textbf{Prof.~Pradyumn Kumar Sahoo} has over two decades of extensive research experience in Applied Mathematics, Cosmology, General Relativity, Modified Gravity Theories, and Astrophysical Objects. He earned his Ph.D. in Mathematics from Sambalpur University, Odisha, India, in 2004. Since joining the Department of Mathematics at BITS Pilani, Hyderabad Campus, in 2009 as an Assistant Professor, he has advanced to the position of Professor and also served as Head of the Department from 2020 to 2024.  

Prof.~Sahoo has been recognized internationally, including being awarded a Visiting Professor Fellowship at Transilvania University of Brasov, Romania, and being ranked among the top 2\% of scientists worldwide in Nuclear and Particle Physics, according to a Stanford University–led survey. His prolific research output includes more than 250 publications in high-impact journals. He has supervised or co-supervised 19 Ph.D. students (11 graduated and 8 ongoing) and guided numerous M.Sc. dissertations.  

He is an active member of the COST Action CA21136: \textit{Addressing observational tensions in cosmology with systematics and fundamental physics}. As a visiting scientist, he has engaged with leading global research centers, including CERN, Geneva. His contributions extend to invited talks at numerous national and international conferences and to fostering collaborations worldwide.  

Prof.~Sahoo has also made significant contributions to BITS Pilani through five sponsored research projects supported by UGC, CSIR, NBHM, ANRF--DST, and DAAD (RISE) Worldwide. Beyond research, he plays a vital role in the academic community, serving as an expert reviewer for SERB and UGC research projects and as an editorial board member of several reputed international journals.

\end{document}